\documentclass[11pt]{article}

\usepackage[margin=1in]{geometry}
\usepackage[utf8]{inputenc}
\usepackage{amsmath}
\usepackage{amsfonts}
\usepackage{bm}
\usepackage{tikz}
\usetikzlibrary{arrows.meta, positioning}
\usepackage{mathtools}
\usepackage{graphicx}
\usepackage[authoryear]{natbib}
\usepackage{import}

\usepackage{standalone}
\usepackage{tabularx}
\usepackage{lscape}
\usepackage{longtable}
\usepackage{array,multirow}
\usepackage{authblk}
\usepackage{tcolorbox}
\usepackage{pgfplots}
\usepackage{tikzscale}
\usepackage{pdfpages}
\usepackage{makecell}
\usepackage{mdframed}
\usepackage{enumitem}%
\usepackage{dsfont}
\usepackage{amsthm}
\usepackage{stmaryrd}
\usepackage{amssymb}
\usepackage{hyperref}
\usepackage{rotating} 

\usepackage{graphicx}
\usepackage{tcolorbox}
\usepackage{tikzscale}
\usetikzlibrary{arrows, automata}

\newcommand{\type}{\mathsf{type}}
\newcommand{\samp}{\mathsf{samp}}

\newcommand{\outlier}{\mathsf{outlier}}
\newcommand{\source}{\mathsf{source}}
\newcommand{\charic}{\mathsf{char}}
\newcommand{\PMA}{\mathsf{PMA}}

\usepackage{soul}
\usepackage{xcolor}

\usepackage{parskip}
\pgfplotsset{compat=1.18}

\mathtoolsset{showonlyrefs}

\begin{document}
\title{{Bayesian Statistical Modeling in Action for Estimation and Forecasting in Low- and Middle-income Countries:
The Case of the Family Planning Estimation Tool}}
\author[1]{Leontine Alkema\thanks{
This work was supported, in whole or in part, by the Gates Foundation (INV-00844 and OPP1192802). Under the grant conditions of the Foundation, a Creative Commons Attribution 4.0 Generic License has already been assigned to the Author Accepted Manuscript version that might arise from this submission. We are thankful to many individuals whose interaction with FPET has contributed to this work. We would also like to thank Jonathan Bearak, John Casterline, Vladimira Kantorova, and Mark Wheldon, for helpful inputs and discussions. Generative AI tools (OpenAI GPT-5-mini; OpenAI GPT-4 and 5) were used for language improvement. Contact: lalkema@umass.edu.}}
\affil[1]{\textit{Department of Biostatistics and Epidemiology, 
University of Massachusetts Amherst, 
USA}}

\author[2] {Herbert Susmann}
\affil[2]{\textit{Division of Biostatistics, Department of Population Health, NYU, 
USA}}

\author[1]{Evan Ray}

\author[3]{Shauna Mooney}
\affil[3]{\textit{Hamilton Institute and Department of Mathematics and Statistics, Maynooth University, Ireland}}
\author[3]{Niamh Cahill}

\author[4]{Kristin Bietsch}
\affil[4]{\textit{Avenir Health, 
USA}}

\author[5]{A.A. Jayachandran}
\affil[5]{\textit{Track20 (India), Avenir Health, India}}

\author[6]{Rogers Kagimu}
\affil[6]{\textit{Division of Health Information Management and Division of Reproductive and Infant Health, Ministry of Health, Uganda}}


\author[4]{Priya Emmart}

\author[7]{Zénon Mujani}
\affil[7]{\textit{Programme National de la Santé Reproductive 
and Tulane International, Democratic Republic of Congo  }} 

\author[8]{Khan Muhammad}
\affil[8]{\textit{United Nations Population Fund, 
Pakistan}}

 \author[9]{Brighton Muzavazi}
\affil[9]{\textit{Track20 and Ministry of Health and Child Care, 
Zimbabwe}}

\author[4]{Rebecca Rosenberg}
\author[4]{John Stover}
\author[4]{Emily Sonneveldt}

\date{
\today}

\maketitle

\abstract{ 
The Family Planning Estimation Tool (FPET) is used in low- and middle-income countries to produce estimates and short-term forecasts of family planning indicators, such as modern contraceptive use and unmet need for contraceptives. Estimates are obtained via a Bayesian statistical model that is fitted to country-specific data from surveys and service statistics data.  The model has evolved over the last decade based on user inputs. 

In this paper we summarize the main features of the statistical model used in FPET and introduce recent updates related to capturing contraceptive transitions, fitting to survey data that may be error prone, and the use of service statistics data. We assess model performance through a validation exercise and find that FPET is reasonably well calibrated. 

We use our experience with FPET to briefly discuss lessons learned and open challenges related to the broader field of statistical modeling for monitoring of demographic and global health indicators.

}

\clearpage 


\section{Introduction}
The elevation of family planning (FP) on the global stage in 2012 with the launch of FP2020 provided an opportunity to create new approaches for monitoring of FP programs (\citep{fp202_2014}). The need for annual estimates to track progress of the initiative led to the development of standard indicators, approaches, and methodologies. For countries, having annual estimates of FP indicators was a priority to better gauge progress towards their own objectives. After the FP2020 initiative concluded, the FP2030 initiative was started with a continued focus on measurement and setting of country-specific commitments (\citep{fp2030}).

Track20, implemented by Avenir Health and funded by the Gates Foundation, is a global family planning project aimed at improving global and country level use of data, see \url{https://www.track20.org/}. The project’s aims include the development and introduction of innovative and user-friendly methodologies, tools, and approaches that build capacity and broaden who can effectively engage with family planning data. Track20 has provided the annual estimates used for FP2020 related measurements and continues to assess progress for FP2030 and the Ouagadougou Partnership (focused on FP in countries in western Africa, see \url{https://www.speakupafrica.org/program/speak-up-africa-website-blurb-for-the-ouagadougou-partnership-op/}). This is done through a bottom-up reporting process led by government monitoring and evaluation (M\&E) officers, effectively linking country and global FP data.

The Family Planning Estimation Tool (FPET) is used by Track20 in low- and middle-income countries to produce estimates and short-term forecasts of FP indicators, such as modern contraceptive use and unmet need for contraceptives. Estimates are obtained via a Bayesian statistical model that is fitted to country-specific data from surveys and service statistics data. The model used in FPET is a country-specific implementation of a global model for FP estimation (\cite{alkema2013mcpr,cahill2018mcpr, kantorova_estimating_2020}). The country-specific implementation, referred to as a local model, was produced to introduce and facilitate in-country usage by the Track20 project. FPET has evolved over the last decade based on usage in the Track20 project and other user inputs and requests. Recent statistical modeling updates include relaxation of strong parametric assumptions made regarding contraceptive use transitions (\cite{susmann_flexible_2025}), updated use of survey data to improve estimation in the presence of data outliers (\cite{alkema_2024dm}), and updated use of service statistics data to better account for uncertainty associated with these data (\cite{mooney_emu}). 

In this paper, we present the current version of the Family Planning Estimation Tool. We first introduce how FPET came about, how it is used in the Track20 project, and how the modeling approach has evolved. The methods section presents the current model set up, highlighting how different advances in statistical modeling were incorporated. We use our experience with FPET to discuss lessons learned and open challenges related to the broader field of statistical modeling for monitoring demographic and global health indicators.

\section{Background}

\subsection{FP indicators} 

We are interested in family planning among women of reproductive age  in different populations of interest. In this paper, we focus on annual FP outcomes at the national level, from 1970 until 2030, for women based on their marital status. We refer to married women as those women who are married or in a formal union \citep{kantorova_estimating_2020}. Unmarried women are the complement of this group and includes women who are separated, divorced, or widowed. 

The FP indicators considered are contraceptive use and unmet need for contraceptives. Contraceptive methods are categorized into modern versus traditional methods. Modern methods of contraception include female and male sterilization, oral hormonal pills, the intra-uterine device, male and female condoms, injectables, the implant (including Norplant; Wyeth-Ayerst, Collegeville, PA, USA), vaginal barrier methods, standard days method, lactational amenorrhea method, and emergency contraception. Traditional methods of contraception include abstinence, the withdrawal method, the rhythm method, douching, and folk methods. Unmet need for family planning among married women is defined as the percentage of women who want to stop or delay childbearing but who are not currently using any method of contraception to prevent pregnancy \citep{bradley_revising_2012}. Also included are women who are currently pregnant or postpartum amenorrheic whose pregnancies were mistimed or unwanted. For unmarried women, sexual activity is considered in the calculation of unmet need, as summarized in \cite{kantorova_estimating_2020}.

FP indicators are combined to summarize demand for FP and demand satisfied with modern methods. Specifically, demand for FP is defined as the sum of modern users, traditional users, and those women unmet need for any method.  Demand satisfied with modern methods, or demand satisfied in short, is defined as the share of women with a demand for FP who use a modern method of contraception.

\subsection{Survey data}\label{sec-surveydatabase}

Survey data on FP are obtained in household surveys in which women answer questions related to marriage, pregnancies, sexual activity, and contraceptive use. For example, contraceptive prevalence data obtained from surveys refers to the percentage of women who report themselves or their partners as currently using at least one contraceptive method of any type (modern or traditional). Household surveys may be conducted by international organizations or by local organizations or governments. Survey programs and categorizations include the Demographic and Health Survey Program (DHS), Performance Monitoring for Action (PMA), UNICEF Multiple Indicator Cluster Surveys (MICS), national surveys, and other surveys. Micro data from each survey can be used to calculate FP indicators for the population-time period covered by the survey, as well as a measure of the uncertainty owed to the survey design, referred to as sampling error. 

\begin{sidewaysfigure}[htbp]
    \centering
    \includegraphics[width=0.9\textwidth]
        {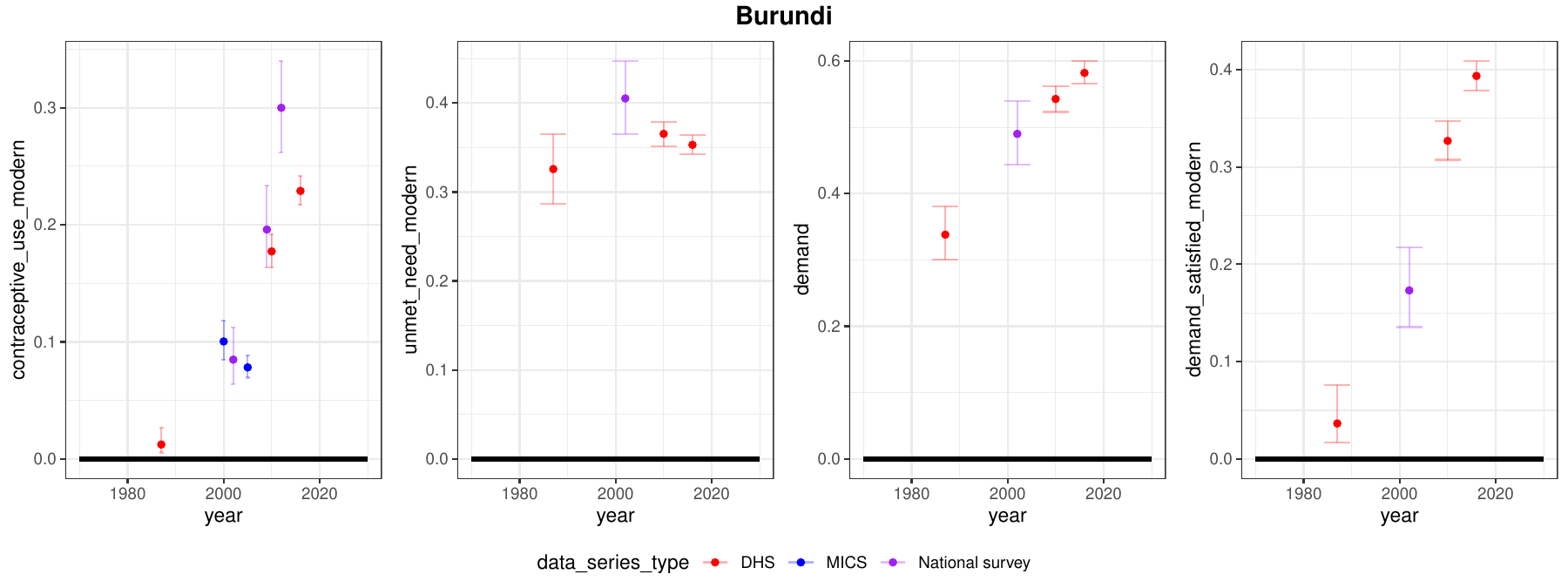}\\
    \includegraphics[width=0.9\textwidth]
        {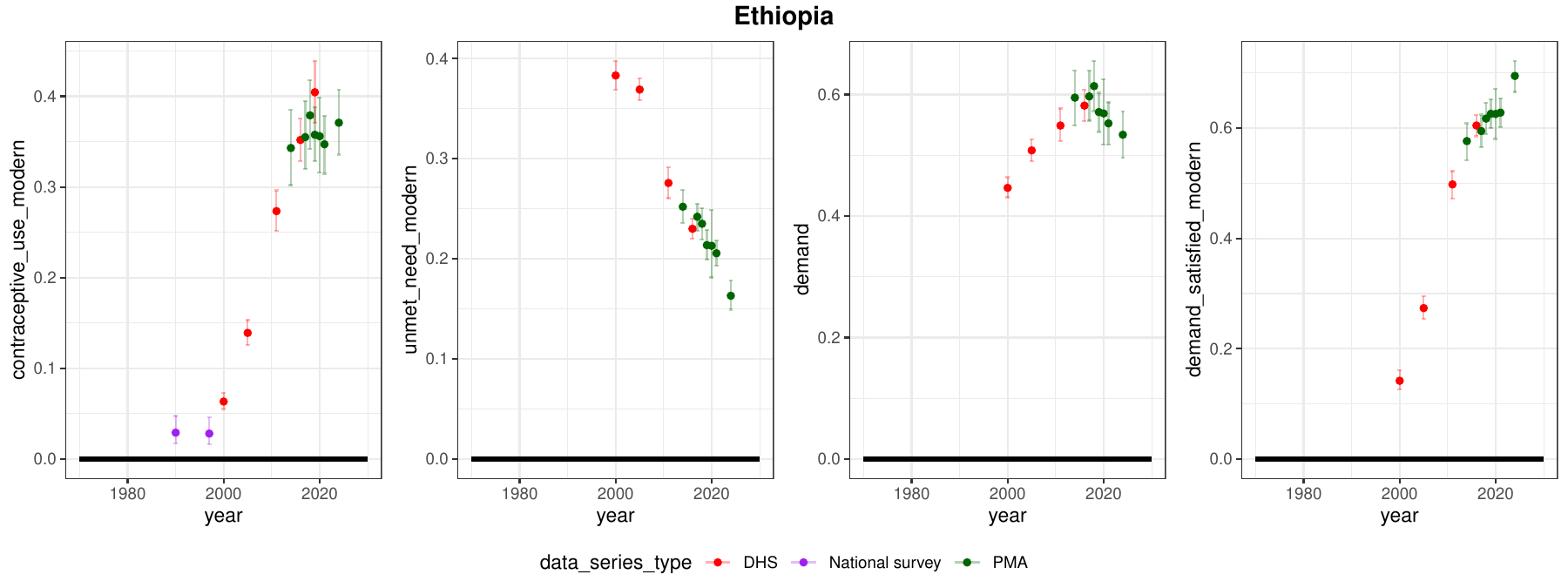}\\
    \caption{\textbf{FP data for married women in Burundi and Ethiopia.} Data are shown in colored dots with 95\% confidence intervals that reflect sampling errors. }
    \label{fig-data}
\end{sidewaysfigure}

Survey data compilation is carried out by the United Nations Population Division (\citep{molitoris_world_nodate}) and the Track20 project. 
The data base contains observations on the compositional vector of FP proportions, i.e., contraceptive use of modern and traditional methods, and unmet need for any method. Demand and demand satisfied can be calculated from these outcomes. Examples of available survey data are shown in Figure~\ref{fig-data} for Burundi and Ethiopia.

\subsection{Why a model is needed to produce FP estimates}

Often survey data alone for some indicator of interest may not be sufficient to inform programs and understand progress. First, data are not necessarily available for all years of interest, e.g., for years in the past, the period after the most recent data point until the current year, and future years. Moreover, data are subject to measurement error, which may be substantial. 

These points are illustrated in Figure~\ref{fig-data}. In Burundi, data are not recent -- its most recent survey is from 2016 -- and the most recent national survey, from 2012, suggests a large fluctuation in modern contraceptive use (mCPR) which is likely to be due to measurement error. In Ethiopia, data from PMA surveys suggest levels that are lower than those observed in the most recent DHS survey. These examples illustrate the need to integrate data sources to produce reliable estimates and forecasts.

\subsection{What FPET does and its usage in the Track20 project}
\begin{figure}[htbp]
    \centering
  \includegraphics[width=0.45\textwidth]{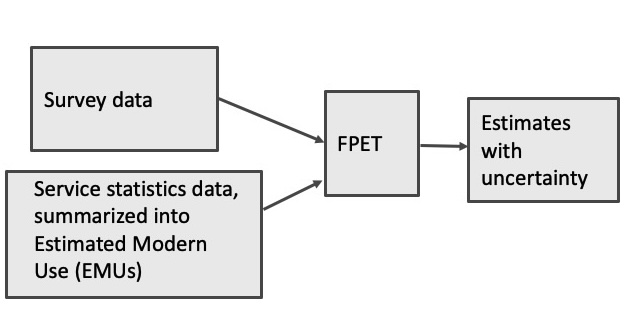}
  \includegraphics[width=0.45\textwidth]{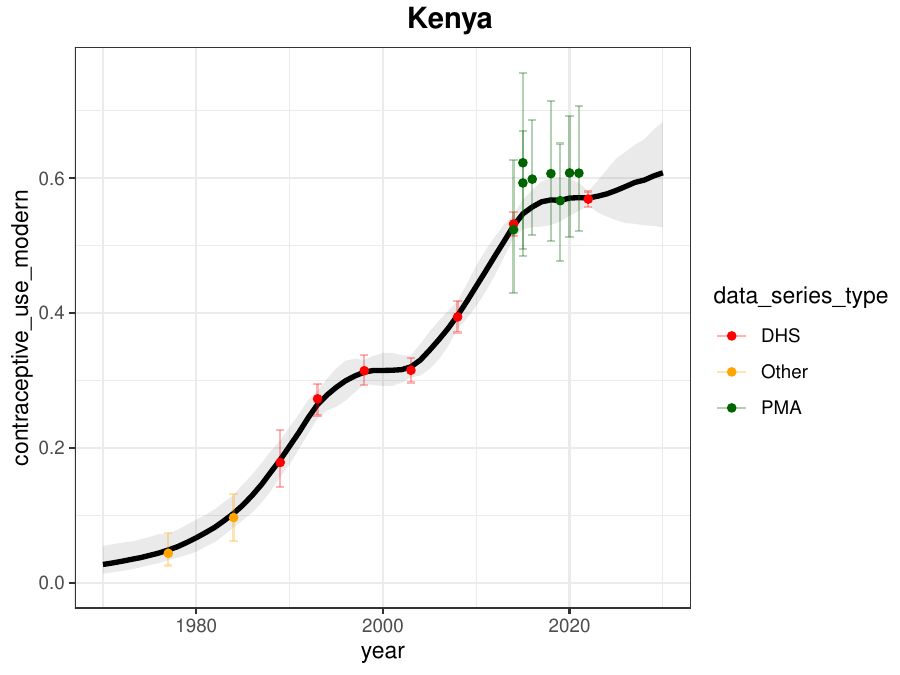}
    \caption{\textbf{The Family Planning Estimation Tool (FPET) produces estimates of FP indicators, based on survey data and observations on estimated modern use (EMUs), obtained from service statistics.} The graph shows survey data on modern contraceptive use for married women of reproductive age in Kenya, together with an FPET fit (point estimates are given by the solid black line, shaded areas represent 90\% credible intervals). }
    \label{fig-fpetoverview}
\end{figure}

FPET produces estimates of FP indicators using available data for a population of interest, see Figure~\ref{fig-fpetoverview}. 
FPET produces estimates of FP indicators using available surveys for a population (e.g., a country or subnational region).   In addition to survey data, service statistics data may also be used to inform the estimates. 
Service statistics data refer to data obtained from routine data collection. These systems may record various FP-related outcomes, including FP commodities distributed to clients or facilities, visits to FP facilities/providers, or FP users.  Track20 developed a tool to calculate an Estimated Modern Use (EMU) indicator from these different types of service statistics data \citep{bietsch_estimated_2025, emutrack20, ss2emu_tool}. EMU estimates have been used to inform mCPR estimates in the absence of recent survey data \citep{cahill2018mcpr, mooney_emu}. 


The Track20 project works to build the capacity of countries to generate and use data to inform their programming, improve impact, and accelerate progress toward their FP goals. One of the pillars of the Track20 approach is the cultivation of a network of M\&E Officers dedicated to increasing the quality and use of FP data. These M\&E Officers work within or in partnership with Ministries of Health in 34 countries. This approach positions governments to be the driving force of FP data instead of relying on implementing partners to produce needed analyses and findings. Track20 provides ongoing on-demand technical support to these countries. This includes an annual training that brings all M\&E Officers together to learn additional analytical skills, engage in south-to-south learning (referring to countries in the global south sharing experiences and learning from one another, see for example, https://unsouthsouth.org/about/about-sstc/), and to identify opportunities for new analyses and tools based on current government challenges. This direct exchange with multiple countries enables Track20 to develop tools that directly speak to current government priorities. It also provides an opportunity for any new tools to be vetted and tried across many countries, providing valuable feedback that improves Track20’s work.

Track20 M\&E officers use FPET to assess progress towards country strategies and commitments. The officers are trained in FPET and use it to produce FP estimates for their country. These FPET estimates are reviewed, vetted, and approved through annual consensus meetings. These meetings are convened by governments and attended by partners and donors, providing a time to reflect on progress and identify any barriers or opportunities moving forward. 

The estimation of FP indicators in FPET has evolved based on user inputs and further model developments. The first version was based on a global modeling approach to estimate and project FP indicators for married women aged 15-49 in all countries in the world using survey data (\cite{alkema2013mcpr}). For in-country usage in the Track20 project, we produced FPET based on a local version of the global model, allowing users to fit the model to data from one country only (\cite{new_levels_2017}). This version expanded upon the global model by allowing for the use of service statistics data.  

Subsequently, FPET evolved to better capture local contexts. Model updates were introduced to improve predictive performance and to better account for survey data quality issues (\cite{cahill2018mcpr}), and to improve the use of service statistics data (\cite{cahill_using_2021}). The population considered was extended from married women to all women, using a local implementation of the global model for estimation among unmarried women (\cite{kantorova_estimating_2020,guranich_fpemlocal_2021}). In 2024, several updates were introduced to further improve FPET’s utility. The next section introduces the FPET model specification based on these updates.

\section{Methods}
\subsection{Overview of workflow}
We start by introducing notation to formalize the problem of producing estimates and forecasts of FP indicators. We are interested in the proportion of women of reproductive age in some population who either use a modern method of contraception, use a traditional method, have an unmet need for any method, or do not have a demand for FP (no need group). We denote these proportions as $\phi_j$ for $j=1,\hdots, 4$ referring to modern use, traditional use, unmet need for any method, and women ``with no need`` respectively. Unmet need for modern methods is defined as the sum of traditional users and women with an unmet need for any method. Demand $\eta_{1}$ is defined as the sum of modern users, traditional users, and unmet need for any method: $\eta_{1} = \phi_1 + \phi_2 + \phi_3$.  Demand satisfied is defined as the ratio of modern users over those with a demand, $\eta_{2} = \phi_1/(\phi_1 + \phi_2 + \phi_3)$. 

For FPET, we are interested in FP indicators by country, for different years, and for women based on their marital status. In terms of notation, when referring to specific population-periods, additional indices are used. In particular, $\phi_{j,c,t,m}$ refers to the $j$-th proportion for country $c$, year $t$, and marital group $m$. 

The workflow for estimating FP indicators for all countries and by marital status, and for producing local models for fitting, is summarized in Figure~\ref{fig-workflow}. We first fit global models separately by marital status; these models describe the evolution of FP indicators over time in each country and are informed by survey data. Details on the global models are provided in the next section.

The main utility of FPET is to allow countries to produce estimates for a specific population by fitting a model to that population's data alone. We call these "local models": models derived from the global models but adapted so they can be fitted using data from a single population. In FPET, local models are implemented by fixing non-country-specific parameters at the point estimates obtained from the global fits. For local models that estimate FP indicators among all women, service statistics data can also be included. Details on local models are introduced in Section~\ref{sec-localmodels}.

\begin{figure}[htbp!]
\begin{tikzpicture}[
    node distance=2.5cm and 2.5cm,
    block/.style={rectangle, draw, fill=green!20, minimum width=4cm, minimum height=1cm, align=center},
    datablk/.style={rectangle, rounded corners=6pt, draw, fill=orange!20, minimum width=3cm, minimum height=1cm, align=center},
    localblk/.style={rectangle, draw, fill=yellow!20, minimum width=3cm, minimum height=1cm, align=center},
    arrow/.style={-{Stealth[scale=1.3]}, thick}
  ]

\node[datablk] (survey) {Survey data};

\node[block, below left=1.6cm and -0.5cm of survey] (globalM) {Global model fitting\\ for married women};
\node[block, below right=1.6cm and -0.5cm of survey] (globalU) {Global model fitting\\ for unmarried women};

\node[datablk, below=1cm of globalM] (service) {Service statistics data};
\node[localblk, below=1cm of globalU] (local) {Local model fitting\\ \small (by marital group, or\\ for both marital groups)};

\draw[arrow] (survey) -- (globalM);
\draw[arrow] (survey) -- (globalU);

\draw[arrow]
  (survey.east)  .. controls ++(3,0) and ++(5,5) .. node[right, 
  ] {Data for 1 country}(local.east);
  
\draw[arrow] (globalM) -- (local);
\draw[arrow] (globalU) -- (local);

\draw[arrow, dashed] (service) -- (local);

\end{tikzpicture}
\caption{\textbf{Workflow for estimating FP indicators by marital status and generating local models for country-level fitting.} Global models (green) are fitted by marital status, using survey data. Local models (yellow) use information from global models and are fitted to survey data from one country and, optionally, service statistics data. }\label{fig-workflow}
\end{figure}

\subsection{FPET global statistical model specification}

\subsubsection{Overview}
To produce estimates and forecasts for FP indicators for each marital group, we made assumptions in the global model about how available data relate to the indicators and how FP indicators may change over time and vary across different populations. More formally, we defined a likelihood function or data model for available data and a process model that defined the parametrization of model parameters. The diagram in Figure~\ref{fig-diagrammodel} illustrates the set up. We introduce the process and survey data models for modern use and unmet need for modern methods below. The appendix includes the specification of the process and data models for traditional use.

\begin{figure}[htbp!]
\begin{tikzpicture}[
    node distance=1.5cm,
    every node/.style={font=\small, align=center},
    block/.style={draw, thick,  minimum width=3cm, minimum height=2cm, inner sep=8pt},
    data/.style={block, fill=green!20},
    process/.style={block, fill=green!20},
    hier/.style={block, fill=green!20},
    prior/.style={block, fill=green!20},
    >={Stealth[width=2.5mm,length=3mm]},
     datablk/.style={rectangle, rounded corners=6pt, draw, fill=orange!20, minimum width=3cm, minimum height=1cm, align=center},
]

\node[data] (data) {
    \textbf{Data model} \\
    $\mathrm{logit}(y_i)| \phi_{c[i], t[i]}, \sigma_i^2 \sim N(\mathrm{logit}(\phi_{c[i], t[i]}), \sigma_i^2)$
};
\node[datablk, left = of data] (survey) {Survey data};
\node[process, below=of data] (process) {
    \textbf{Process model} \\
$\begin{aligned}
 & g_1(\eta_{c,t}) = \begin{cases}
  \Omega_c,  &t = t^*,\\
g_1(\eta_{c,t-1}) + f(\eta_{c,t-1},\lambda_c,\beta_c) + \varepsilon_{c,t}, & t > t^,\\
g_1(\eta_{c,t+1}) - f(\eta_{c,t+1},\lambda_c,\beta_c) - \varepsilon_{c,t+1}, & t < t^*,
\end{cases}\\
& \varepsilon_{c,t}| \rho_{\varepsilon}, \varepsilon_{c,t-1},\sigma_\varepsilon^2 \sim N\big(\rho_{\varepsilon}\varepsilon_{c,t-1},\sigma_\varepsilon^2\big)
\end{aligned}$
};

\node[hier, right=of data, below = of process] (hier) {
    \textbf{Hierarchical distributions} \\
    on transition model parameters $\bm{\beta}_c$, $\lambda_c$, $\Omega_c$
};


\draw[->, thick] (process) -- node[left, font=\footnotesize] {Specification for $\phi$ \\ (derived from $\eta$s)} (data);

\draw[->, thick] (hier) -- (process);
\draw[->] (survey) -- (data);

\end{tikzpicture}
\caption{\textbf{Overview of global model specification.} The process model specifies how FP indicators vary over time and differ across populations. The data model specifies how observed survey data relates to FP indicators. Hierarchical distributions are used for country-specific transition model parameters.}\label{fig-diagrammodel}
\end{figure}

\subsubsection{Process model: How do FP indicators vary over time and differ across populations?}

\begin{figure}[htbp]
    \centering
   \includegraphics[width=\textwidth]{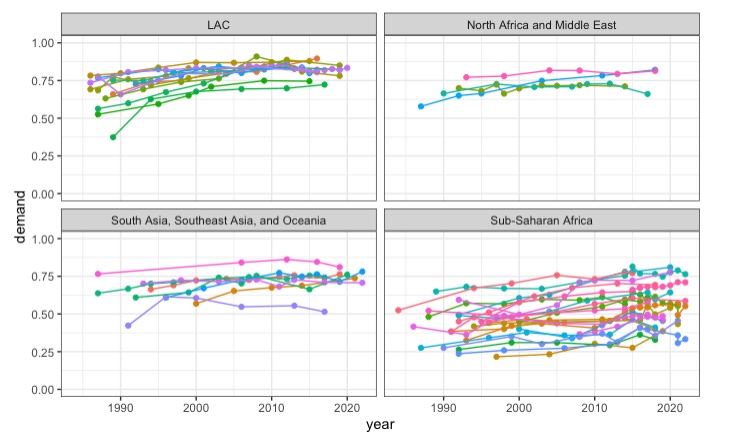}\\
   \includegraphics[width=\textwidth]{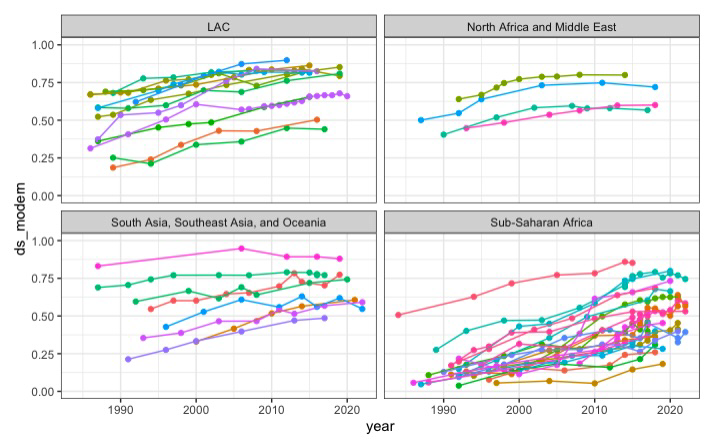}
    \caption{\textbf{Observed data on demand and demand satisfied for modern methods for selected countries over time. } Observations refer to prevalence among married women of reproductive age. Countries are included in the selected main regions if they have more than 4 observations.  }
    \label{fig-transitioneda}
\end{figure}

\paragraph{Capturing contraceptive transitions}  
In the process model, we consider how FP indicators change over time. The main goal in FPET is to capture long-term and short-term changes. Over longer time periods, the main assumptions underlying FPET are based on capturing a contraceptive transition, introduced by  \cite{alkema2013mcpr, cahill2018mcpr} and \cite{susmann_flexible_2025}. 
Specifically, over longer time periods, we assume that demand for contraceptives increases from low to high levels, where increases start slowly at low demand, accelerate once growth has been initiated, and then rates of change slow down as demand reaches high levels, which we refer to as an asymptote. Similarly, we assume that demand satisfied with modern methods also follows this type of increase.  Illustrations of observed trends globally, that support these assumptions, are given in Figure~\ref{fig-transitioneda}. Based on combining the two transitions, we obtain modern contraceptive use (mCPR) and unmet need for modern methods. Resulting estimates show a typical pattern in unmet need whereby unmet need increases at the start of a transition, when supply is lagging demand, followed by a decrease once supply catches up with demand. 

Transitions in FP indicators are modeled using flexible Bayesian hierarchical B-spline \textit{transition models} (\cite{susmann_flexible_2025}).
In this type of model, the rate of change in an indicator of interest is parameterized using a systematic component and a smoothing component, where the systematic component captures the expected long-term transition from low to high levels of the indicator while the local smoothing component captures deceleration or acceleration as indicated by the data. We use a Bayesian transition model to capture the increase from low levels to an asymptote in demand and demand satisfied with modern methods. 

The remainder of this section introduces the set up of a transition model for one indicator $\eta_{c,t}$. The full specification for the transition models for demand and demand satisfied, for women by marital status, is given in Appendix Section~\ref{sec-app-trad}. 

\paragraph{Bayesian B-spline transition model}
A transition model for indicator of interest  $\eta_{c,t}$ is specified as follows:
\begin{align} 
 g_1(\eta_{c,t}) &= \begin{cases}
 \Omega_c, & t = t^*, \\
 g_1(\eta_{c,t-1}) + f(\eta_{c,t-1}, {\lambda}_c, {\beta}_c) + \varepsilon_{c,t}, & t > t^*, \\
 g_1(\eta_{c,t+1}) - f(\eta_{c,t+1},{\lambda}_c, {\beta}_c) - \varepsilon_{c,t + 1}, & t < t^*,
\end{cases} \label{eq-transitionfunction}
\end{align}
where function \(g_1() = \Phi^{-1}()\) transforms the indicator to the
inverse-probit scale and \(\Omega_c\) is the level of the indicator on
that transformed scale at a fixed reference year \(t^*\).
The \textit{transition function} $f(\eta_{c,t},{\lambda}_c, {\beta}_c)$ captures the expected rate of change of the logit-transformed indicator and the stochastic smoothing component $\varepsilon_{c,t}$ captures deviations away from the expected rate of change. 

\begin{figure}[htbp]
    \centering
    \includegraphics[width=0.45\textwidth]{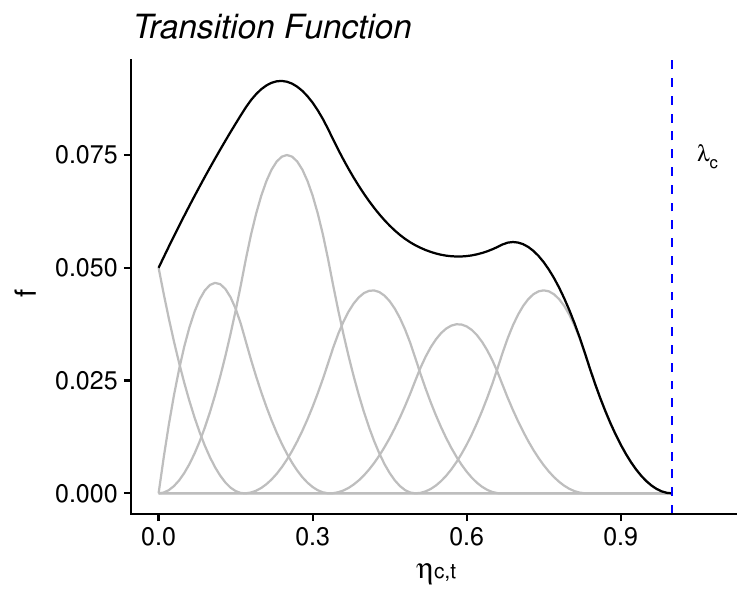}    
  \includegraphics[width=0.45\textwidth]{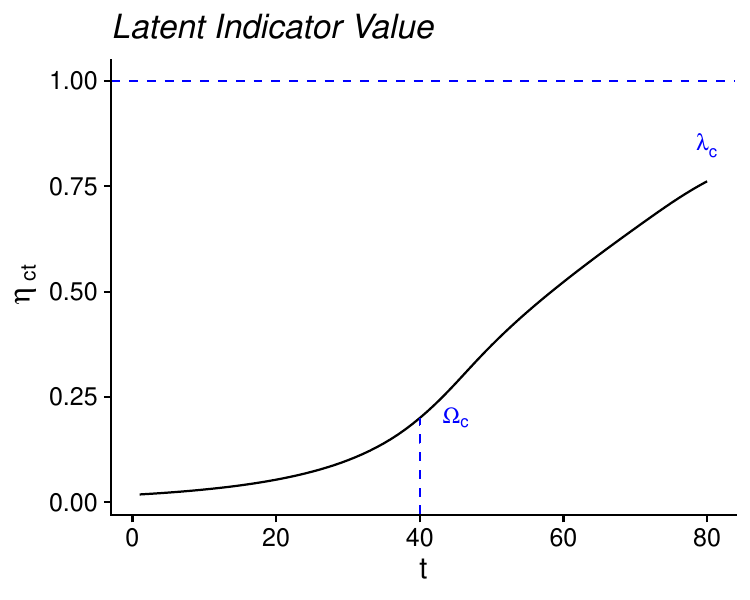}    
    \caption{\textbf{ Illustration of the transition model specification for indicator $\eta_{c,t}$. }  
    Left: Rate of change (black) as a function of the $\eta_{c,t}$, parameterized using B-splines. The gray lines illustrate a scaled version of the non-zero spline basis functions. Right: Transition of $\eta_{c,t}$ from low to high levels, based on the rate of change from plot on left. Figure adapted from \cite{susmann_flexible_2025}.}
    \label{fig-splines}
\end{figure}

The specification of the transition function $f(\eta_{c,t},{\lambda}_c, {\beta}_c)$ follows that of \cite{susmann_flexible_2025}, see Figure~\ref{fig-splines}.
The transition function is parameterized using B-splines, as follows:
\begin{align}
    f(\eta_{c,t},{\lambda}_c, {\beta}_c) = \sum_{j=1}^{J} \bm\beta_{c, j} B_{j}\left(\eta_{c,t}/\lambda_c \right),
\end{align}
where $\bm{\beta}_c$ is the vector of splines parameters, $B_{j}(\cdot)$ refers to the $j$th spline basis function with normalized domain $[0,1]$, and $\lambda_c$ is the asymptote. An illustration of this transition function is given in Figure~\ref{fig-splines}.  

The smoothing terms $\varepsilon_{c,t}$ represent deviations from the smooth transition functions, and are incorporated to capture short-term accelerations or stalls as suggested by the data. The smoothing terms are assigned an AR(1) model:
$$\varepsilon_{c,t}|\varepsilon_{c,t-1}, \rho_{\varepsilon}, \sigma_\varepsilon \sim N(\rho_{\varepsilon}\cdot \varepsilon_{c,t-1}, \sigma_{\varepsilon}^2),$$
with unconditional standard deviation $\text{sd}(\varepsilon_{c,t})= \sigma_\varepsilon \slash \sqrt{1 - \rho_{\varepsilon}^2}$. Using an AR(1) smoothing term in the expression for the rate of change 
implies that the model for $\eta_{c,t}$ itself can be seen as an ARIMA(1,1,0) model on the logit scale with time-varying drift term  $f(\eta_{c,t - 1}, \bm{\beta}_c, \lambda_c)$.

\paragraph{Sharing information across populations using hierarchical models}
Our goal is to estimate and forecast FP indicators from 1970 until 2030 or beyond for all countries globally. However, a country’s data may not be sufficient to estimate FP for the time period of interest. For example, and as illustrated in Figure~\ref{fig-data}, some countries may have very limited data on FP. For countries that are at the start of their transition, country data alone do not allow for estimating its asymptotes. 

To overcome data limitations, we fit a global model in which information is shared across populations using hierarchical models. For each model, we group countries based on geo-political characteristics into clusters and subclusters of countries, following the groupings proposed by \cite{bearak_unintended_2020}. In addition, for the modeling of FP among unmarried women, we further extend the groupings to account for differences in cultural norms related to FP and sexual activity among unmarried women, following the approach by \cite{kantorova_estimating_2020}. 

In the transition model, three sets of model parameters are given hierarchical priors:
\begin{enumerate}
\item asymptote $\lambda_{c}$ controlling the limiting value of an indicator;
\item spline coefficients $\bm{\beta}_{c,j}$  determining the transition function $f$; and
\item $\Omega_{c}$, the level of an indicator in a reference year (on the transformed scale).
\end{enumerate}
For a subset of parameters, hierarchical priors are assigned to transformations of the model parameters, to constrain parameters to appropriate ranges. Full specification of hierarchical models are given in the Appendix. Here we summarize the set up using general notation and give one example. 

The general hierarchical model set up is as follows. Let $\theta_\gamma^{(l)}$ denote the value of one of the transformed parameters for a geographic unit indexed by $\gamma$ at hierarchical level $l = 0, 1, \hdots, L$. For national-level estimation, level $l$ starts at the global level with $l=0$ and ends at the country level with $l=L$. Additional levels may include the clusters or subclusters of countries. For each parameter, we adopt the following setup: 
\begin{align*}
\theta_\gamma^{(l)} \mid \theta^{(l-1)}_{\gamma^\prime}, \sigma^{(l)}_\theta &\sim N\left(\theta^{(l-1)}_{\gamma^\prime}, \sigma^2_{l, \theta}\right), \text{ for } l > 1.
\end{align*}
Here, $\theta^{(l-1)}_{\gamma^\prime}$ is the parameter value at one level higher in the hierarchy for the region $\gamma^\prime$ that contains unit $\gamma$. At the top level of the hierarchy, $l = 0$, only a single parameter is estimated, representing an intercept that applies to all lower level geographic units. The parameter $\sigma_{l,\theta}$ for $l > 1$ describes variability of values of the parameter $\theta$ across geographic units at level $l$ of the hierarchy. Priors are defined for the intercept and the variance terms:
\begin{align*}
  \theta^{(0)} &\sim N\left(s_{0, \theta}\cdot m_{0, \theta}, s_{0, \theta}^2\right),\\
    \sigma_{l,\theta} &\sim N^{+}\left(0, w_{0, \theta}^2\right),
\end{align*} 
where $s_{0,\theta}\cdot m_{0, \theta}$ and $s_{0, \theta}$ refer to the prior mean and standard deviation of $\theta^{(\mathrm{global})}$, and $w_{0, \theta}$ refers to the prior standard deviation
for $\sigma_{l,\theta}$. 

Specific settings for all parameters are given in the Appendix. As an example, we consider the parameter for the level of transformed demand among married women in reference year $t^*$, given by $\Omega_c$. 
No transformation is used for this parameter because $\Omega$ is already unconstrained. The hierarchical levels are given by world, cluster, subcluster, and country. The model set up is as follows, naming the hierarchical levels: 
\begin{align*}
  \Omega_c^{(\mathrm{country})} \mid \Omega_{s[c]}^{(\mathrm{subcluster})}, \sigma_{ \mathrm{country},\Omega}^2 &\sim N(\Omega_{s[c]}^{(\mathrm{subcluster})}, \sigma_{ \mathrm{country},\Omega}^2), \label{eq-countryhier} \\
  \Omega_s^{(\mathrm{subcluster})} \mid \Omega_{w[s]}^{(\mathrm{cluster})}, \sigma_{ \mathrm{subcluster},\Omega}^2 &\sim N(\Omega_{w[s]}^{(\mathrm{cluster})}, \sigma_{\mathrm{subcluster},\Omega}^2), \nonumber \\
  \Omega_w^{(\mathrm{cluster})} \mid \Omega^{(\mathrm{global})}, \sigma_{ \mathrm{cluster}, \Omega}^2 &\sim N(\Omega^{(\mathrm{global})}, \sigma_{ \mathrm{cluster}, \Omega}^2), \\
  \sigma_{l, \Omega} &\sim N^{+}(0, 2^2), \text{ for } l=1,2,3\\
  \Omega^{(\mathrm{global})} &\sim N(0, 3^2).
\end{align*}

\subsubsection{Data model for survey data}

The survey data model specifies the relation between the observed survey data and the FP indicators that are estimated, such that the overall model can be  fitted to the survey data while accounting for different sources of error. 

The survey data base contains observations on the compositional vector of FP proportions, i.e., contraceptive use of modern and traditional methods, and unmet need for any method (see Section~\ref{sec-surveydatabase}). For model fitting, we consider three proportions: mCPR, the ratio of unmet need for modern methods out of non-modern-users, and traditional users out of those with an unmet need for modern methods. 

All observed proportions may be subject to various errors. First, all observations are subject to sampling error arising from the survey design. In addition, other non-systematic errors may occur which we will refer to as measurement errors. Measurement errors may occur due to errors introduced during data collection or compilation. The extent of such errors may vary based on survey characteristics. For example for mCPR estimation, some survey programs have been found to provide more variable measurements of mCPR than others (\cite{alkema2013mcpr}). Extreme errors may also occur, as we saw in the mCPR example for Burundi in Figure~\ref{fig-data}, resulting in outlying observations due to measurement error. Finally, errors may occur if the sampled population differs from the population of interest. For mCPR estimation, such errors are typically absent from recently collected data by international programs but may be present in older data or data from national or other survey programs. Examples include variation in age groups (\cite{alkema2013mcpr,kantorova_estimating_2020}). When using survey data, we need to account for data errors, to produce estimates that are aligned with observations that are very certain and deemed to be close to the true value, while smoothing over outlying data points that are very uncertain. 

To model the relation between FP indicators and an observed proportion, accounting for possible errors in the observed proportion, we use the normal-with-optional-shrinkage (NOS) class of data models (\cite{alkema_2024dm}). This class of data models accounts for different types of errors that may occur. These errors include sampling errors and differences in observational uncertainty based on survey characteristics. In addition, the NOS data model uses shrinkage priors to 
produce estimates that are robust to outlying observations. 

We summarize the main set up here for one proportion, i.e., mCPR. Additional details are given in the Appendix (Section~\ref{app-datamodel}). 

With slight misuse of notation for ease of exposition, let $y_i$ refer to one observed proportion and let $\phi_{c[i], t[i]}$ refer to the true latent indicator for the corresponding country-year. For example, let $y_i$ refer to observed mCPR and $\phi_{c[i], t[i]}$ to true mCPR. The total error associated with $y_i$, on the logit-transformed scale, is decomposed into type-specific errors:
\begin{align}
    E_i &= \mathrm{logit}(y_i) - \mathrm{logit}(\phi_{c[i], t[i]}),\\
    & = E_i^{(\samp)} + \mathbb{I}(d[i] \in \mathcal{D})E_{i}^{(\source)} + \mathbb{I}(i \in \mathcal{C})E_{i}^{(\charic)} + \mathbb{I}(i \in \mathcal{I})E_i^{(\outlier)}, 
\end{align}
where sampling error is given by $E_i^{(\samp)}$, source-type specific measurement error by $E_{i}^{(\source)}$, $E_{i}^{(\charic)}$ captures differences if the sampled population differs from the population of interest, and finally, observation-specific error for outliers is given by $E_i^{(\outlier)}$. Indicator functions indicate whether the error term is assigned to each observation.

In the NOS data model applied to FP indicators, each type-specific error term is assumed to be normally distributed with mean zero and standard deviation $\sigma_i^{(\type)}$: 
\begin{align}
    E_i^{(\type)}|\sigma_i^{(\type)}  &\sim N\left(0, \sigma_i^{(\type)2}\right). \label{eq-errors}
\end{align}
For the sampling error $E_i^{(\samp)}$, the associated variance is $\sigma_i^{(\samp)2} = s_i^2$, where $s_i$ is the sampling error on the logit-scale. Source-type errors capture that errors from some programs may be larger than errors from other programs:
$\sigma_i^{(\source)} = \tilde{\sigma}_{d[i]}^{(\source)}$,
where $d[i] \in \{ 1, \dots, D \}$ indexes the survey source of observation $i$, i.e., DHS, MICS, PMA, National survey, or other survey. 
The term $E_i^{(\charic)}$ with variance $\sigma^{(\charic)2}$ is added if a survey did not capture the entire population of interest, for example, if ever-married women were interviewed as compared to married women. The subset of observation indices for which the population surveyed differs from that of interest, $\mathcal{C}$, is obtained from survey meta data.  


The final  error term, referred to as outlier error, is designed to handle detection of outlier observations and smoothing over them. Outlier detection and smoothing is accomplished by assigning shrinkage priors to the outlier error terms (\cite{piironen2017horseshoe}), to shrink outlier errors towards zero for most observations, only allowing large errors when the observation is extremely different from the trend. The outlier error standard deviation term $\sigma_i^{(\outlier)}$  is parameterized using a scale parameter $\tau$ that is shared across observations, and an observation-specific scale parameter $\gamma_i$,  $\sigma_i^{(\outlier)}= \sqrt{\frac{\tau^2 \vartheta^{2} \gamma_{i}^{2}}{\vartheta^{2} + \tau^2 \gamma_{i}^{2}}}$.
The local scale parameter $\gamma_{i}$ is assigned a heavy-tailed half-Cauchy prior, $\gamma_{i} \sim C^+(0, 1)$, to allow individual errors to escape from shrinkage to zero. 

Based on the expression for the error terms in Eq.~\eqref{eq-errors}, and assuming independence across errors for a single observation, the marginal density for individual observations is given by 
\begin{eqnarray}
\mathrm{logit}(y_i)|\phi_{c[i],t[i]}, \sigma_i \sim N(\mathrm{logit}(\phi_{c[i], t[i]}), \sigma_i^2), \label{dm-normal} \end{eqnarray}
with $\sigma_i^2 = s_i^2 + \mathbb{I}(d[i] \in \mathcal{D})\sigma_{i}^{(\source)2} + \mathbb{I}(i \in \mathcal{C})\sigma^{(\charic)2} + \mathbb{I}(i \in \mathcal{I})\sigma_i^{(\outlier)2}$. We assume observations are conditionally independent except for data collected by the PMA program, as results from consecutive PMA surveys may be correlated because of repeated sampling of individuals across survey rounds. For PMA data series for a country-marital group, we use the multivariate normal specification:
\begin{align}
    \bm{z}|\bm{\phi}, \bm{\Sigma} \sim N(\mathrm{logit}(\bm{\phi}), \bm{\Sigma}), 
\end{align}
where $\bm{z}$ refers to the vector of logit-transformed observations, $\mathrm{logit}(\bm{\phi})$ refers to the respective mean vector of $\bm{z}$, and $\bm{\Sigma}$ to its associated covariance matrix. We set $\Sigma_{j,l} = Cov(z_j, z_l) = \sigma_j \cdot \sigma_l \cdot \rho_{\PMA}^{|t[j] - t[l]|}$, where  $\rho_{\PMA}$ refers to the autocorrelation in PMA data errors. 

\paragraph{When to (not) include outlier error terms?}
Outlier error terms are included only for observations with indices $\mathcal{I}$. This subset is introduced to avoid automatically smoothing over high-quality observations that are outlying but are deemed to capture true fluctuations in FP indicators. 

A data classification algorithm is used to produce a set of reference survey data points that are, by default, not deemed to be outlying, as follows. For each country, we first label observations as possibly outlying if there are data quality concerns or other documented substantive information to suggest they may be subject to substantial error. For example, DHS observations collected before 1990 may be outlying because of possible misclassification of women who were pregnant or post-partum. A survey may be classified as being possibly outlying if issues with the survey process have been documented. 
Subsequently, for each country, we check if we can identify a reference source category among the left over points, which, as the name suggests, provides a reference in the estimation exercise and is assumed to be not subject to outlier errors. This category is given by DHS data, if available. If no DHS data are available, data from national surveys or other survey programs are used, with the choice based on the source type with the most observations after 1990 in that country. 
If the algorithm does not result in a reference category, i.e., if there are no non-possibly-outlying observations that are from DHS, national, or other surveys, then the population does not have a reference category and all observations are possibly outlying. 

The default classification of possibly outlying points is used in global model fitting for estimation of long term trends. In local model fitting, users may update the classification to reflect external information about data quality.

\subsubsection{Global model fitting and computation}

Global model fitting refers to estimating all model parameters of the combined process and survey data model, as summarized in Figure~\ref{fig-diagrammodel}. Global model fitting occurs separately by marital group, as summarized in Figure~\ref{fig-workflow}. 

To estimate  mCPR and unmet need for modern methods, per marital group, we use two steps. In step 1A, we estimate global and world region parameters governing long-term trends in demand and demand satisfied. In step 1B, parameters governing  short-term fluctuations and data quality are estimated. The motivation for fitting sequentially in two steps is to avoid possible issues identifying long-term trends. Details are provided in Appendix Section~\ref{sec-details1a1b}. 

The global model was implemented in the \texttt{Stan} programming language \citep{stan2023, cmdstanr2022}, and all analyses were conducted using the \texttt{R} statistical computing environment \citep{r2022}. Code is available in the open-source R package \texttt{fpet2}, see \cite{fpetrpackage}. The package's website contains detailed instructions for global model fitting, see \url{https://alkemalab.github.io/fpet2/articles/workflow_globalnationalfitting.html}.

\subsection{Local modeling of FP indicators}\label{sec-localmodels}

The main utility of FPET is as a tool for countries to produce estimates for a particular population of interest, by fitting a model to data from that population alone. As summarized in Figure~\ref{fig-workflow}, we use what we refer to as ``local models``: models that are derived from global (hierarchical) models but adapted such that they can be fitted to the data from one population alone.

For FPET, we provide an implementation of local models through fixing non-country-specific parameters at point estimates from the global fit. The set up follows the set up introduced in the \texttt{localhierarchy} R package \citep{alkema_2025localhierarchy}. This set up is introduced in the next section. In local models, countries may want to use service statistics data to inform FP estimates. This is introduced in Section~\ref{sec-serviceestats}.  

A web-based application of the local model with a user-friendly interface is available at \url{https://fpet.track20.org/}.

\subsubsection{Local model specification and implementation}

For local modeling, we use the specification and implementation provided in the \texttt{localhierarchy} R package \citep{alkema_2025localhierarchy}.  For national estimation using the FPET model, we start by fitting the global model as explained in the previous section, using a global data base and estimating all model parameters, including data quality parameters and means and variance of hierarchical distributions. In the derived local model (see Figure~\ref{fig-local}), parameters that are not country-specific are fixed using parameter estimates (posterior means) from the global model fit. 

Local modeling is implemented in the R package \texttt{fpet2}  (\citep{fpetrpackage}). The package uses functionality from \texttt{localhierarchy} to pass estimates of non-country-specific parameters from global models to local models. Local models are available for estimation of one marital group only, and for estimating both marital groups and calculating all-women estimates. 

\begin{figure}[htbp!]
\begin{center} 
\begin{tikzpicture}[
    node distance=1.5cm,
    every node/.style={font=\small, align=center},
    block/.style={draw, thick,  minimum width=3cm, minimum height=2cm, inner sep=8pt},
    data/.style={block, fill=yellow!20},
    process/.style={block, fill=yellow!20},
    hier/.style={block, fill=yellow!20},
    prior/.style={block, fill=green!20},
    >={Stealth[width=2.5mm,length=3mm]},
     datablk/.style={rectangle, rounded corners=6pt, draw, fill=orange!20, minimum width=3cm, minimum height=1cm, align=center},
]

\node[data] (data) {
    \textbf{Data model} \\
   
    $\mathrm{logit}(y_i)| \phi_{c[i], t[i]} \sim N(\mathrm{logit}(\phi_{c[i], t[i]}), \hat{\sigma}_i^2),$

};
\node[datablk, left = of data] (survey) {Survey data \\ for one country};
\node[process, below=of data] (process) {
    \textbf{Process model} \\
$\begin{aligned}
g_1(\eta_{c,t}) &= \begin{cases}
\Omega_c, & t = t^*,\\
g_1(\eta_{c,t-1}) + f(\eta_{c,t-1},\lambda_c,\beta_c) + \varepsilon_{c,t}, & t > t^,\\
g_1(\eta_{c,t+1}) - f(\eta_{c,t+1},\lambda_c,\beta_c) - \varepsilon_{c,t+1}, & t < t^*,\\
\end{cases}\\
\varepsilon_{c,t}|  \varepsilon_{c,t-1}&\sim N\big(\hat{\rho}_{\varepsilon}\varepsilon_{c,t-1},\hat{\sigma}_\varepsilon^2\big)
\end{aligned}$
};
 \node[hier, below = of process] (hier) {
    \textbf{Priors on transition model parameters}
   $\bm{\beta}_c$, $\lambda_c$, $\Omega_c$ \\ Priors are derived from point estimates of hierarchical mean and \\variance  parameters. 
   In general notation, with $l$ referring to the country level:\\
  $  \begin{aligned}
\theta_\gamma^{(l)} &\sim N(\hat{\theta}^{(l-1)}_{\gamma^\prime}, \hat{\sigma}^2_{l, \theta}), 
\end{aligned}$
\\
where $\hat{\theta}^{(l-1)}_{\gamma^\prime}$ refers to the point estimate at one level higher in the hierarchy.
}; 
\node[prior, right=of data] (prior) {
    Point estimates \\
    from global model\\ 
    ($\hat{\sigma}$'s, $\hat{\rho}$'s, $\hat{\theta}^{(l-1)}$'s)
};

\draw[->, thick] (process) -- node[left, font=\footnotesize] {Specification\\ for $\phi$ \\ (derived from $\eta$s)} (data);

\draw[->] (survey) -- (data);
\draw[->] (prior) -- (data);
\draw[->] (prior) -- (process);
\draw[->] (prior) -- (hier.east);

\draw[->, thick] (hier) -- (process);

\end{tikzpicture}
\caption{\textbf{Overview of local model specification.} The local model specification follows that of the global model (Figure~\ref{fig-diagrammodel}), with parameters that are not country-specific fixed at the estimates obtained from the global model fit. }\label{fig-local}
\end{center}
\end{figure}

\subsubsection{Use of service statistics data}\label{sec-serviceestats}

Service statistics data are used in FPET to inform mCPR estimates. 
There are four types of FP service statistics data: the number of contraceptive commodities distributed to clients, the number of contraceptive commodities distributed to facilities, the number of times clients interacted with a provider for contraceptive services, and the number of current contraceptive users of any method. Track20 developed a tool to calculate an Estimated Modern Use (EMU) indicator from these different types of service statistics data \citep{bietsch_estimated_2025, emutrack20, ss2emu_tool}. EMU estimates have been used in FPET to inform mCPR estimates in the absence of recent survey data \citep{cahill2018mcpr}. Here we summarize the approach used in the most recent version of FPET. Additional details can be found in \cite{mooney_emu}.

To reduce risk of bias, we use observed EMU rates of
change to inform mCPR estimates, as opposed to EMU level estimates. The motivation for using rates of change comes from difficulties capturing some contraceptive methods in specific EMUs or country settings. For example, in some countries, EMU estimates of IUD use based on service statistics data do not align with what is expected from survey data. In other settings, some methods cannot be adequately captured, such as sterilization when using commodities supplied to clients data, or condom use. When methods are missing, EMUs are biased but changes in EMUs, i.e., increases or decreases, are not necessarily biased. We note that percentage point changes in mCPR are the sum of method-specific percentage point changes, hence observed changes in mCPR do not depend on methods that do not change with time. This explains the rationale for using rates of change in EMUs in FPET: the risk of bias is lower for rates of change as compared to levels in settings where some methods may be difficult/impossible to capture in service statistics data.

When including rates of change of EMUs in FPET, we account for observation-specific uncertainty derived during EMU calculation, as well as uncertainty based on a country-type specific assessment of error. The data model is given by:
\begin{align}
\Delta EMU_{d,c,t}|\sigma_{EMU,d,c,t} \sim N(\Delta \phi_{c, t, m},  \sigma_{EMU,d,c,t}^2), 
\end{align}
where $\Delta EMU_{d,c,t}$ refers to the observed change in EMU  for data type $d$, country $c$, and year $t$. $\Delta \phi_{c, t, m}$ refers to the change in modern contraceptive use for the same country-period for marital group $m$, where $m$ refers to all women or married women. Error variance $\sigma_{EMU,d,c,t}^2$ is given by 
\begin{align}
\sigma_{EMU,d,c,t}^2 = s_{EMU,d,c,t}^2 + \sigma_{EMU, d, c}^2,
\end{align}
with $s_{EMU,d,c,t}^2$ the pre-calculated variance based on uncertainty in inputs and adjustments, and $\sigma_{EMU, d, c}^2$ a country-type specific variance. We obtain $s_{EMU,d,c,t}^2$ based on a Monte Carlo simulation, using uncertainty in inputs associated with the adjustments of EMUs to capture the private share (\cite{mooney_emu1}).  The country-type variance is estimated hierarchically:

\begin{align}
\log(\sigma_{EMU, d, c})|\mu_{EMU, d}, \sigma_{EMU} \sim N(\mu_{EMU, d}, \sigma_{EMU}^2),
\end{align}
with type-specific mean $\mu_{EMU, d}$, 
and overall variance  $\sigma_{EMU}^2$, both of which are hyperparameters, estimated hierarchically using training data from all countries. Through this set-up, the country-type variance is based on global type-specific variance, as well as country data on the difference between change in EMU and survey-informed change in mCPR. As a result, the total variance associated with an EMU-based rate of change will be smaller for countries in which EMUs are more closely tracking survey-based estimates of mCPR.

\clearpage 

\section{Results}
In this section, we first present illustrative findings that show country-specific fits to highlight modeling aspects. We then discuss results from a model validation exercise, followed by illustrative findings in local modeling when including service statistics data.  Results for all countries are given in Appendix Figure~\ref{fig-app-results}. 

\subsection{Illustrative findings}
\begin{sidewaysfigure}[htbp]
    \centering
    \includegraphics[width=0.9\textwidth]
        {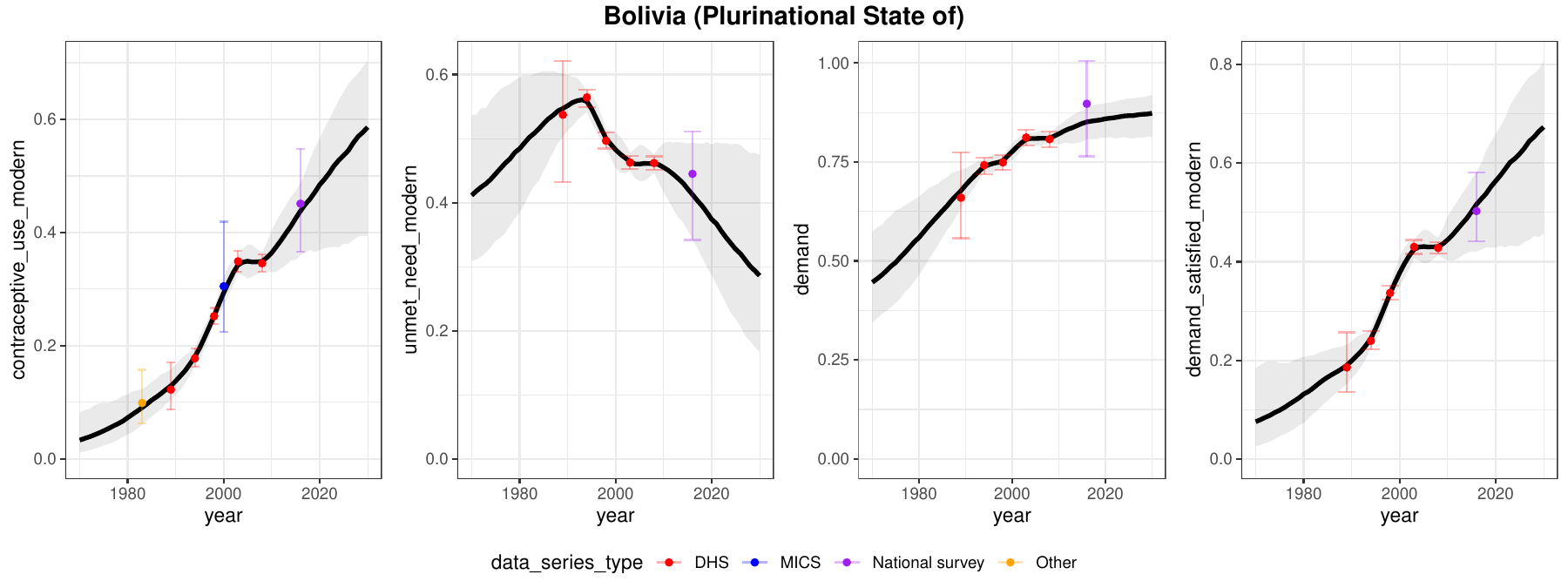}\\
    \includegraphics[width=0.9\textwidth]
        {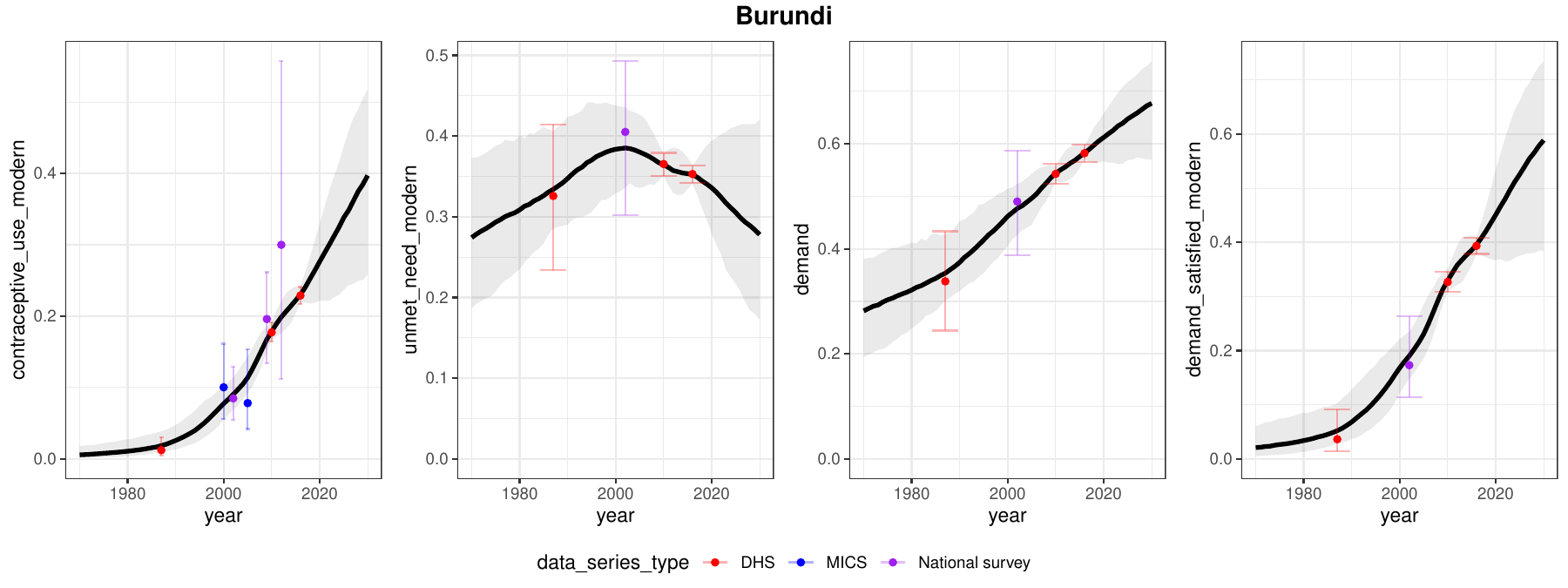}\\
     \caption{\textbf{Data, estimates, and forecasts from 1970 until 2030 for FP indicators for married women in Bolivia and Burundi.} Graphs show mCPR, unmet need for modern methods, demand, and demand satisfied with modern methods. Survey data are shown in colored dots with 95\% confidence intervals that reflect total error variance. FPET estimates are shown in black, with grey shaded areas representing 90\% credible intervals.}
    \label{fig-transitionexamples}
\end{sidewaysfigure}

\begin{sidewaysfigure}[htbp]
    \centering
    \includegraphics[width=0.9\textwidth]
        {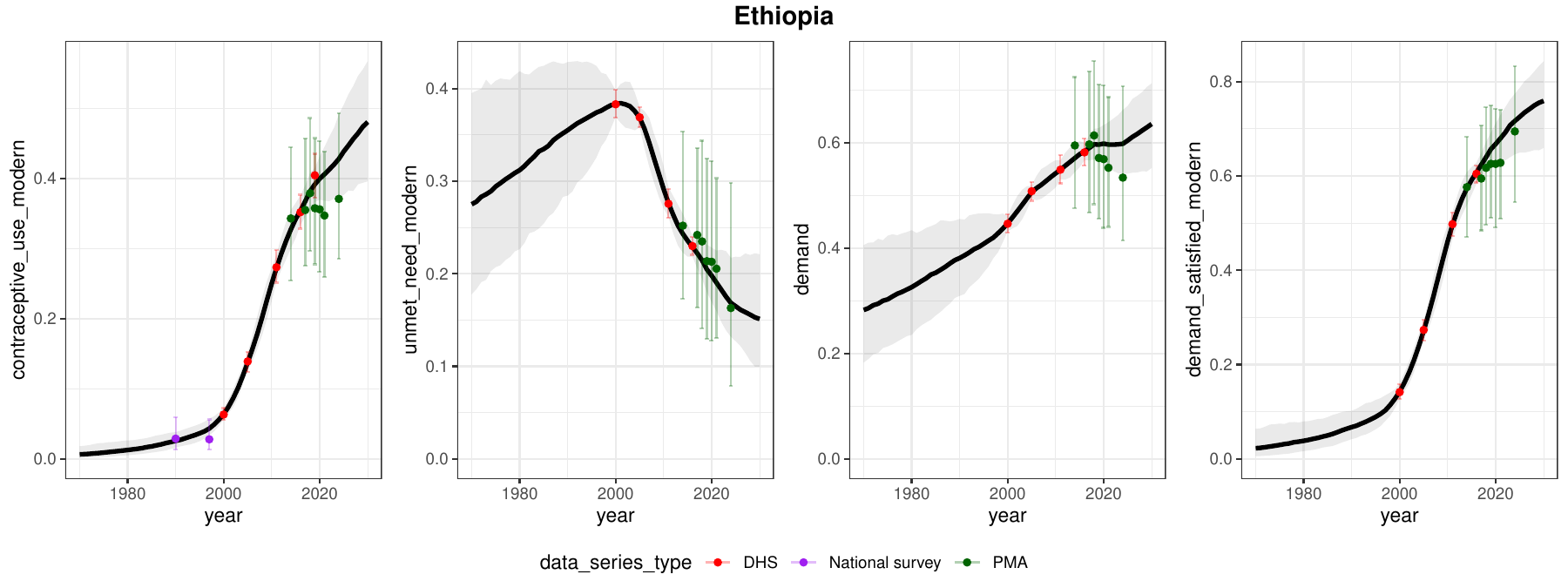}\\
    \includegraphics[width=0.9\textwidth]
        {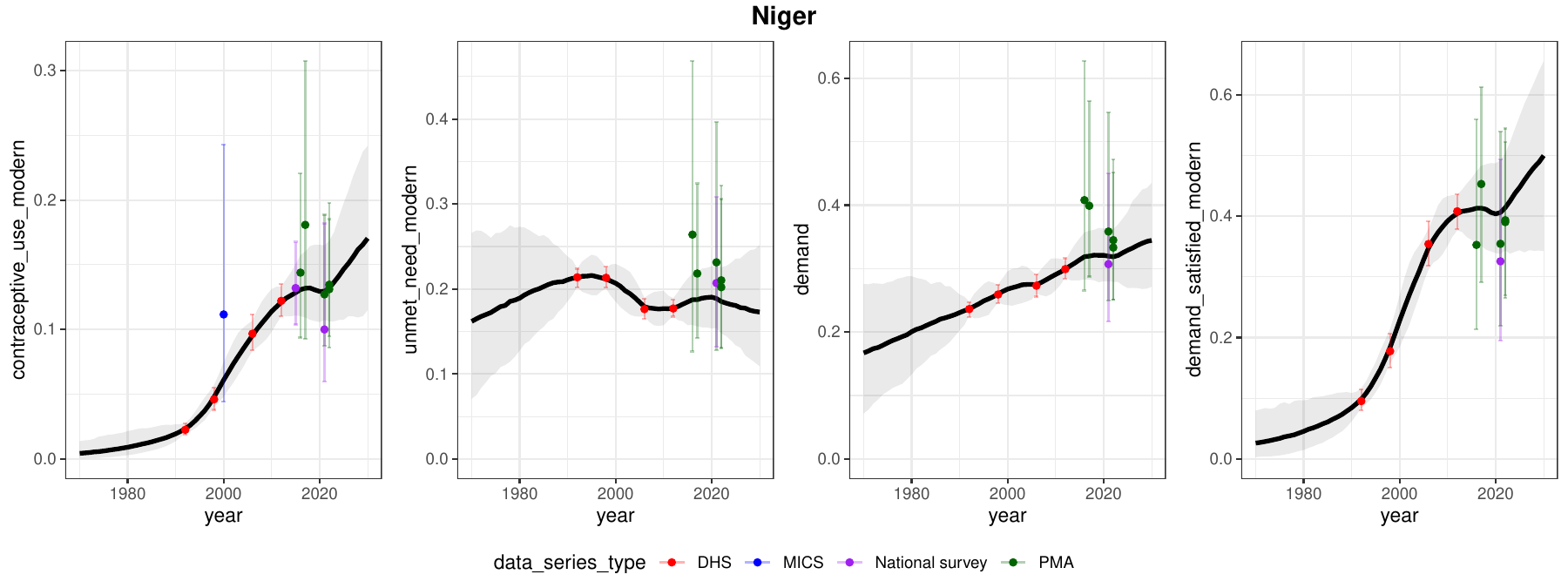}\\
     \caption{\textbf{Data, estimates, and forecasts from 1970 until 2030 for FP indicators for married women in Burundi and Ethiopia.} Graphs show mCPR, unmet need for modern methods, demand, and demand satisfied with modern methods. Survey data are shown in colored dots with 95\% confidence intervals that reflect total error variance. FPET estimates are shown in black, with grey shaded areas representing 90\% credible intervals.}
    \label{fig-transitionexamples2}
\end{sidewaysfigure}

To illustrate different contraceptive transitions and the effect of hierarchical modeling, model fits for married women in Bolivia, Burundi, Ethiopia, and Niger are shown in Figures~\ref{fig-transitionexamples} and \ref{fig-transitionexamples2}. The figures illustrate how the model set up, using transition functions, allows for estimation and forecasting in different settings. 

Results for Burundi and Bolivia illustrate how the model can produce estimates and forecasts for all years, even if data are limited for recent years. The uncertainty associated with FP indicators increases with the forecast horizon. 

In Burundi, the most recent national survey indicates that mCPR is high, as compared to DHS observations in surrounding years. The FPET fit for Burundi treats the national survey observation as implausibly high and smooths over it, i.e., the model does not shrink the outlier error for the national survey in Burundi based on information in surrounding years. If substantive information on mCPR in Burundi would suggest that the national survey accurately captured a temporary uptake, the outlier error could be removed in a local model fit, to produce estimates that reflect that reality. 

In Ethiopia, FPET estimates are close to the most recent DHS, accounting for correlation in the PMA data series, that trend lower. 

Niger illustrates a setting with limited recent FP data: the absence of recent DHS surveys — which typically have smaller sampling errors and are considered less susceptible to non‑sampling errors — means analysts must rely on PMA and national surveys that carry greater uncertainty. FPET’s role in this context is to integrate those uncertain sources and produce estimates with uncertainty that accurately reflect the available evidence. Consequently, FPET estimates for Niger are substantially more uncertain for recent years because of these data limitations.

\clearpage

\subsection{Model validation}
Model development and checking took place in different stages. To summarize final performance, we carried out a validation exercise in which we fit FPET to a training data set that excluded survey data that had been collected in or after 2020. The goal was to check how well the estimates and predictions based on the training set, i.e. estimates and forecasts that would have been produced in 2020, aligned with: (1) the estimates based on the full data set; and (2) the held-out survey data reported for years after 2019. We calculated summary measures based on prediction errors and calibration.

The first set of measures is focused on the comparison between point estimates and credible intervals (CIs) from the training and full data set. We calculate prediction errors in the FPET point predictions, referring to the difference between the FPET point prediction for a given year and its updated estimate. Here the year 2024 is chosen, to reflect a forecast horizon of 5 years based on the training data that excludes data in and after 2020. We also check how often the point prediction falls outside of the CIs based on all data. We expect these errors to be small and for point predictions to fall inside updated intervals at at least the nominal level. Results are given in Table~\ref{tab-val2020} and suggest that FPET
is reasonably well calibrated. For example, the median absolute error in mCPR among all women is 1.7\%.  Similarly, the point predictions that would have been constructed in 2020 generally fall within the credible intervals constructed using all data at nominal levels or above, see Table~\ref{tab-calibration}. We do note a  negative bias of -1.8\% for predicting mCPR among married women (meaning that FPET overpredicted the mCPR levels). This type of bias was also present in the previously-used FPET model, the updated model has improved the bias from -2.6 to -1.8\% (see appendix Section~\ref{sec-app-validation}). 

\begin{table}[htbp]
\begin{center}
\centering
\begin{tabular}{llrrrr}
  \hline
indicator & married? & MedE & MedAE & prop $<1$\% & prop $<5$\% \\ 
  \hline
mCPR & all & -1.1 & 1.7 & 0.4 & 0.8 \\ 
  mCPR & no & -0.4 & 1.4 & 0.4 & 0.9 \\ 
  mCPR & yes & -1.8 & 2.4 & 0.3 & 0.7 \\ 
  unmet modern & all & 0.5 & 1.2 & 0.5 & 1.0 \\ 
  unmet modern & no & -0.1 & 0.6 & 0.7 & 1.0 \\ 
  unmet modern & yes & 1.1 & 1.6 & 0.3 & 0.9 \\ 
   \hline
\end{tabular}
\caption{\textbf{Summary of prediction errors (in percentage) for the year 2024}. An error refers to the difference between the estimate for the indicator based on the full data set and the indicator based on the training data, for the year 2024 (5 year forecast horizon). A positive (negative) error indicates that the prediction based on data up to 2018 underpredicted (overpredicted) the target value.
Table columns: MedE = median error, MedAE = median absolute error,  prop $<$ 1 or 5 \% = proportion of errors less than 1 or 5 percentage points.}
    \label{tab-val2020}
    \end{center}
\end{table}

\begin{table}[htbp]
    \centering
   \begin{tabular}{llrr}
  \hline
indicator & marital\_status & below 90\% CI & above 90\% CI \\ 
  \hline
mCPR & married & 8.3 & 1.4 \\ 
  mCPR & unmarried & 5.6 & 1.4 \\ 
  unmet & married & 0.0 & 1.4 \\ 
  unmet & unmarried & 1.4 & 0.0 \\ 
   \hline
\end{tabular}
    \caption{\textbf{Percentage of estimates for 2024 based on data prior to 2020 that fall outside their respective credible intervals based on all data.}. 
Table columns: below or above 90\% CI  = percentage of estimates that fall below or above the 90\% CI.}
    \label{tab-calibration}
\end{table}

The second set of measures is focused on the differences between left-out survey data and point predictions. We do not necessarily expect these differences to be small for all survey data: we only expect small differences for survey data that is subject to small sampling and non-sampling errors. Results for observations from DHS surveys in sub-Saharan Africa, where such errors are generally small, are shown in Table~\ref{tab-dhs}. We see that results are  comparable to the results in Table~\ref{tab-val2020}.

\begin{table}[htbp]
    \centering
\begin{tabular}{llrrrr}
  \hline
indicator & married? & MedE & MedAE & prop $<1$\% & prop $<5$\% \\ 
  \hline
mCPR & no & -0.5 & 2.4 & 0.1 & 0.8 \\ 
  mCPR & yes & -1.7 & 2.1 & 0.2 & 0.7 \\ 
  unmet modern & no & 0.1 & 0.9 & 0.6 & 1.0 \\ 
  unmet modern & yes & 1.3 & 2.1 & 0.3 & 0.9 \\ 
   \hline
\end{tabular}
    
    \caption{\textbf{Summary of prediction errors (in percentage)  for left-out DHSs in sub-Saharan Africa}. An error refers to the difference between a left-out DHS observation and the estimate for the indicator based on the validation data. A positive (negative) error indicates that the prediction based on data up to but excluding 2020 underpredicted (overpredicted) the target value.
Table columns: MedE = median error, MedAE = median absolute error,  prop $<$ 1 or 5 \% = proportion of errors less than 1 or 5 percentage points.}
    \label{tab-dhs}
\end{table}

\subsection{Including service statistics data}
\begin{figure}[htbp]
    \centering
    \includegraphics[width=\textwidth]{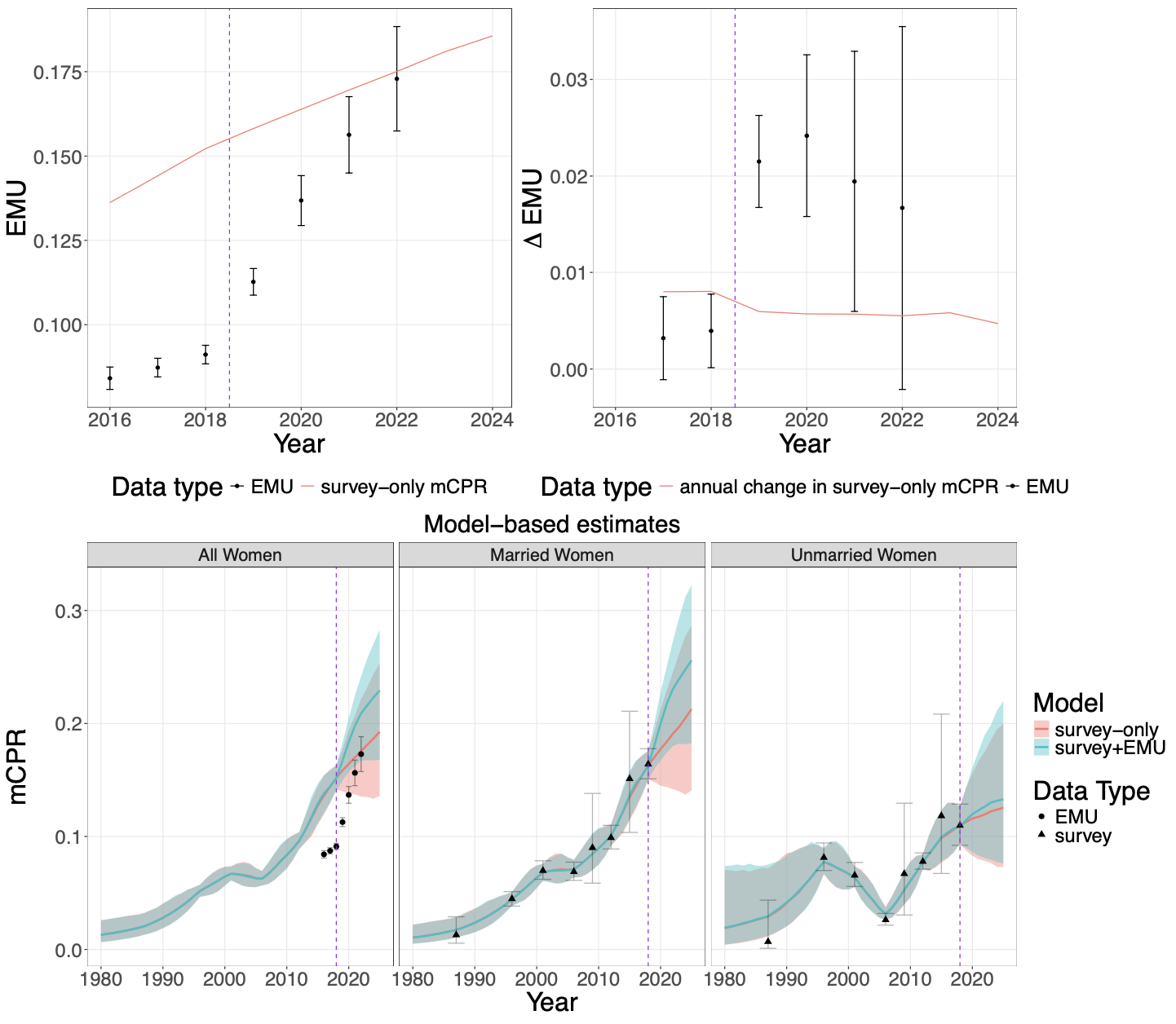}
    \caption{\textbf{Illustration of the effect of including EMUs when estimating mCPR.} The top row shows observed EMUs (black) for the level (left) and rate of change (right). The vertical lines indicate the 95\% confidence intervals based on $s_{EMU,d,c,t}$ alone. Estimates of mCPR that are based on survey data alone are added in red.  The purple dashed line indicates the most recent survey year. The top row illustrates that the EMU-based rates of change suggest an acceleration after the most recent survey year, with rates of change that are higher than survey-based predictions. 
    Plots in bottom row illustrate the effect of including the EMUs in FPET estimates: the acceleration in EMUs past the most recent survey results in faster mCPR growth, as compared to survey-only estimates. }
    \label{fig-emus}
\end{figure}

The effect of including EMUs is illustrated in a case study in Figure~\ref{fig-emus}\footnote{EMU data are country-owned and not publicly available. We illustrate EMU data for an anonimized population.}. The top row shows information on mCPR (left) and rates of change in mCPR (right), based on EMUs (black) and survey-based estimates of mCPR (red). The black vertical lines indicate the 95\% confidence intervals for mCPR based on individual EMUs and their respective $s_{EMU,d,c,t}$ alone. The EMU-based rates of change suggest an acceleration after the most recent survey year, with rates of change that are higher than survey-based predictions. Plots in the bottom row illustrate the effect of including the EMUs in FPET estimates: the high rate of growth in EMUs past the most recent survey results in faster estimated mCPR growth, as compared to survey-only estimates. 
  
\cite{mooney_emu} assess whether the inclusion of EMUs improves predictions for years after the most recent year with survey data. They find that EMUs improve predictive performance. The inclusion of EMUs is found to be particularly helpful in settings with stalls or accelerations that occurred after the most recent survey year.

\section{Discussion}
In this paper, we introduced the Family Planning Estimation Tool. We summarized the Bayesian hierarchical transition modeling  approach underlying FPET and introduced recent updates related to capturing contraceptive transitions, fitting to survey data that may be error prone, and the use of service statistics data. A validation exercise suggested that FPET is reasonably well calibrated for predicting FP outcomes. We presented country-level examples to illustrate FPET’s main utilities: (1) reconciling data from multiple sources with varying levels of uncertainty, and (2) producing backcasts, nowcasts, and forecasts in data-limited settings.

Based on our experience with FPET, we believe that local tools are a key to success for producing locally-relevant estimates.  In our set-up, we derived a local tool from a global model, to allow users to fit the model to their own data. At a minimum, these tools allow for production of estimates based on data from a particular setting alone. It is typically easier to get buy-in for a set of estimates that is produced in-house, based on a fit to data that has been vetted, as opposed to estimates that have been produced in a top-down approach. But the use and usefulness of local models go well beyond the production of just one set of estimates. Having a model that can be run annually builds a mechanism for data review, of surveys but most importantly service statistics.  Track20’s experience suggests that countries have improved service statistics over the project because they are reviewing them annually, usually in a consensus meeting with many stakeholders, and it prompts follow-up action. Local tools also enable sensitivity analyses of various forms: models can be fitted to specific data sets or based on changing model assumptions. These kinds of fits can be helpful for analysts to better understand the data and model. Track20 has successfully used this approach in consensus meetings, as well as in additional technical meetings with various government and non-government stakeholders. For example, in India and Kenya, the tool has been used to analyze data outliers and inform decisions related to data exclusion. In summary, by enabling in-house production of estimates and additional assessments, local tools empower governments and move away from a commonly used approach of externally produced global health estimates.

The use of Bayesian models enabled the development and evolution of FPET. The Bayesian framework provides flexibility in translating assumptions into probabilistic models and allows for uncertainty to be accounted for, and propagated across various model components. Active development of computational approaches and Bayesian modeling software, such as those by the Stan team, have resulted in reduced computation times and expanded the types of models that can be fitted. More generally, these developments, as well as the sharing of open-source code bases for Bayesian modeling in various initiatives are greatly increasing the availability of Bayesian models for a range of demographic and global health indicators. 

We conclude by commenting on the role of the DHS program in FP monitoring and the implications of its recent termination. The DHS has been a cornerstone of FP monitoring in many countries, providing household survey data that inform FPET estimates and a wide range of other analyses (see, for example, \citep{unpd_dhs2025}). Following the DHS program’s recent termination as part of the suspension/termination of several USAID‑funded programs under the new U.S. administration (e.g., see \citep{khaki_when_2025}), the need to combine data from multiple sources and to develop new data systems is more important than ever. In a data landscape that excludes DHS, available sources include surveys conducted by other programs (e.g., MICS, PMA), country‑level surveys, and routine service‑statistics systems. Although each of these sources may have limitations, as discussed in this paper, approaches such as FPET offer a standardized framework for reconciling heterogeneous data and producing coherent FP estimates. Sustained investment in alternative data sources, transparent methods, and country capacity will be critical to maintain robust FP monitoring in the absence of DHS.

\section{Data availability}
Data and code are included in the fpet2 R package (\citep{fpetrpackage}), available at \url{https://github.com/AlkemaLab/fpet2}. Example use is illustrated on the package’ website (see \url{https://alkemalab.github.io/fpet2}).

\bibliography{bibliography}

\clearpage

\section{Appendix}

\subsection{Process model specification for all proportions}
Main outcomes of interest are proportions $\phi_{j,c,t,m}$ for category $j=1,\hdots, 4$ (modern use, traditional use, unmet need for any method, and women ``with no need`` respectively), country $c = 1, \hdots, C$, year $t = 1970, \hdots, 2030$, and marital group $m$ referring to married and unmarried women. 

For each marital group, we specify models for demand and demand satisfied (detailed in the next section) and for the ratio of traditional use to unmet need for modern methods (detailed in Section~\ref{sec-app-trad}). Demand $\eta_{1,c,t,m}$ is defined as the sum of modern users, traditional users, and unmet need for any method: $\eta_{1,c,t,m} = \phi_{1,c,t,m} + \phi_{2,c,t,m} + \phi_{3,c,t,m}$.  Demand satisfied is defined as the modern users over demand, $\eta_{2,c,t,m} = \phi_{1,c,t,m}/(\phi_{1,c,t,m} + \phi_{2,c,t,m} + \phi_{3,c,t,m})$. The ratio of traditional use to unmet need for modern methods is defined as $\eta_{3,c,t,m} = \phi_{2,c,t,m}/(\phi_{2,c,t,m} + \phi_{3,c,t,m})$. 

Based on the models for demand, demand satisfied, and the traditional use ratio, the four proportions are obtained as follows:
\begin{eqnarray*}
\phi_{1,c,t,m} \text{ (modern use)} &=& \eta_{2,c,t,m} \cdot \eta_{1,c,t,m},\\
\phi_{2+3,c,t,m}\text{ (unmet need for modern methods)} &=&  \eta_{1,c,t,m}  - \phi_{1,c,t,m},\\
\phi_{2,c,t,m} \text{ (traditional use)} &=& \eta_{3,c,t,m} \cdot \phi_{2+3,c,t,m},\\
\phi_{3,c,t,m} \text{ (unmet need for any method)} &=& \phi_{2+3,c,t,m} - \phi_{2,c,t,m},\\
\phi_{4,c,t,m}  \text{ (no need)} &=& 1 - \eta_{1,c,t,m}. 
\end{eqnarray*}

\subsection{Model specification for demand and demand satisfied}

\subsubsection{Process model specification for demand and demand satisfied}\label{specification-of-a-transition-model}
We model demand and demand satisfied, separately by marital group (4 models total). As specified in the main text, a transition model for one indicator of interest $\eta_{c,t}$ (dropping additional indices for notational convenience) is specified as follows:

\begin{align} 
 g_1(\eta_{c,t}) &=\begin{cases}
 \Omega_c, & t = t^*, \\
 g_1(\eta_{c,t-1}) + f(\eta_{c,t-1}, {\lambda}_c, {\beta}_c) + \varepsilon_{c,t}, & t > t^*, \\
 g_1(\eta_{c,t+1}) - f(\eta_{c,t+1},{\lambda}_c, {\beta}_c) - \varepsilon_{c,t + 1}, & t < t^*,
\end{cases} , \label{eq-process_app} \\
    f(\eta_{c,t},{\lambda}_c, {\beta}_c) &= \sum_{j=1}^{J} \bm\beta_{c, j} B_{j}\left(\eta_{c,t}/\lambda_c \right),\label{eq-process_app2}\\
    \varepsilon_{c,t}|\varepsilon_{c,t-1}, \rho_{\varepsilon}, \sigma_\varepsilon &\sim N(\rho_{\varepsilon}\cdot \varepsilon_{c,t-1}, \sigma_{\varepsilon}^2),\label{eq-process_app3}
\end{align}
where function \(g_1() = \Phi^{-1}()\) transforms the indicator to the
inverse-probit scale and \(\Omega_c\) is the level of the indicator on
that transformed scale at a fixed reference year \(t^*\) (the year 2004 is used as a reference year). 
\emph{Transition function} \(f(\eta_{c,t - 1}, {\beta}_c, \lambda_c)\)
captures the expected rate of change of the logit-transformed indicator and is parameterized
using B-splines. The stochastic smoothing component \(\varepsilon_{c,t}\) captures
deviations away from the expected rate of change. 

Process model parameters (one set for each of the four models) are given by:
\begin{itemize}
    \item \({\beta}_c\): vector of country-specific splines parameters;
    \item \(\lambda_c\): country-specific asymptote;
    \item $\Omega_c$: country-specific level (on transformed scale) in reference year $t^*$;
    \item $\varepsilon_{c,t}$: country-year deviation terms; 
    \item $\rho_{\varepsilon}$ and $\sigma_\varepsilon$: global AR-parameters.
\end{itemize}

For all four transition models, the priors on the AR-parameters are given by
\begin{align*}
    \rho_{\varepsilon} &\sim N_{(0,1)}(0, 0.5),\\
    \sigma_\varepsilon &\sim N^+(0, 0.2).
\end{align*}

Country-specific transition parameters $\lambda_{c}, \bm{\beta}_{c,j}, \Omega_{c}$ are given hierarchical priors. To work with parameters that are unconstrained, the following transformations are used:
\begin{align*}
    \lambda_c &= \lambda_c^{(low)} + (1- \lambda_c^{(low)})\Phi\left(\tilde{\lambda}_c\right),\\ 
\beta_{c,k} &= \beta_{c}^{(low)} + (\beta_{c}^{(up)}  - \beta_{c}^{(low)})\cdot \text{inv-logit}(\tilde{\beta}_{c,k}),
\end{align*}
where the \(\tilde{\cdot}\) is added to refer to the transformed and unconstrained parameter. These transformations result in $\lambda_c$ being bounded between $\lambda_c^{(low)}$ and $1$, and the $\beta_{c,k}$s are constrained to be between \(\beta_{c}^{(low)}\)
and \(\beta_{c}^{(up)}\). The bounds differ across the four transition models and are given in Table~\ref{tb-bounds}. 

Each of the three (unconstrained) parameters \({\Omega}_c\),
\(\tilde{\lambda}_c\), and \(\tilde{\beta}_{c,k}\), are modeled hierarchically. The hierarchical groupings vary by
marital group and parameter and are given in Table~\ref{tb-groupings}. For each parameter, priors for global intercepts and standard deviation terms are given by
\begin{align*}
  \theta^{(0)} &\sim N(s_{0, \theta}\cdot m_{0, \theta}, s_{0, \theta}^2),\\
    \sigma_{l,\theta} &\sim N^{+}(0, w_{0, \theta}^2),
\end{align*} 
where $s_{0,\theta}\cdot m_{0, \theta}$ and $s_{0, \theta}$ refer to the prior mean and standard deviation of $\theta^{(\mathrm{global})}$, and $w_{0, \theta}$ refers to the prior standard deviation
for $\sigma_{l,\theta}$. Prior parameters (\(m_{0, \mu}, s_{0, \mu}, w_{0, \mu}\))
also differ across indicators and marital groups and are given in Table~\ref{tb-priorparam}. Variances for the transformed asymptote and spline coefficients are constrained to decrease with decreasing hierarchical level (i.e., the variability across countries within a region is assumed to be smaller than the variability of regions in the world). 

\begin{table}[htbp!]
\begin{tabular}{|l|l|l|l|l|c|} \hline
Group & Indicator & Parameter & Setting & Value \\ \hline
married & Demand satisfied & $\beta_{c,k}$ & Lower bound & 0.01 \\
married & Demand satisfied & $\beta_{c,k}$ & Upper bound & 0.5 \\
married & Demand satisfied & $\lambda_c$ & Lower bound & 0.1 \\
married & Demand & $\beta_{c,k}$ & Lower bound & 0.01 \\
married & Demand & $\beta_{c,k}$ & Upper bound & 0.5 \\
married & Demand & $\lambda_c$ & Lower bound & 0.1 \\
unmarried & Demand satisfied & $\beta_{c,k}$ & Lower bound & 0.01 \\
unmarried & Demand satisfied & $\beta_{c,k}$ & Upper bound & 0.5 \\
unmarried & Demand satisfied & $\lambda_c$ & Lower bound & 0.05 \\
unmarried & Demand & $\beta_{c,k}$ & Lower bound & 0.01 \\
unmarried & Demand & $\beta_{c,k}$ & Upper bound & 0.5 \\
unmarried & Demand & $\lambda_c$ & Lower bound & 0.01 \\\hline
\end{tabular}
\caption{Bounds used for transition model parameters.}\label{tb-bounds}
\end{table}

\begin{table}[htbp!]
\begin{tabular}{|@{}|l|l|l|p{6cm}|}\hline
Group & Indicator & Parameter & Hierarchical groupings \\\hline
married & Demand satisfied & $\beta_{c,k}$ & cluster, subcluster, iso \\
married & Demand satisfied & $\Omega_c$ & cluster, subcluster, iso \\
married & Demand satisfied & $\lambda_c$ & cluster, iso \\
married & Demand & $\beta_{c,k}$ & cluster, subcluster, iso \\
married & Demand & $\Omega_c$ & cluster, subcluster, iso \\
married & Demand & $\lambda_c$ & cluster, iso \\
unmarried & Demand satisfied & $\beta_{c,k}$ & clusterandsa0\_unmarried, iso \\
unmarried & Demand satisfied & $\Omega_c$ & regional2\_unmarried, iso \\
unmarried & Demand satisfied & $\lambda_c$ & clusterandsa0\_unmarried, iso \\
unmarried & Demand & $\beta_{c,k}$ & clusterandsa0\_unmarried, iso \\
unmarried & Demand & $\Omega_c$ & is\_unmarried\_sexual\_activity,
hier\_regional\_unmarried, iso \\
unmarried & Demand & $\lambda_c$ & is\_unmarried\_sexual\_activity,
hier\_regional\_unmarried, iso \\ \hline
\end{tabular}
\caption{Hierarchical groupings used for transition model parameters.}\label{tb-groupings}
\end{table}

\begin{table}[htbp!]
\begin{tabular}{|@{}|l|l|l|l|l@{}|}\hline
Group & Indicator & Parameter & Setting & Value \\\hline
married & Demand satisfied & $\beta_{c,k}$ & Prior SD $s_0$ & 3 \\
married & Demand satisfied & $\beta_{c,k}$ & Prior SD $w_0$ & 2 \\
married & Demand satisfied & $\beta_{c,k}$ & Prior mean $m_0$ & -1 \\
married & Demand satisfied & $\Omega_c$ & Prior SD $s_0$ & 3 \\
married & Demand satisfied & $\Omega_c$ & Prior SD $w_0$ & 2 \\
married & Demand satisfied & $\Omega_c$ & Prior mean $m_0$ & 0 \\
married & Demand satisfied & $\lambda_c$ & Prior SD $s_0$ & 3 \\
married & Demand satisfied & $\lambda_c$ & Prior SD $w_0$ & 2 \\
married & Demand satisfied & $\lambda_c$ & Prior mean $m_0$ & 0 \\
married & Demand & $\beta_{c,k}$ & Prior SD $s_0$ & 3 \\
married & Demand & $\beta_{c,k}$ & Prior SD $w_0$ & 2 \\
married & Demand & $\beta_{c,k}$ & Prior mean $m_0$ & -1 \\
married & Demand & $\Omega_c$ & Prior SD $s_0$ & 3 \\
married & Demand & $\Omega_c$ & Prior SD $w_0$ & 2 \\
married & Demand & $\Omega_c$ & Prior mean $m_0$ & 0 \\
married & Demand & $\lambda_c$ & Prior SD $s_0$ & 3 \\
married & Demand & $\lambda_c$ & Prior SD $w_0$ & 2 \\
married & Demand & $\lambda_c$ & Prior mean $m_0$ & 0 \\
unmarried & Demand satisfied & $\beta_{c,k}$ & Prior SD $s_0$ & 3 \\
unmarried & Demand satisfied & $\beta_{c,k}$ & Prior SD $w_0$ & 2 \\
unmarried & Demand satisfied & $\beta_{c,k}$ & Prior mean $m_0$ & -1 \\
unmarried & Demand satisfied & $\Omega_c$ & Prior SD $s_0$ & 3 \\
unmarried & Demand satisfied & $\Omega_c$ & Prior SD $w_0$ & 2 \\
unmarried & Demand satisfied & $\Omega_c$ & Prior mean $m_0$ & 0 \\
unmarried & Demand satisfied & $\lambda_c$ & Prior SD $s_0$ & 3 \\
unmarried & Demand satisfied & $\lambda_c$ & Prior SD $w_0$ & 2 \\
unmarried & Demand satisfied & $\lambda_c$ & Prior mean $m_0$ & 0 \\
unmarried & Demand & $\beta_{c,k}$ & Prior SD $s_0$ & 3 \\
unmarried & Demand & $\beta_{c,k}$ & Prior SD $w_0$ & 2 \\
unmarried & Demand & $\beta_{c,k}$ & Prior mean $m_0$ & -1 \\
unmarried & Demand & $\Omega_c$ & Prior SD $s_0$ & 3 \\
unmarried & Demand & $\Omega_c$ & Prior SD $w_0$ & 2 \\
unmarried & Demand & $\Omega_c$ & Prior mean $m_0$ & 0 \\
unmarried & Demand & $\lambda_c$ & Prior SD $s_0$ & 3 \\
unmarried & Demand & $\lambda_c$ & Prior SD $w_0$ & 2 \\
unmarried & Demand & $\lambda_c$ & Prior mean $m_0$ & 0 \\
\hline
\end{tabular}
\caption{Prior parameters used for transition model parameters.}\label{tb-priorparam}
\end{table}

\clearpage

\subsubsection{Data model specification for modern use and unmet need for modern methods}\label{app-datamodel}

We consider two proportions: mCPR and the ratio of unmet need for modern methods out of non-modern-users. For a single observed proportion $y_i$, the likelihood  is given by 
\begin{align}
\mathrm{logit}(y_i)|\phi_{c[i],t[i]}, \sigma_i &\sim& N(\mathrm{logit}(\phi_{c[i], t[i]}), \sigma_i^2), \label{eq-dm_app1}\\
\sigma_i^2 &=& s_i^2 + \mathbb{I}(d[i] \in \mathcal{D}) \tilde{\sigma}_{d[i]}^{(source)2} + \mathbb{I}(i \in \mathcal{C})\sigma^{(\charic)2} + \mathbb{I}(i \in \mathcal{I})\sigma_i^{(\outlier)2}, \label{eq-dm_app2}
\end{align}
where $\sigma_i^{(\outlier)}= \sqrt{\frac{\tau^2 \vartheta^{2} \gamma_{i}^{2}}{\vartheta^{2} + \tau^2 \gamma_{i}^{2}}}$, with local scale parameter $\gamma_{i}$ assigned a heavy-tailed half-Cauchy prior:
$$\gamma_{i} \sim C^+(0, 1),$$
to allow individual errors to escape from shrinkage. 

For PMA data series for a country-marital group, we use the multivariate normal specification:
\begin{align}
    \bm{z}|\bm{\phi}, \bm{\Sigma} \sim N(\mathrm{logit}(\bm{\phi}), \bm{\Sigma}), 
\end{align}
where $\bm{z}$ refers to the vector of logit-transformed observations, $\mathrm{logit}(\bm{\phi})$ refers to the respective mean vector of $\bm{z}$, and $\bm{\Sigma}$ to its associated covariance matrix. We set $\Sigma_{j,l} = Cov(z_j, z_l) = \sigma_j \cdot \sigma_l \cdot \rho_{\PMA}^{|t[j] - t[l]|}$, where  $\rho_{\PMA}$ refers to the autocorrelation in PMA data errors. 

Priors on the data quality parameters are the same across proportions and marital groups. The priors are as follows
 \begin{align*}
 \tilde{\sigma}_d^{(source)} &\sim N^+(0, 0.5^2),\\
 \sigma^{(\charic)} &\sim N^+(0, 0.5^2),\\
    \tau &\sim C^+(0, 0.04^2),\\
   \vartheta &\sim N^+(0,1),\\
  \rho_{\PMA} &\sim U(0,1).
   \end{align*}

\subsubsection{Model fitting for estimation of modern use and unmet need for modern methods}\label{sec-details1a1b}
Global model fitting refers to estimating all model parameters of process and survey data model combined, separately by marital group. To estimate  mCPR and unmet need for modern methods, we use two steps. In the first step, we estimate global and world region parameters governing long-term trends in demand and demand satisfied. In the second step, parameters governing  short-term fluctuations and data quality are estimated. 

\paragraph{Step 1A: Estimate long term trends}
To estimate long-term trends, we first fit a global model with transition functions without the AR(1) deviation terms. This step results in the estimation of transition functions that represent smooth long-term trends, for geographic groups at different levels of the hierarchies used.  When fitting the model, the data model is as described in the previous section. Non-sampling errors are included for all source types, including DHS data. The data classification algorithm is used to assess which observations may be possibly outlying. 

\textit{Step 1A model equations:} For the process model, the set up in Equations \eqref{eq-process_app}-\eqref{eq-process_app2} are used with $\varepsilon_{c,t} = 0$ for all $c, t$. For the data model, Equations~\eqref{eq-dm_app1} and \eqref{eq-dm_app2} are used with source-type variance $\tilde{\sigma}_{d}$ included for all source types $d$, including DHS. 

\paragraph{Step 1B: Estimate parameters governing country-specific variations in transition functions, short-term fluctuations, and data quality}
In the second step of model fitting, the full process model is fitted (including AR(1) smoothing terms), to estimate all parameters governing country-specific variations in process functions, as well as data quality. The fitting is informed by the estimates of the long term trends from step 1A. 

\textit{Step 1B model equations:} The process model is defined by Equations \eqref{eq-process_app}-\eqref{eq-process_app3}, using estimated transition functions at the level above the country level as fixed inputs in model fitting. Specifically, in the hierarchical model for transition model country parameters $\lambda_c, \beta_c, \Omega_c$, mean parameters are fixed at point estimates from step 1A: 
\begin{align*}
\theta_\gamma^{(l)} \mid \sigma^{(l)}_\theta &\sim N\left(\hat{\theta}^{(l-1)}_{\gamma^\prime}, \sigma^2_{l, \theta}\right), \text{ for $l$ referring to the country level},
\end{align*}
where $\hat{\theta}^{(l-1)}_{\gamma^\prime}$ refers to the point estimate at one level higher in the hierarchy. Continuing with the example of parameter $\Omega_c$, the density used in step 1B is as follows: 
\begin{align*}
  \Omega_c^{(\mathrm{country})} \mid \sigma_{ \mathrm{country},\Omega}^2 &\sim N\left(\hat{\Omega}_{s[c]}^{(\mathrm{subcluster})}, \sigma_{ \mathrm{country},\Omega}^2\right), 
\end{align*} 
where $\hat{\Omega}_{s[c]}^{(\mathrm{subcluster})}$ was estimated in step 1A. 

In global model fitting to survey data, the data model as given by Equations~\eqref{eq-dm_app1} and \eqref{eq-dm_app2} is used. In step 1B, DHS data are not assigned non-sampling errors; source-type variance $\tilde{\sigma}_{d}$ is included for all source types $d$ except for DHS. The set of possibly outlying points, denoted by $\mathcal{I}$, initially comes from step 1A and is then expanded with observations identified as outliers relative to the long-term trend estimated in step 1A. Concretely, for each observation we compute the difference from the step‑1A long‑term trend estimate and add to $\mathcal{I}$ those observations with absolute errors  in the top 10\%. The rationale is that for FP indicators, extreme fluctuations are more likely to be data errors. By flagging observations that are extreme compared with the smooth long‑term trend, we allow these points to be treated as possibly outlying in the final model fitting (step 1B) so they do not affect the global estimates of parameters governing short-term fluctuations.

\subsection{Modeling traditional use}\label{sec-app-trad}
\subsubsection{Process model for traditional use}
We denote the ratio of traditional use to unmet need for modern methods by $\eta_3$, i.e., $\eta_3 = \phi_2/(\phi_2 + \phi_3)$. We assume an integrated autoregressive process of order 1 (ARI(1)) for this ratio at the country level, using an inverse-probit transform:
\begin{align} 
    \Phi^{-1}(\eta_{3,c,t}) = \Phi^{-1}(\eta_{3,c,t-1})  + \varepsilon_{3,c,t},
\end{align}
where $\varepsilon_{3,c,t} \sim AR(1)$ defined as:
$$\varepsilon_{3,c,t}|\varepsilon_{3,c,t-1}, \rho_{3, \varepsilon}, \sigma_{3,\varepsilon} \sim N(\rho_{3, \varepsilon}\cdot \varepsilon_{3,c,t-1}, \sigma_{3,\varepsilon}^2).$$
The specification for $\eta_{3,c,t}$ is completed through defining its level in a reference year $t^*$:
\begin{align} 
     \Phi^{-1}(\eta_{3,c,t^*}) = \Omega_{3,c}.
\end{align}
The level is assigned a hierarchical prior, following the same set up as used for the $\Omega_c$'s in the transition models for demand and demand satisfied. The groupings and prior parameters are given in Tables~\ref{tb-groupingstrad} and \ref{tb-priorparamtrad}.

\begin{table}[htbp!]
\begin{tabular}{|@{}|l|l|l|p{6cm}|}\hline
Group & Indicator & Parameter & Hierarchical terms \\\hline
married & Trad/Unmet & $\Omega_c$ & cluster, subcluster, iso \\
unmarried & Trad/Unmet & $\Omega_c$ & clusterandsa0\_unmarried, iso \\\hline
\end{tabular}
\caption{Hierarchical groupings used for transformed level of the ratio of traditional use to unmet need. }\label{tb-groupingstrad}
\end{table}

\begin{table}[htbp!]
\begin{tabular}{|@{}|l|l|l|l|l@{}|}\hline
Group & Indicator & Parameter & Setting & Value \\\hline
married & Trad/Unmet & $\Omega_c$ & Prior SD $s_0$ & 3 \\
married & Trad/Unmet & $\Omega_c$ & Prior SD $w_0$ & 2 \\
married & Trad/Unmet & $\Omega_c$ & Prior mean $m_0$ & 0 \\
unmarried & Trad/Unmet & $\Omega_c$ & Prior SD $s_0$ & 3 \\
unmarried & Trad/Unmet & $\Omega_c$ & Prior SD $w_0$ & 2 \\
unmarried & Trad/Unmet & $\Omega_c$ & Prior mean $m_0$ & 0 \\\hline
\end{tabular}
\caption{Prior parameters used for for transformed level of the ratio of traditional use to unmet need.}\label{tb-priorparamtrad}
\end{table}

\subsubsection{Survey data model for traditional use}
We use an NOS data model on the ratio of (logit-transformed) traditional use over those with an unmet need $y_{i,2}/(y_{i,2} + y_{i,3})$. If data on unmet need for any method are not available but traditional use is, we use  $y_{i,2}/(1-y_{i,1})$.

\subsubsection{Model fitting for traditional use}
We estimate global model parameters related to traditional use by fitting the complete global model, including the process model for all indicators, to the survey data on all indicators. In this model fit, non-country-specific process model parameters related to the modeling of demand and demand satisfied, and data model parameters related to mCPR and unmet need for modern methods, are fixed at estimates from the global model fit to these data.

\subsection{Model validation: comparison between FPET2 and FPET1}\label{sec-app-validation}

We compare the validation results for the current version of FPET, referred to as FPET2, with those for the previously used version of FPET, referred to as FPET1. The results for FPET1 are in Tables~\ref{tab-val2020fpet1} and  \ref{tab-dhsfpet1}. 

The main takeaway is that FPET2 does a little better than FPET1 for mCPR for married women in terms of bias and absolute errors for predicting for the year 2024. This result is aligned with the motivation of updating FPET1, which was to improve married-women mCPR estimates. For predicting unmet need for modern methods for married women, and predicting mCPR and unmet need for unmarried women, we find that FPET1 and FPET2 performance is more comparable, without systematic improvements of one over the other.

\begin{table}[htbp]
\centering
\begin{tabular}{|l|l|rrrr|l|}
  \hline
indicator & married? & MedE & MedAE & prop $<1$\% & prop $<5$\% \\ 
  \hline
mCPR & all & -1.6 & 2.5 & 0.3 & 0.8 & fpet1 \\ 
  mCPR & all & -1.1 & 1.7 & 0.4 & 0.8 & fpet2 \\ \hline
  mCPR & no & -0.3 & 1.4 & 0.4 & 0.8 & fpet1 \\ 
  mCPR & no & -0.4 & 1.4 & 0.4 & 0.9 & fpet2 \\ \hline
  mCPR & yes & -2.6 & 4.0 & 0.2 & 0.6 & fpet1 \\ 
  mCPR & yes & -1.8 & 2.4 & 0.3 & 0.7 & fpet2 \\ \hline
  unmet modern & all & 0.7 & 0.9 & 0.5 & 1.0 & fpet1 \\ 
  unmet modern & all & 0.5 & 1.2 & 0.5 & 1.0 & fpet2 \\ \hline
  unmet modern & no & -0.0 & 0.8 & 0.7 & 1.0 & fpet1 \\ 
  unmet modern & no & -0.1 & 0.6 & 0.7 & 1.0 & fpet2 \\ \hline
  unmet modern & yes & 1.0 & 1.6 & 0.3 & 0.9 & fpet1 \\ 
  unmet modern & yes & 1.1 & 1.6 & 0.3 & 0.9 & fpet2 \\ 
   \hline
\end{tabular}
\caption{\textbf{Summary of prediction errors (in percentage) for the year 2024}. An error refers to the difference between the estimate for the indicator based on the full data set and the indicator based on the training data, for the year 2024 (5 year forecast horizon). A positive (negative) error indicates that the prediction based on data up to but excluding 2020 underpredicted (overpredicted) the target value.
Table columns: MedE = median error, MedAE = median absolute error, prop\_less2perc = proportion of errors less than 2 percentage point, prop $<$ 1 or 5 \% = proportion of errors less than 1 or 5 percentage points, fpet refers to the model version.}
    \label{tab-val2020fpet1}
\end{table}

\begin{table}[htbp]
    \centering
\begin{tabular}{|l|l|rrrr|l|}
  \hline
indicator & married? & MedE & MedAE & prop $<1$\% & prop $<5$\% & fpet\\ 
  \hline
mCPR & no & -0.7 & 2.3 & 0.1 & 0.8 & fpet1 \\ 
  mCPR & no & -0.5 & 2.4 & 0.1 & 0.8 & fpet2 \\ \hline
  mCPR & yes & -1.8 & 2.5 & 0.2 & 0.7 & fpet1 \\ 
  mCPR & yes & -1.7 & 2.1 & 0.2 & 0.7 & fpet2 \\ \hline
  unmet modern & no & -0.5 & 1.5 & 0.4 & 1.0 & fpet1 \\ 
  unmet modern & no & 0.1 & 0.9 & 0.6 & 1.0 & fpet2 \\ \hline
  unmet modern & yes & 0.9 & 2.0 & 0.3 & 0.9 & fpet1 \\ 
  unmet modern & yes & 1.3 & 2.1 & 0.3 & 0.9 & fpet2 \\ 
   \hline
\end{tabular}
    \caption{\textbf{Summary of prediction errors (in percentage)  for left-out DHSs in sub-Saharan Africa}. An error refers to the difference between a left-out DHS observation and the estimate for the indicator based on the validation data. A positive (negative) error indicates that the prediction based on data up to but excluding 2020 underpredicted (overpredicted) the target value.
Table columns: MedE = median error, MedAE = median absolute error,  prop $<$ 1 or 5 \% = proportion of errors less than 1 or 5 percentage points.}
    \label{tab-dhsfpet1}
\end{table}

\clearpage

\subsection{Results for all FP2030 countries}

\begin{figure}[h]   \centering
   \caption{\textbf{Data, estimates, and forecasts from 1970 until 2030 for FP indicators for married women for all FP2030 countries.} Graphs show mCPR, unmet need for modern methods, demand, and demand satisfied with modern methods. Survey data are shown in colored dots with 95\% confidence intervals that reflect total error variance. FPET estimates are shown in black, with grey shaded areas representing 90\% credible intervals.} \label{fig-app-results}
\end{figure}



\section*{Supplementary material}

## Algeria married

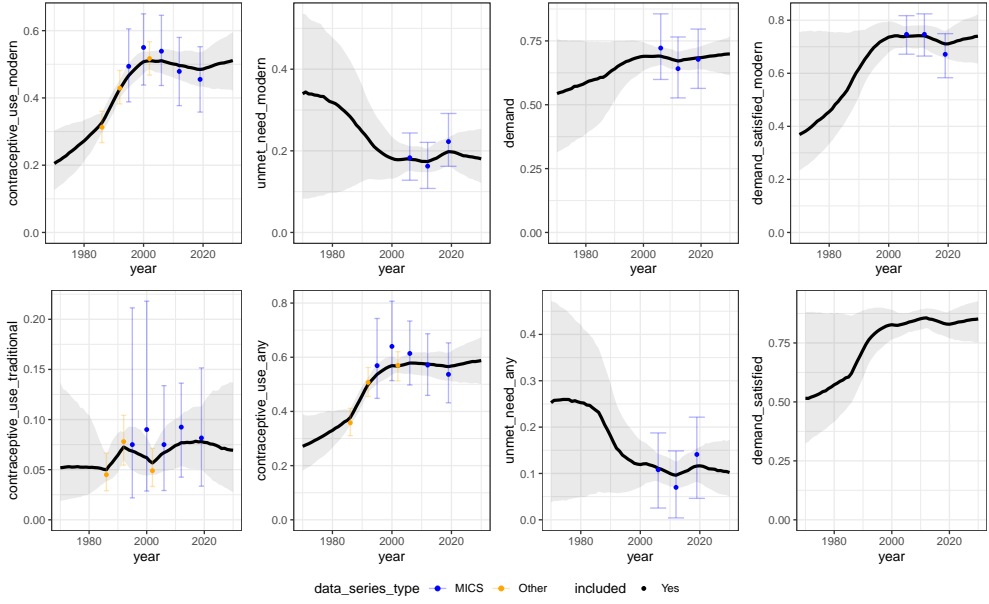

## Algeria unmarried

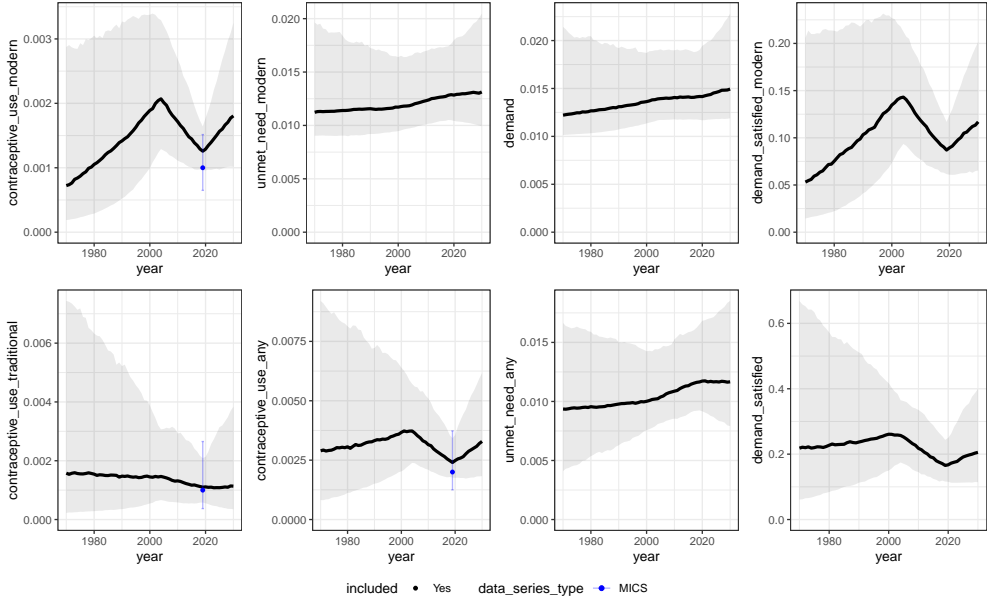

## Algeria all

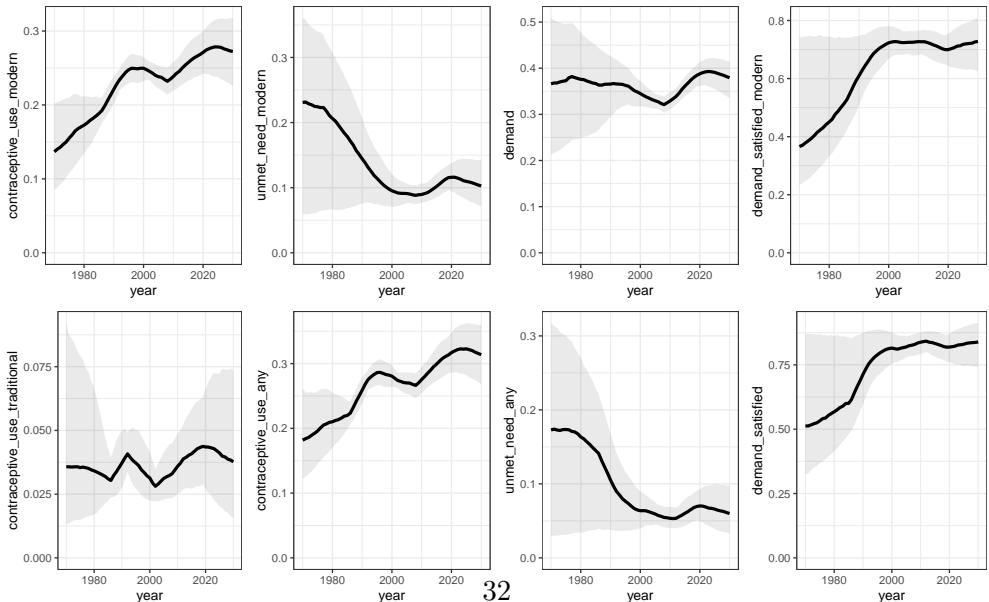

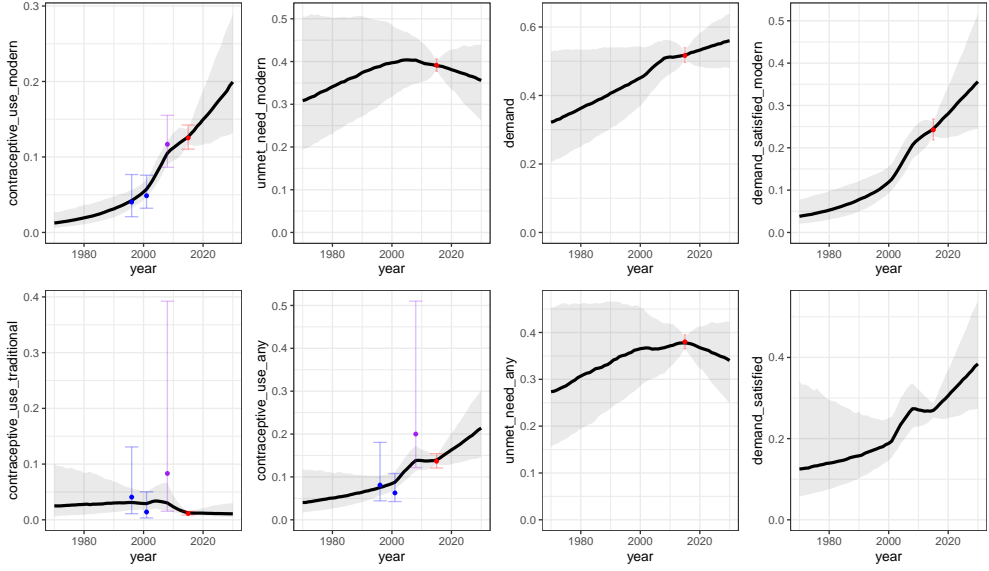

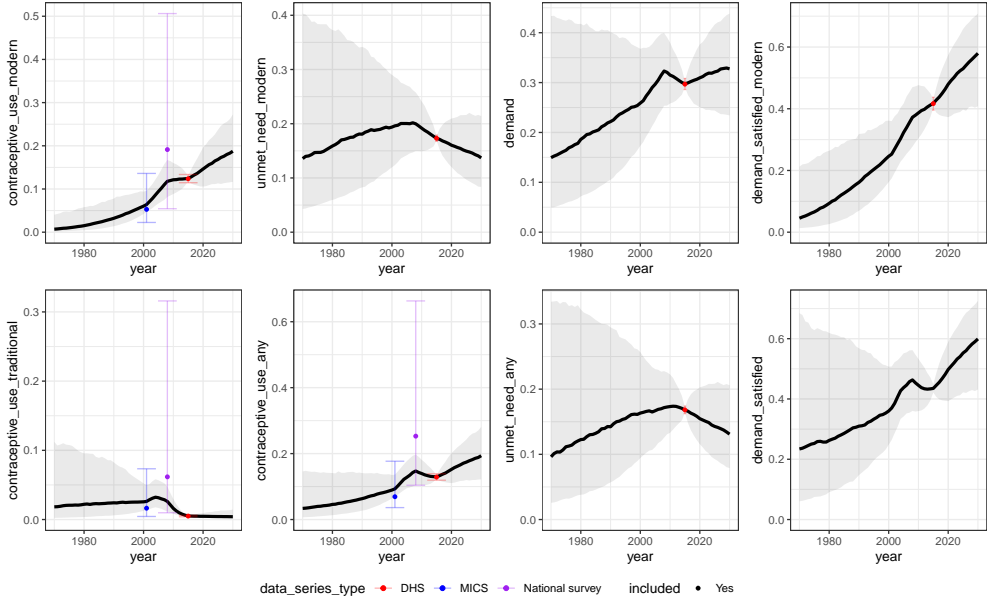

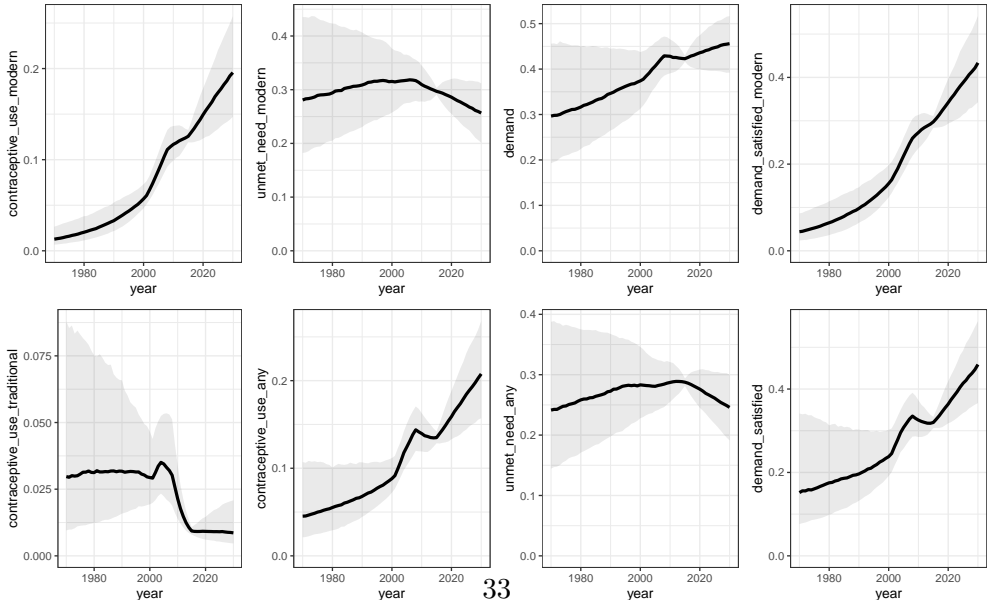

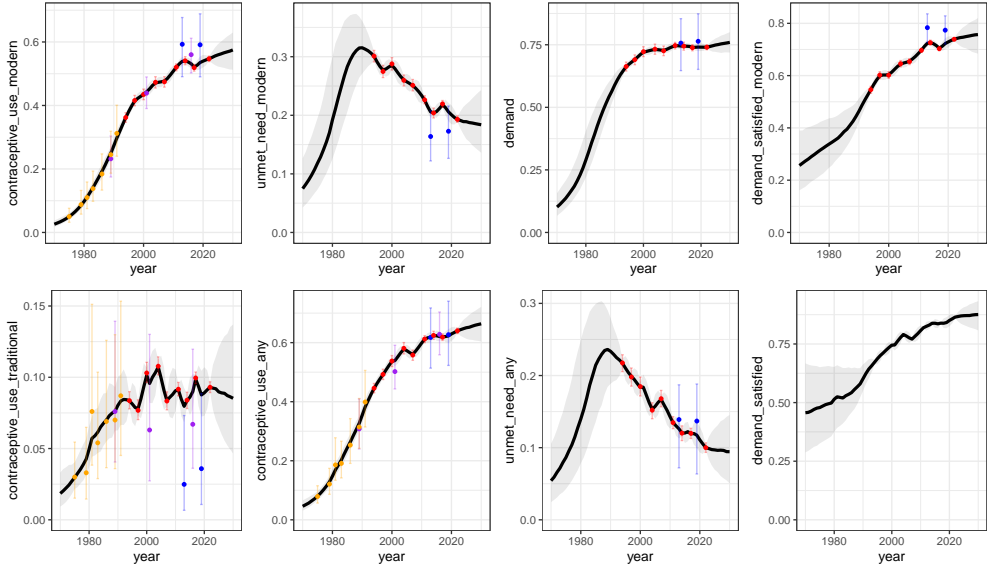

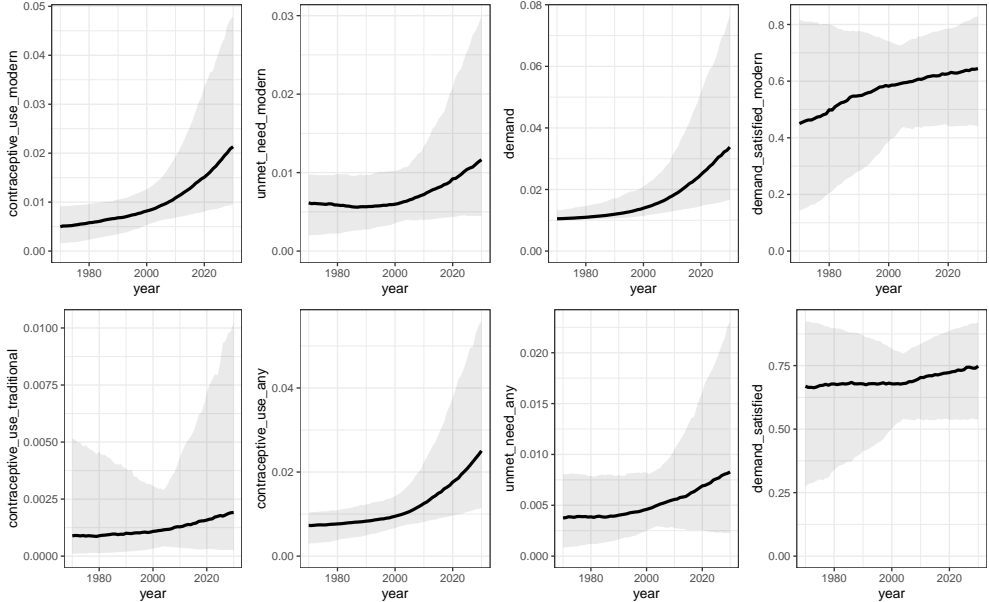

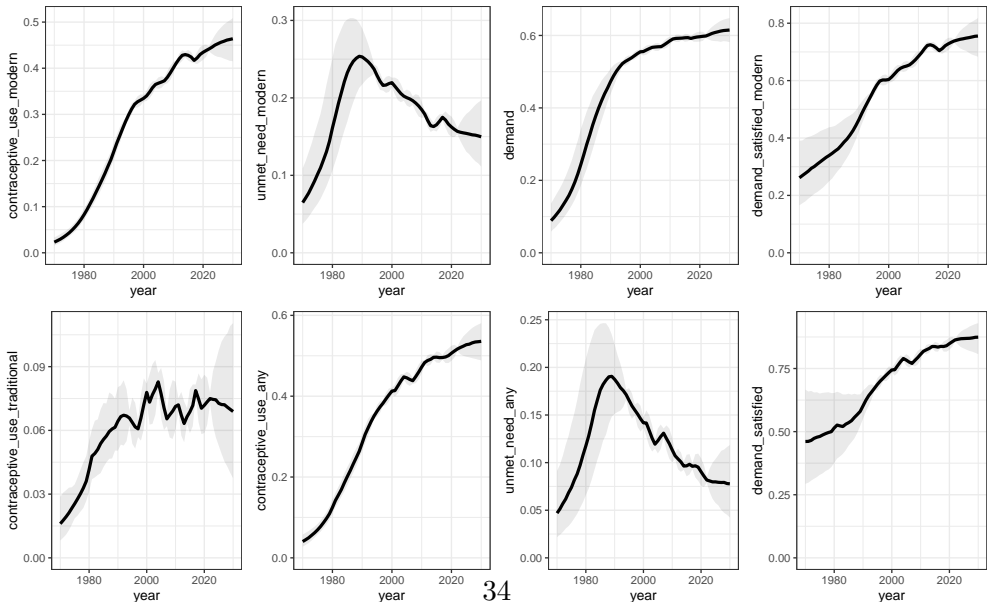

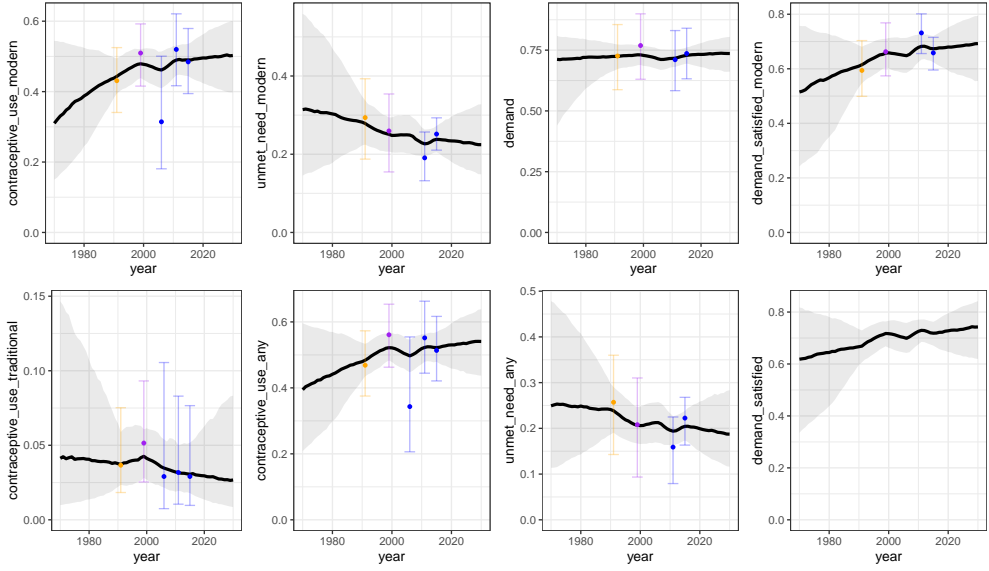

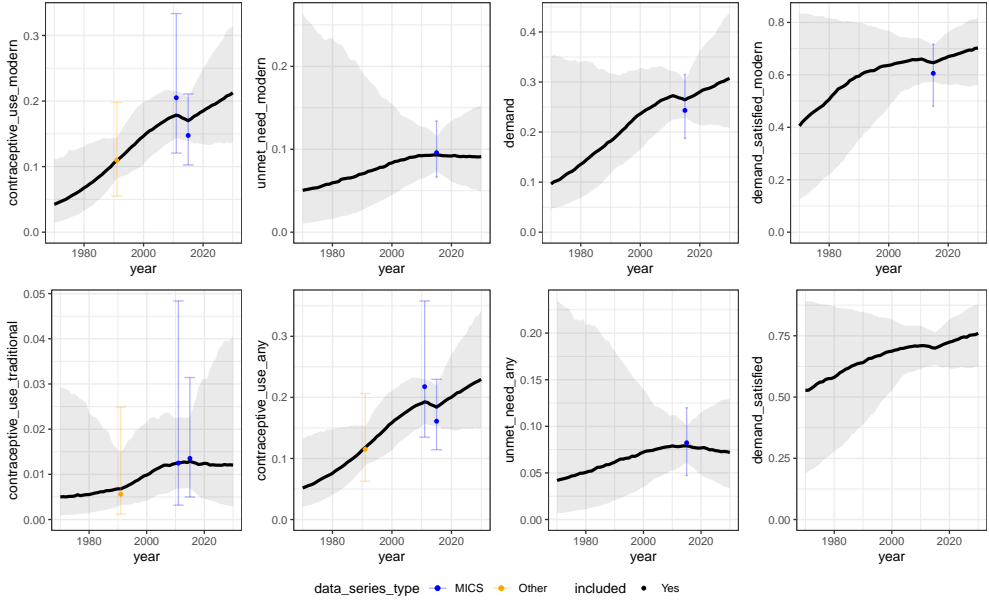

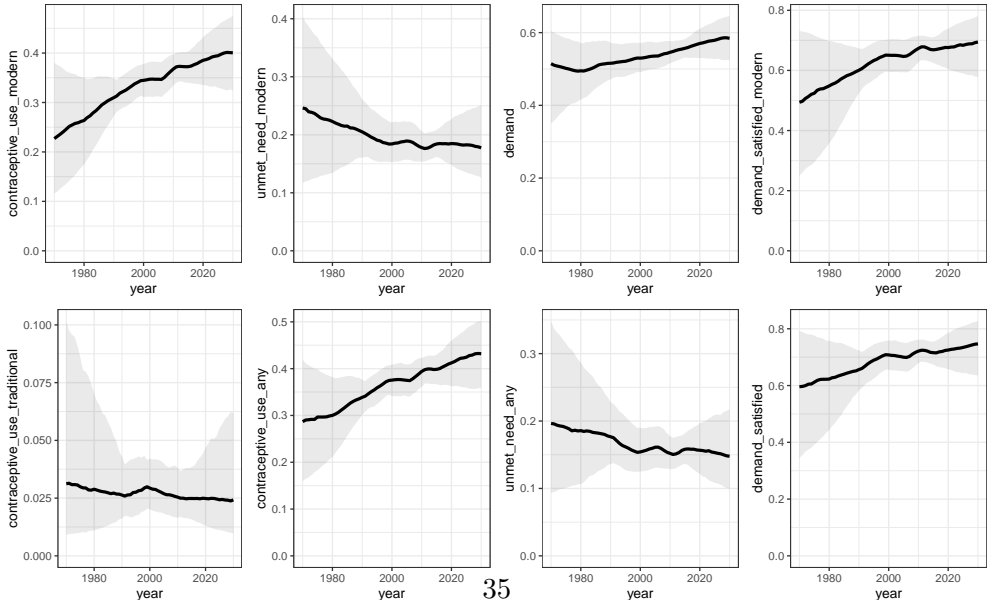

35

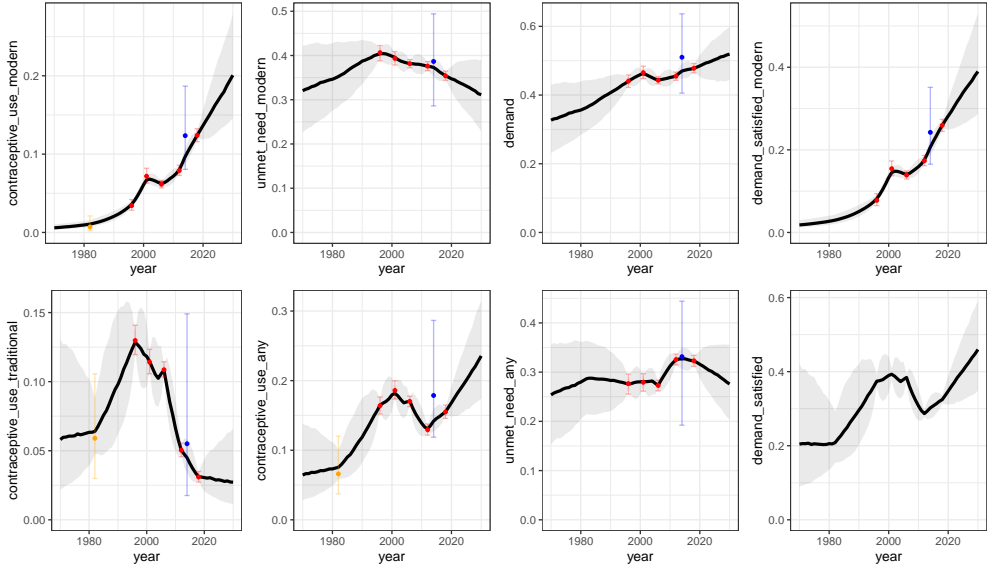

**Benin married**

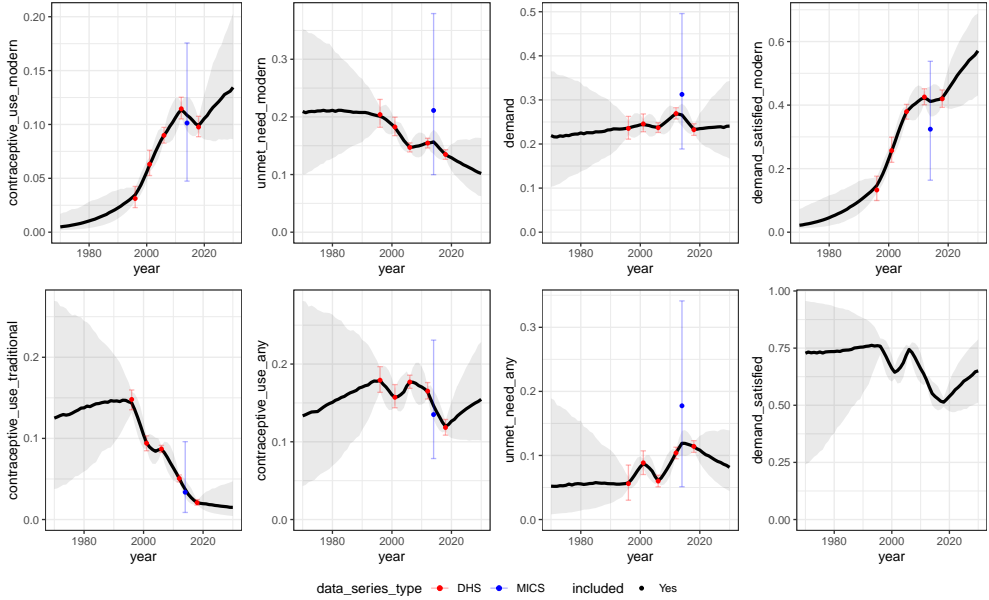

**Benin unmarried**

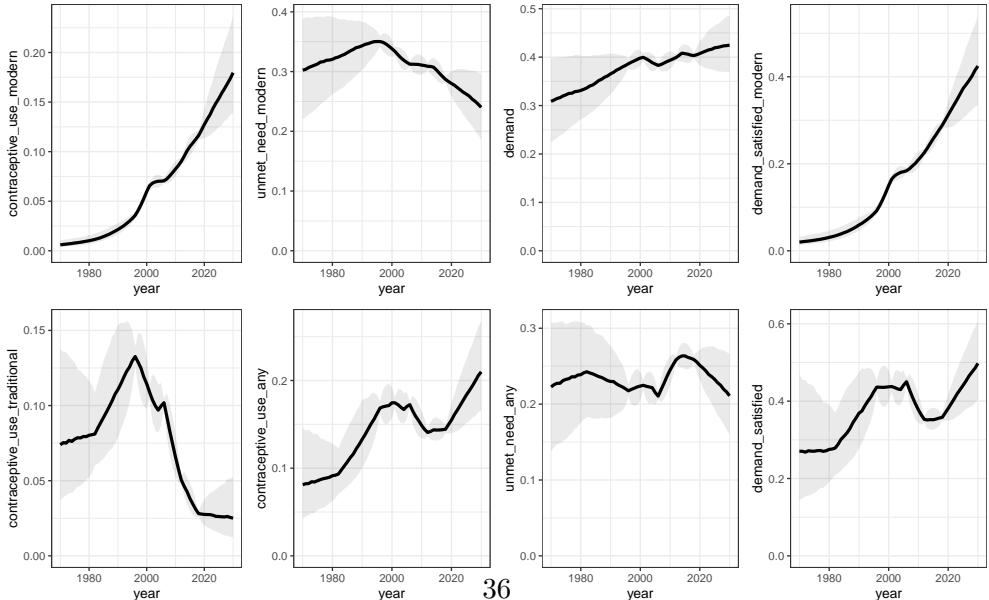

**Benin all**

36

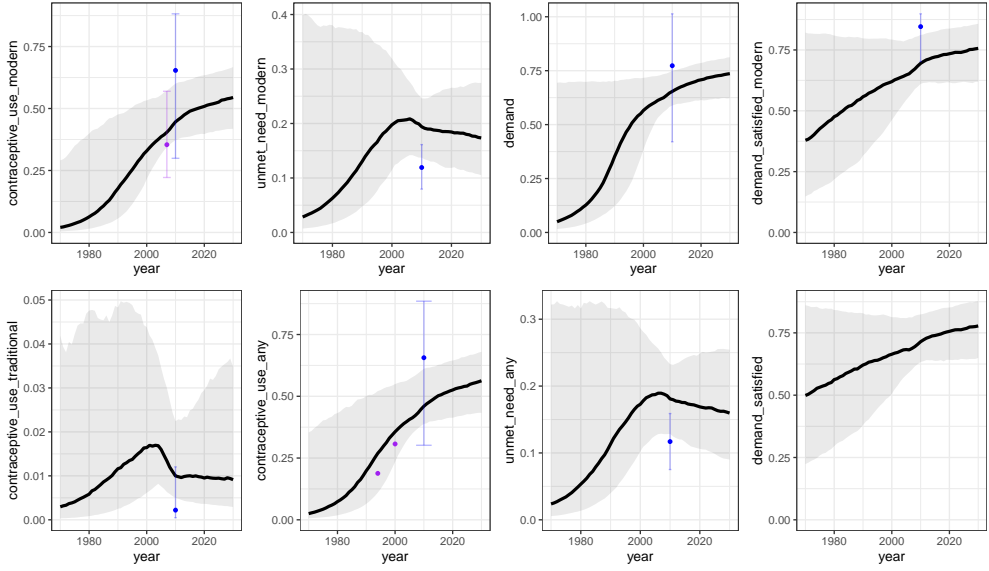

**Bhutan married**

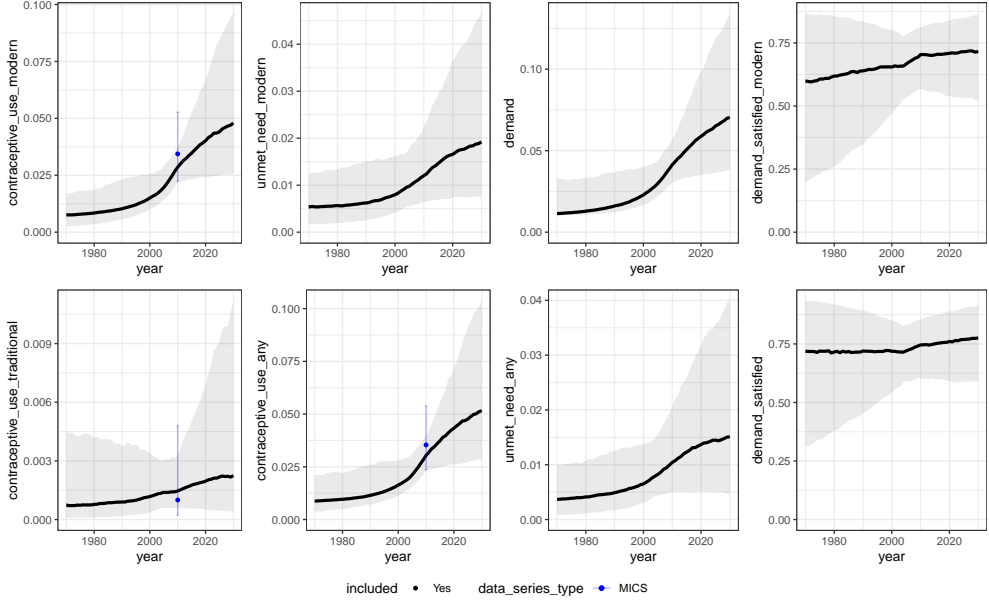

**Bhutan unmarried**

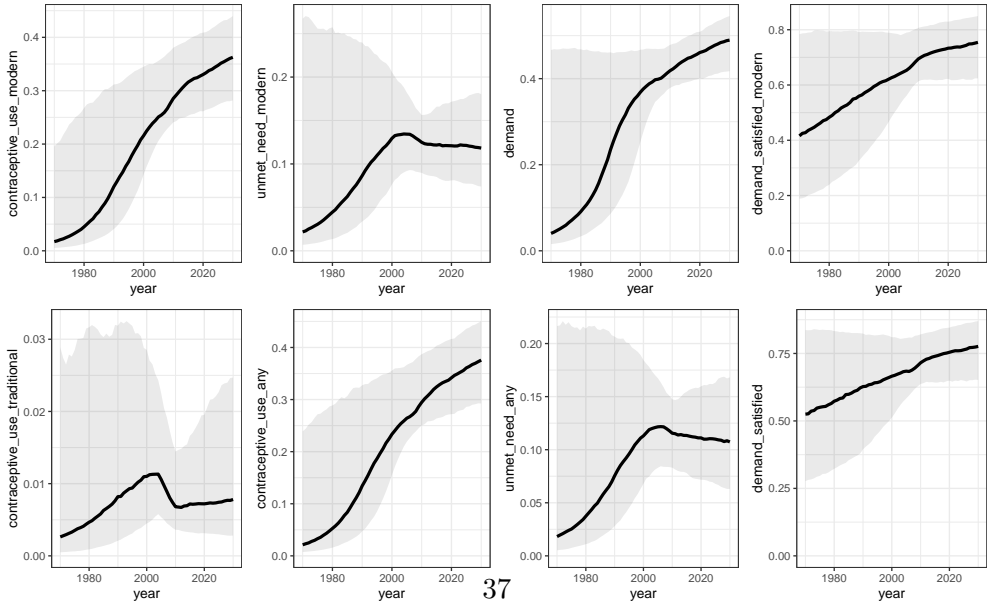

**Bhutan all**

37

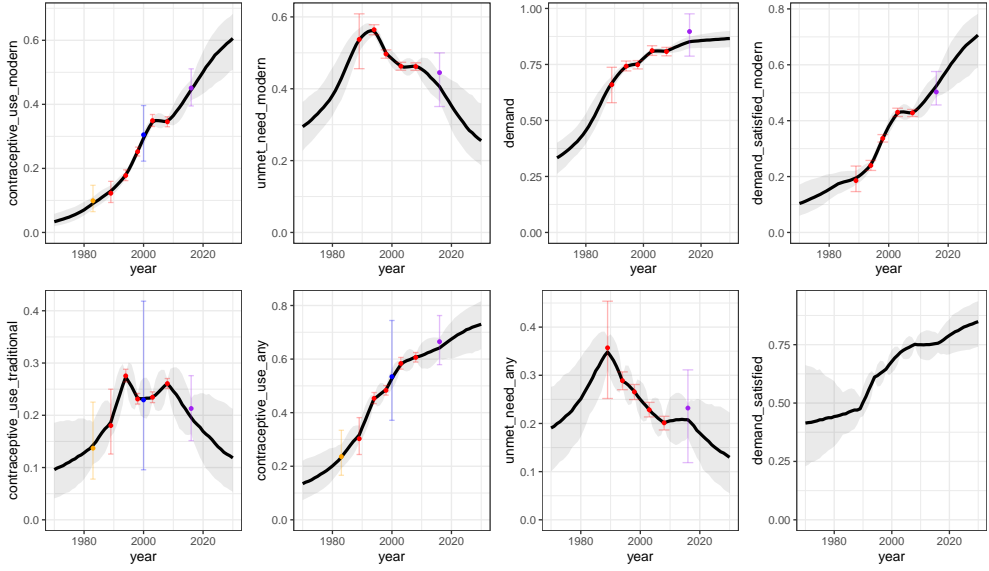

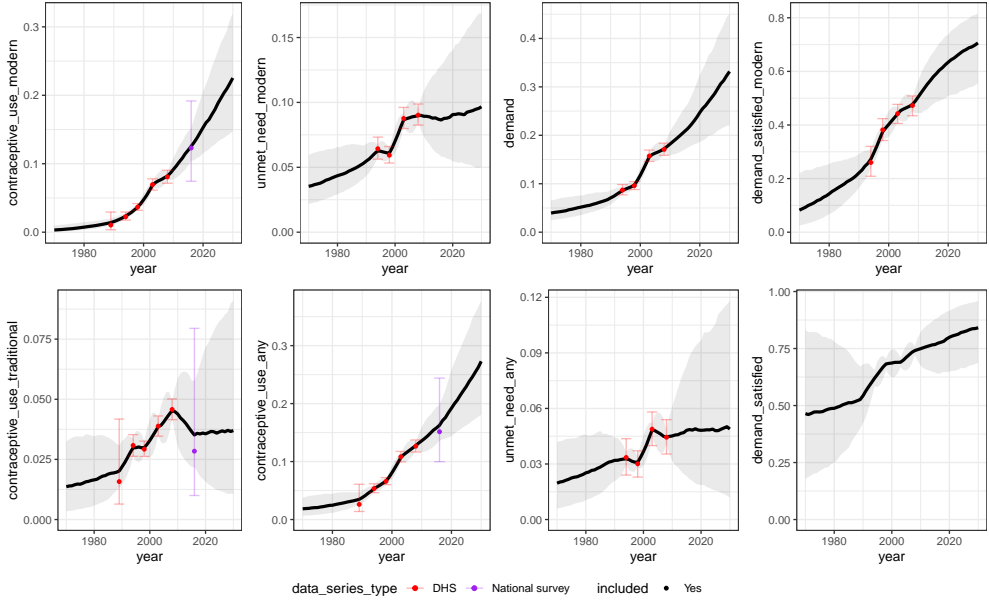

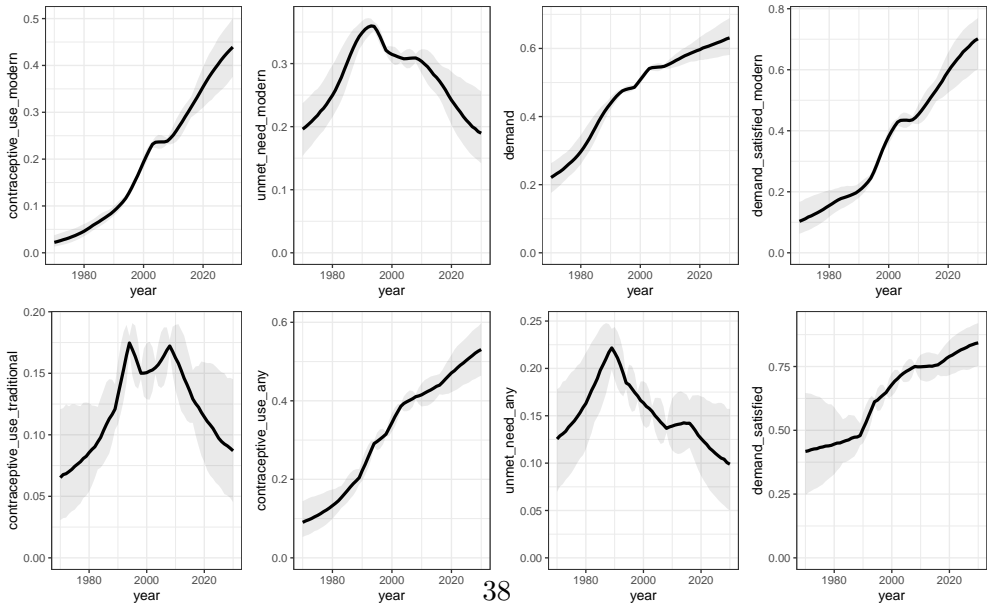

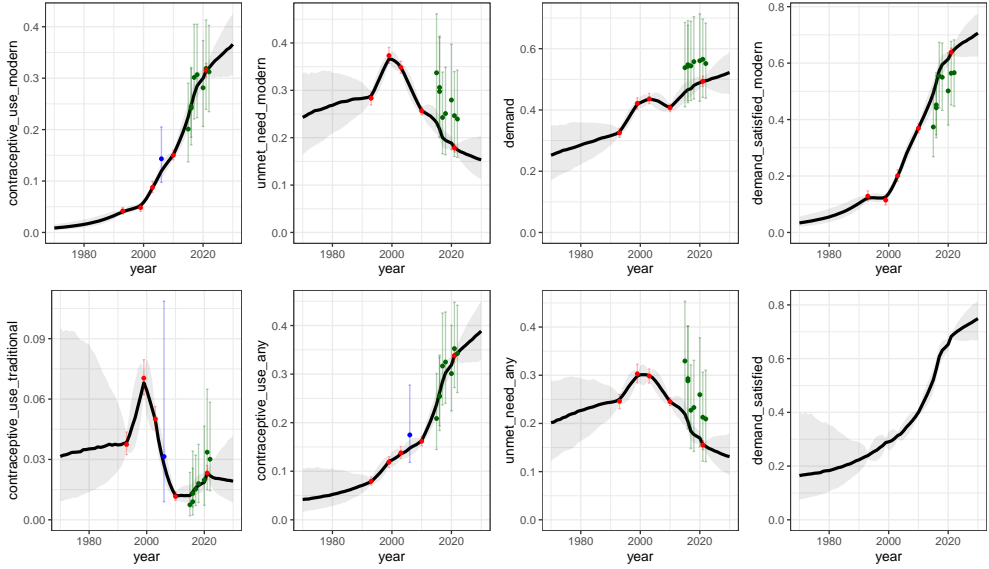

**Burkina_Faso married**

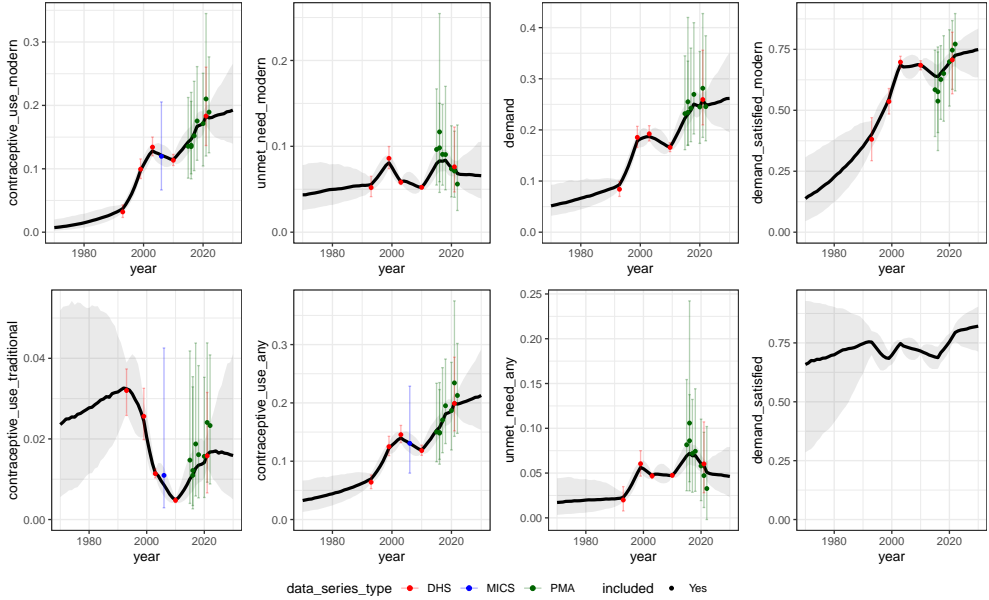

**Burkina_Faso unmarried**

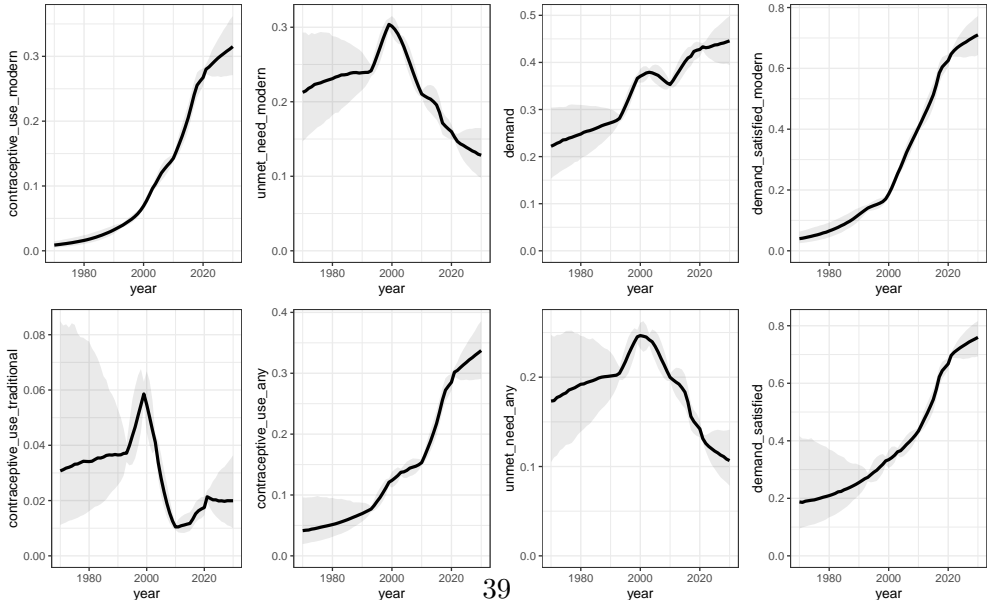

**Burkina_Faso all**

39

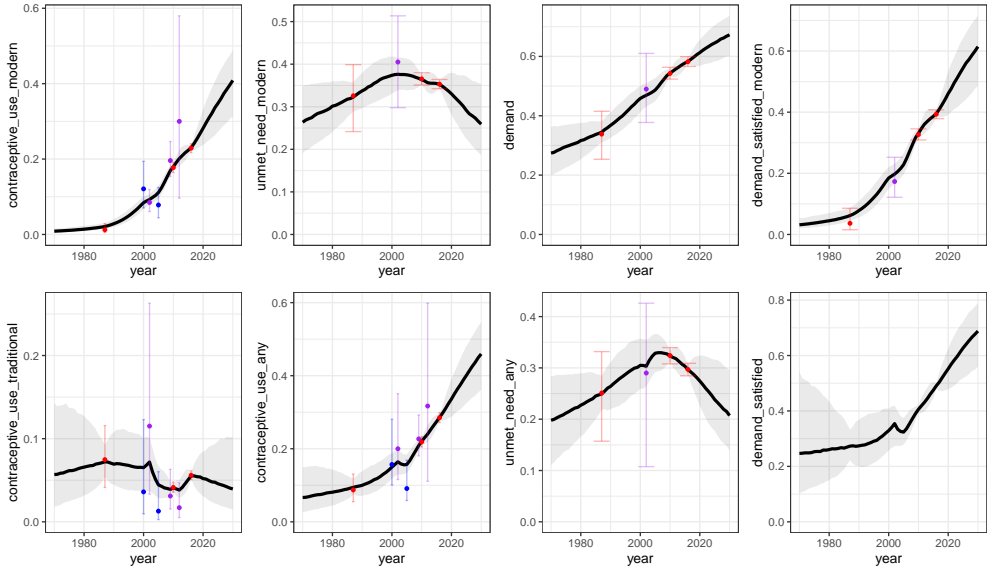

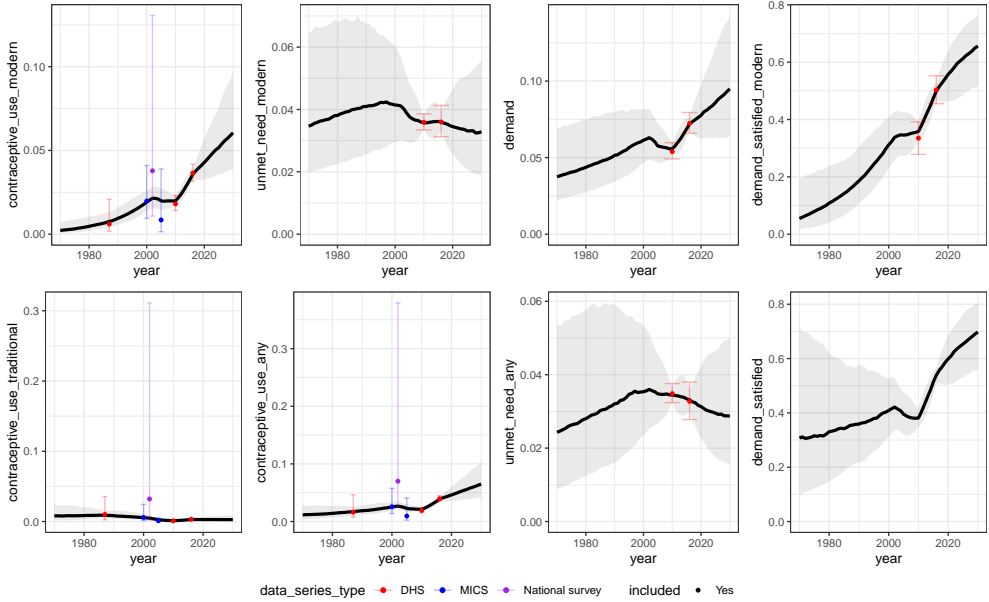

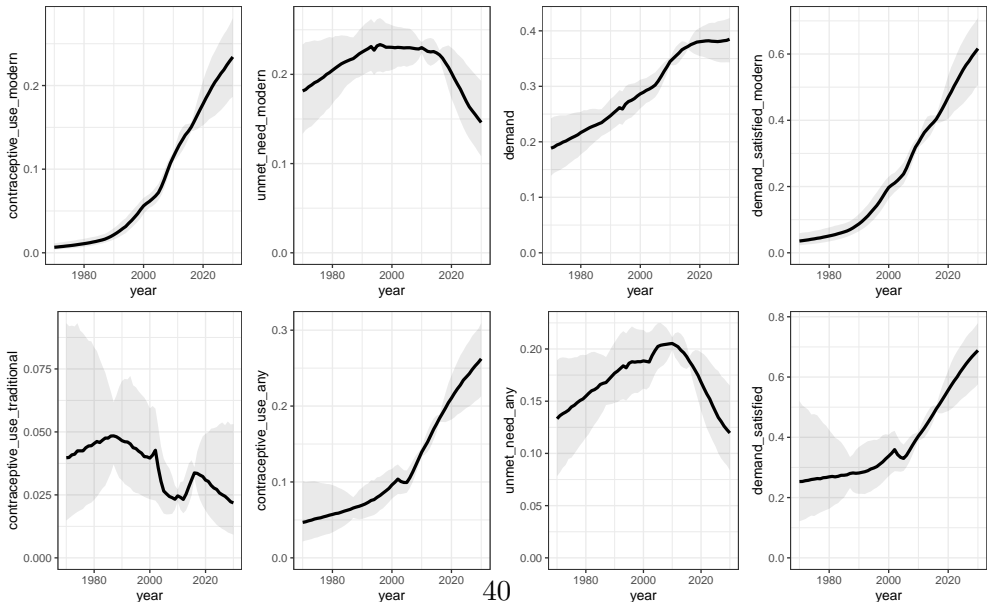

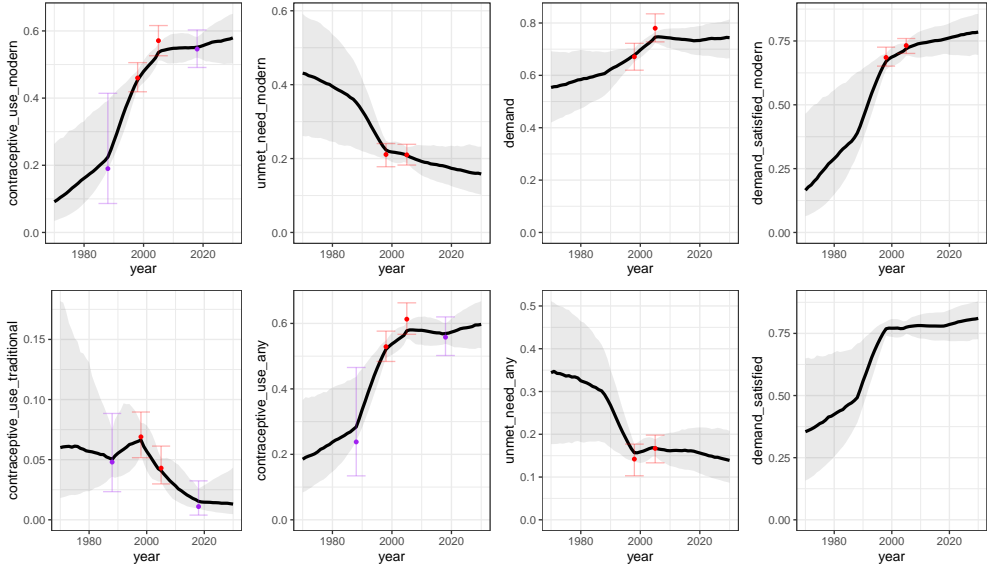

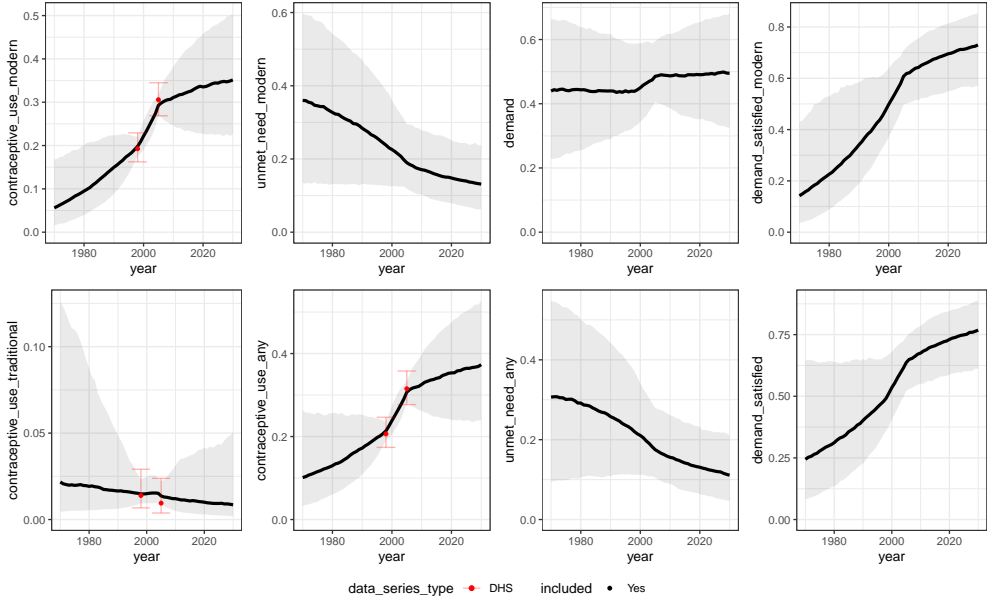

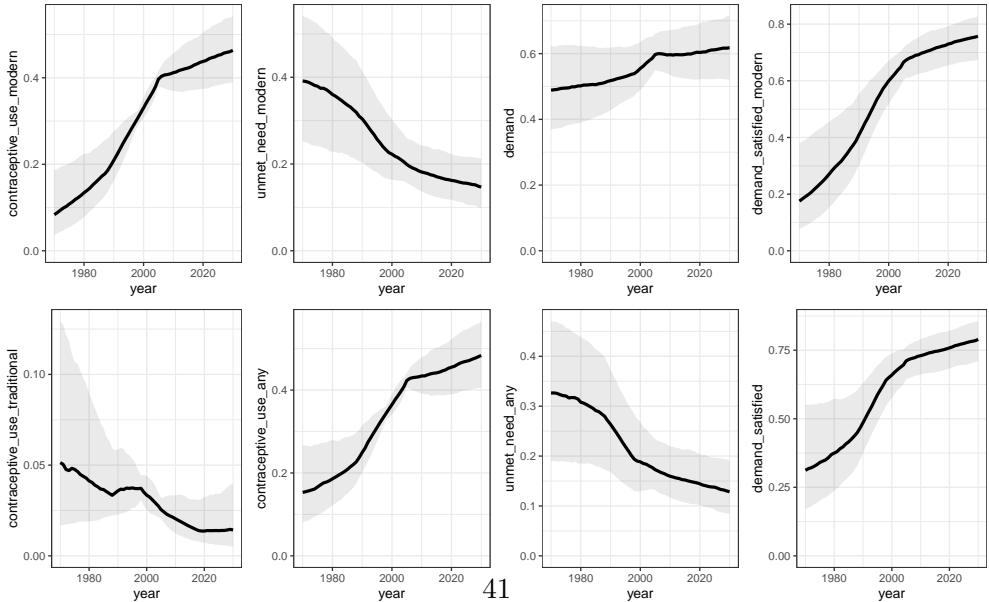

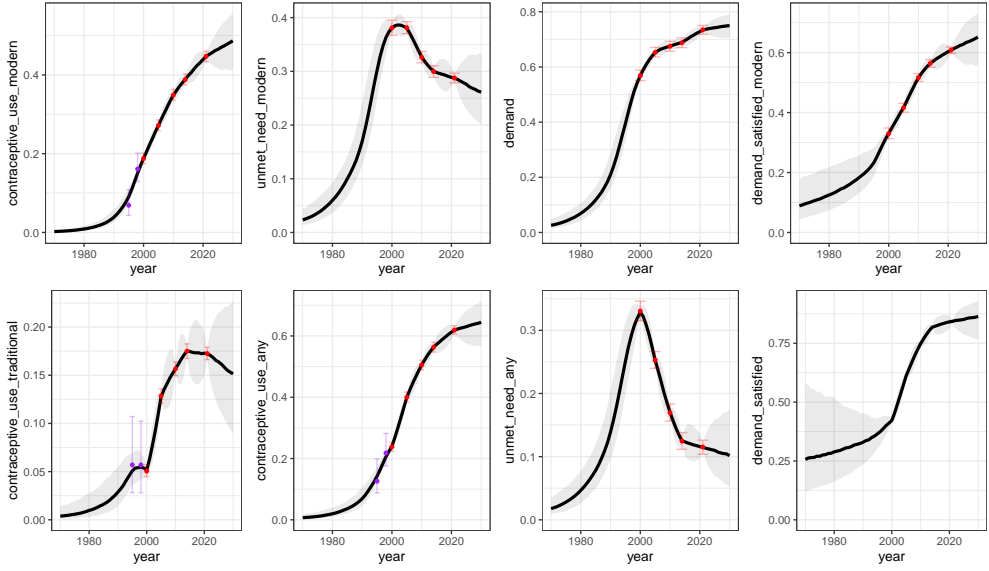

**Cambodia married**

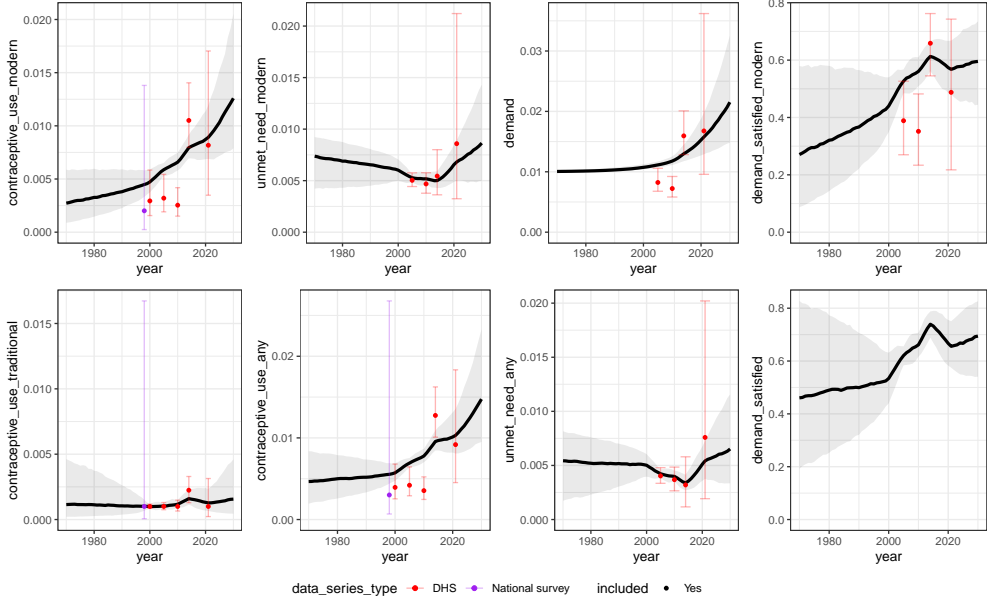

**Cambodia unmarried**

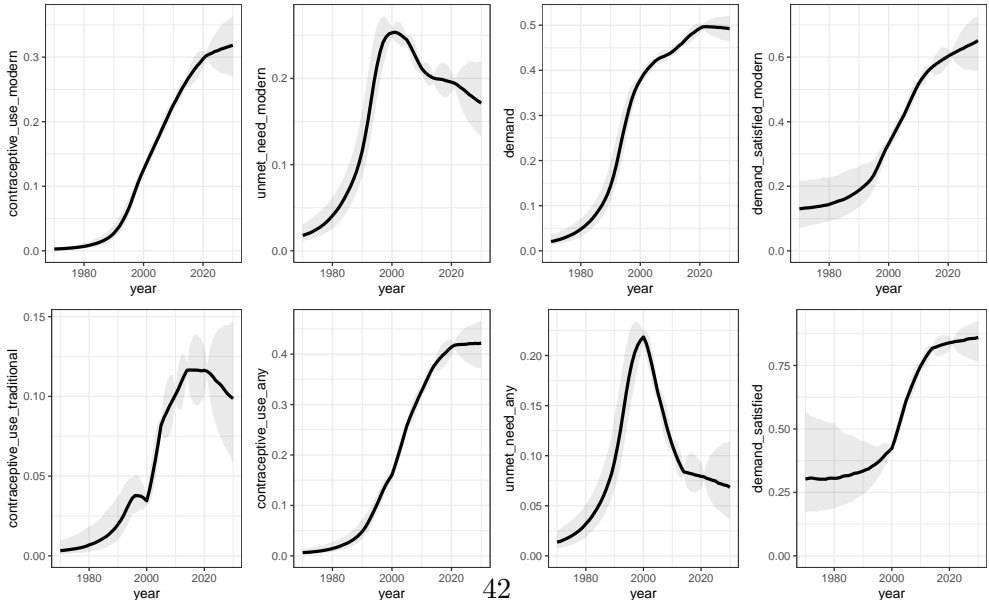

**Cambodia all**

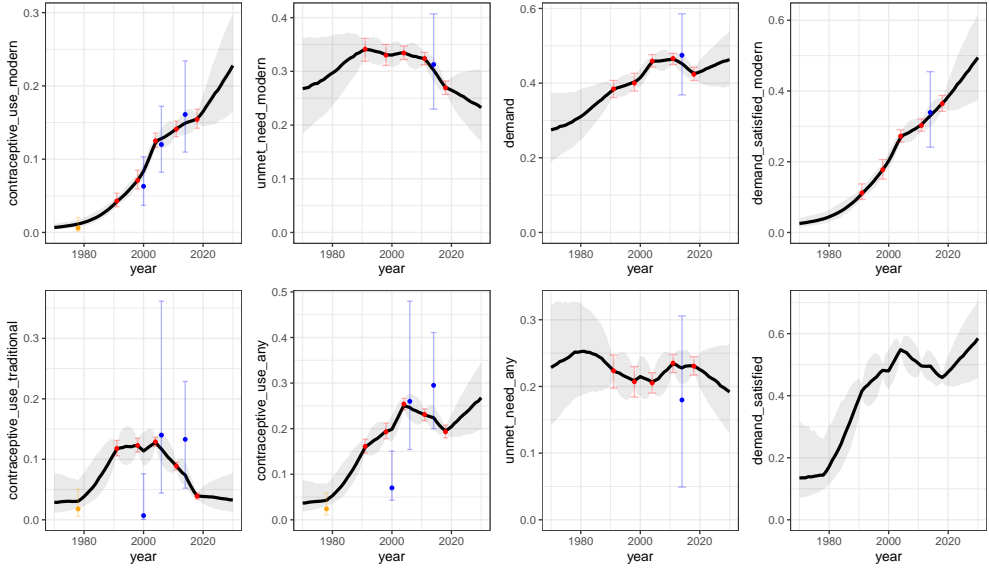

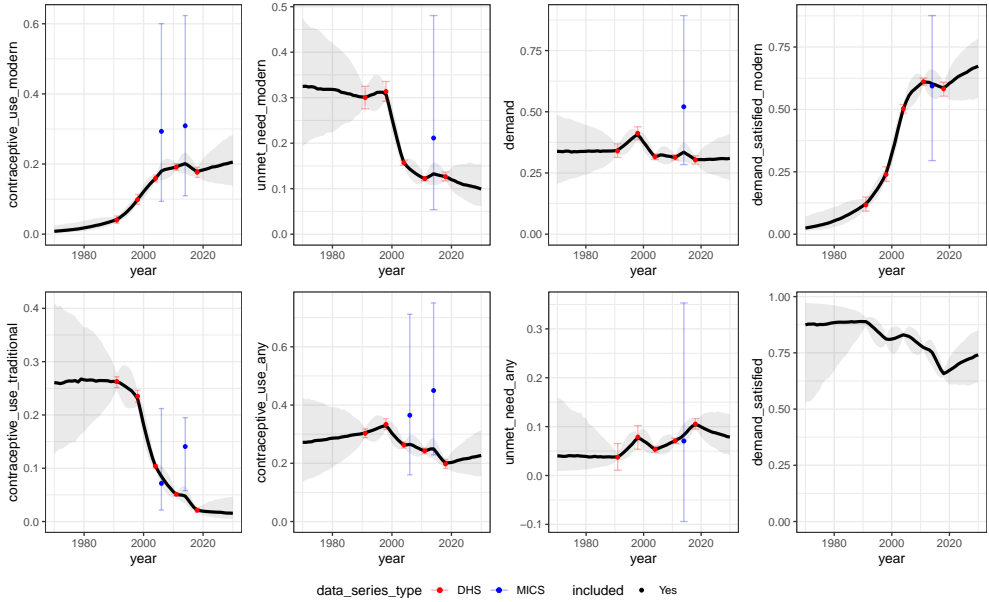

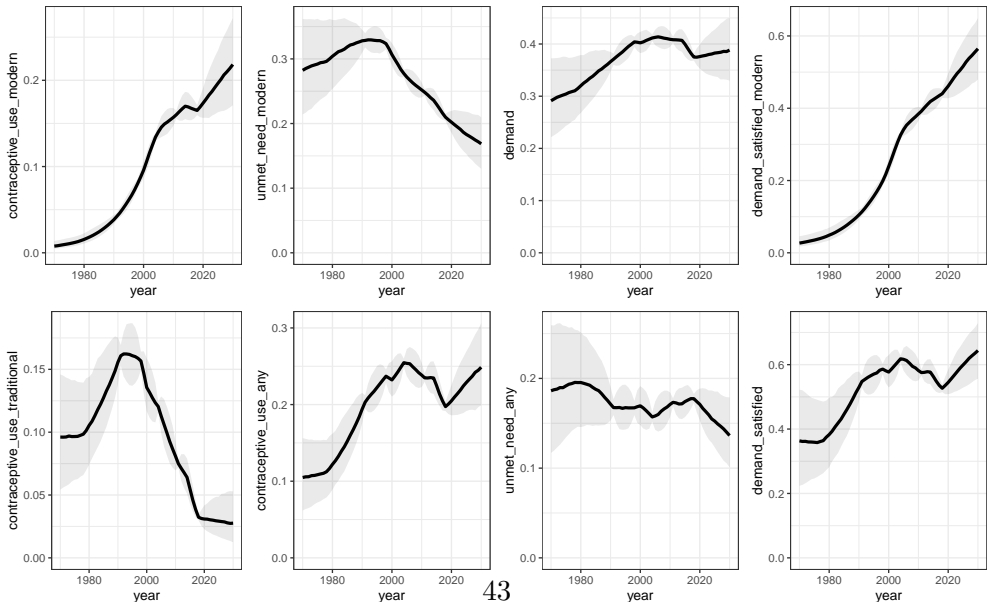

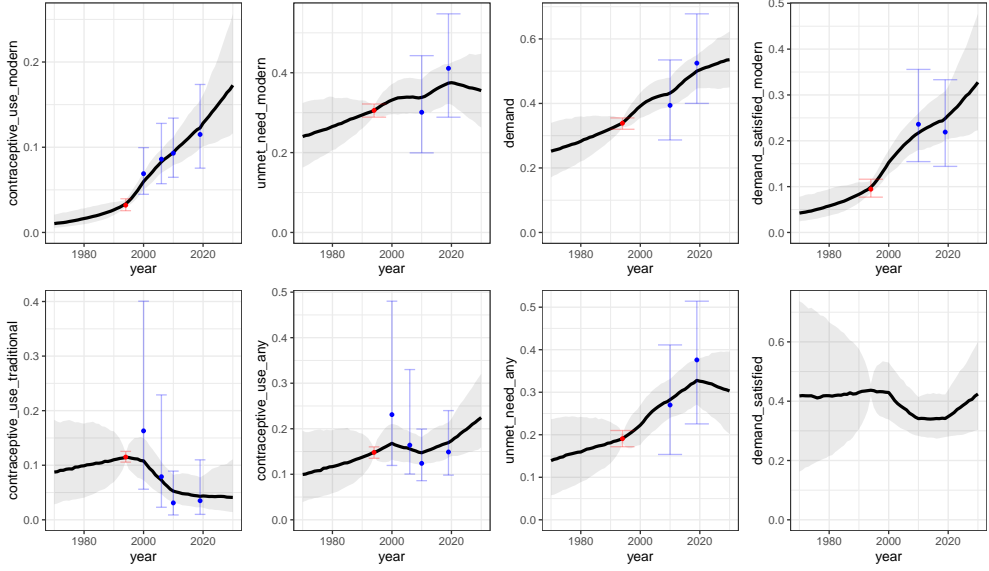

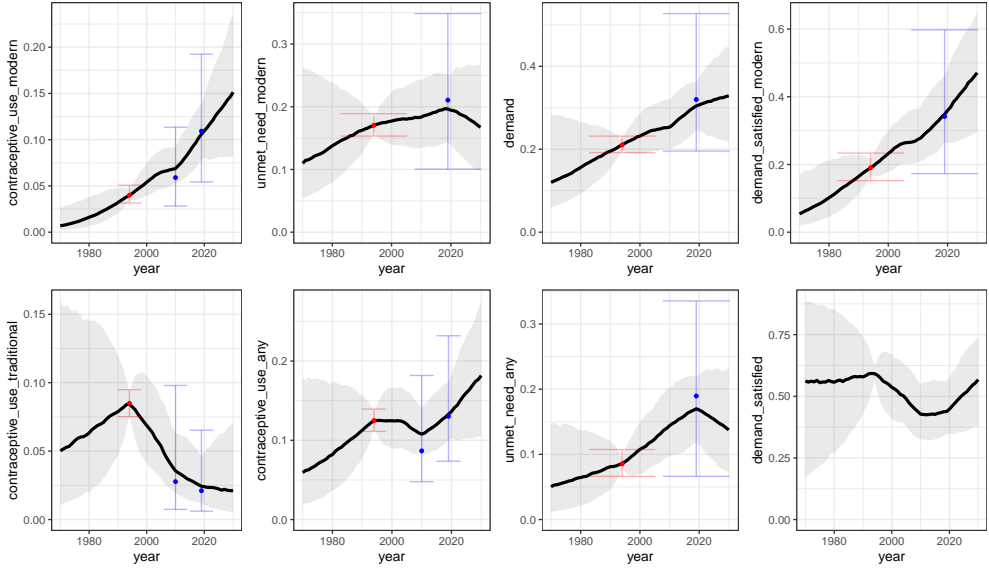

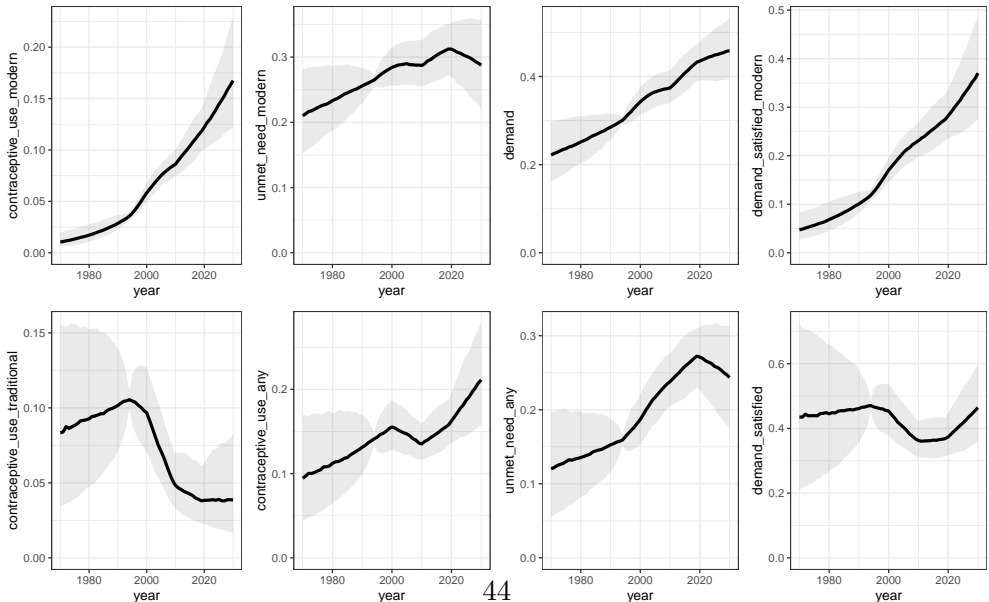

**Chad married**

**Chad unmarried**

**Chad all**

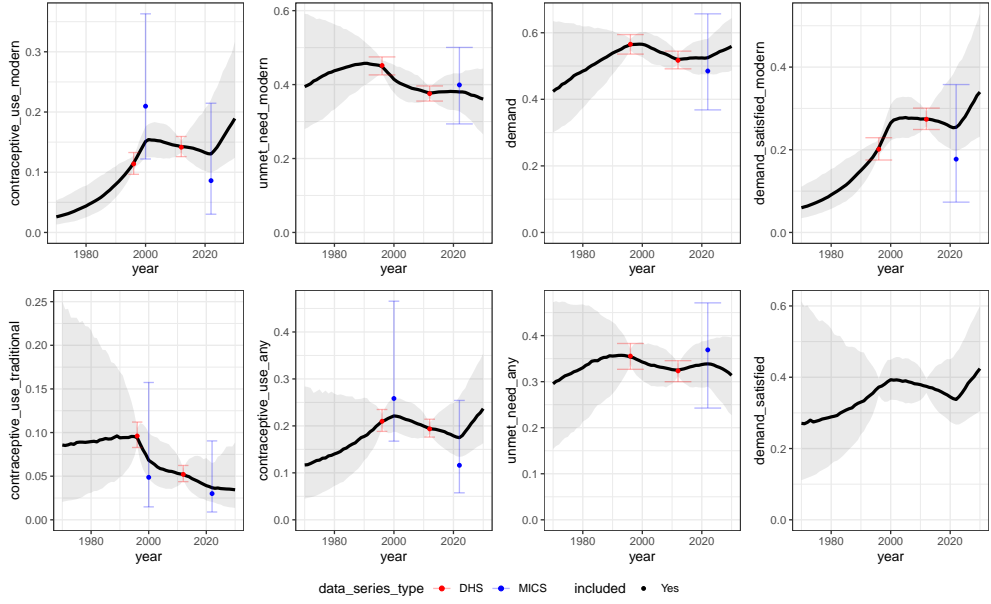

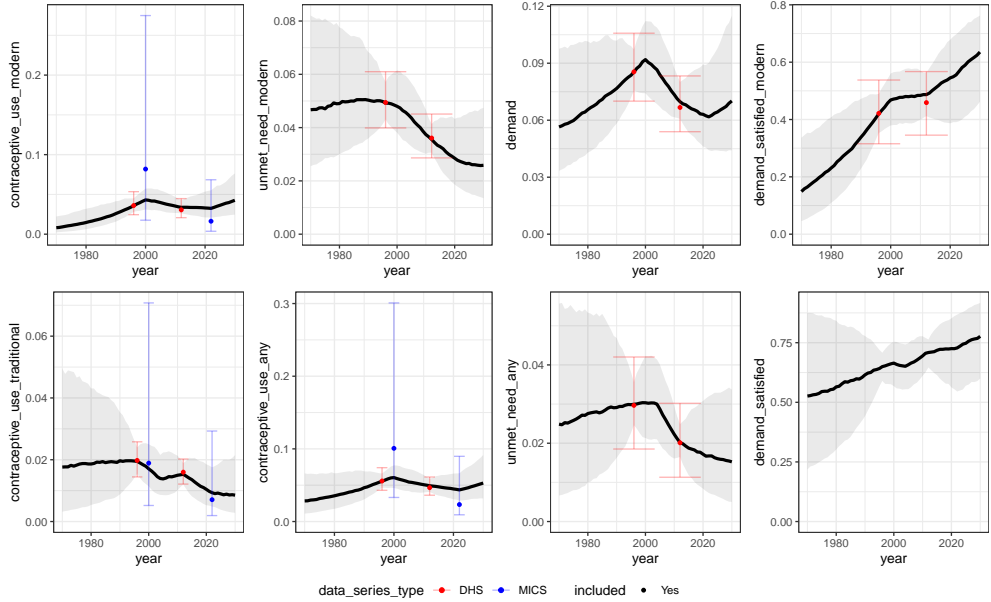

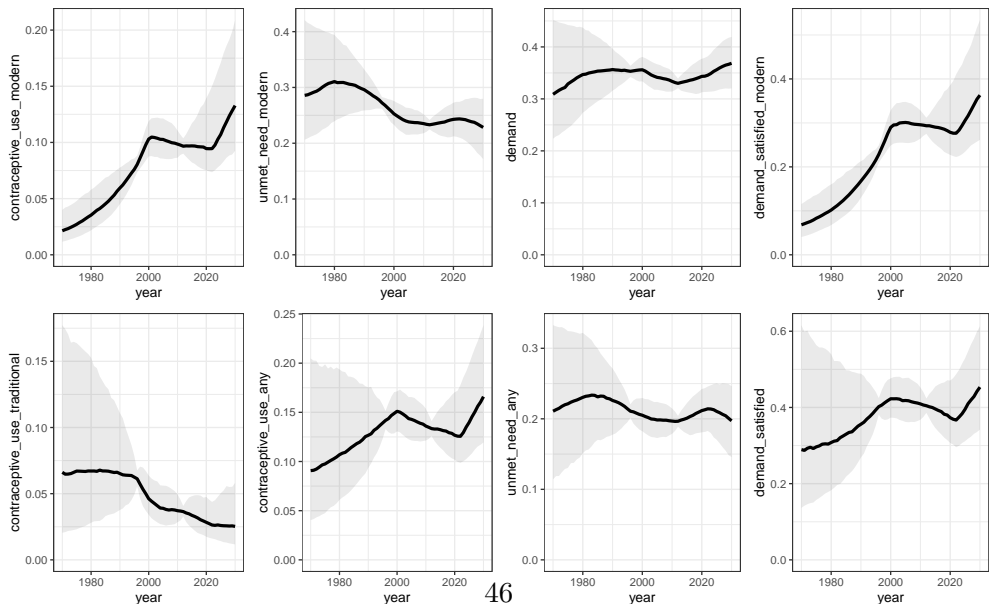

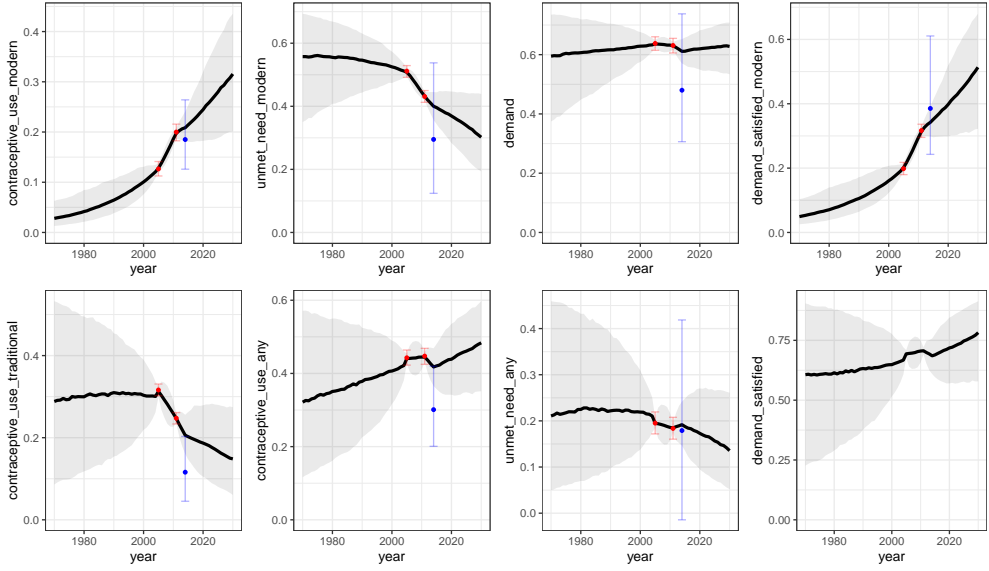

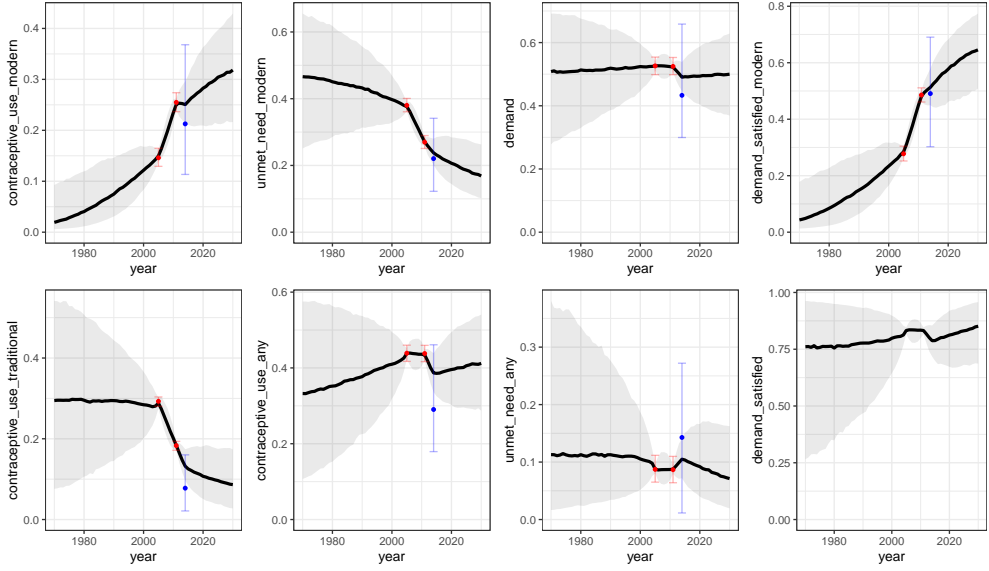

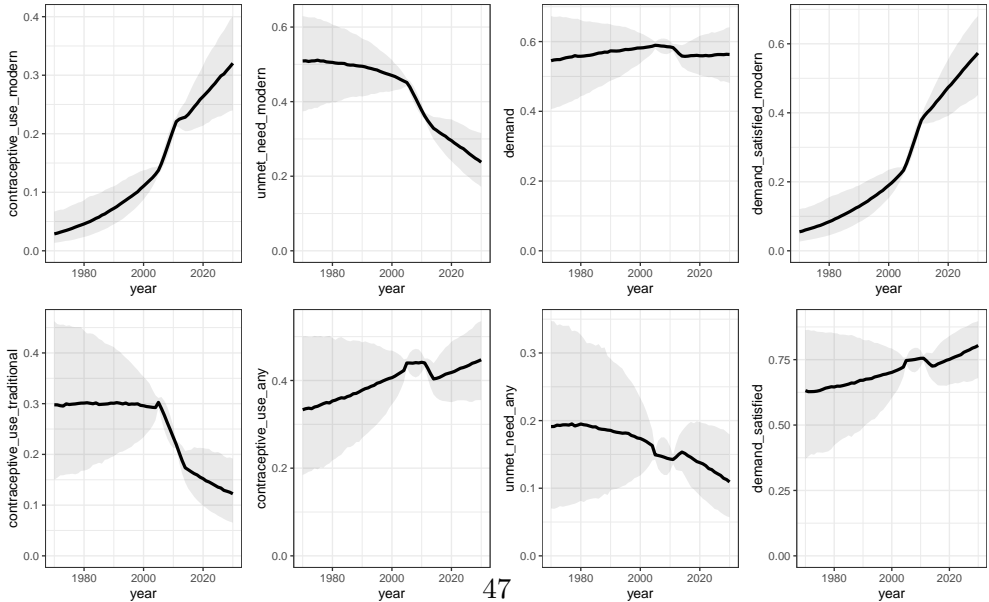

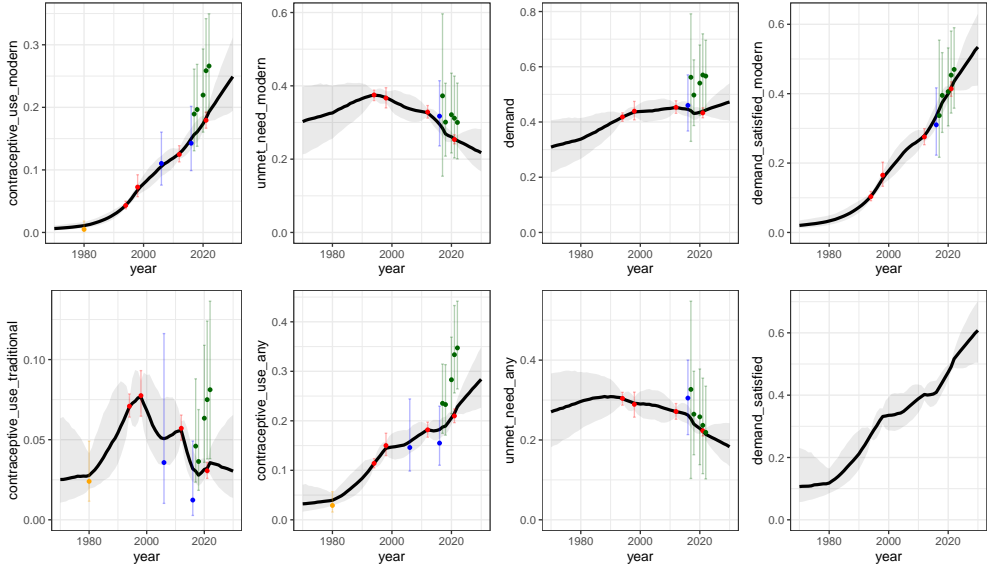

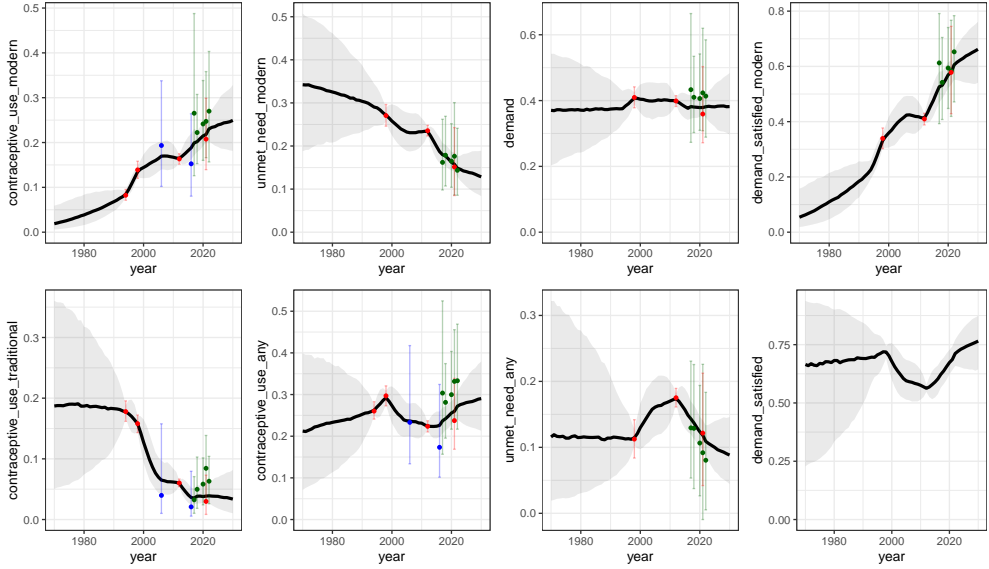

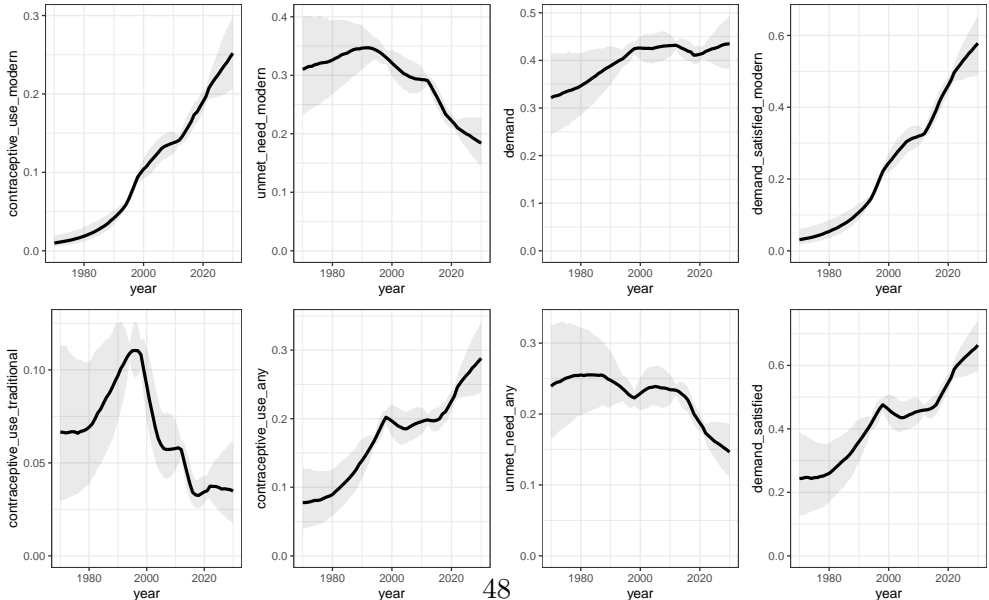

## Democratic_Peoples_Republic_of_Korea married

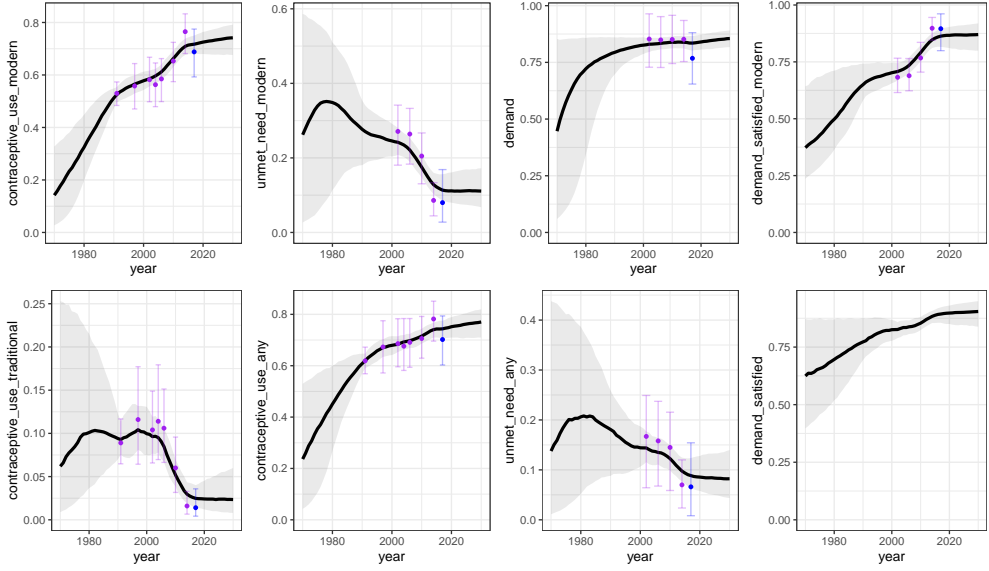

## Democratic_Peoples_Republic_of_Korea unmarried

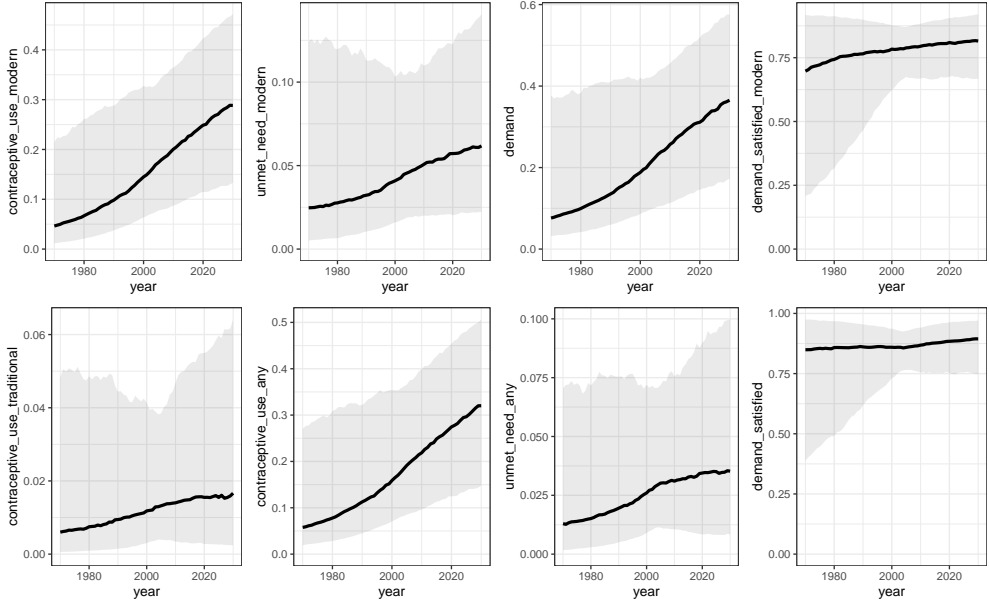

## Democratic_Peoples_Republic_of_Korea all

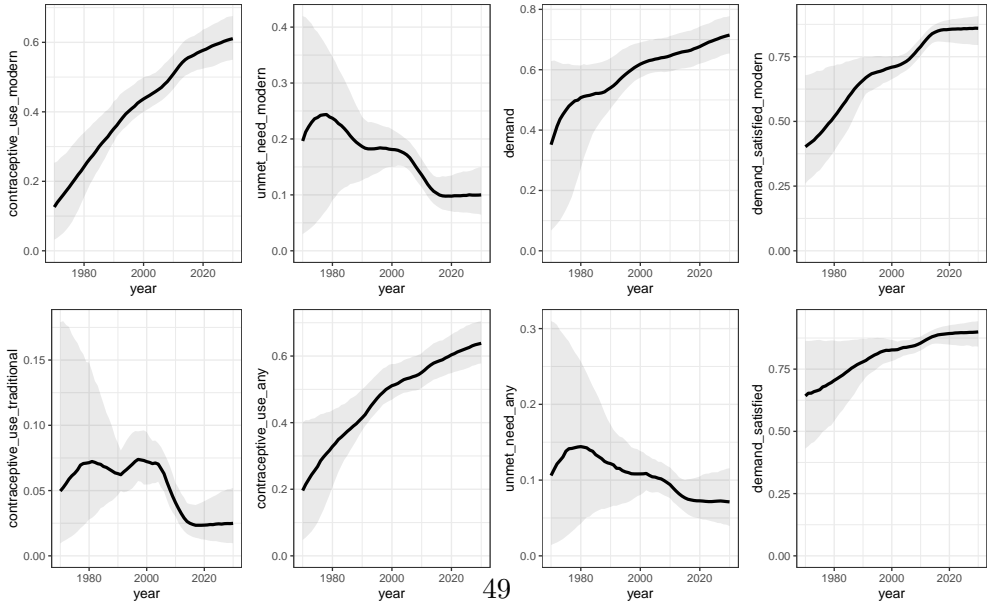

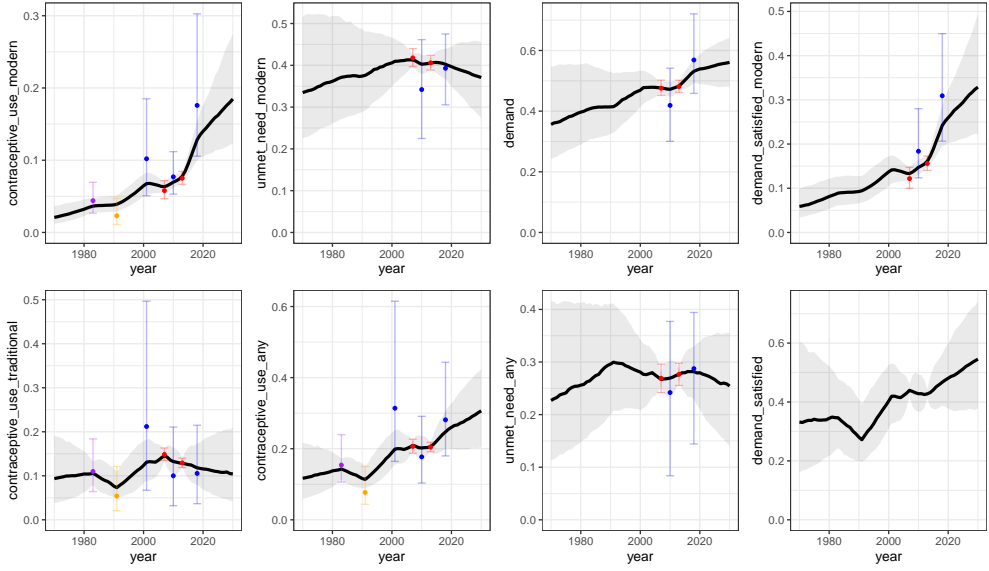

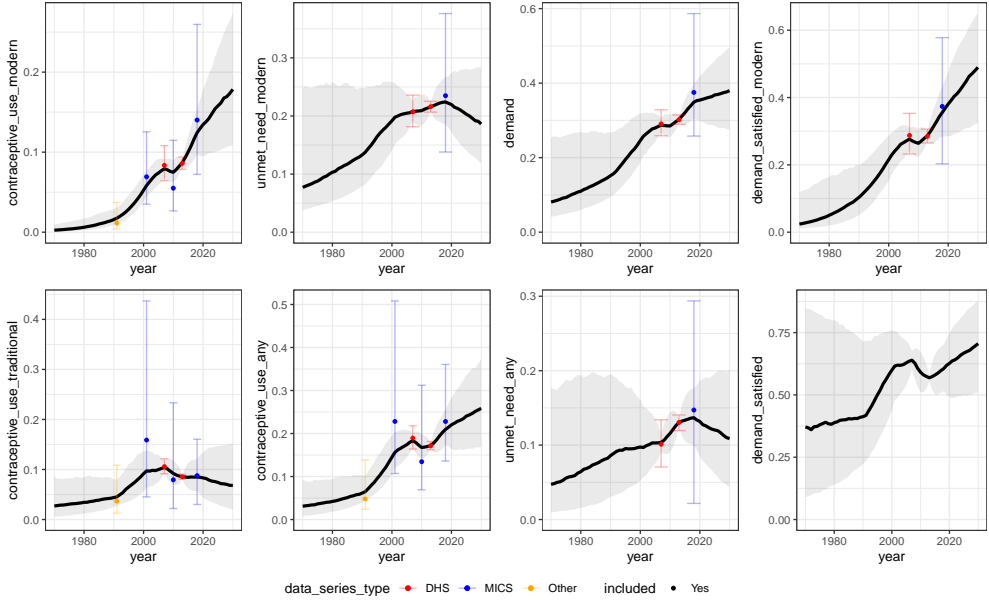

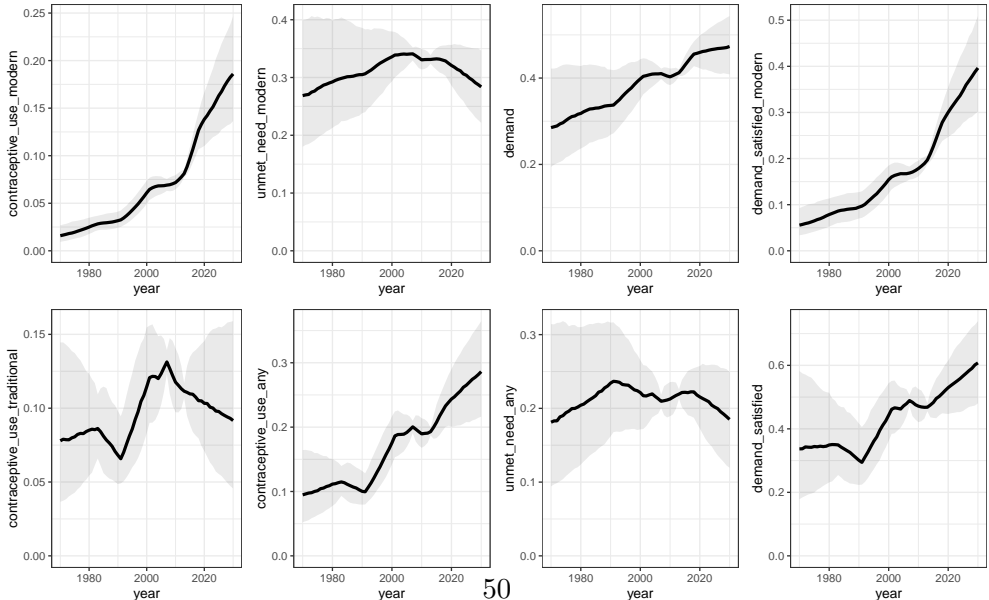

50

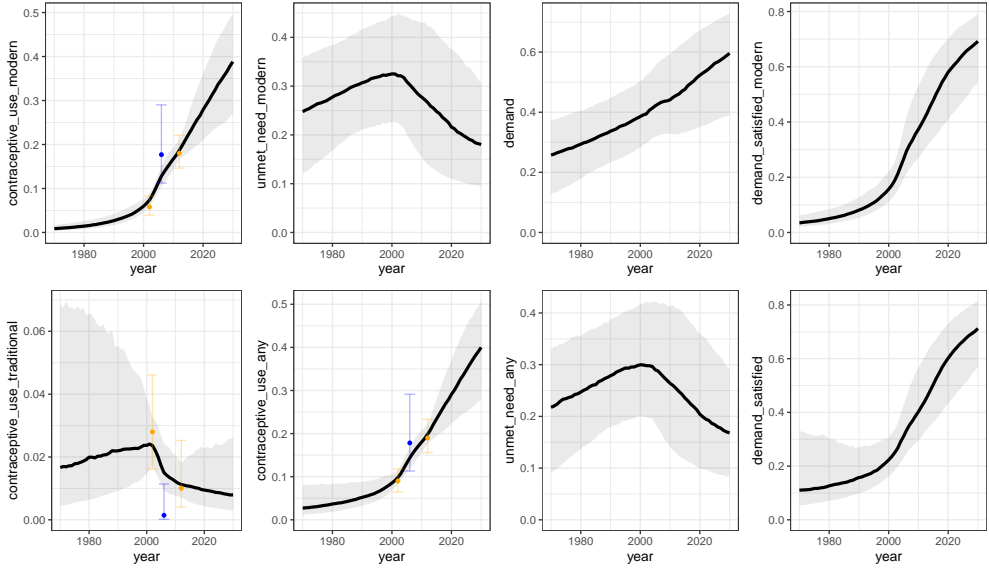

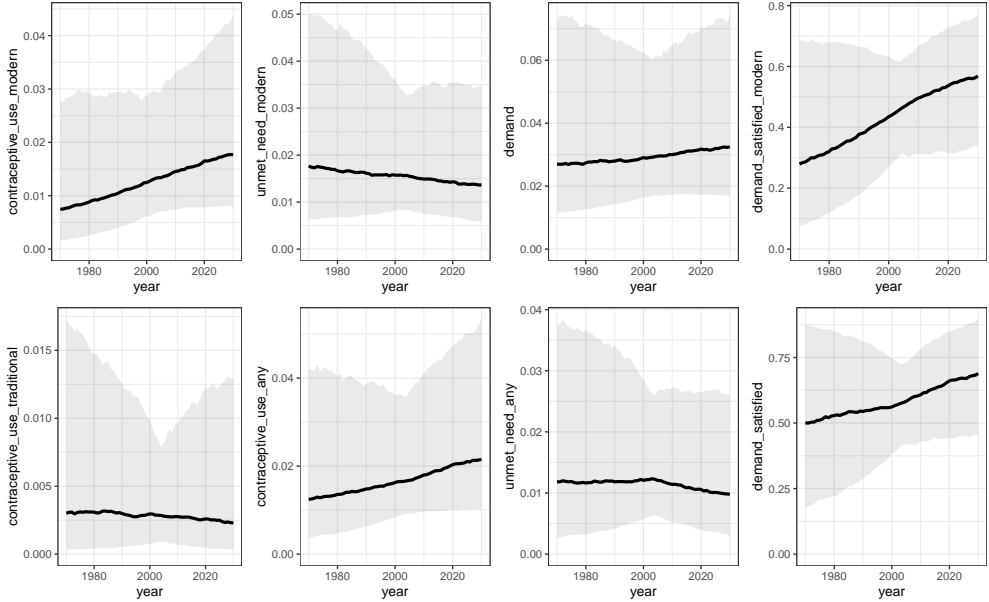

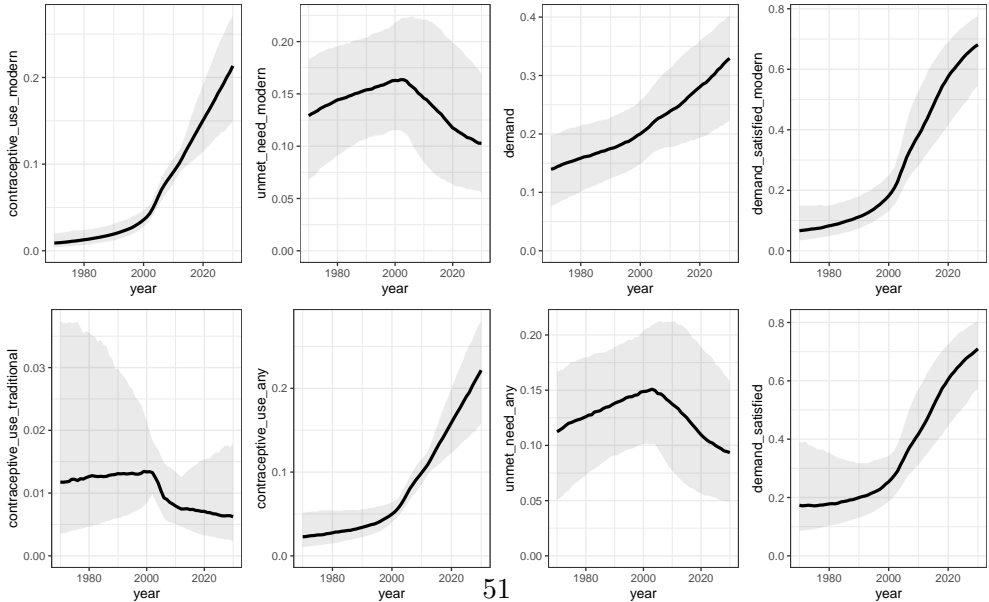

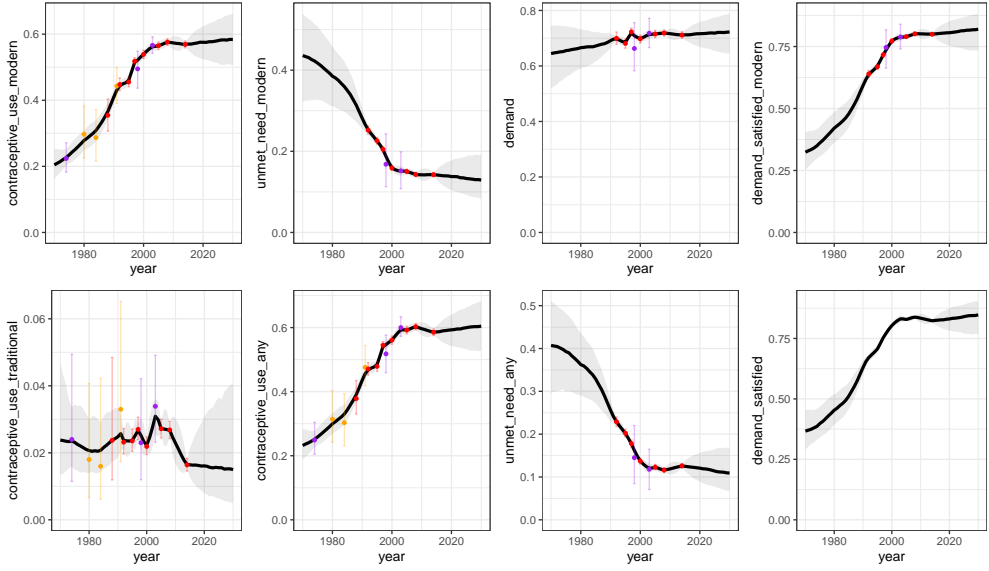

**Egypt married**

included ● Yes    data_series_type ● DHS ● National survey ● Other

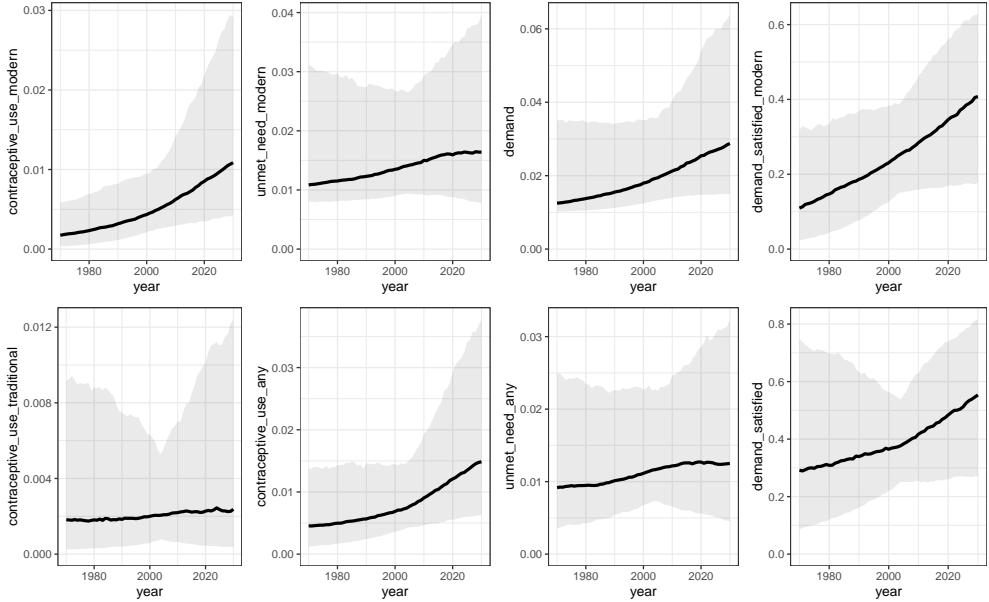

**Egypt unmarried**

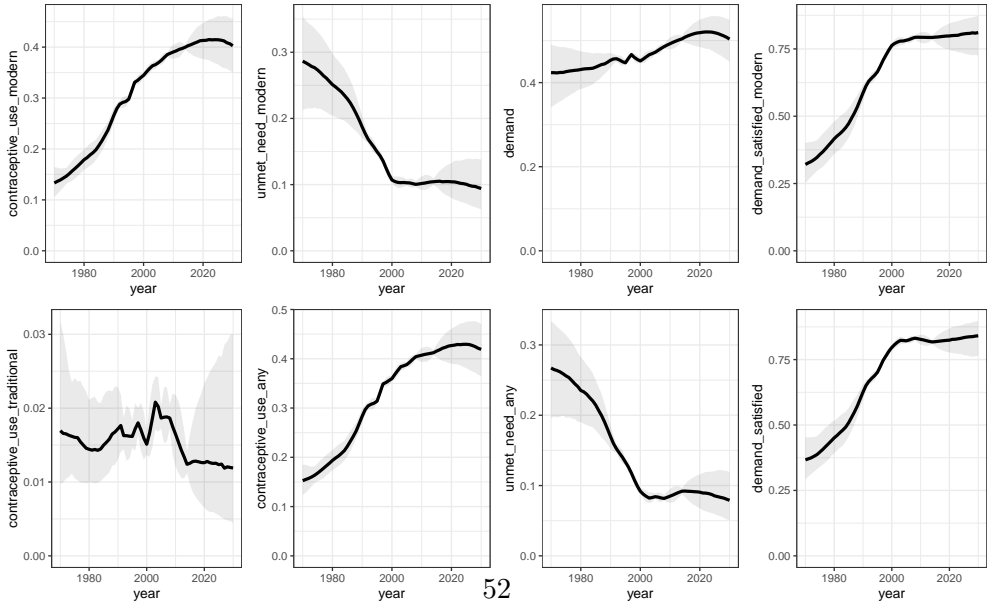

**Egypt all**

52

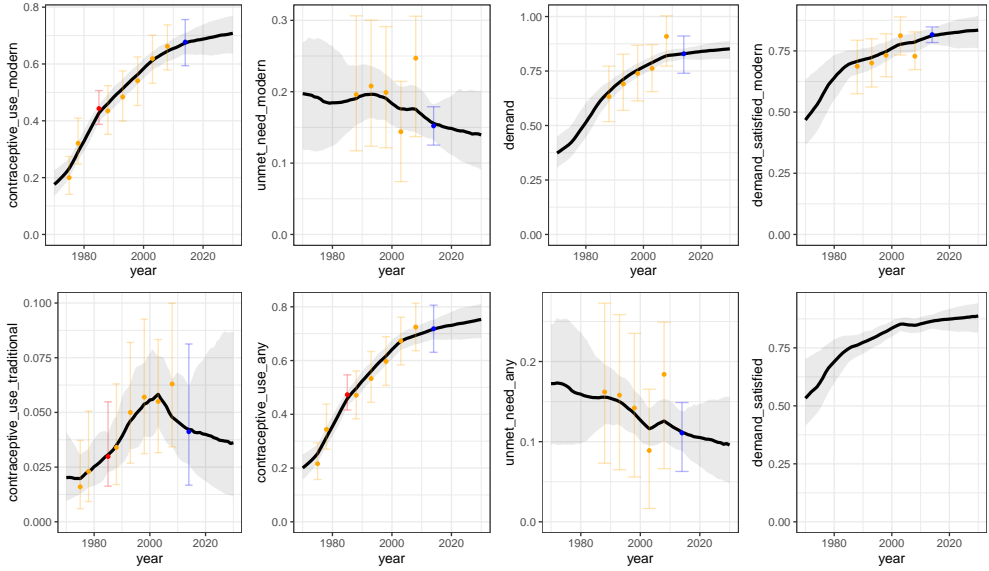

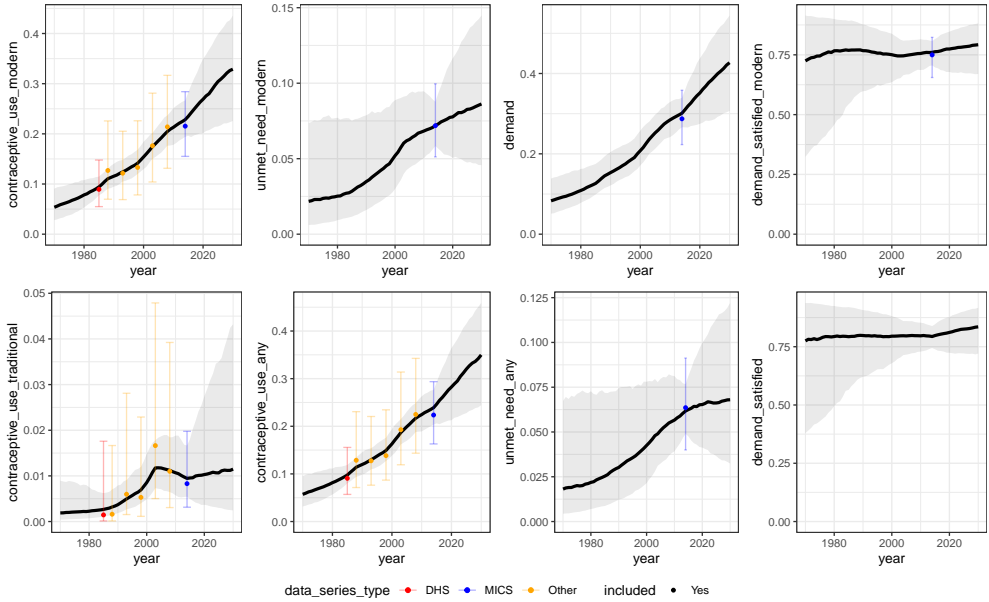

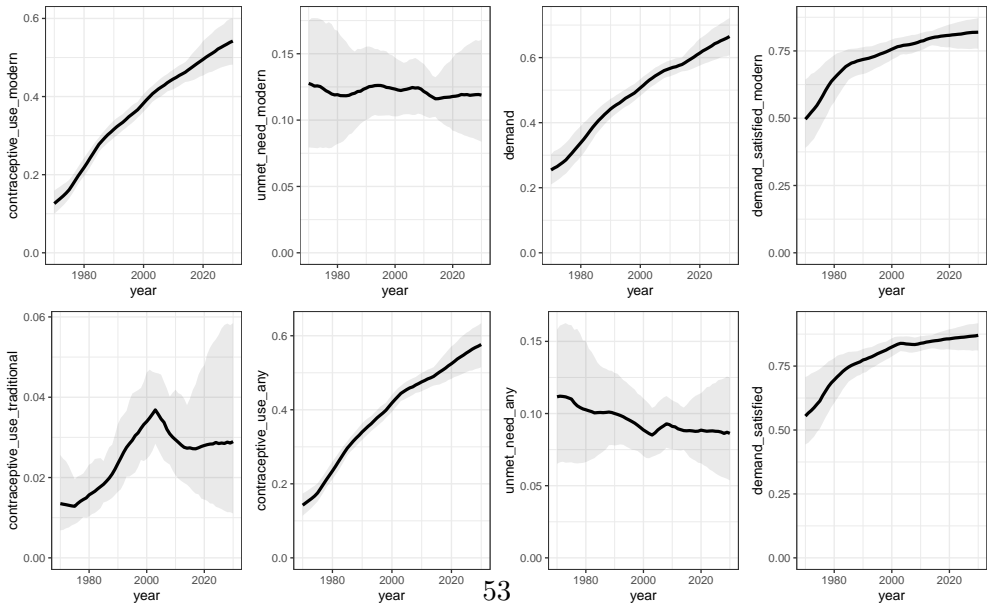

Eritrea married



Eritrea unmarried



Eritrea all

54

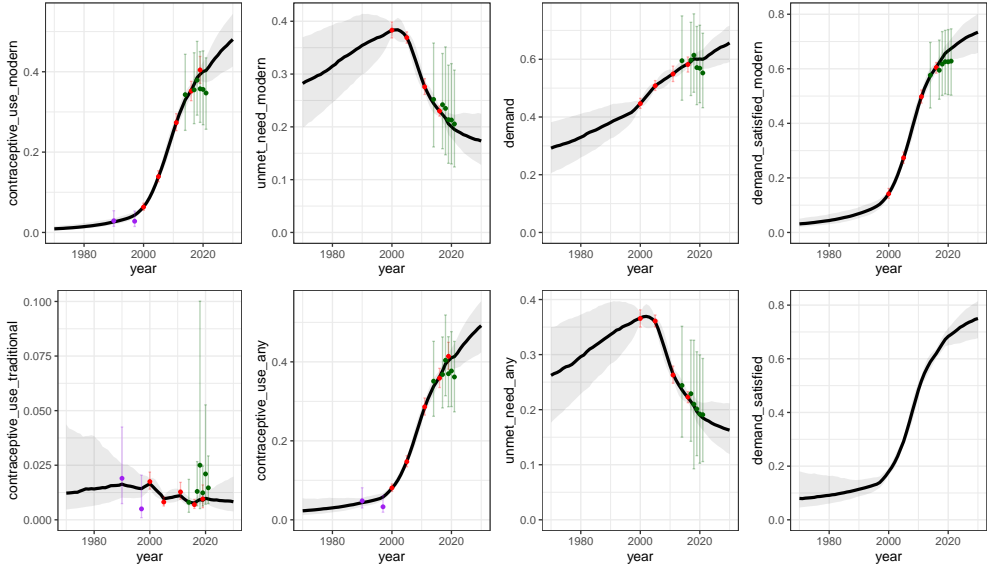

**Ethiopia married**

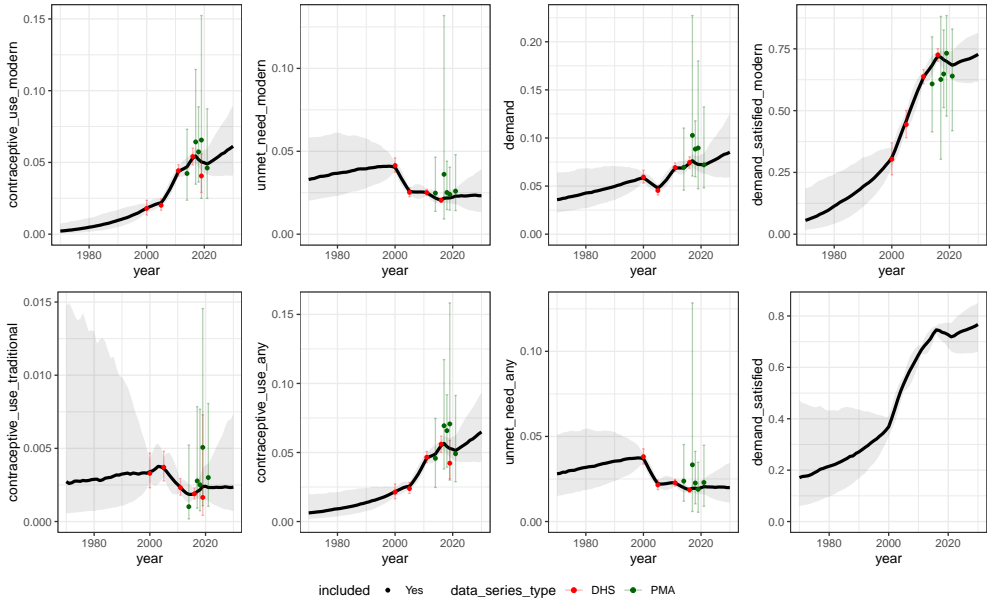

**Ethiopia unmarried**

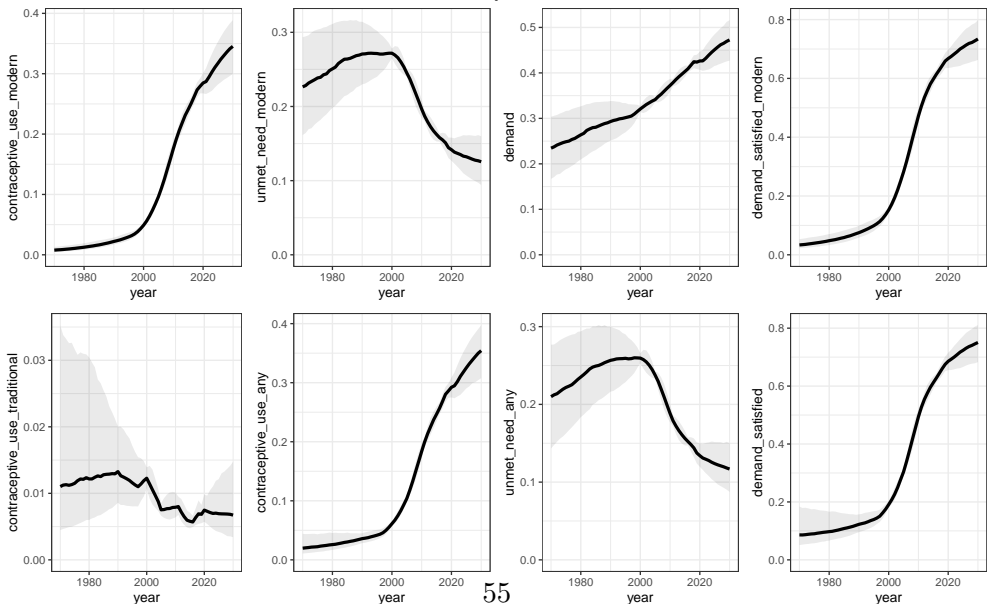

**Ethiopia all**

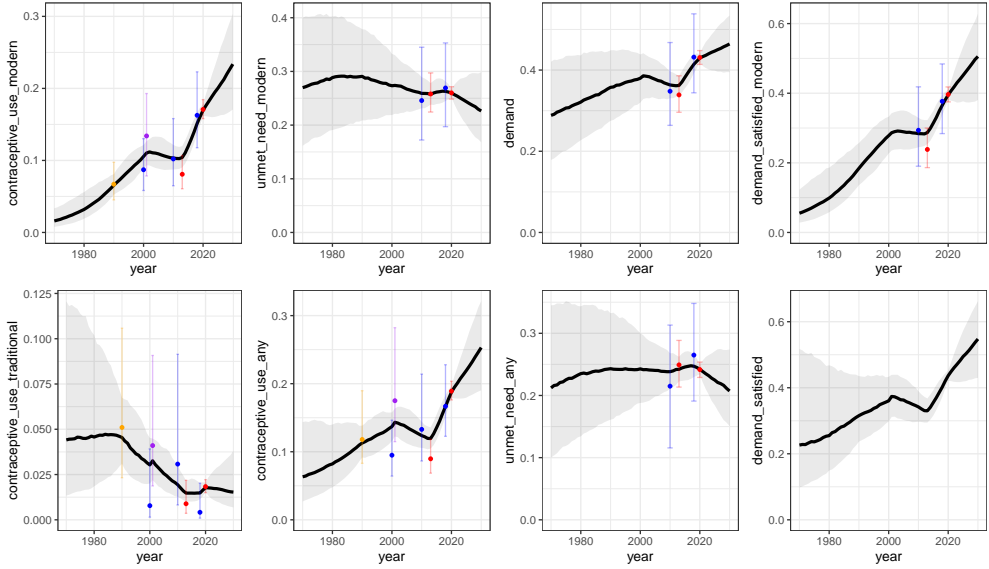

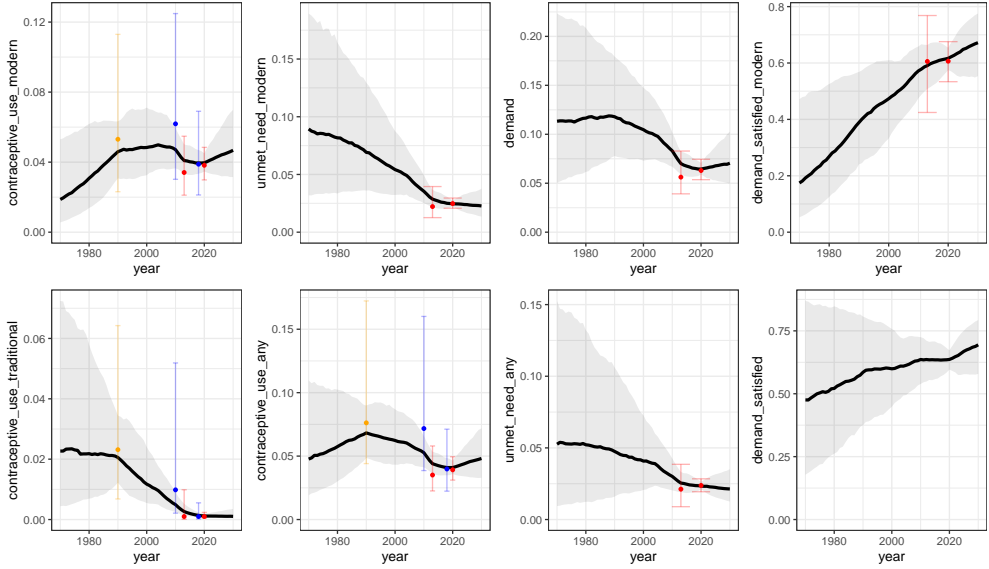

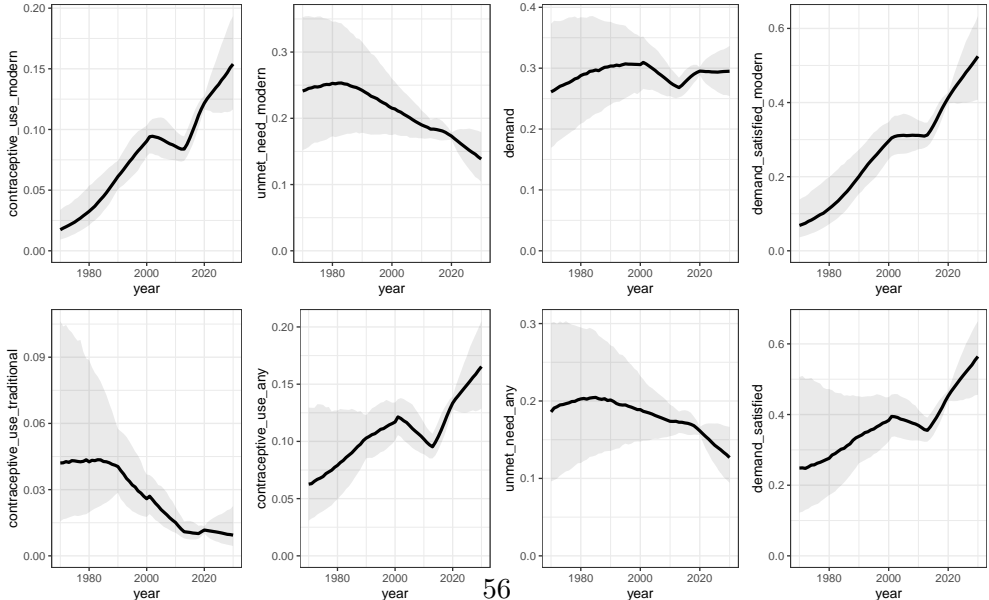

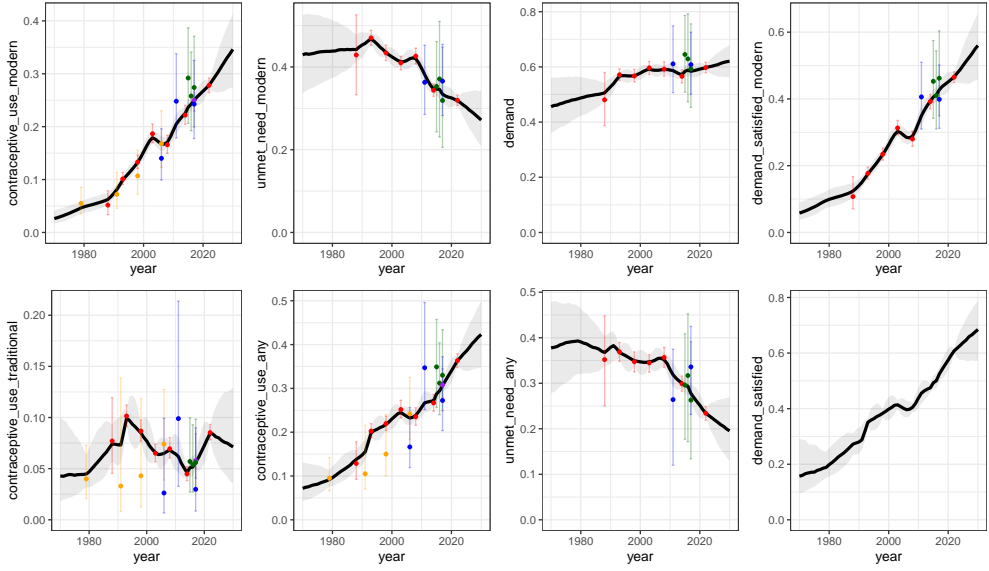

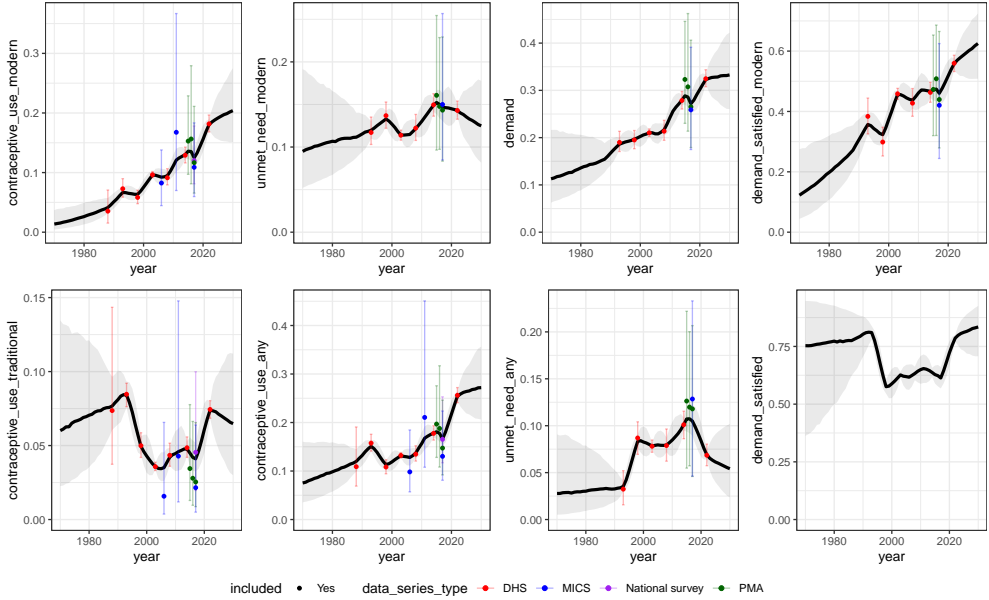

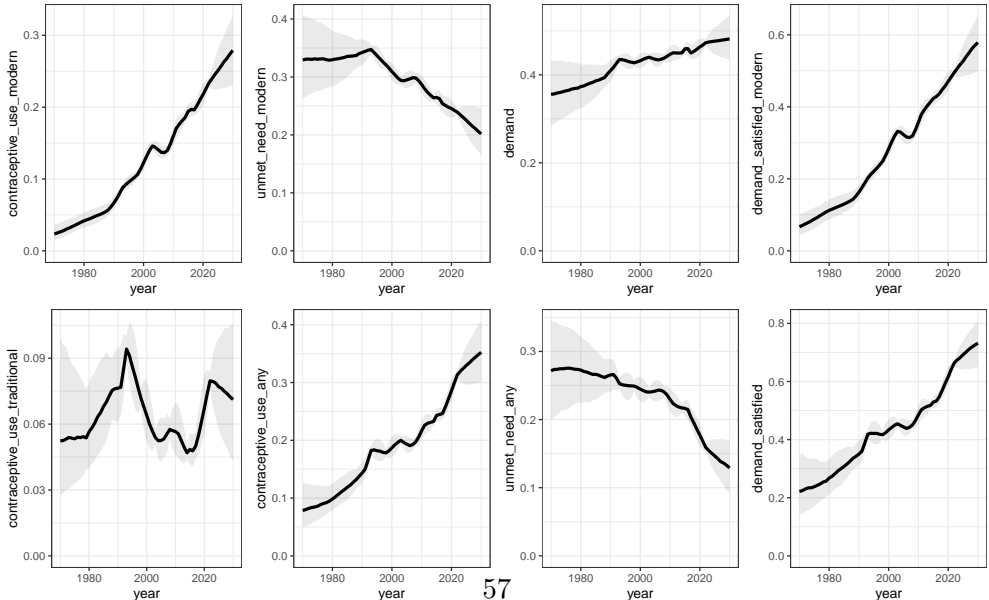

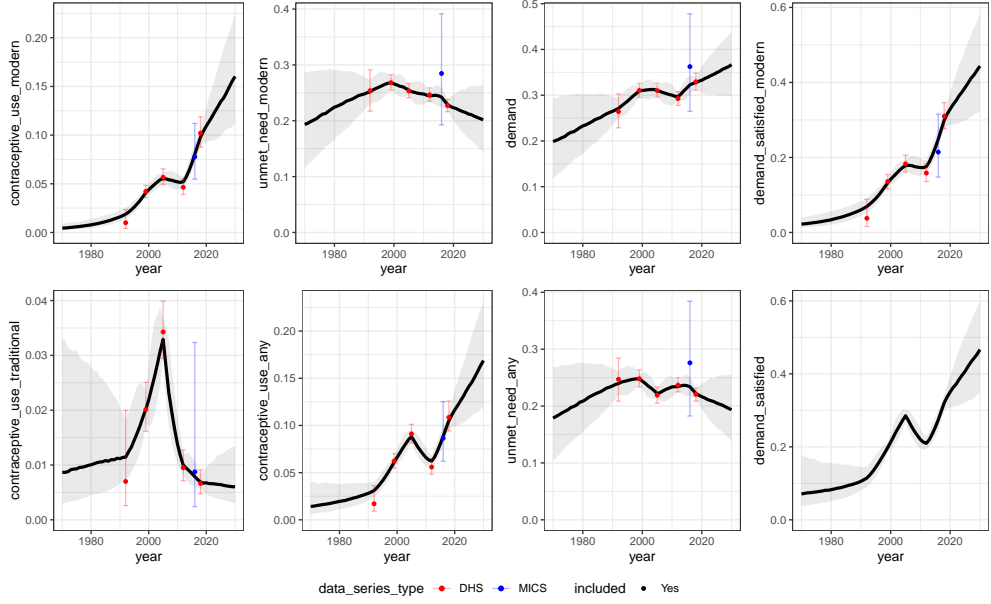

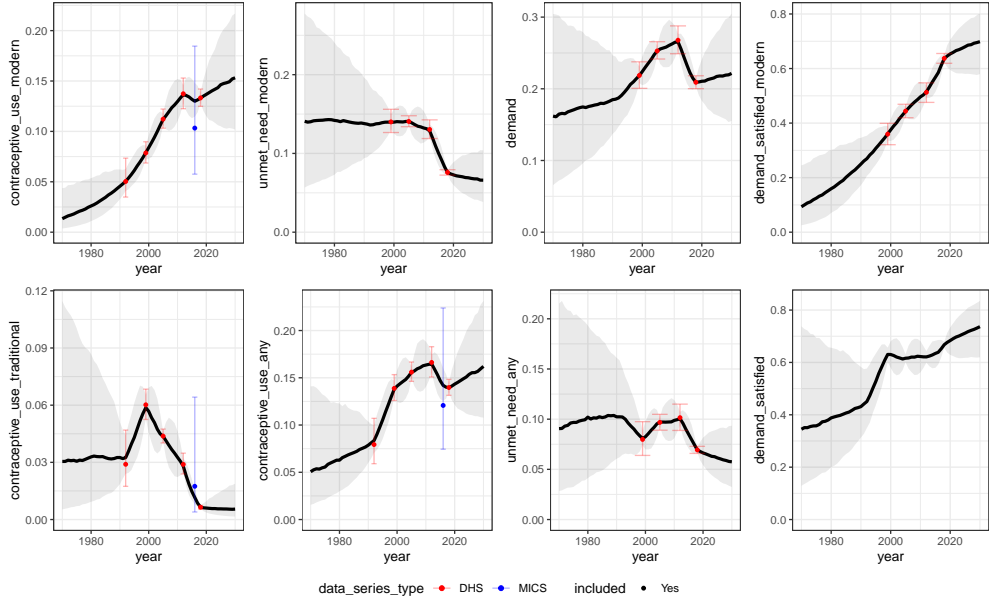

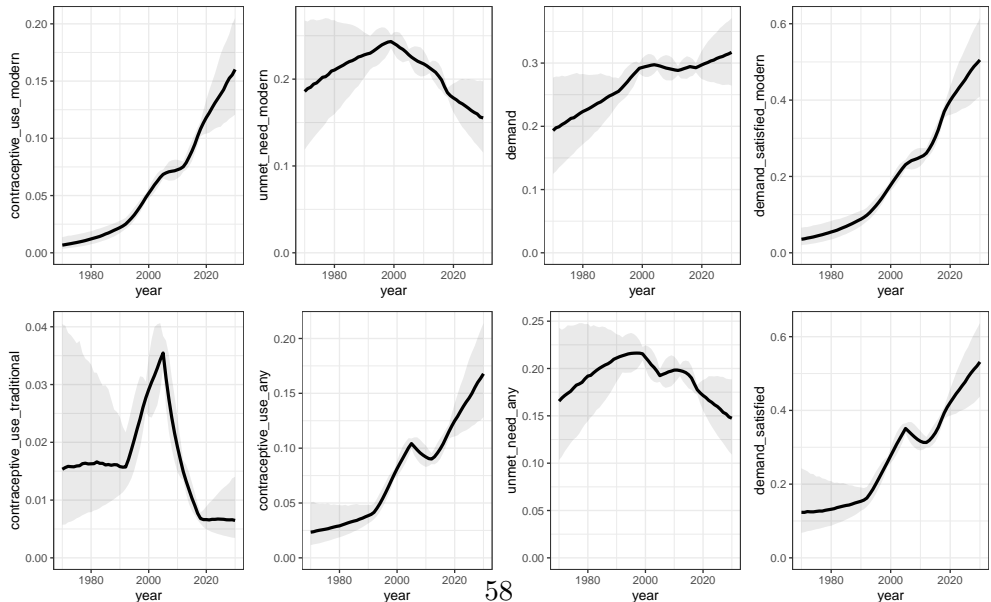

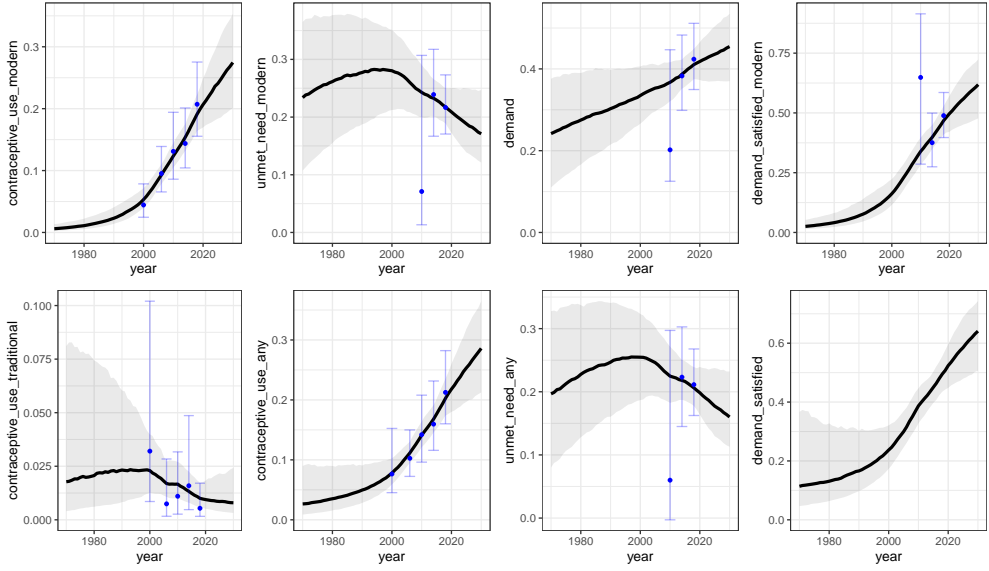

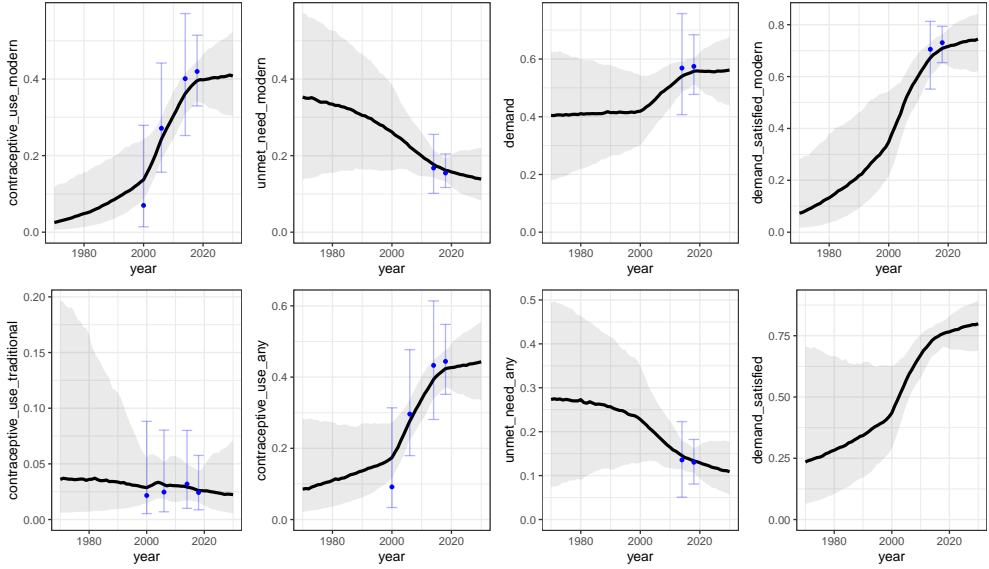

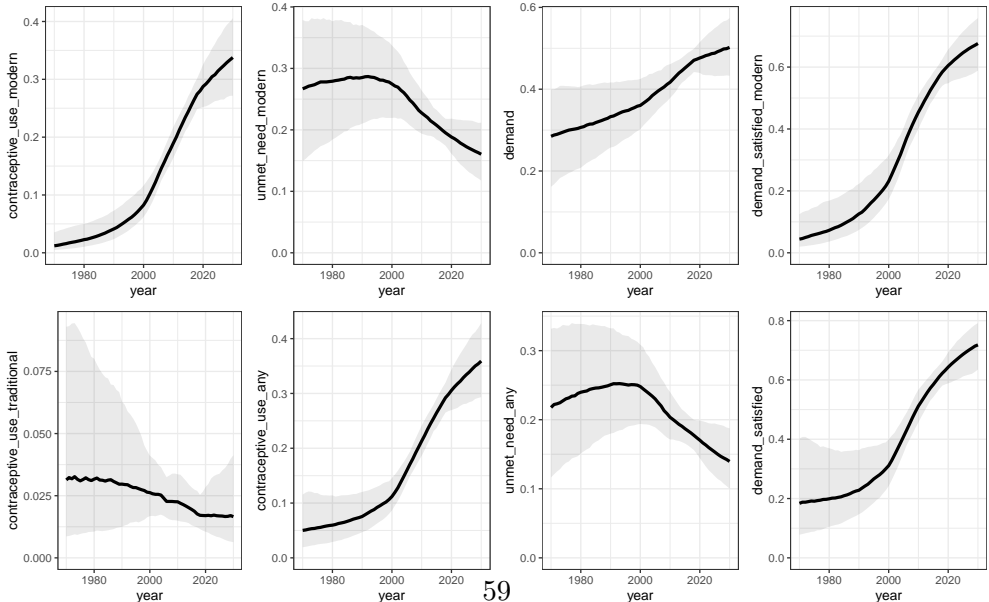

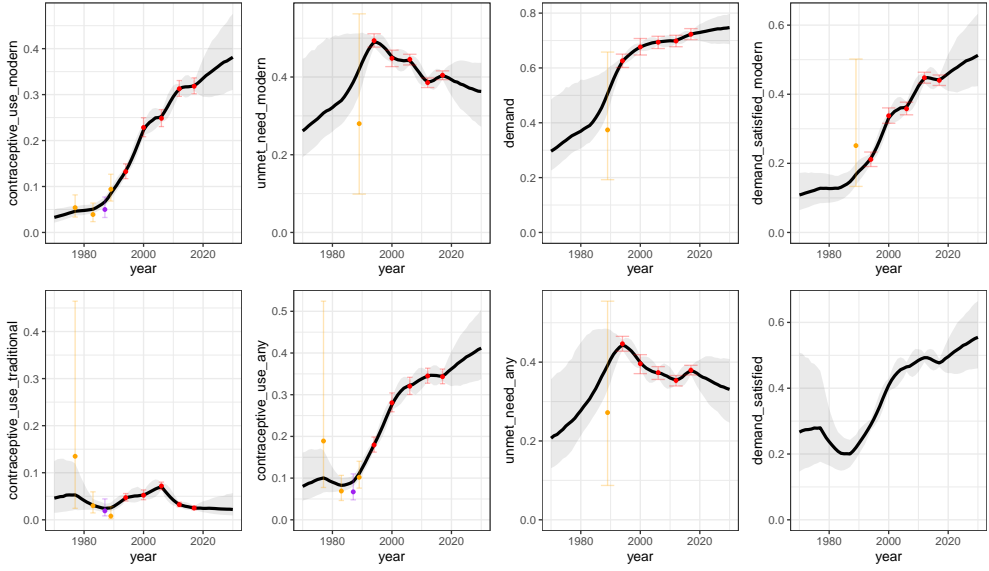

**Haiti married**

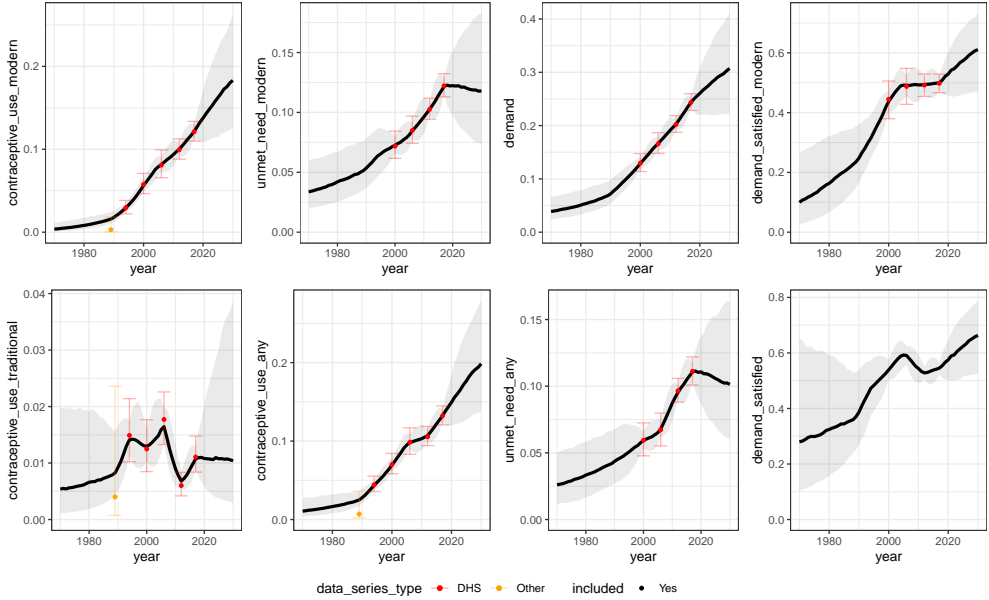

**Haiti unmarried**

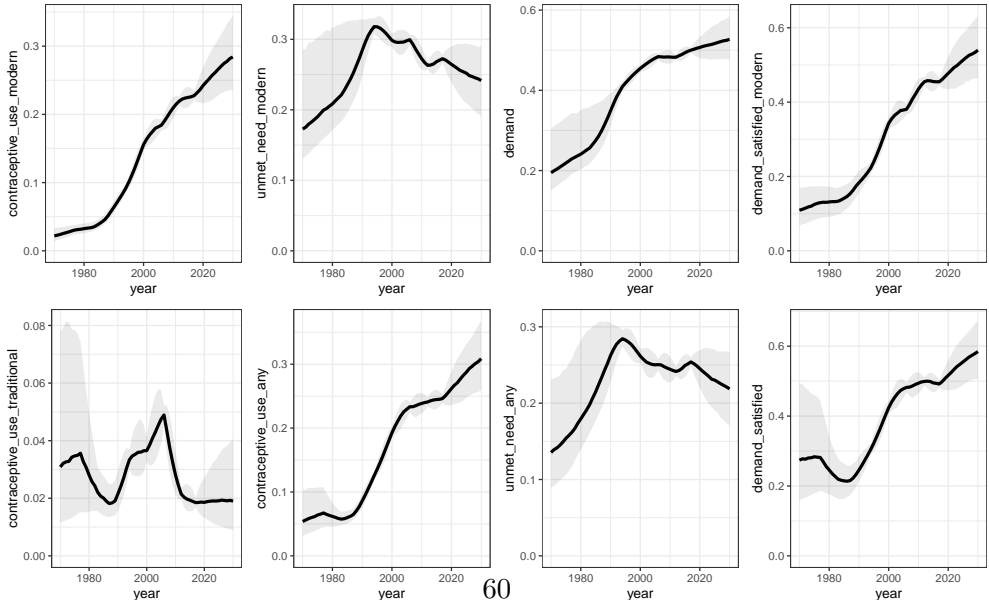

**Haiti all**

60

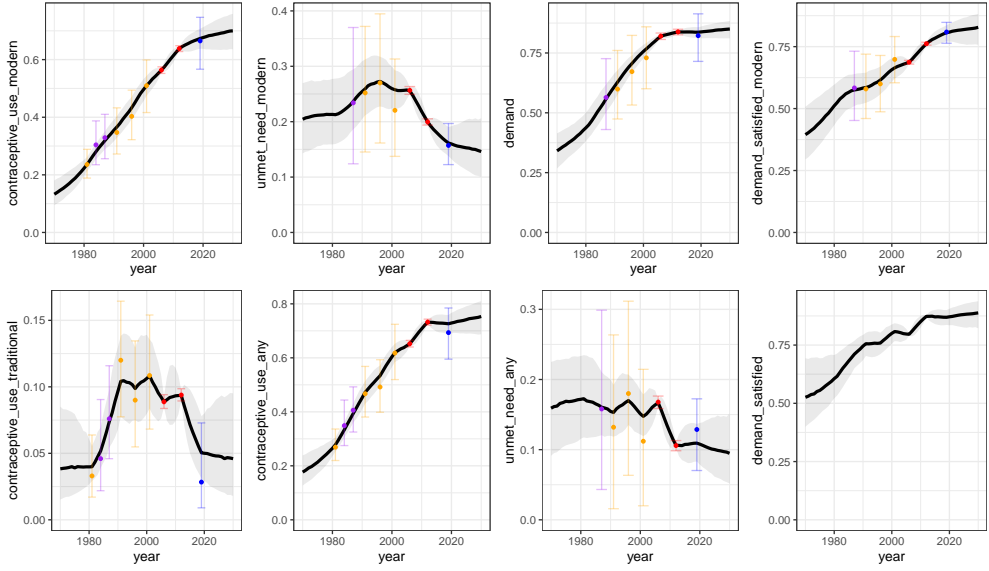

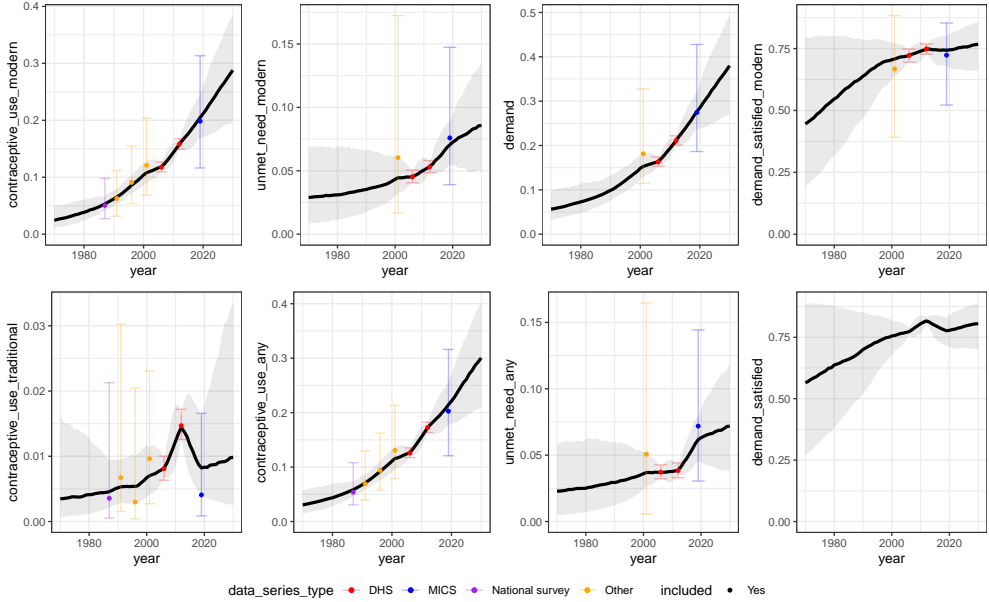

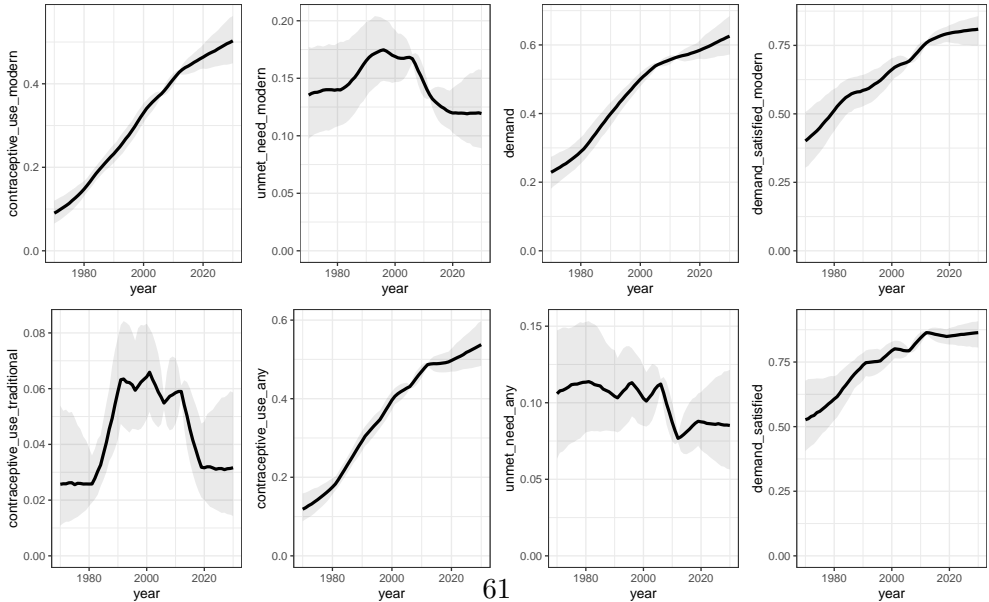

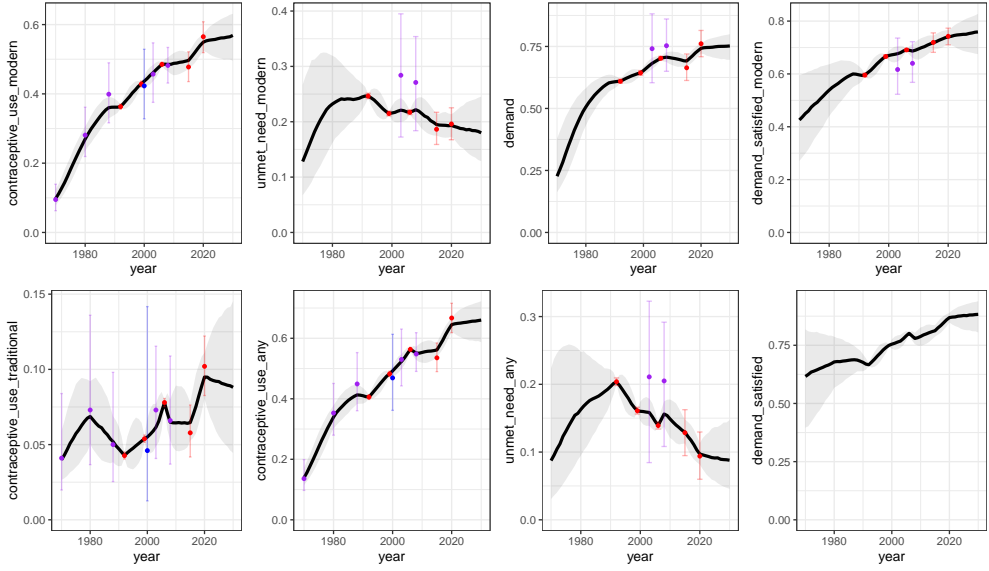

**India married**

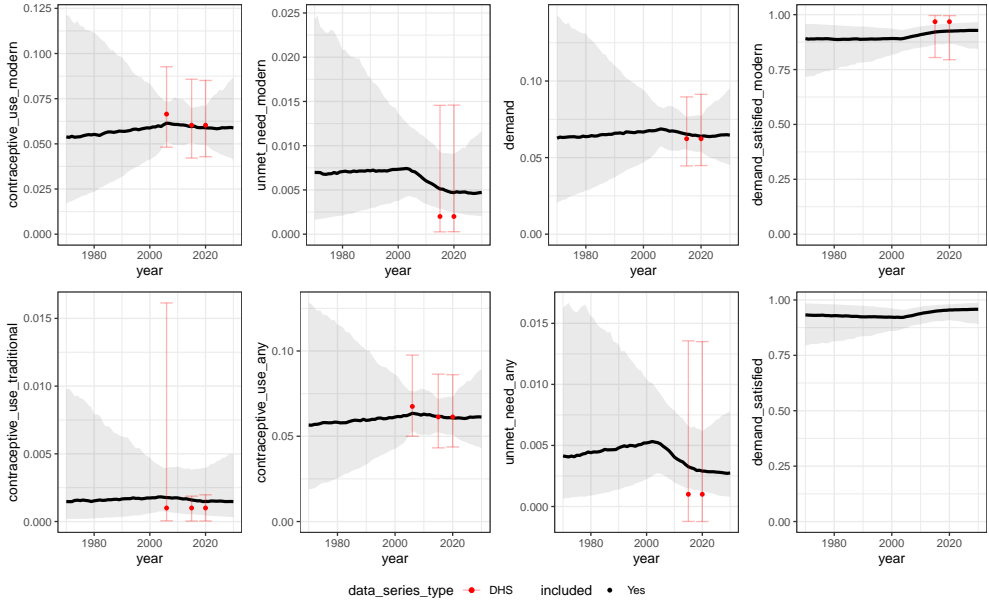

**India unmarried**

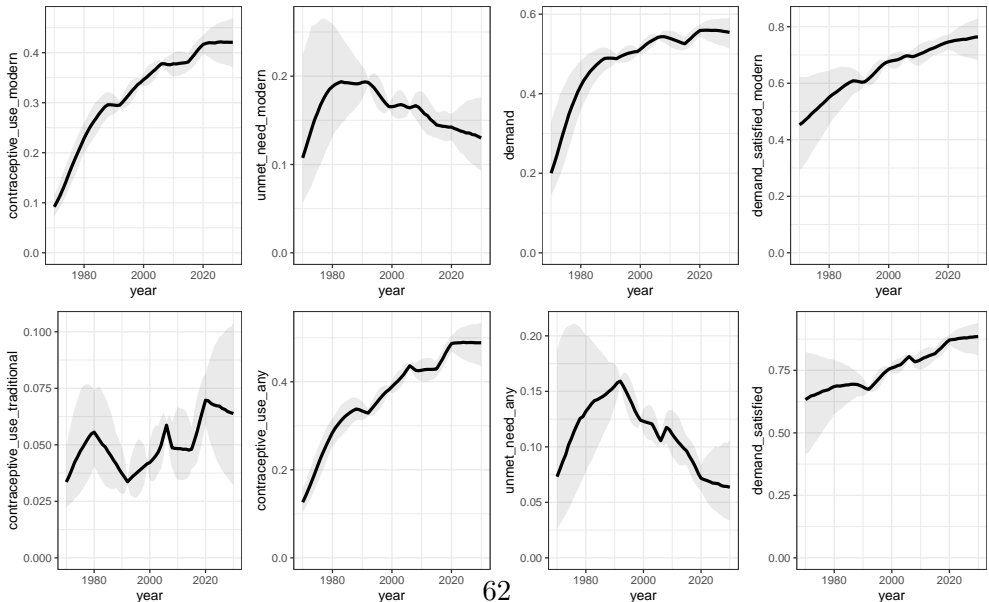

**India all**

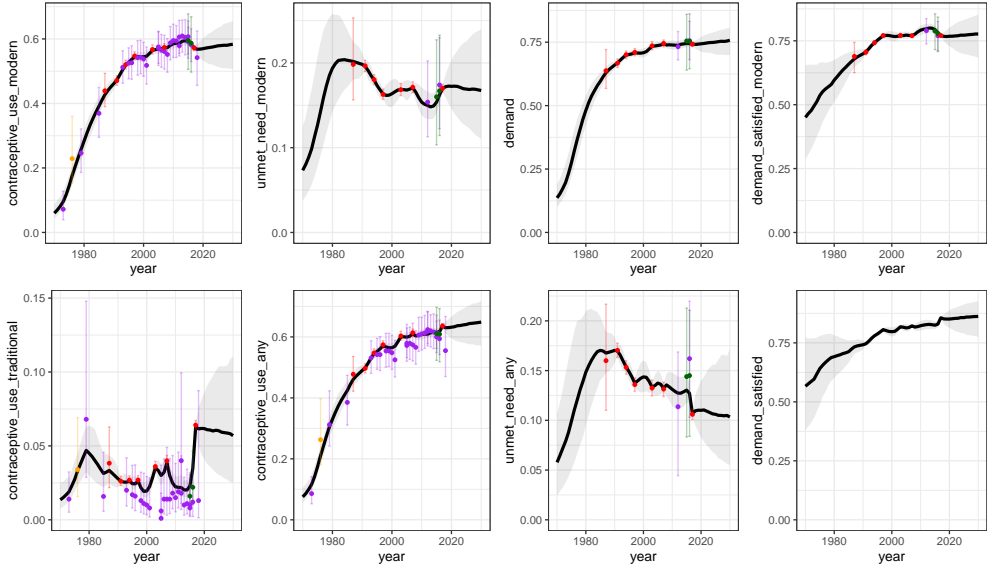

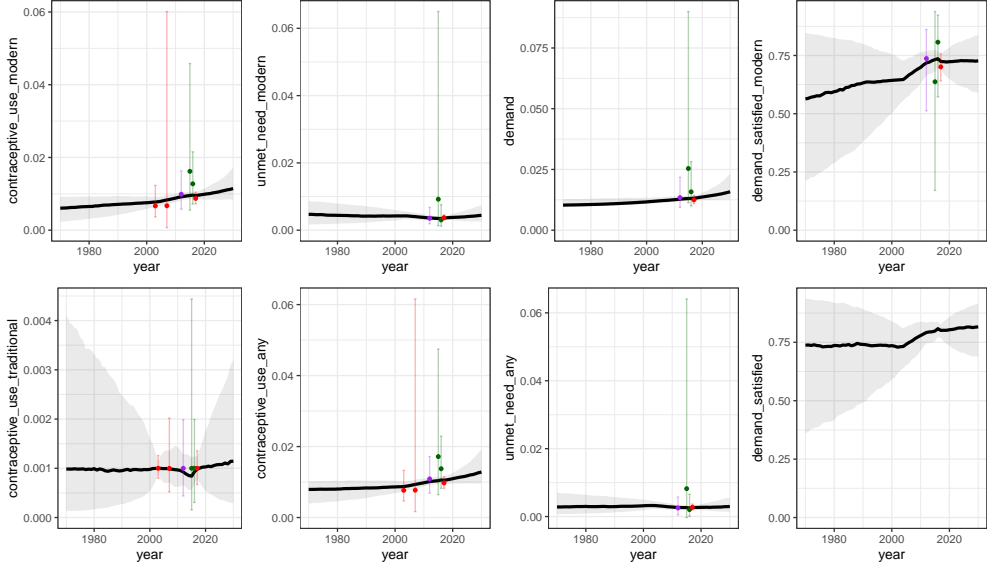

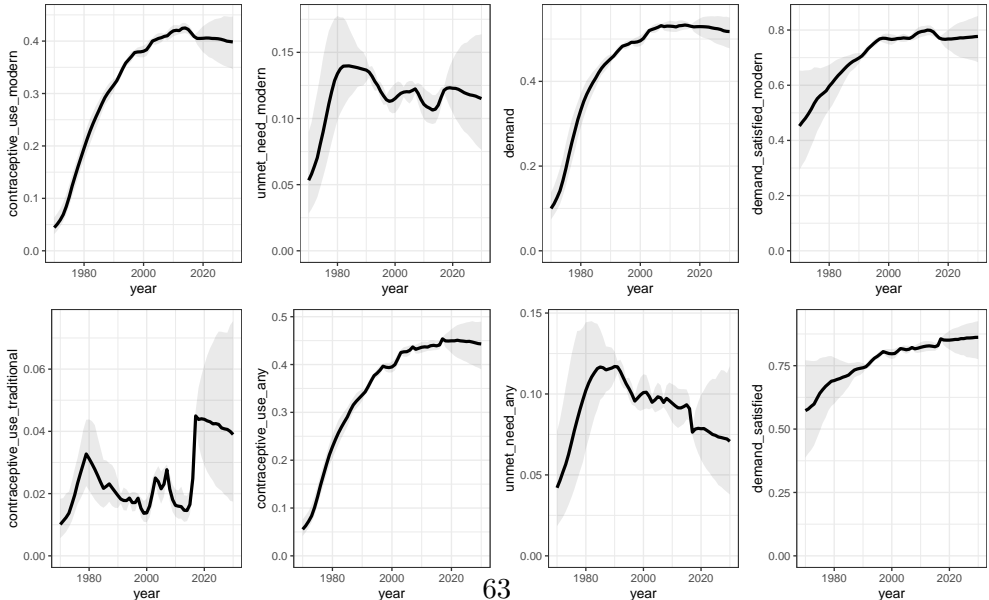

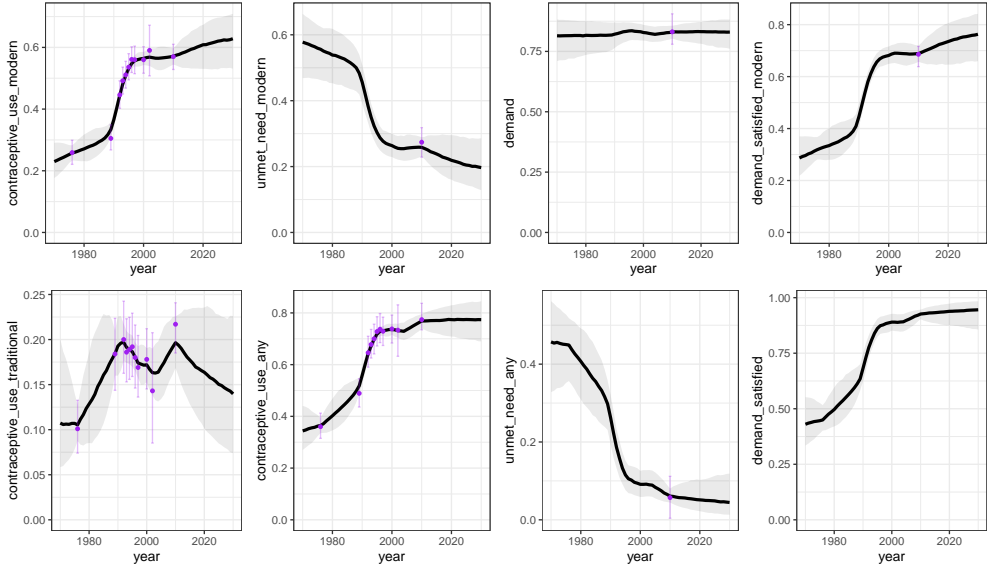

**Iran_Islamic_Republic_of married**

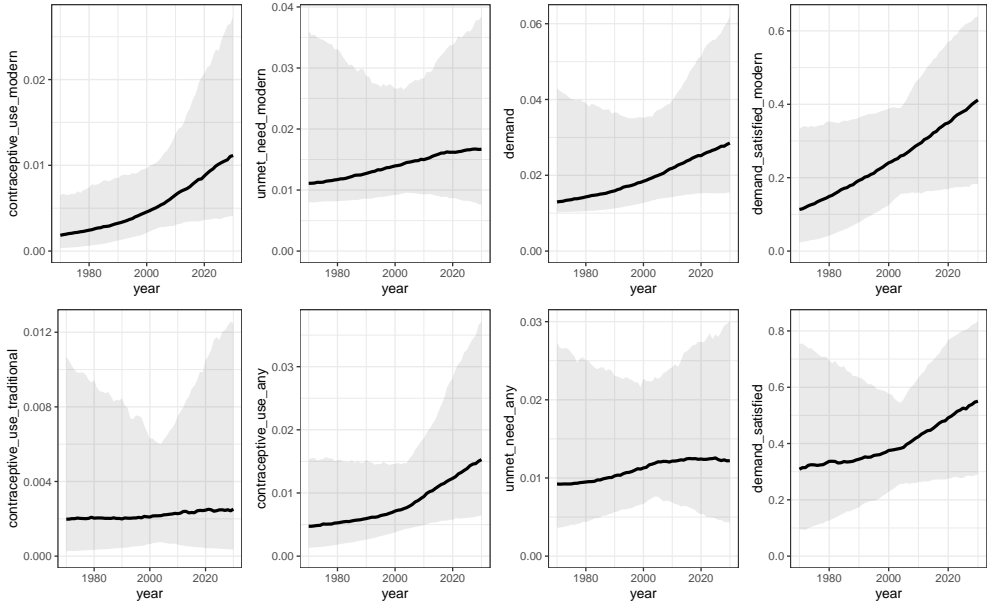

**Iran_Islamic_Republic_of unmarried**

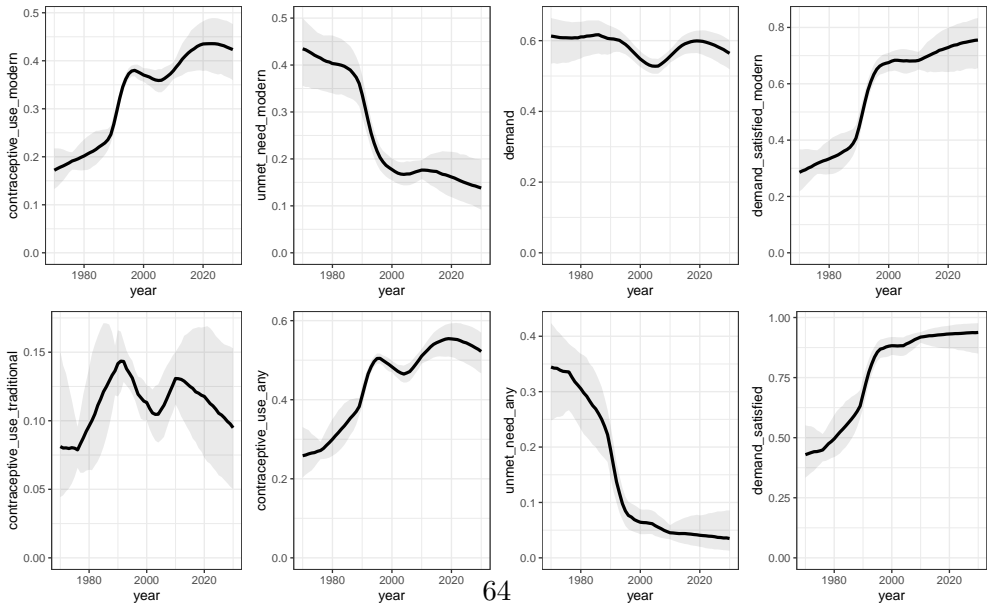

**Iran_Islamic_Republic_of all**

64

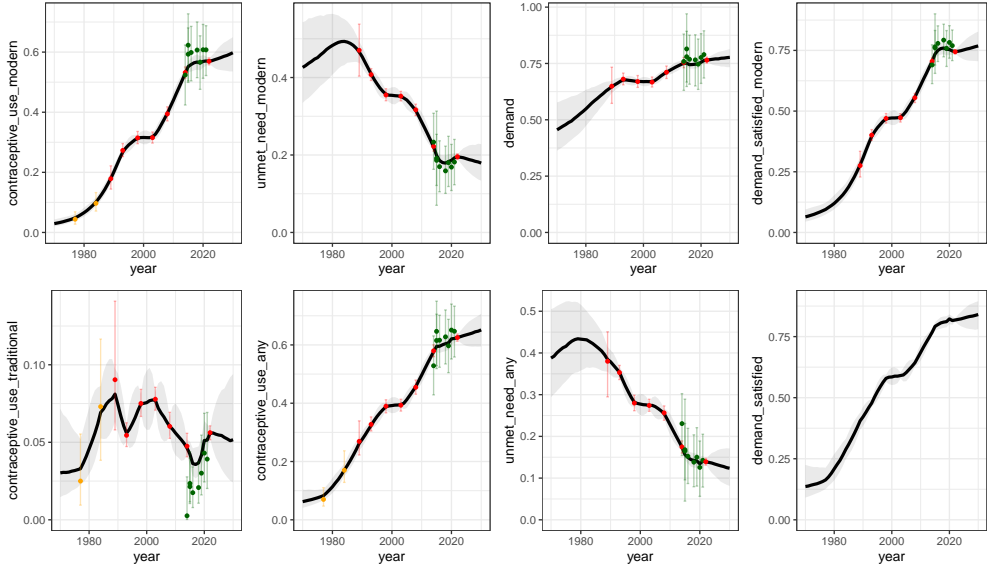

**Kenya married**

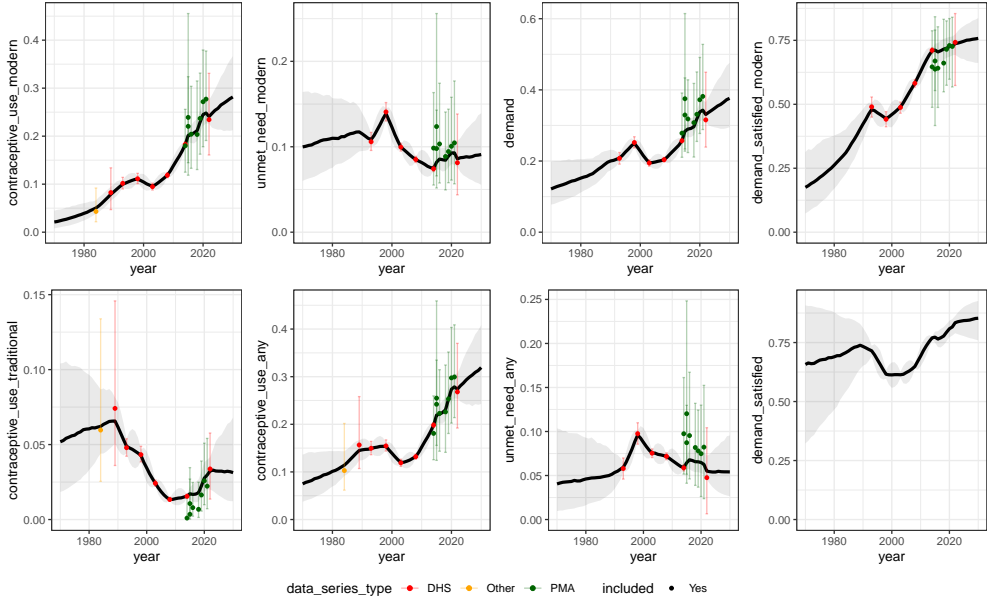

**Kenya unmarried**

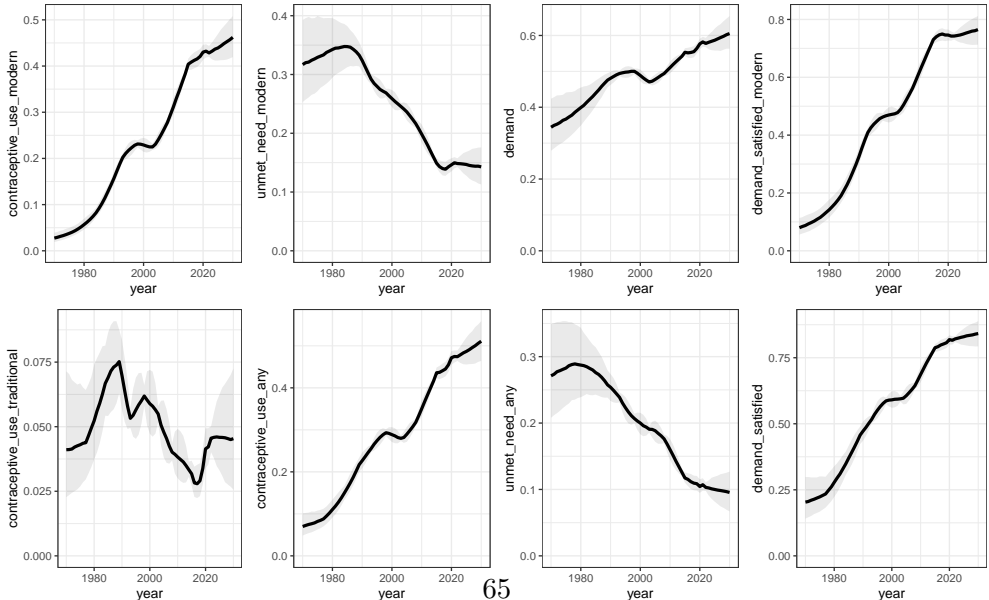

**Kenya all**

65

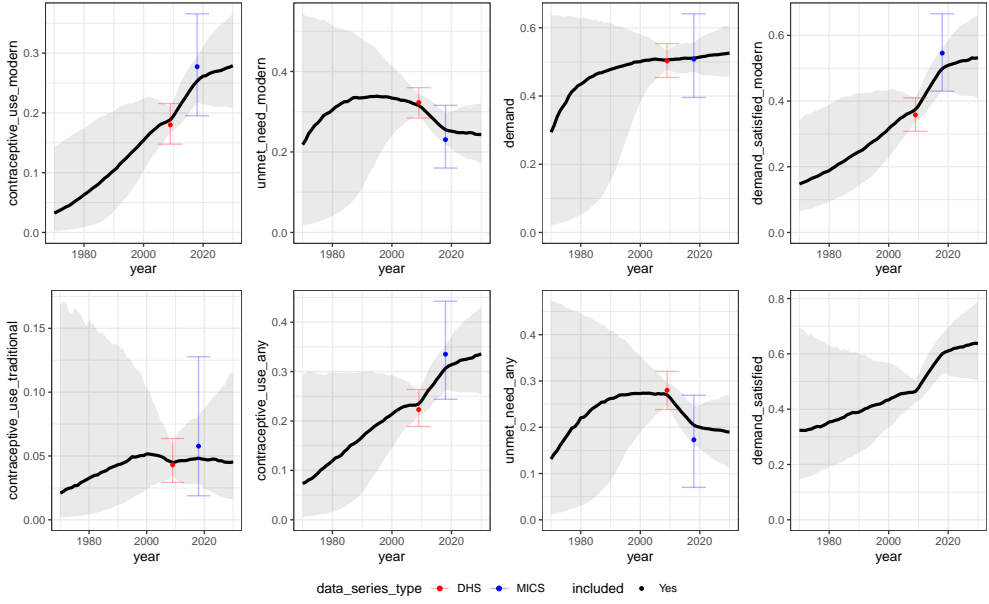

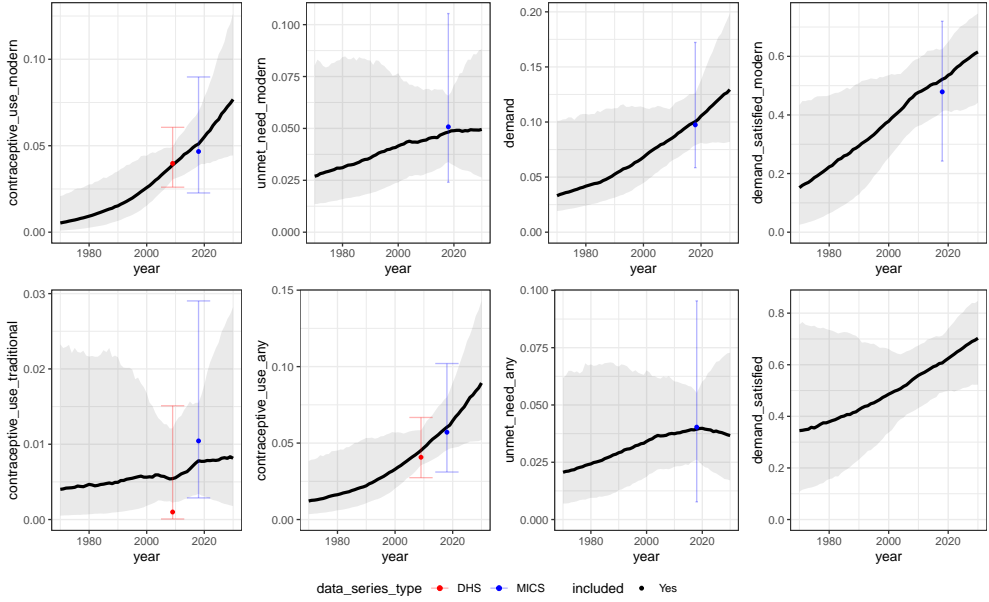

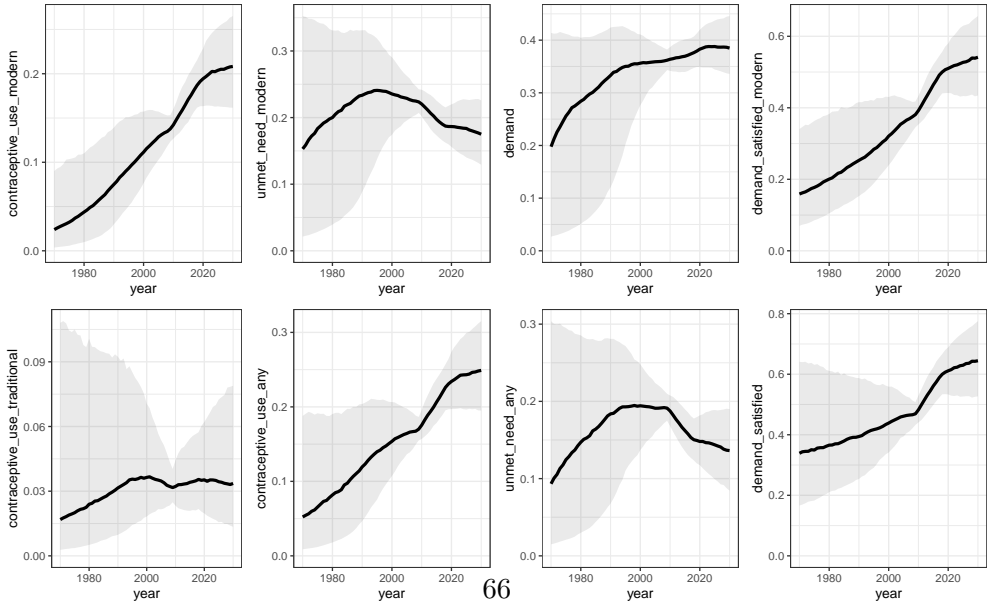

Kyrgyzstan married



Kyrgyzstan unmarried



Kyrgyzstan all

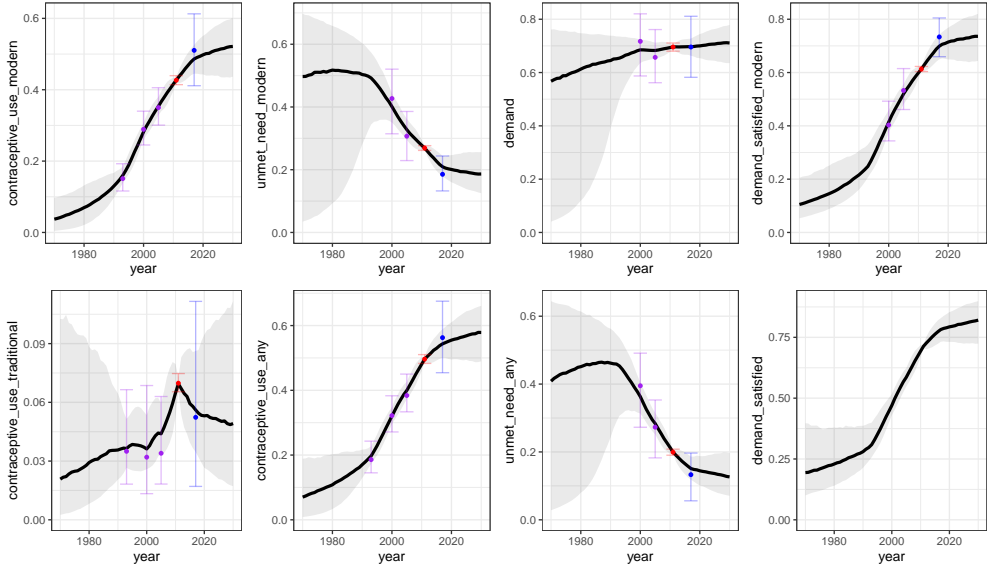

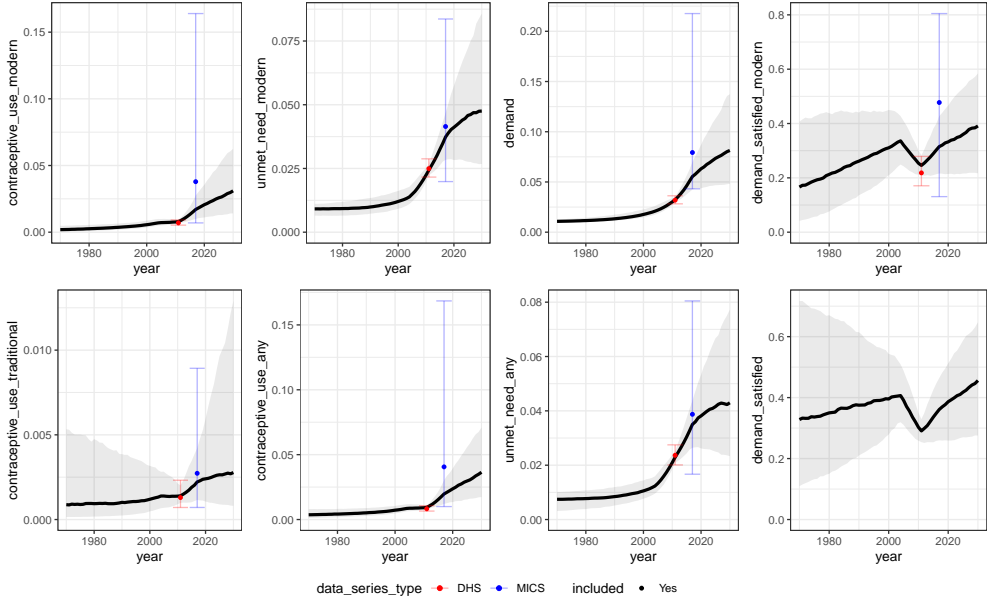

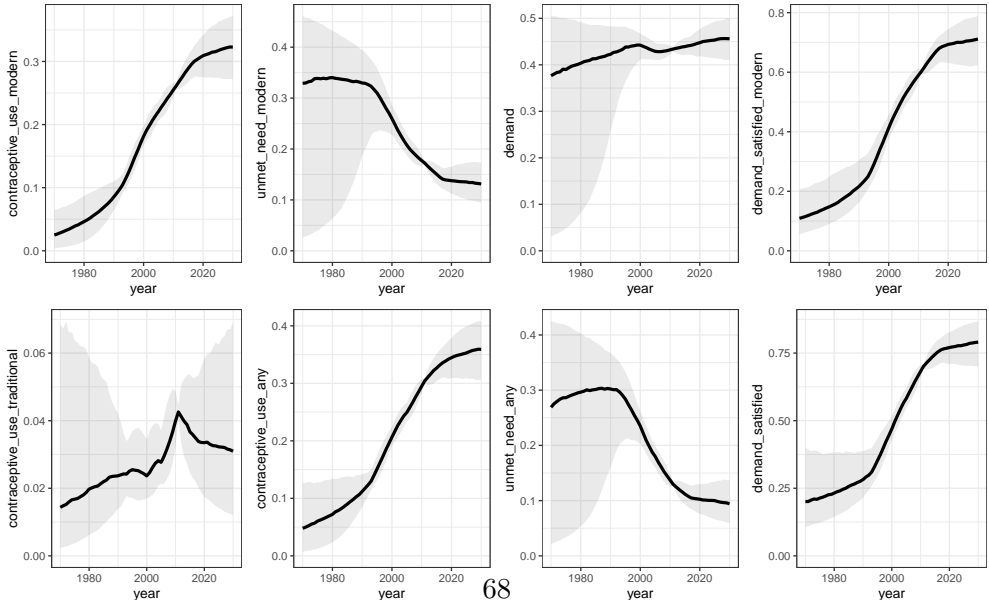

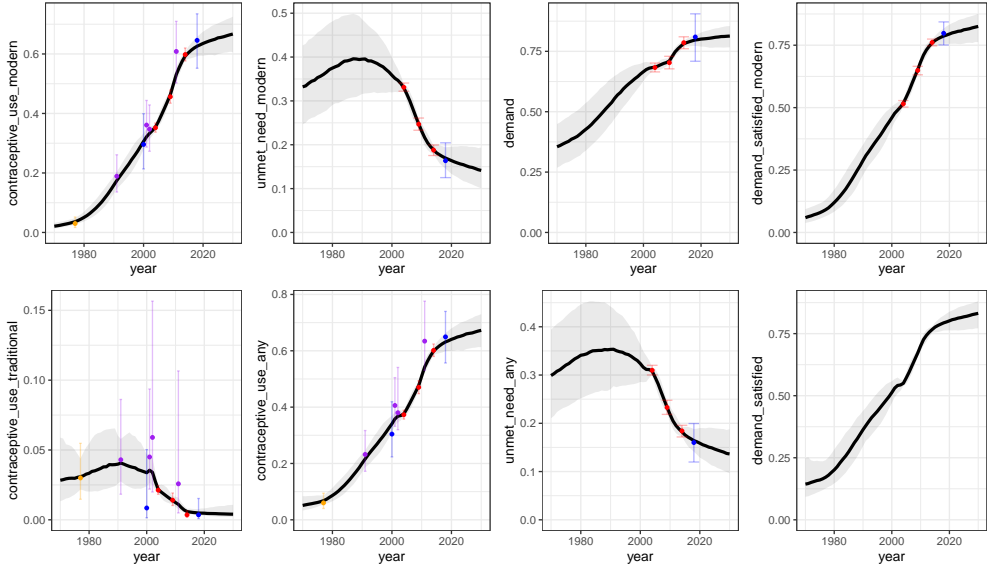

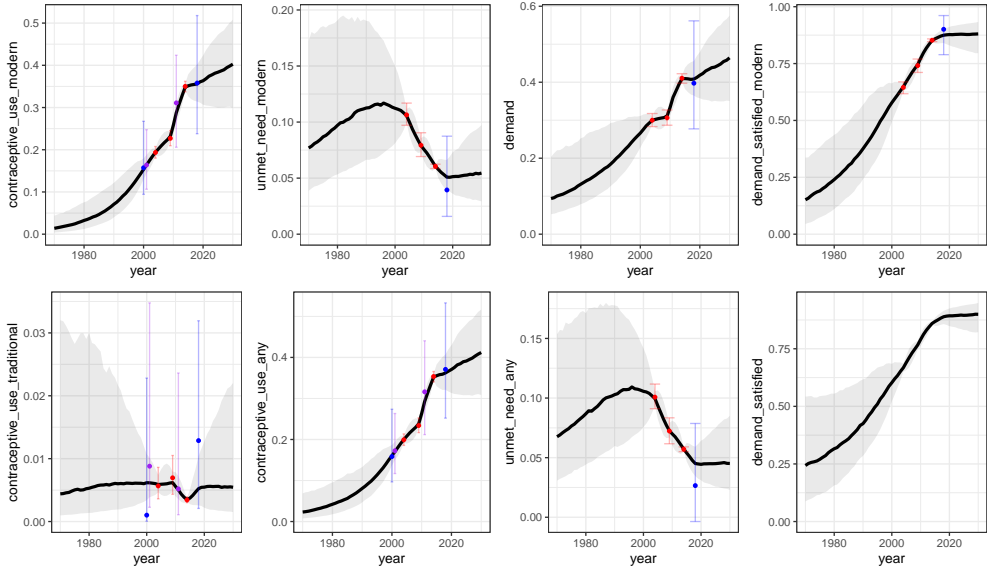

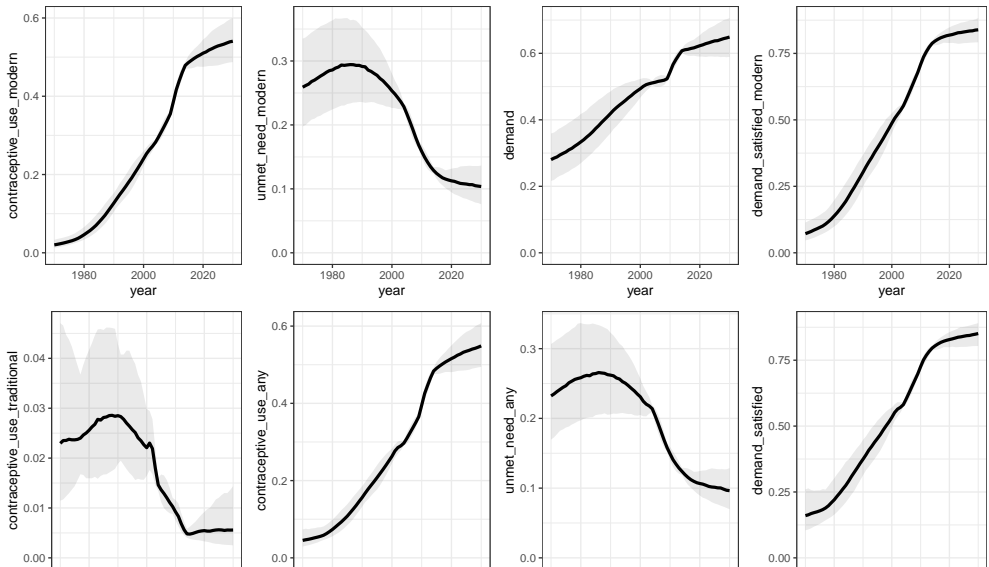

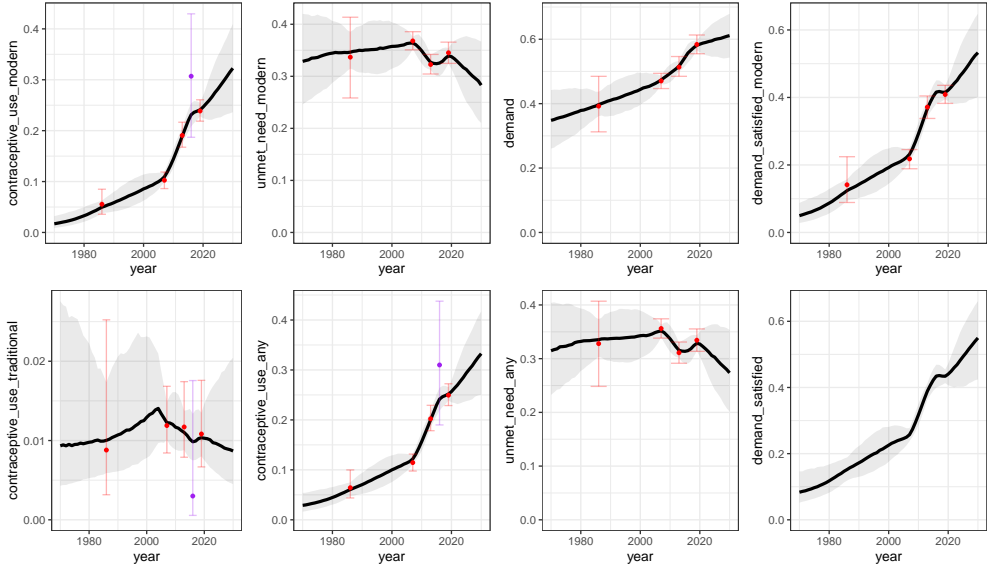

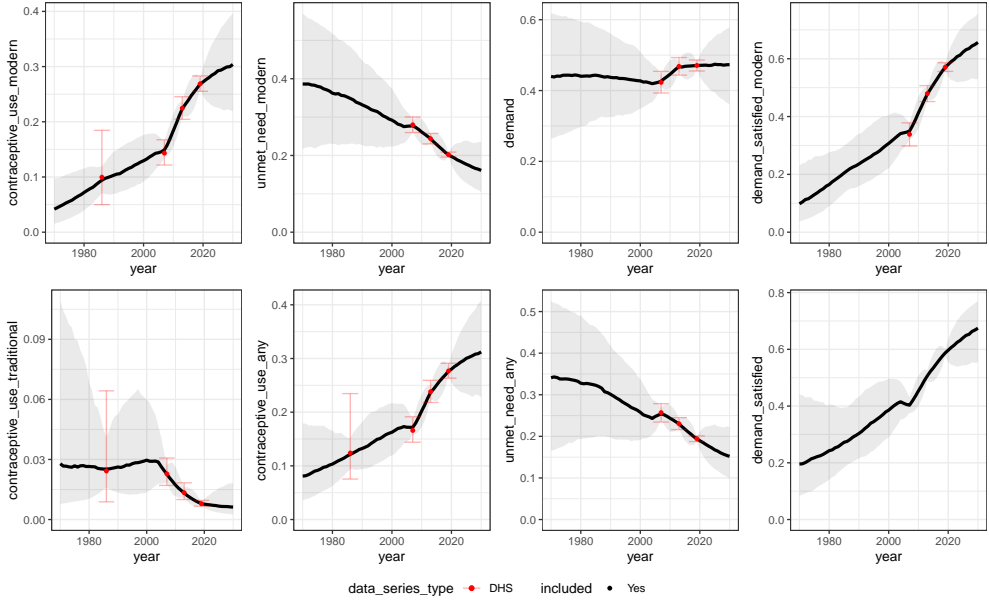

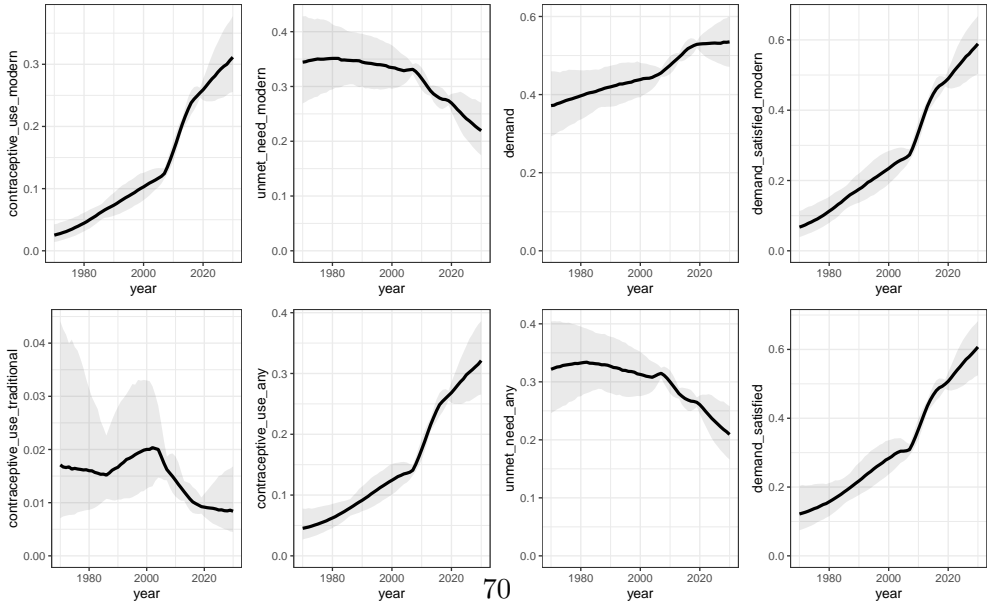

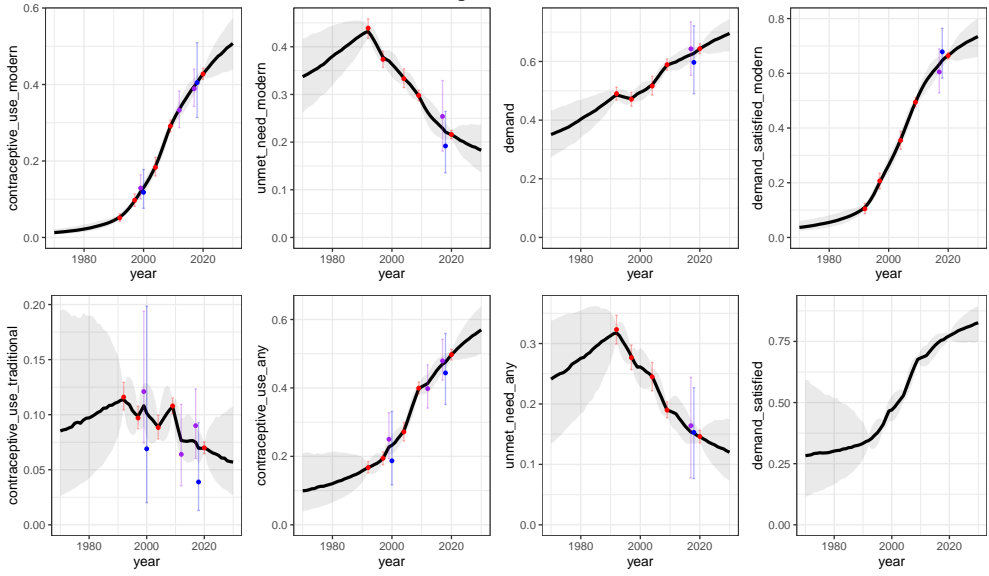

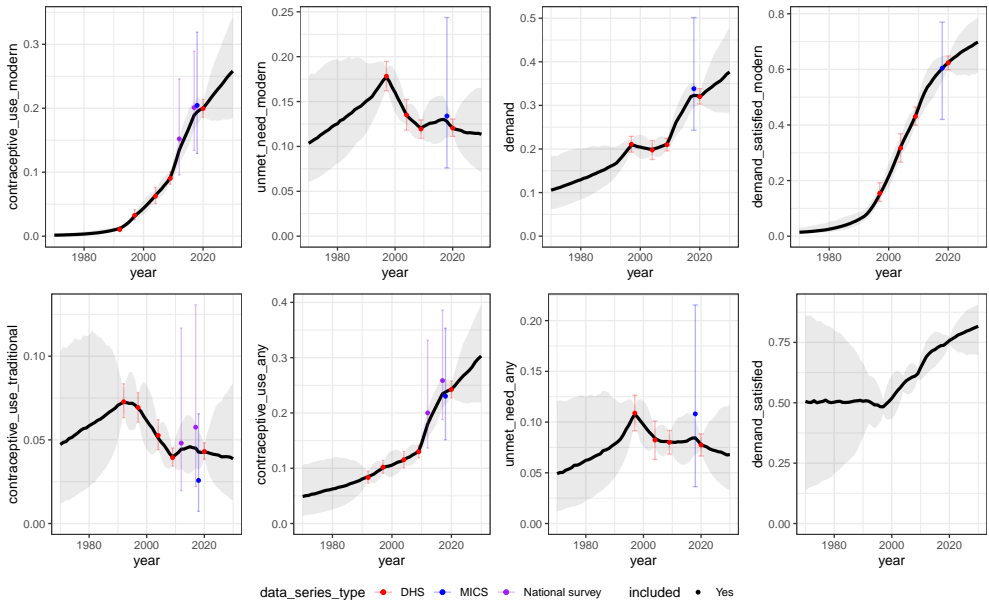

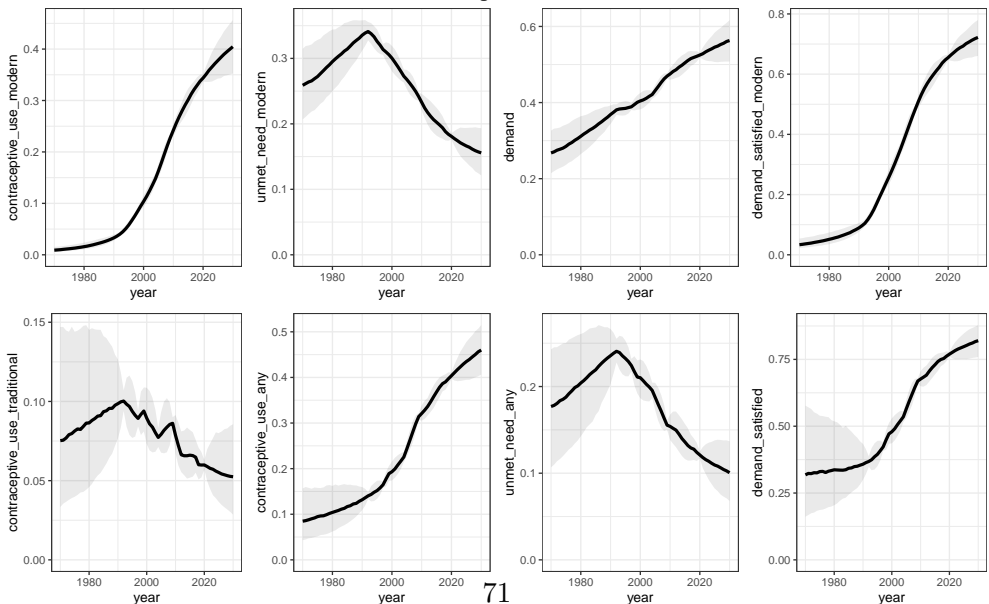

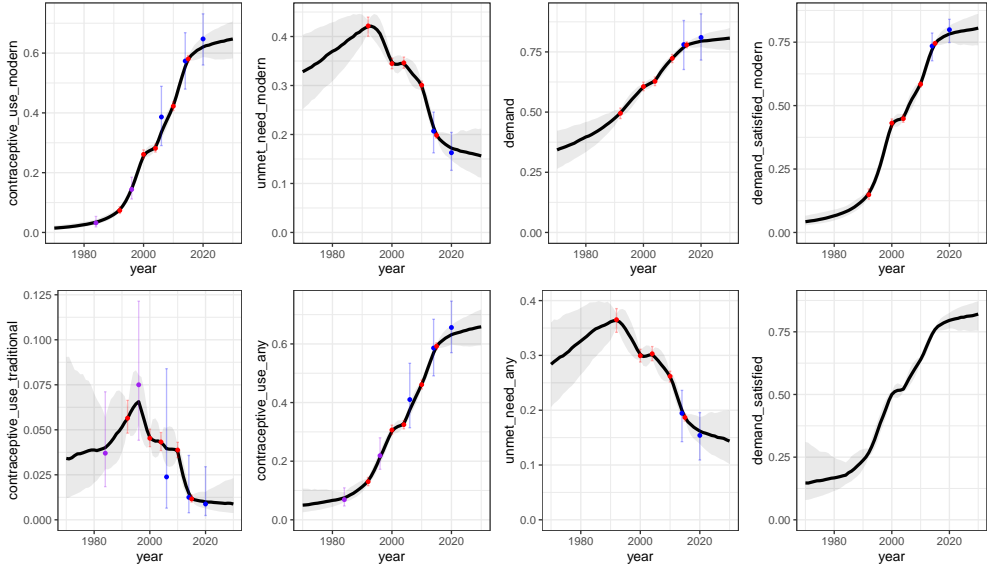

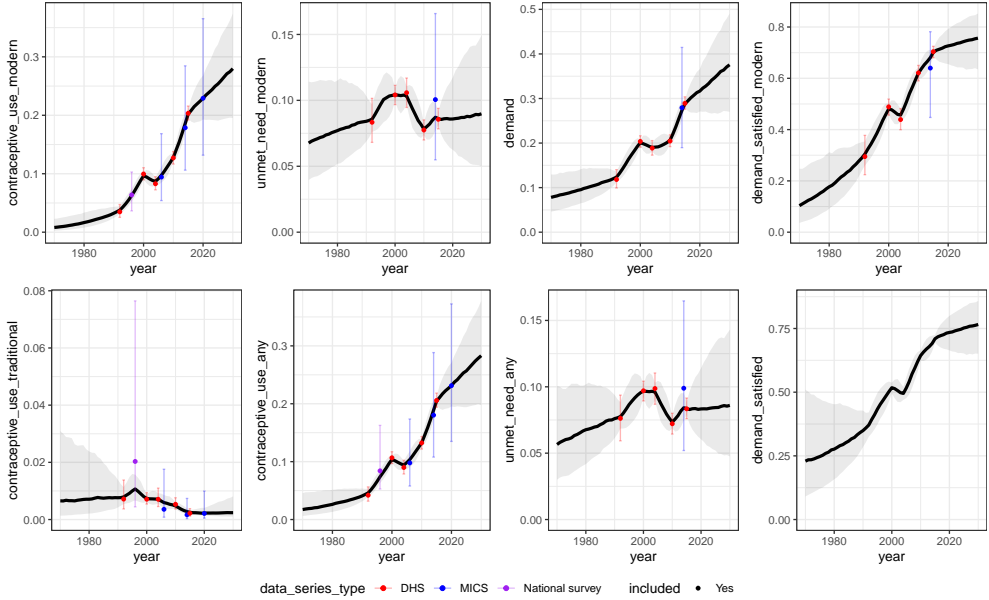

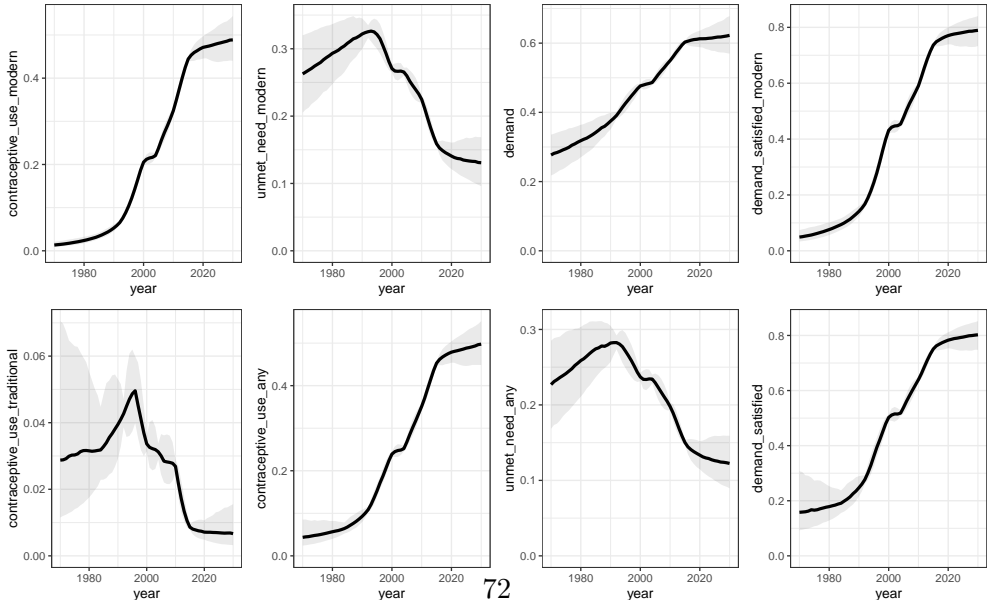

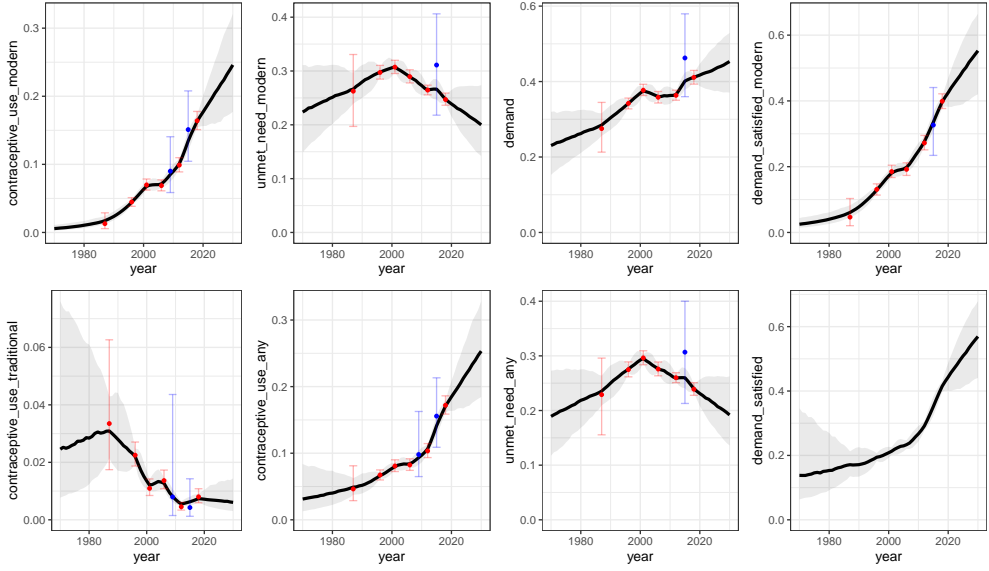

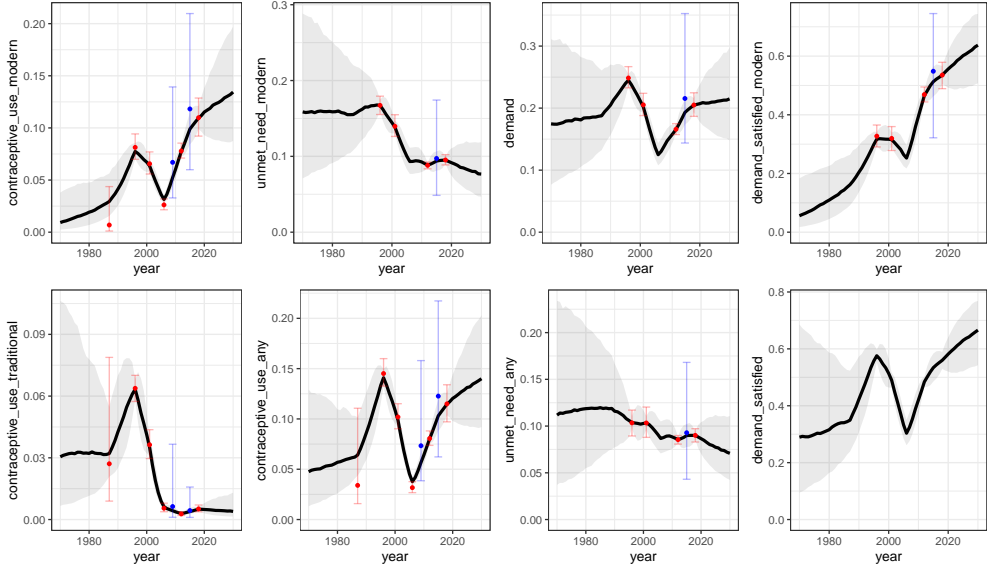

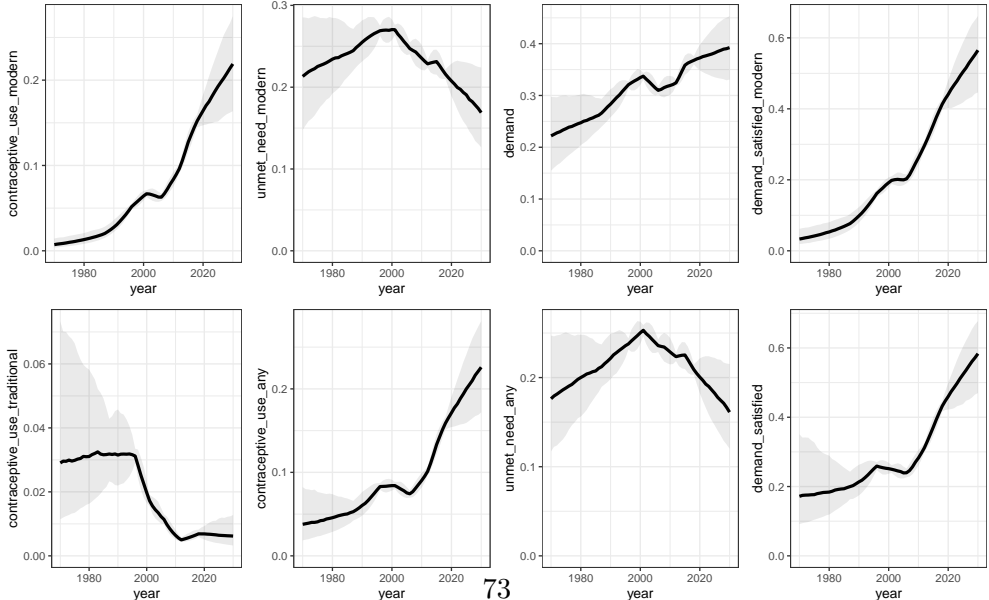

73

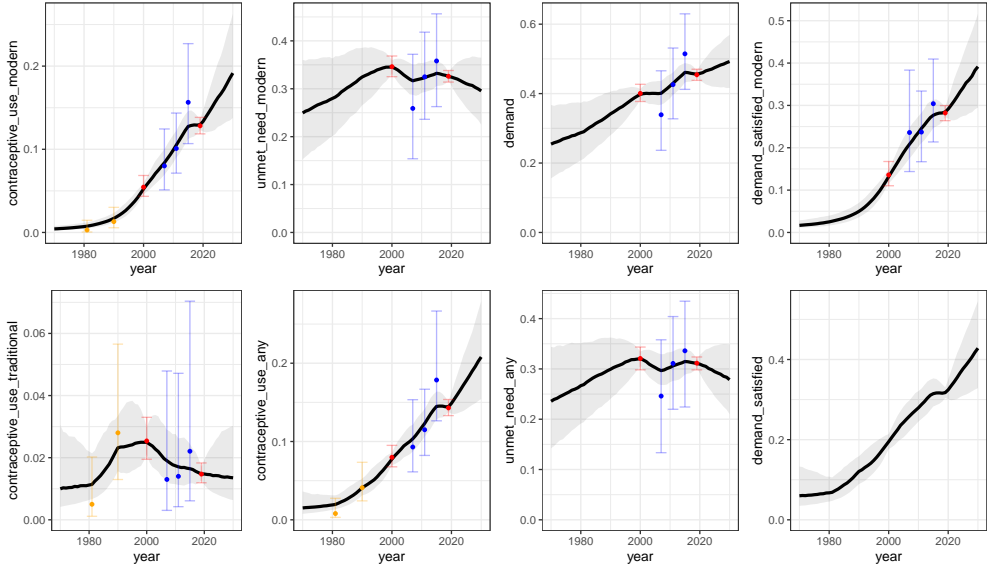

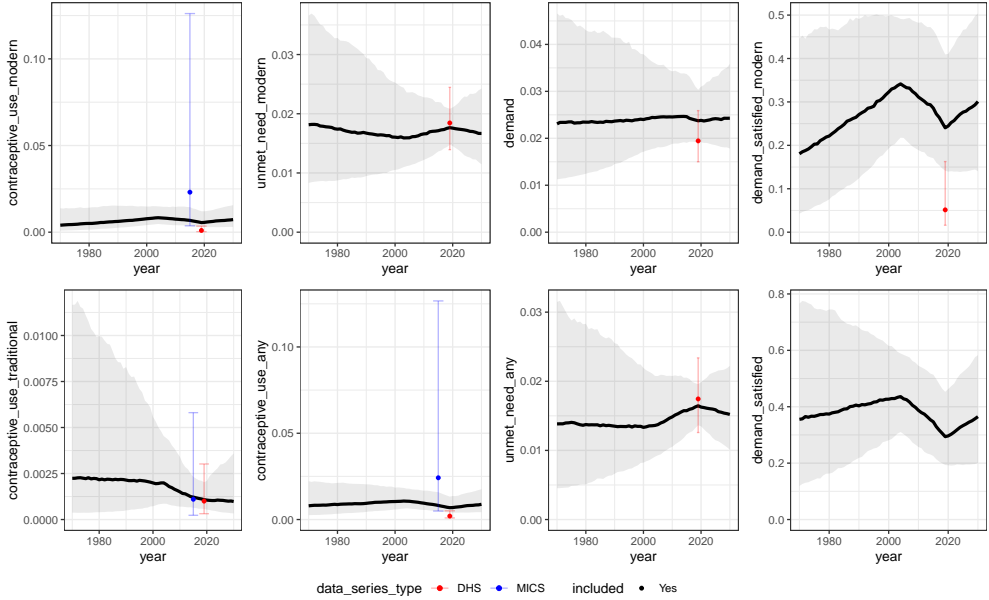

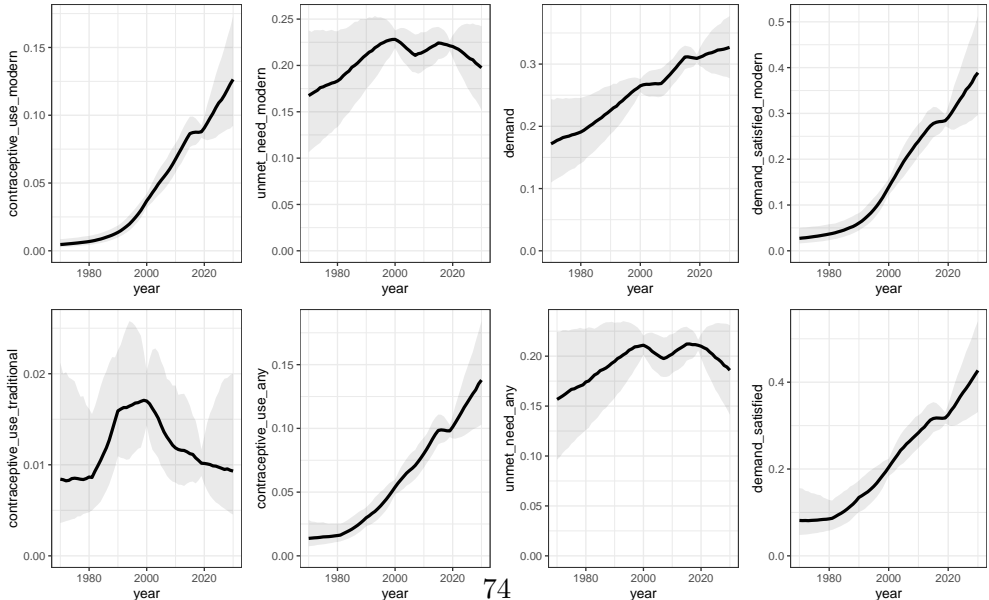

74

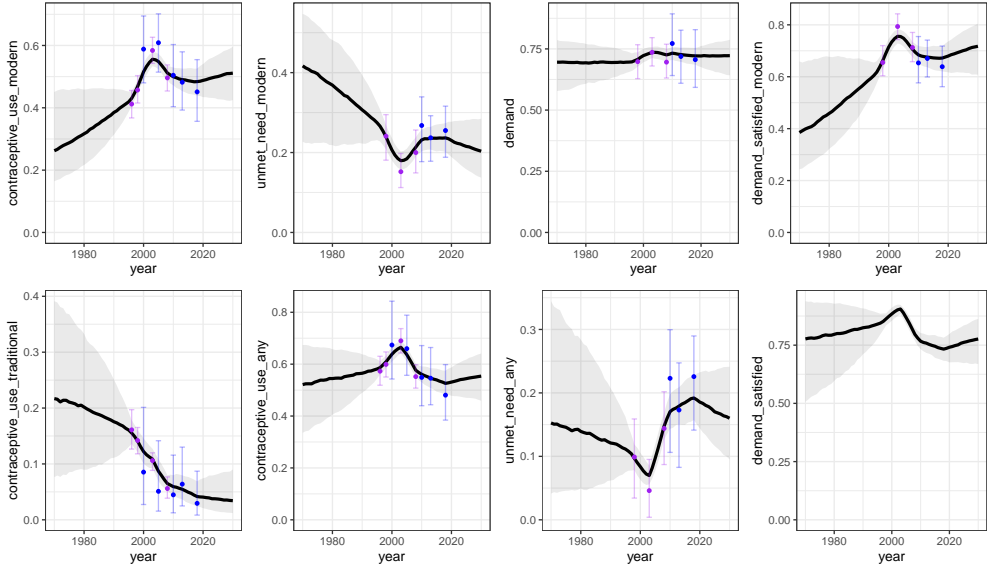

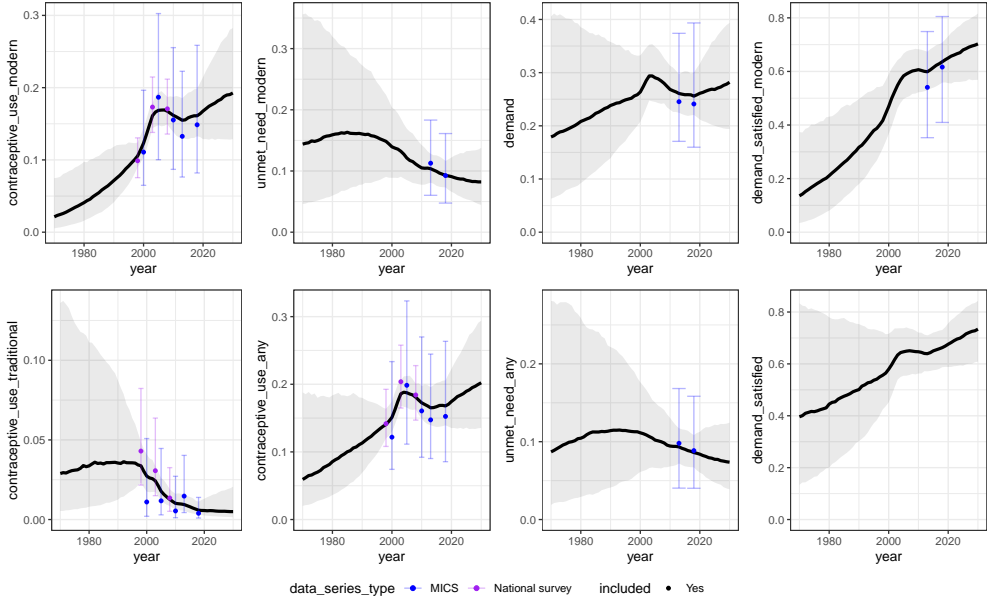

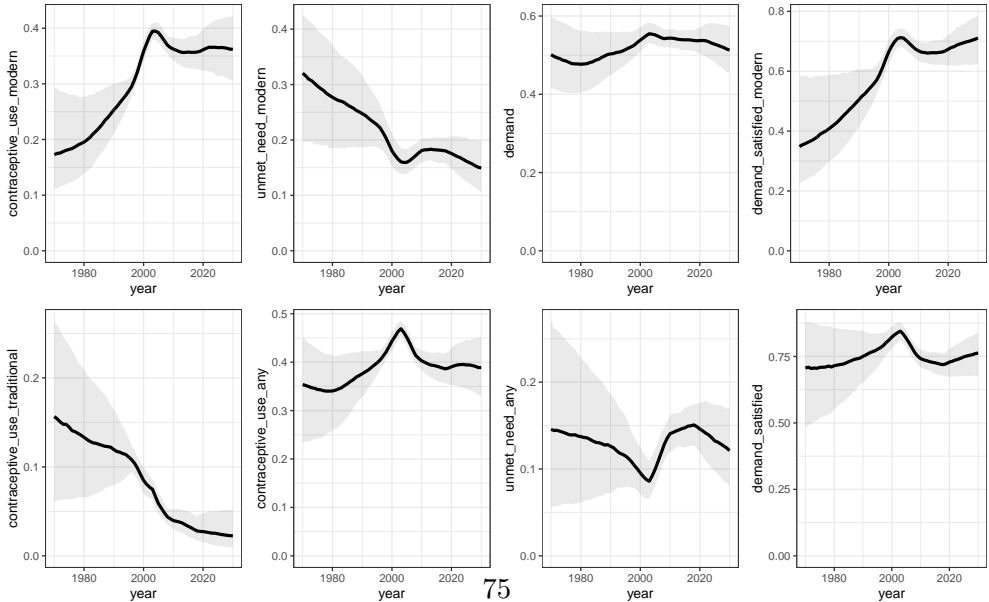

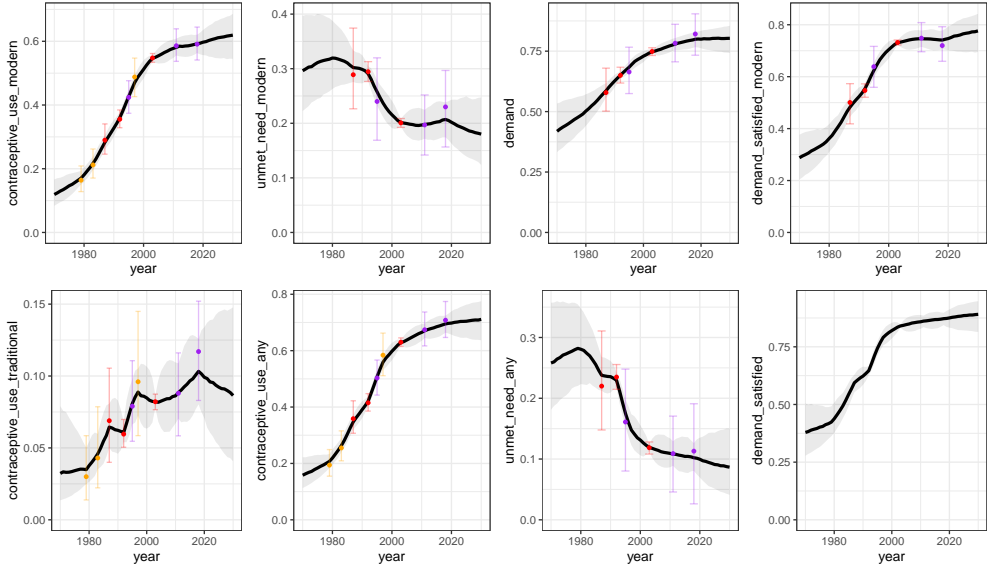

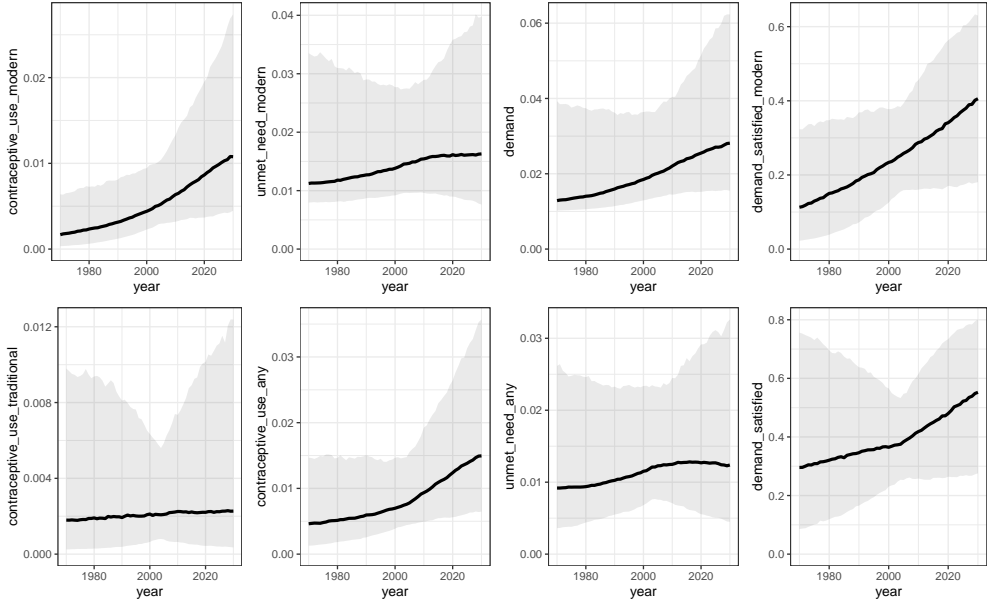

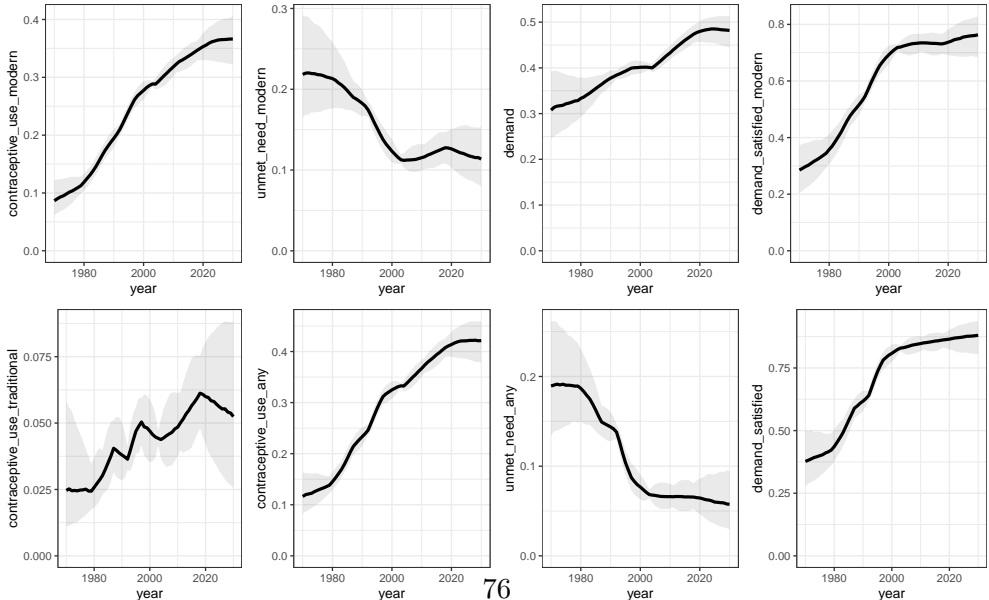

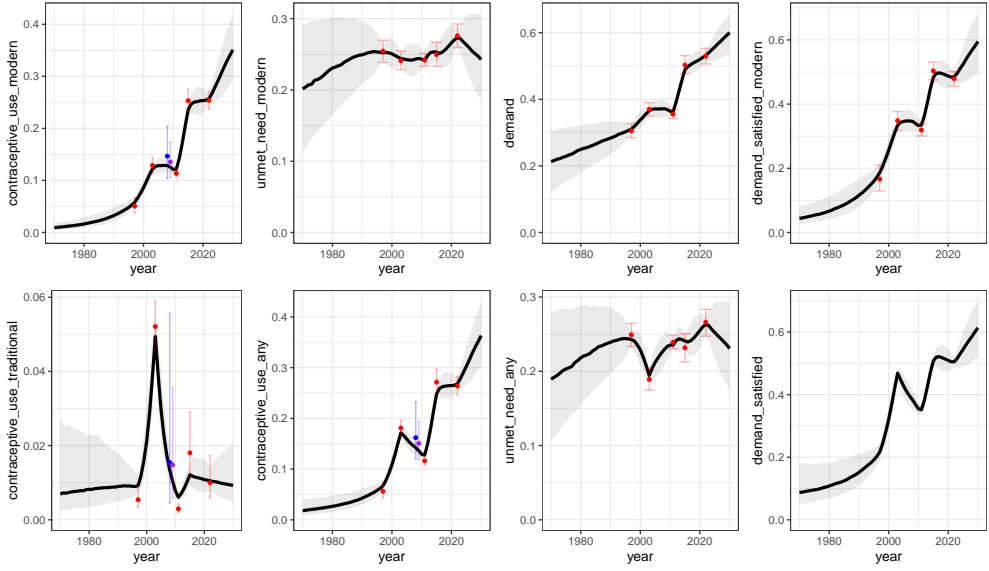

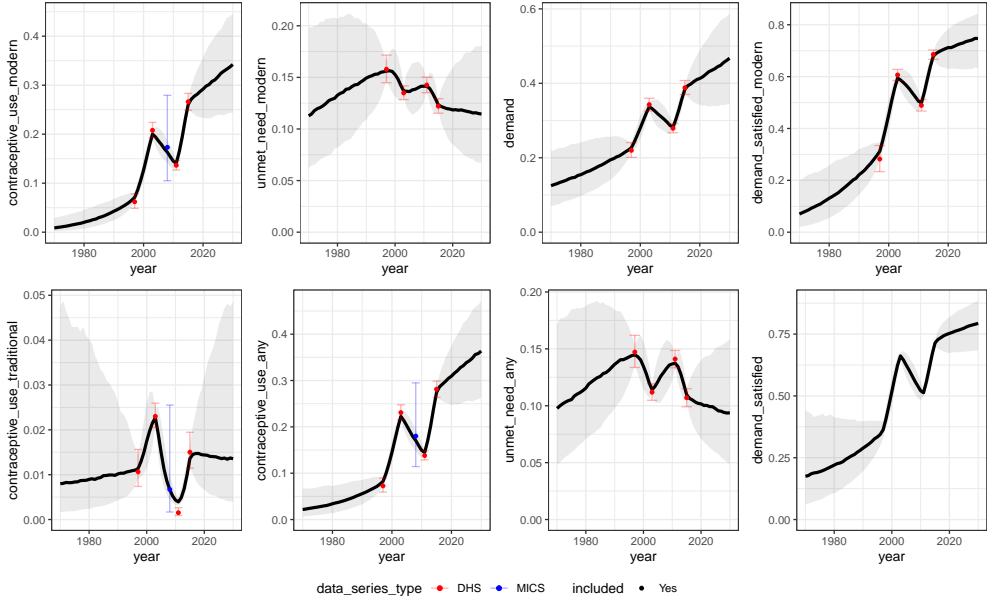

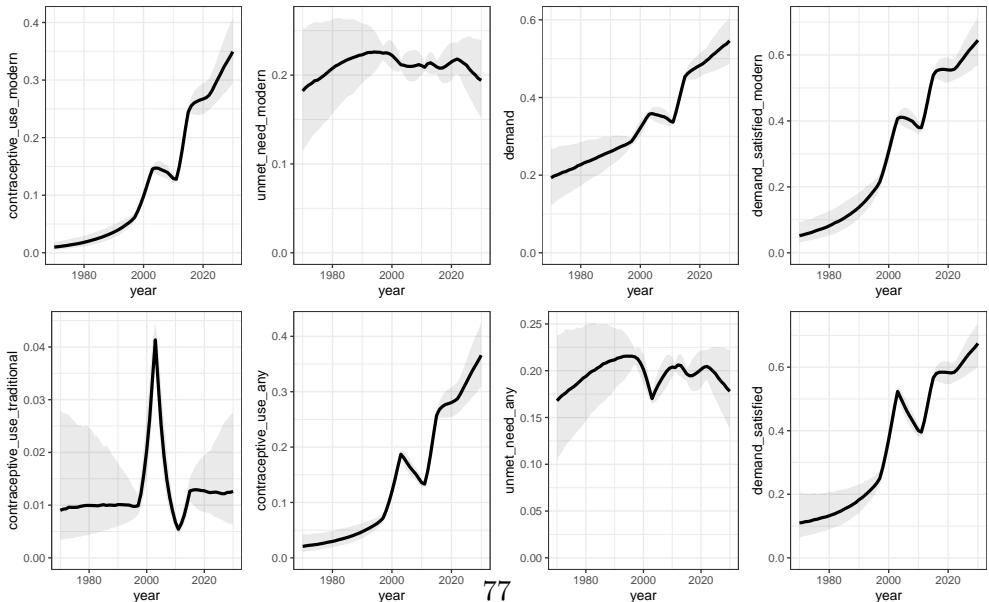

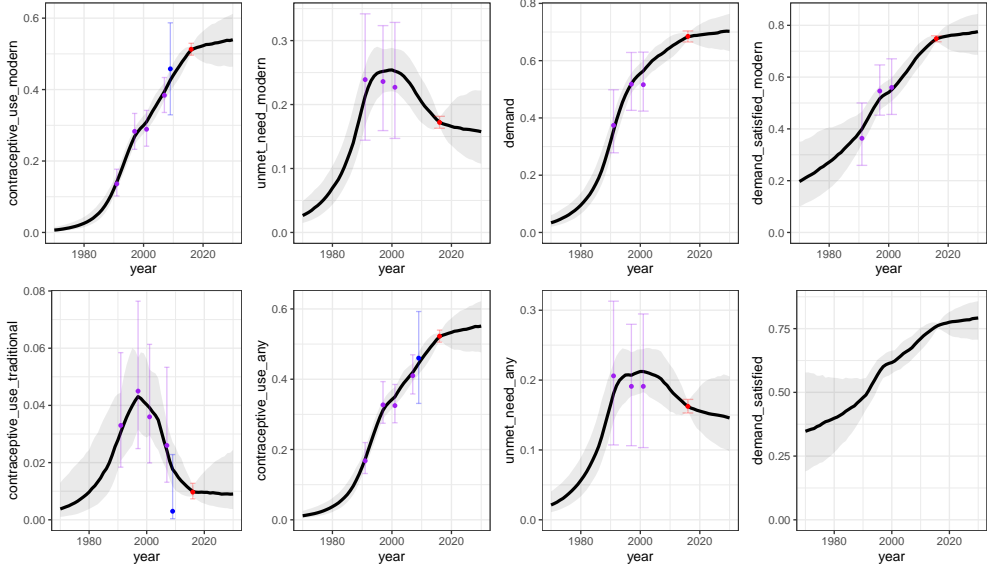

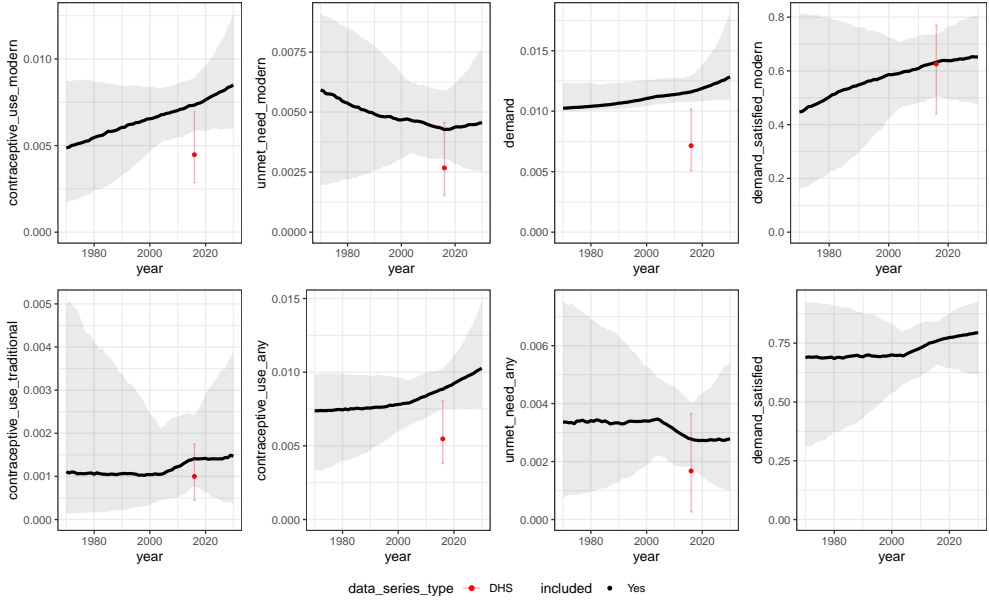

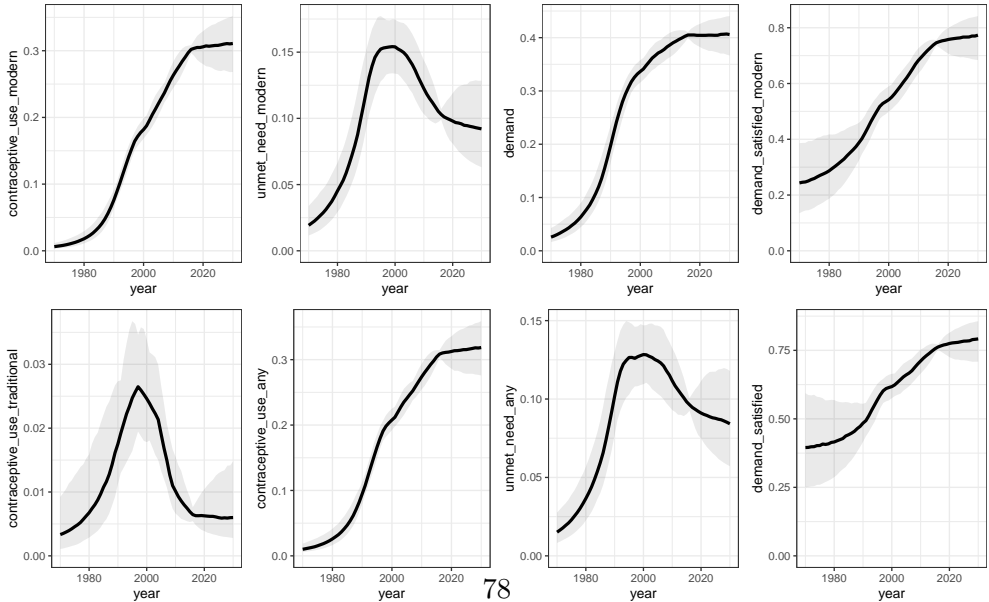

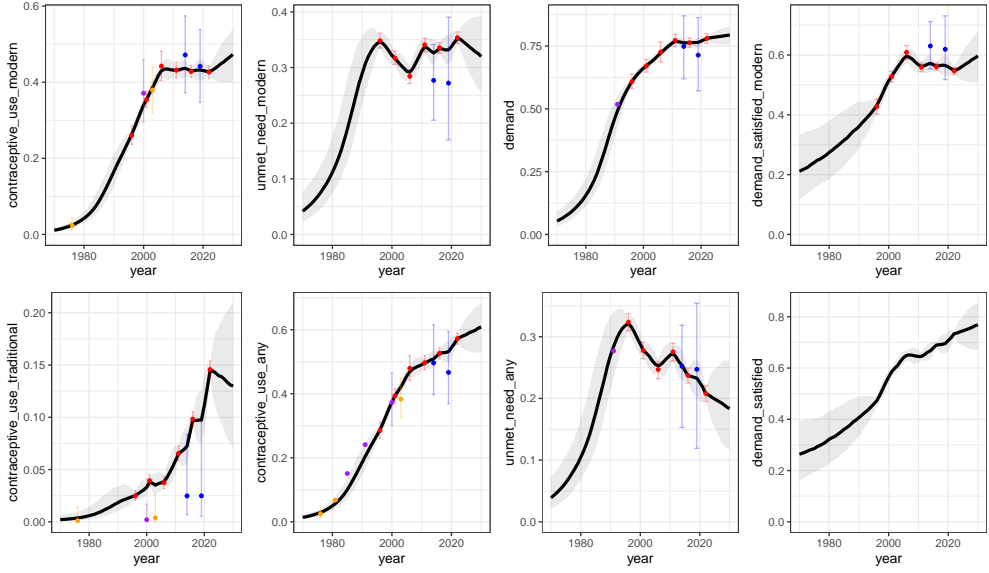

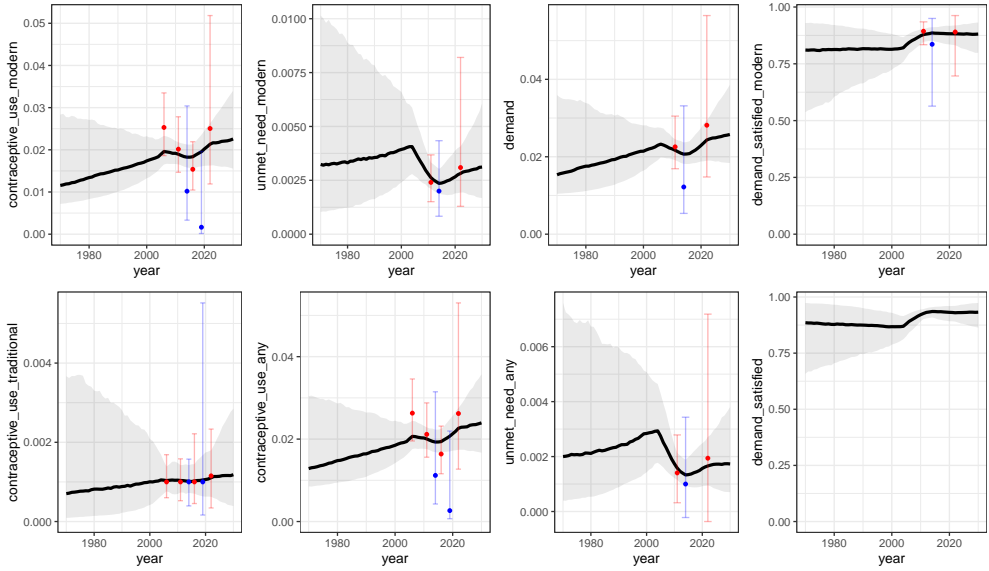

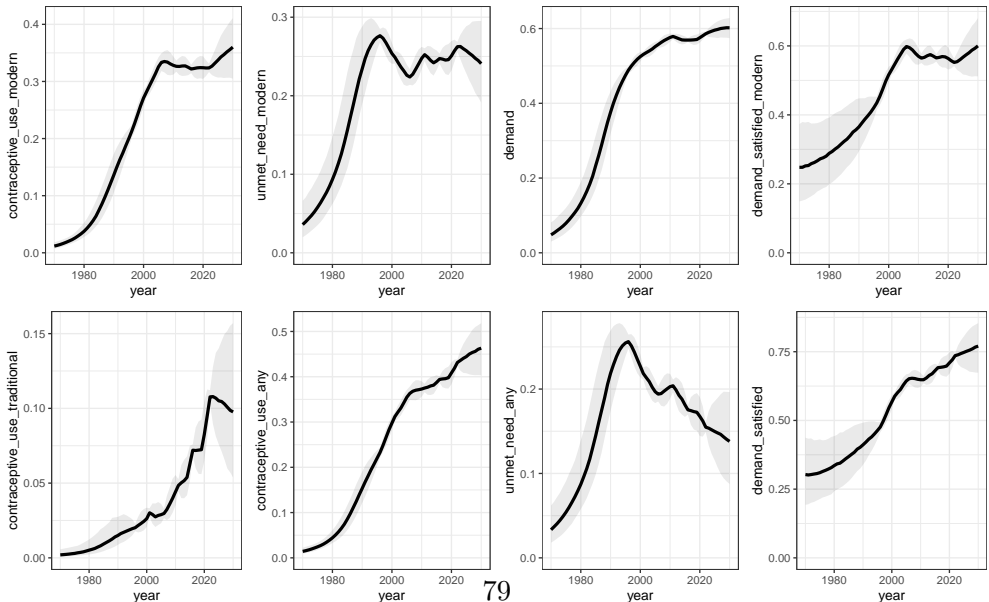

79

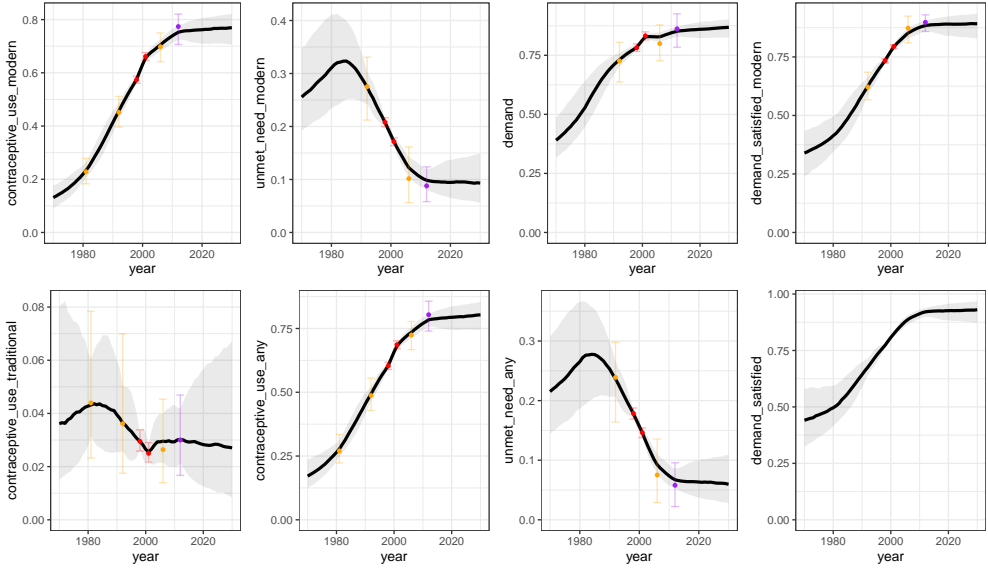

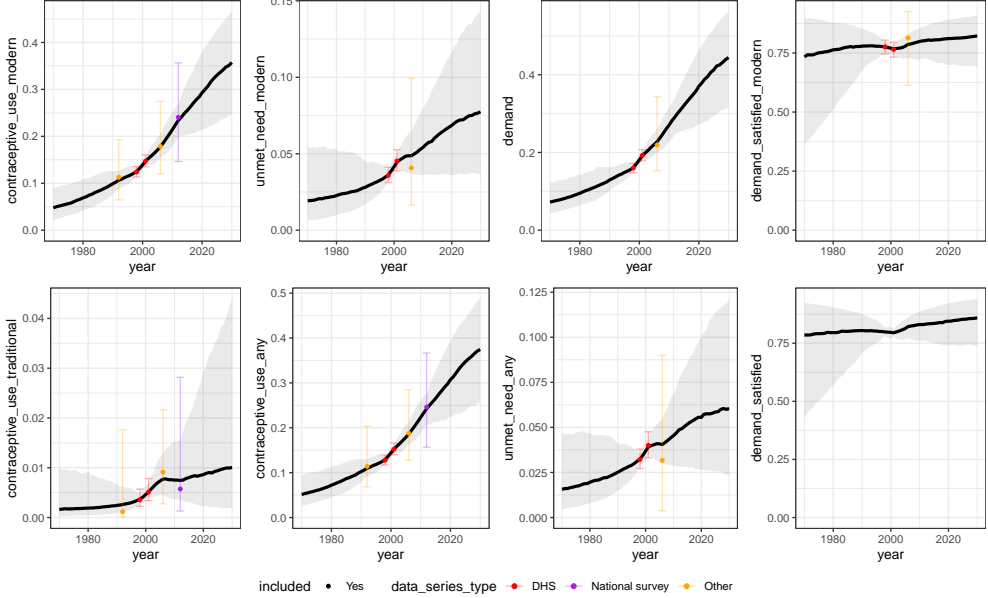

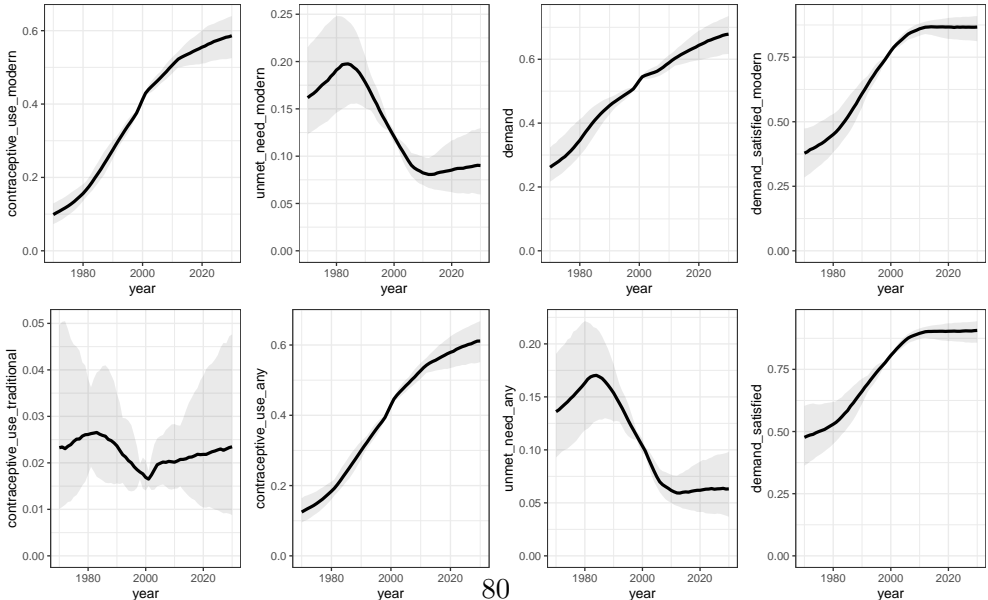

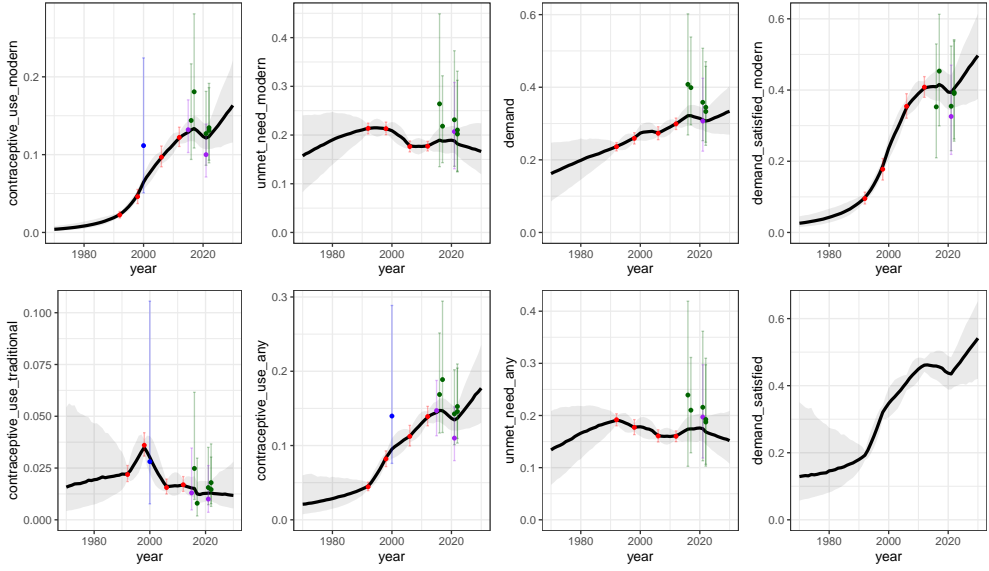

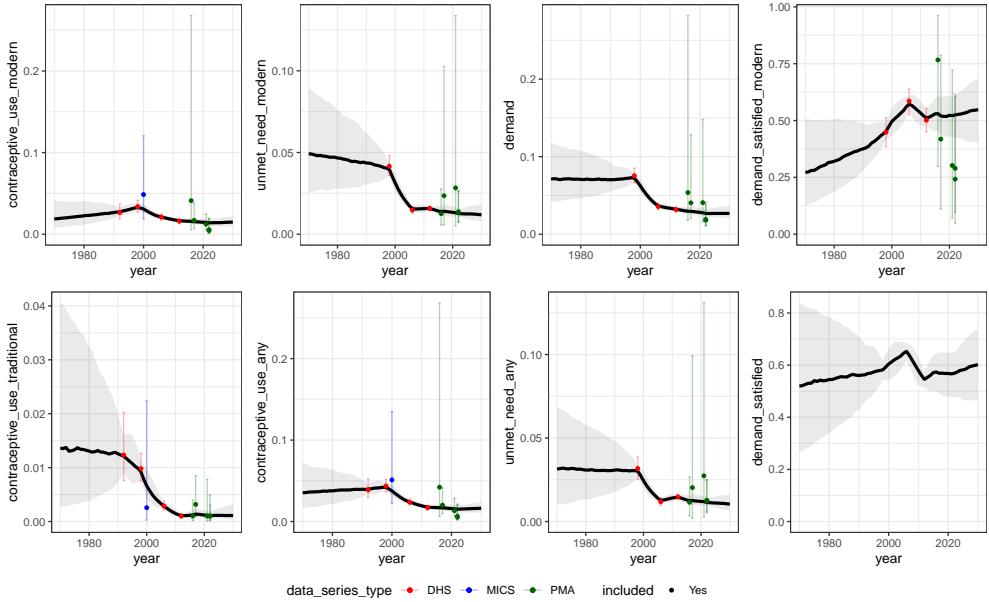

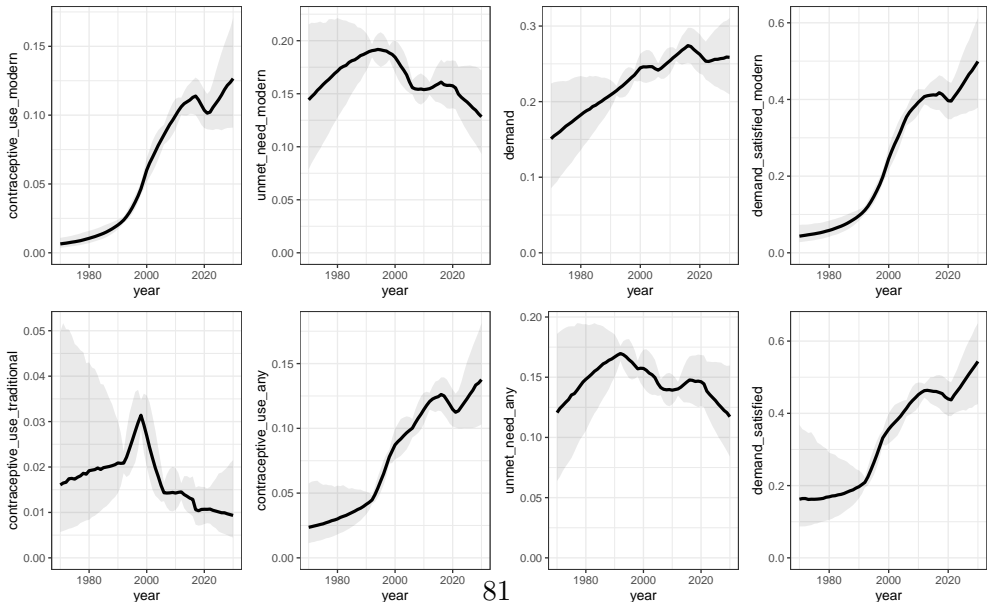

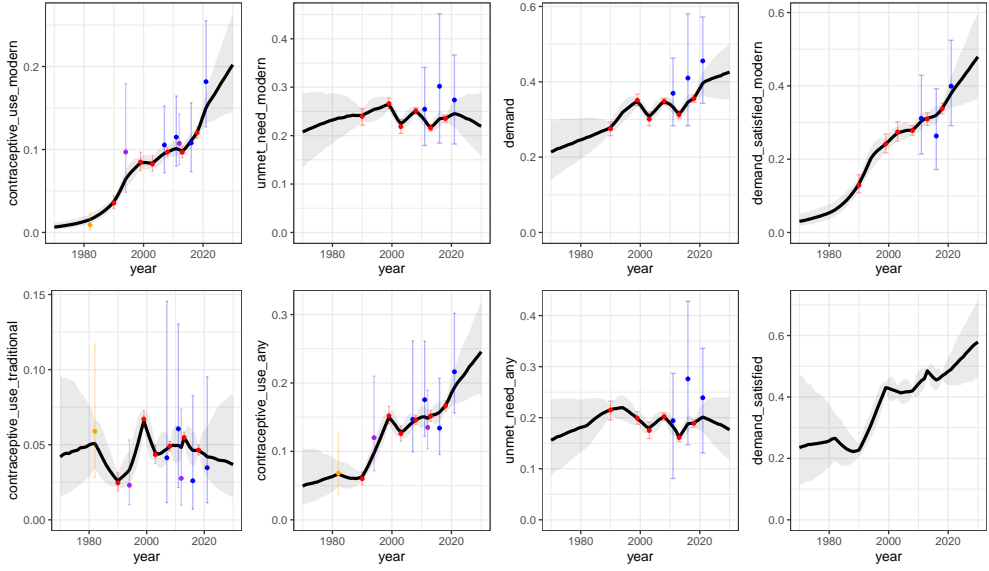

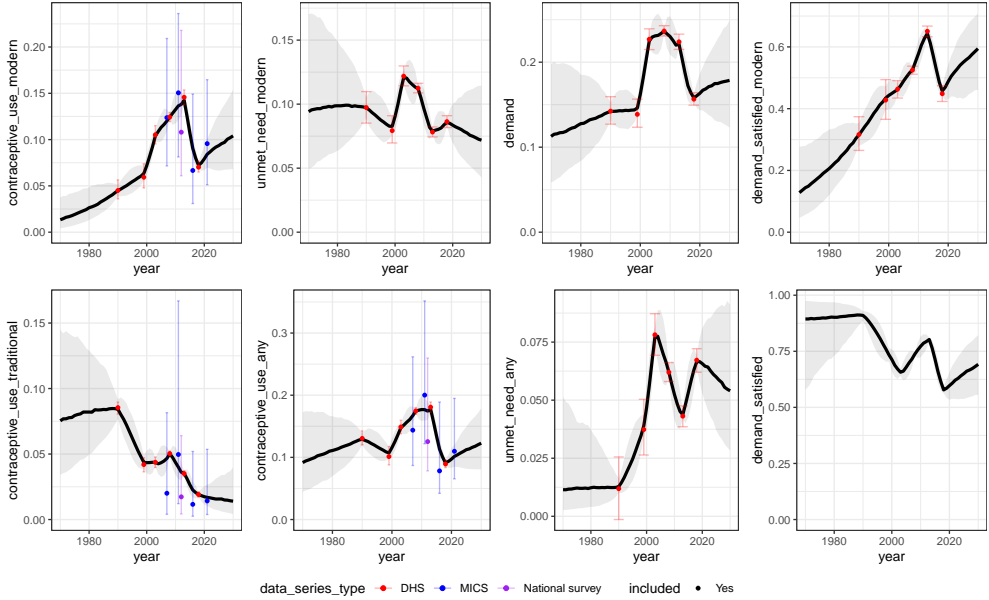

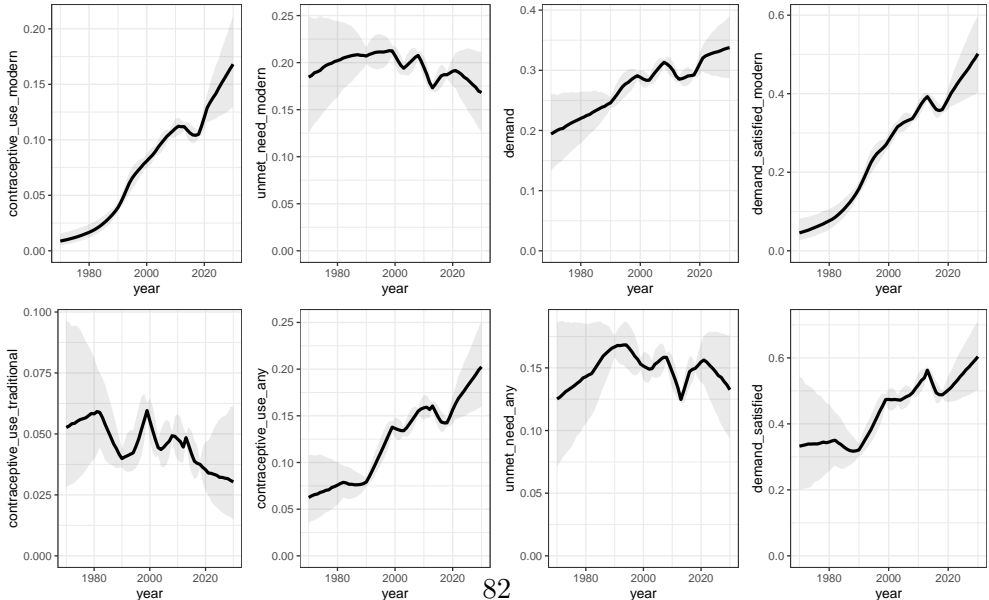

**Pakistan married**

**Pakistan unmarried**

**Pakistan all**

83

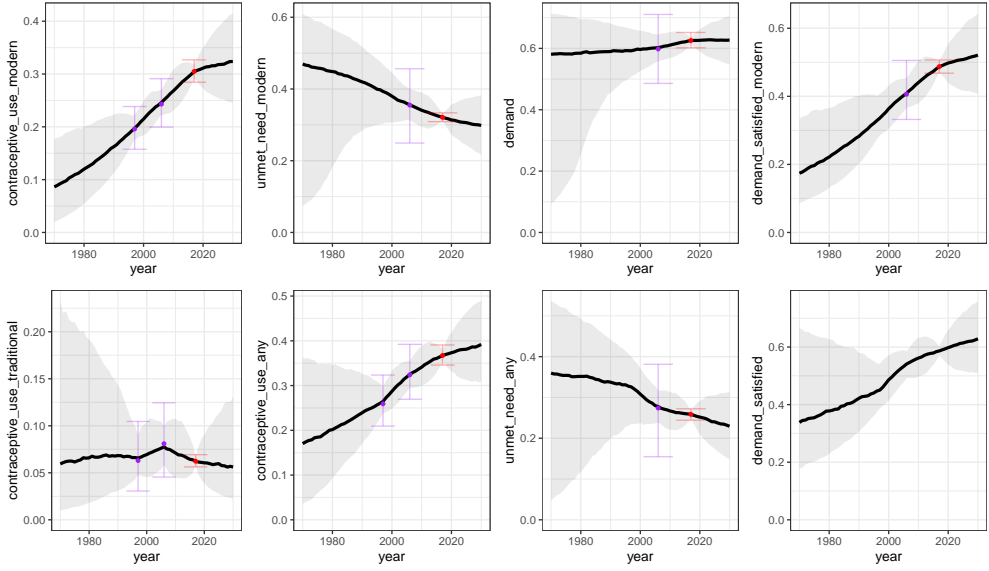

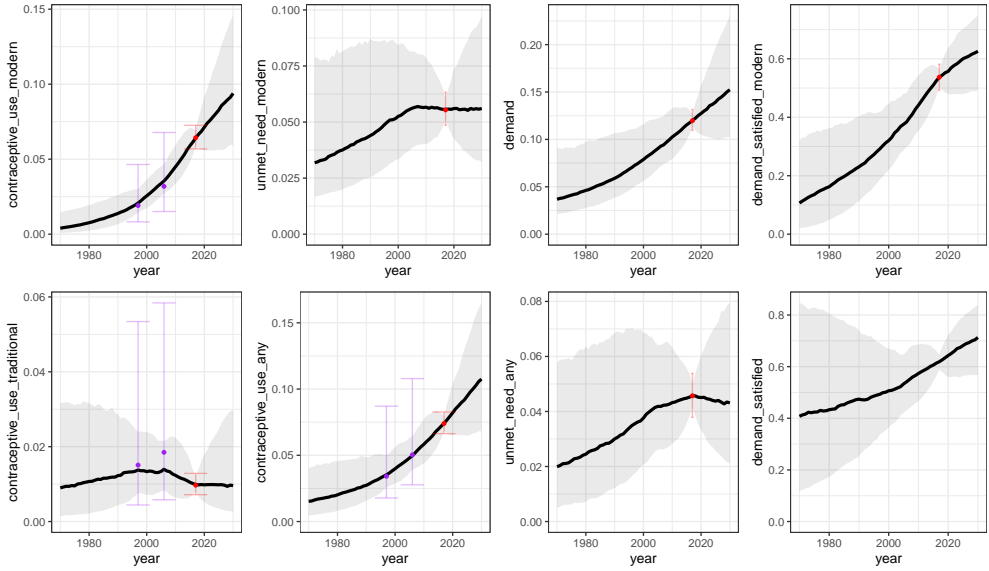

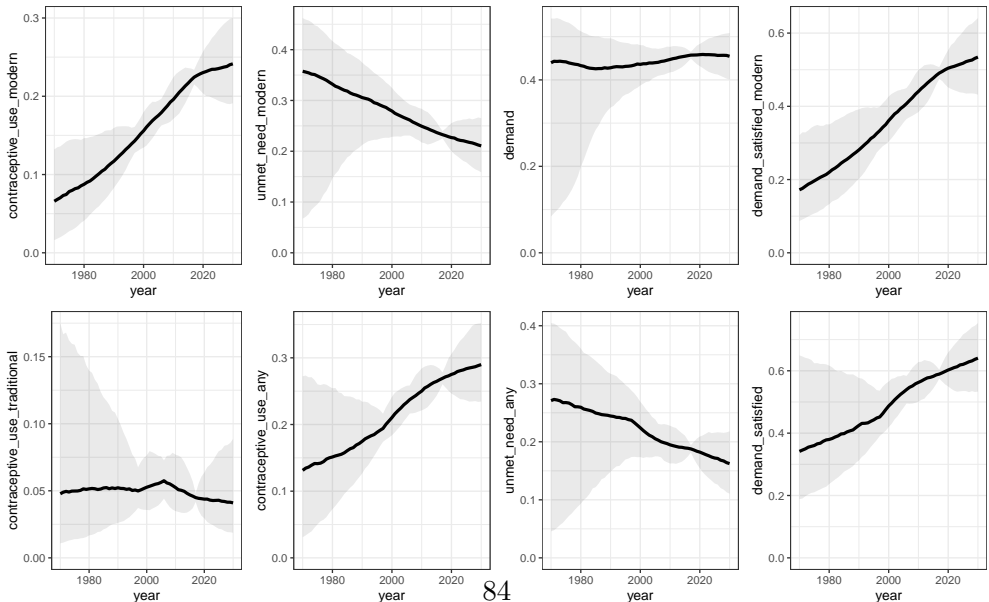

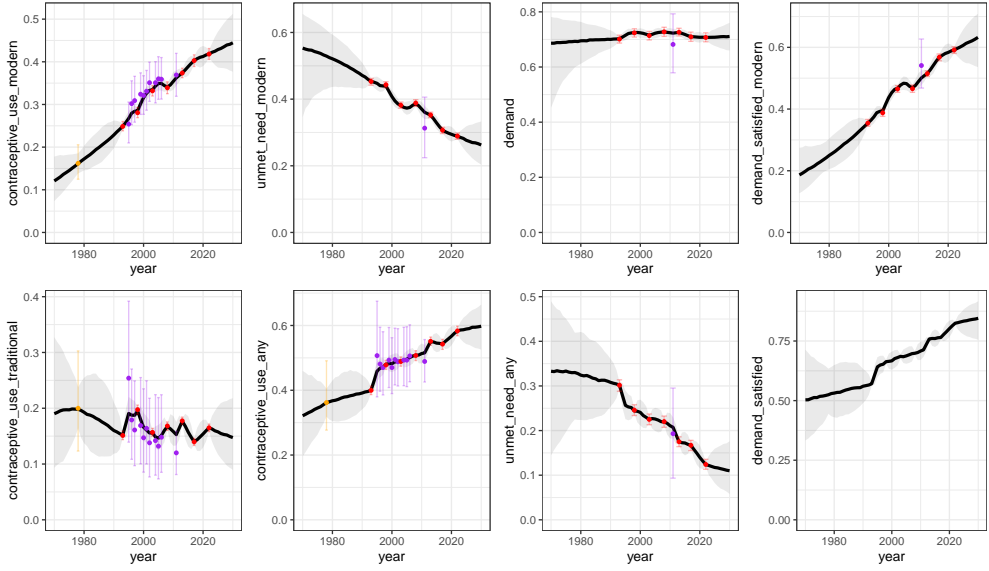

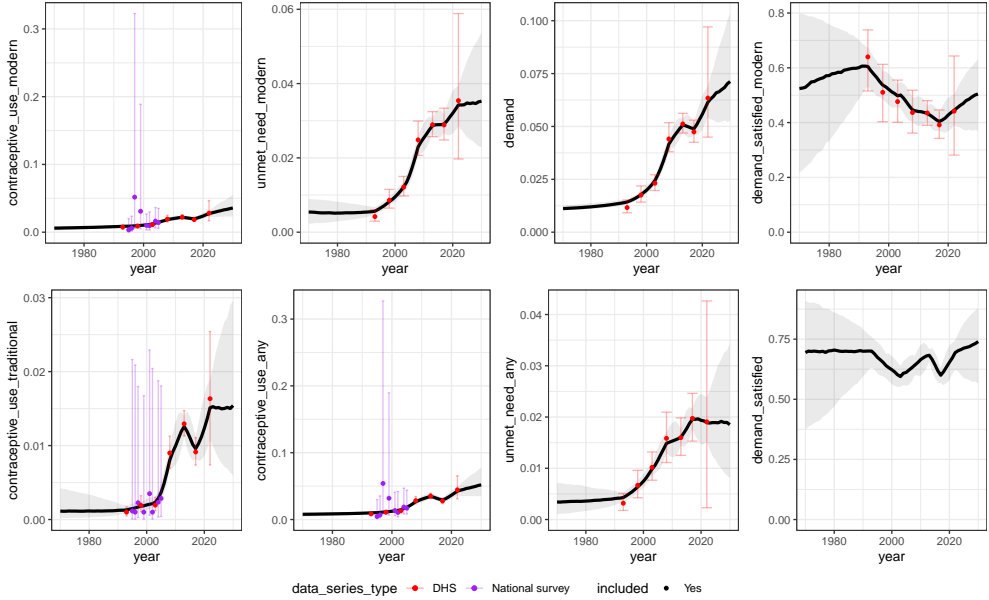

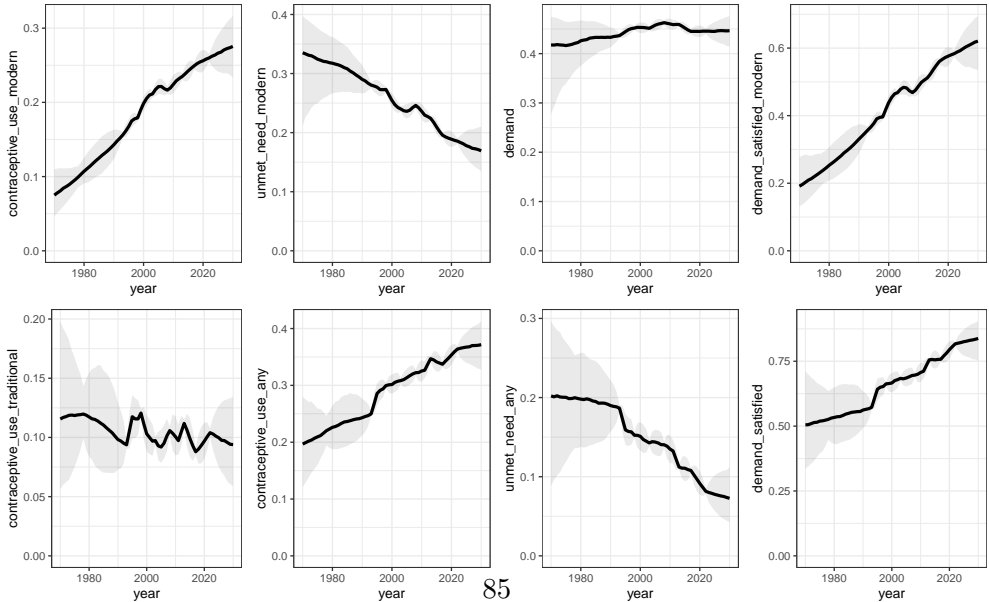

**Rwanda married**

included • Yes    data_series_type ◆ DHS ◆ National survey ◆ Other

**Rwanda unmarried**

data_series_type ◆ DHS ◆ National survey    included • Yes

**Rwanda all**

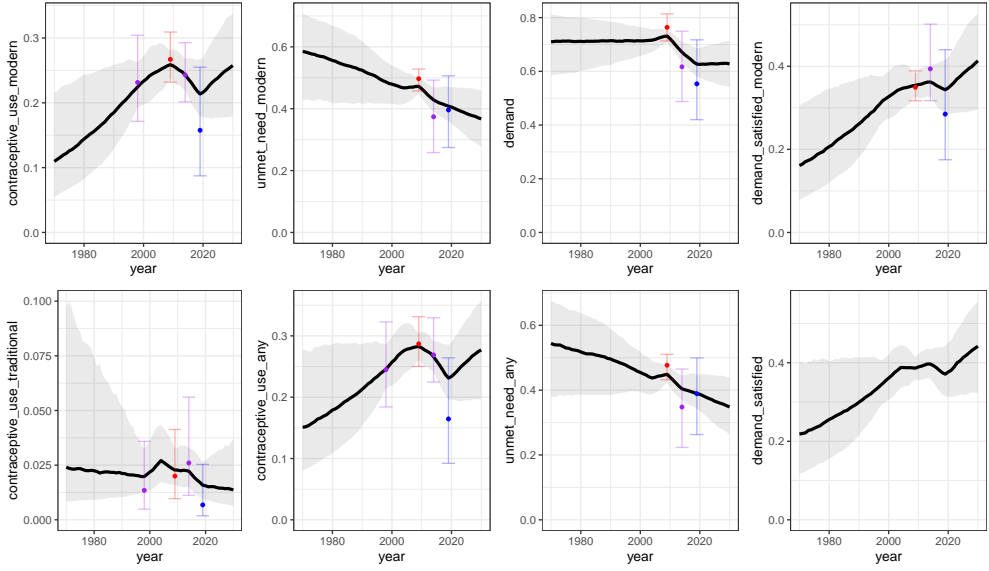

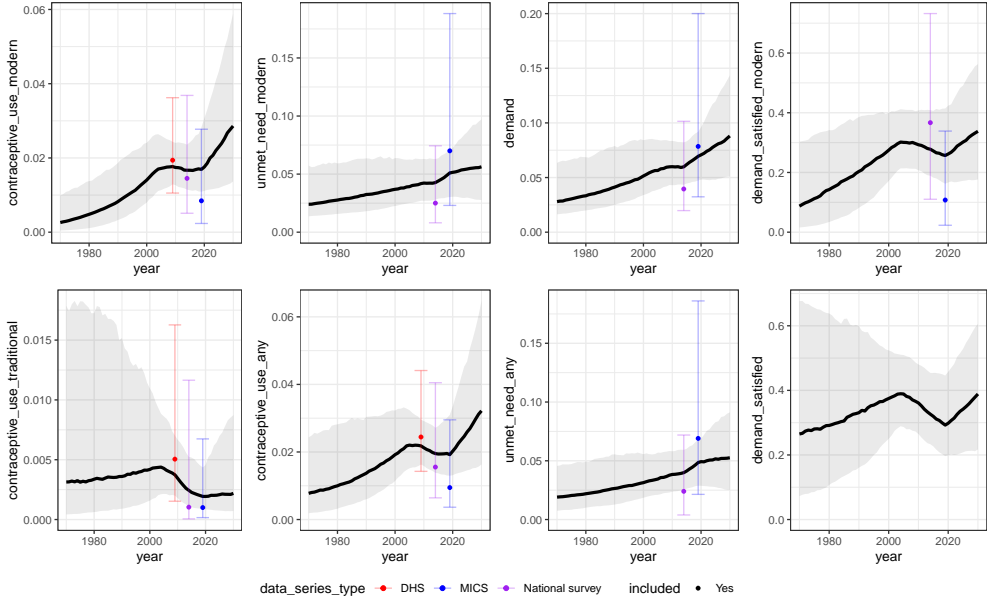

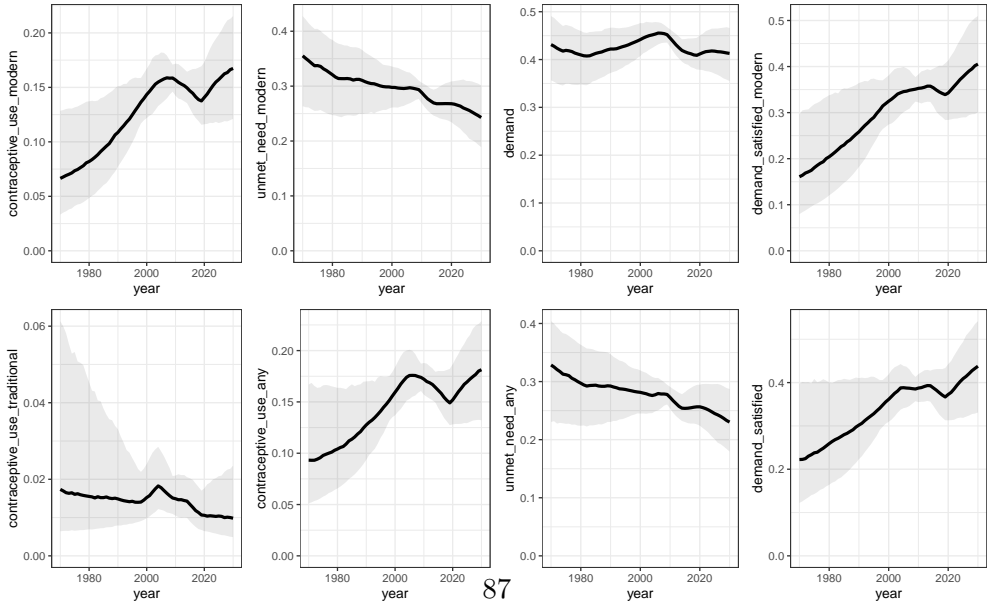

87

**Sao_Tome_and_Principe married**

**Sao_Tome_and_Principe unmarried**

**Sao_Tome_and_Principe all**

data_series_type ● DHS ● MICS    included ● Yes

88

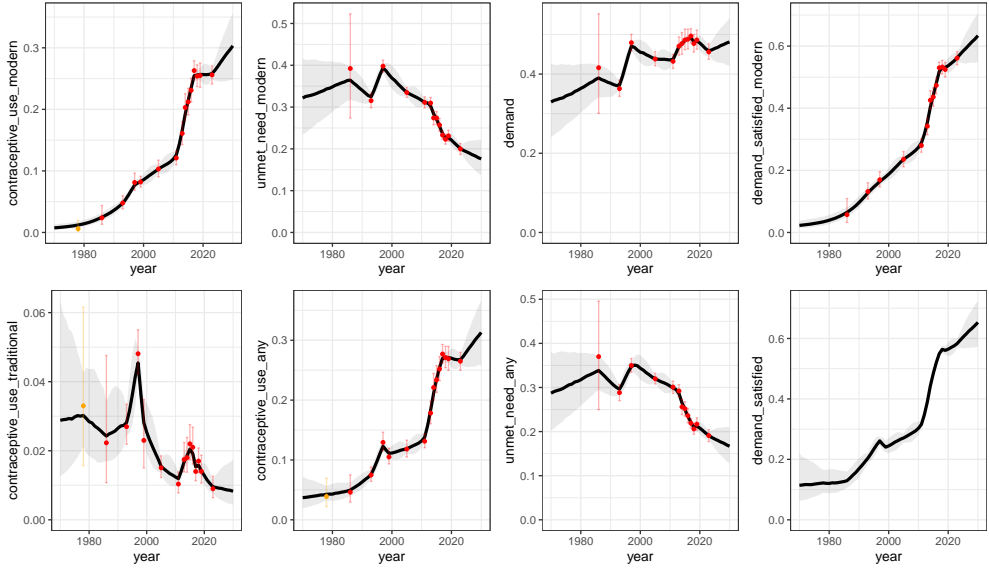

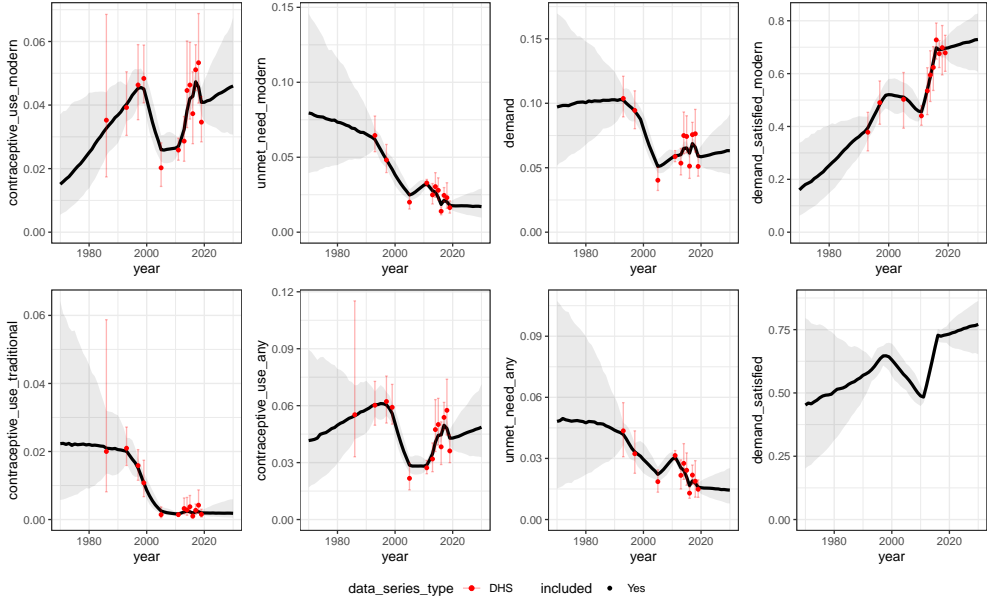

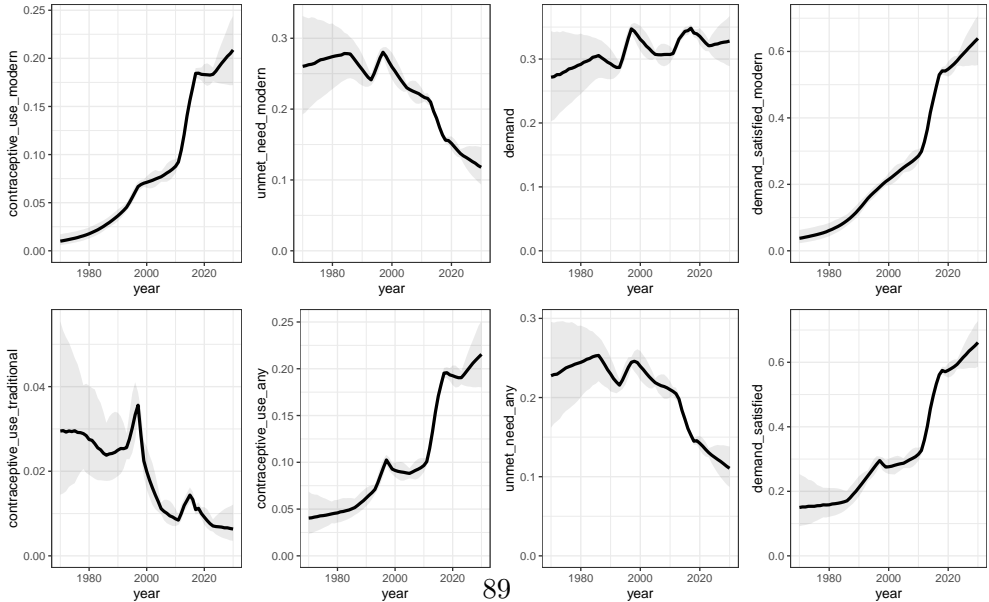

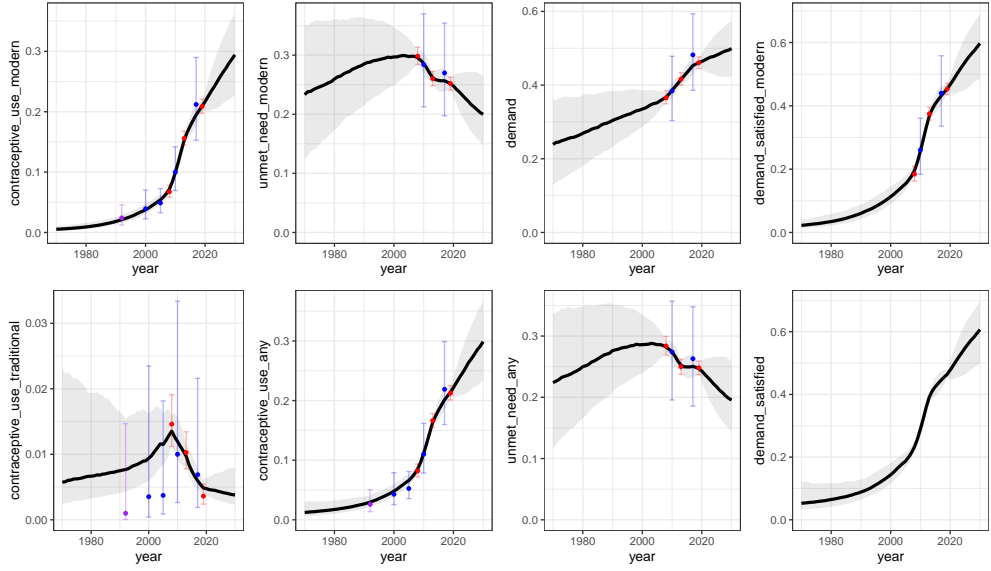

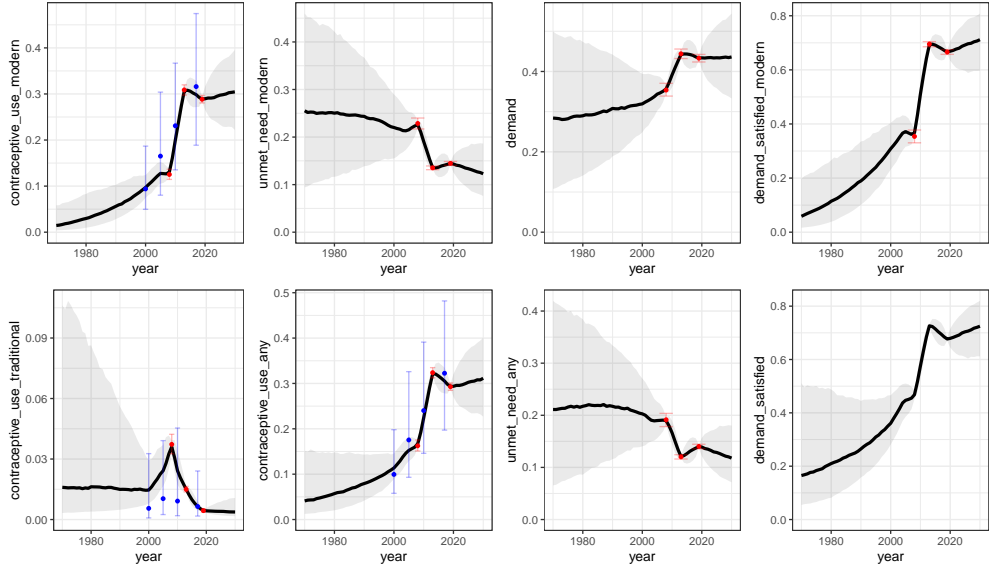

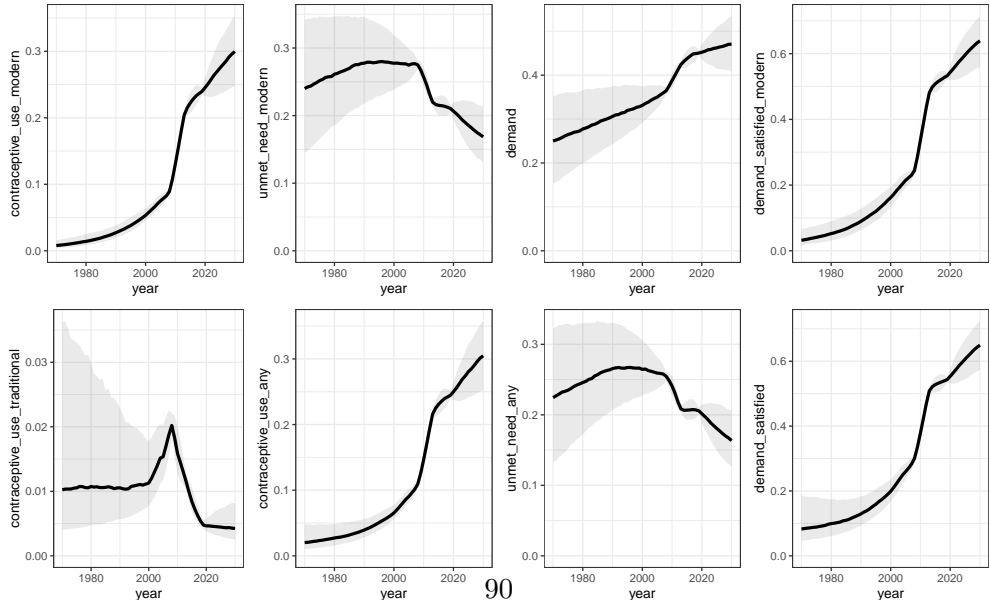

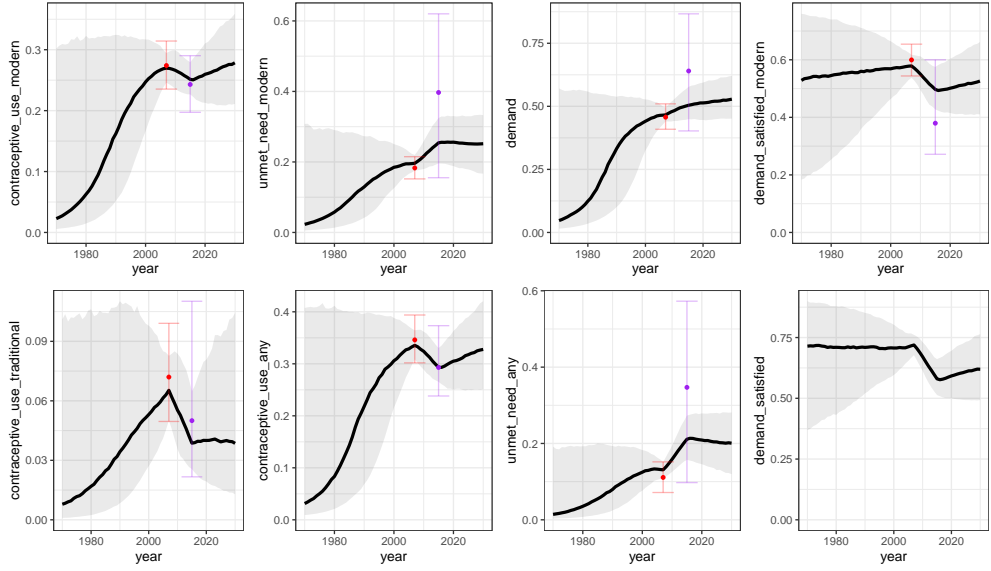

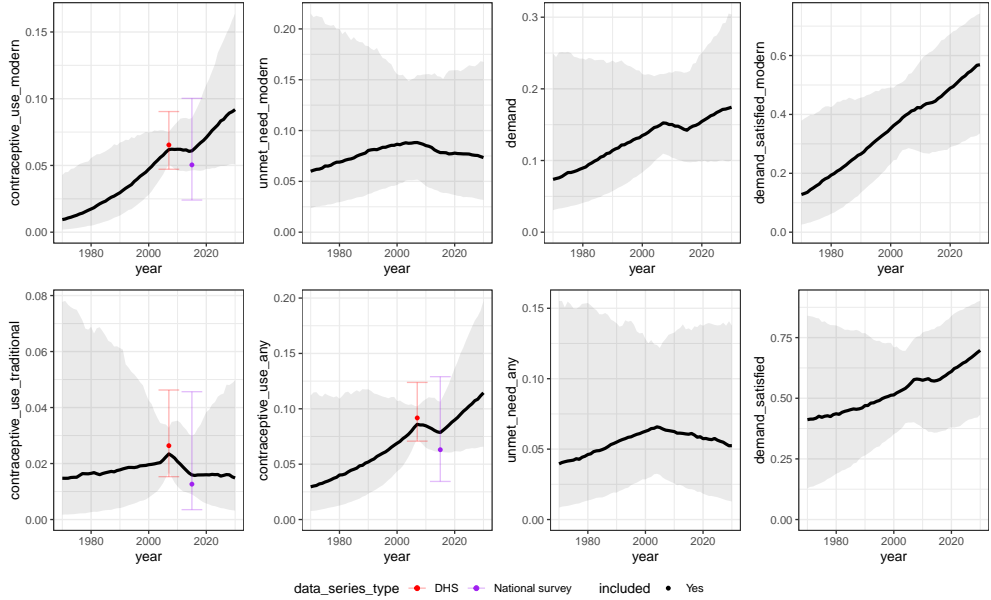

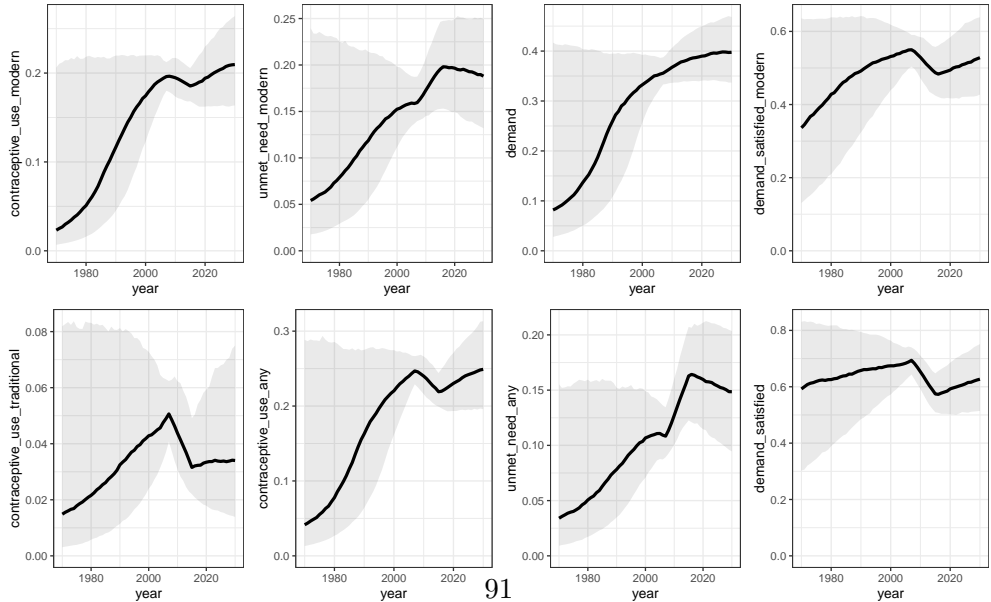

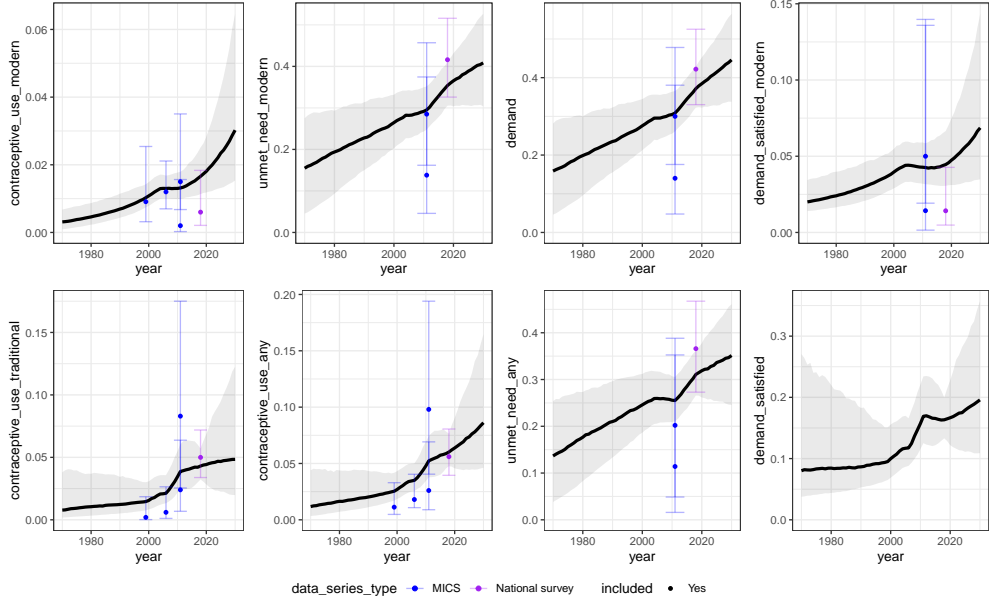

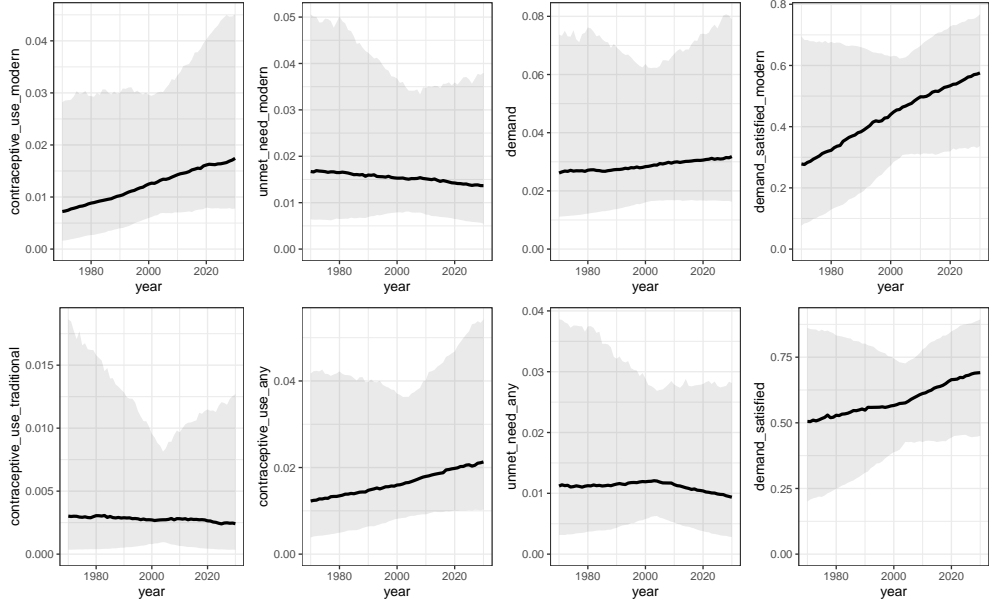

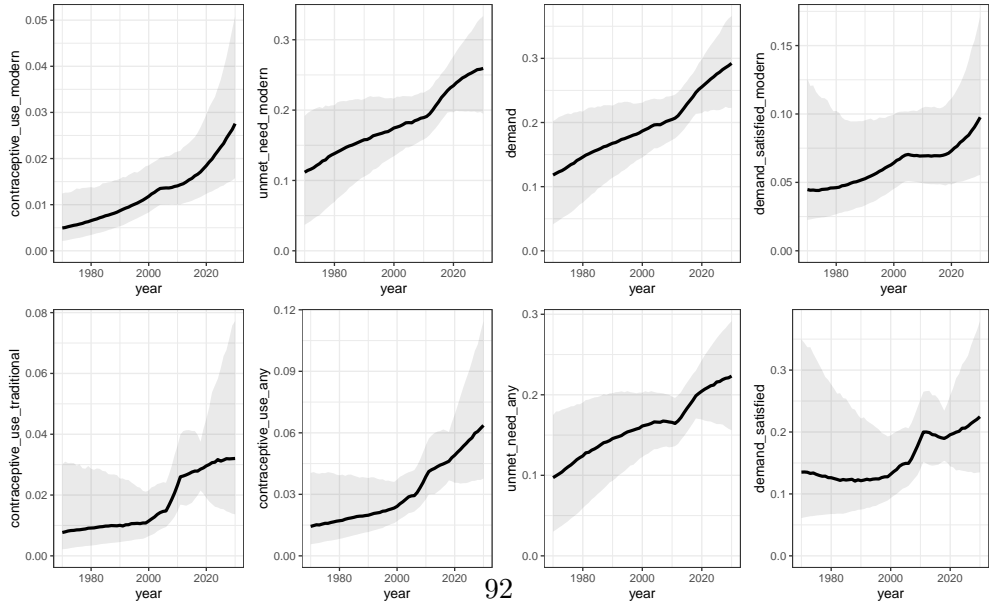

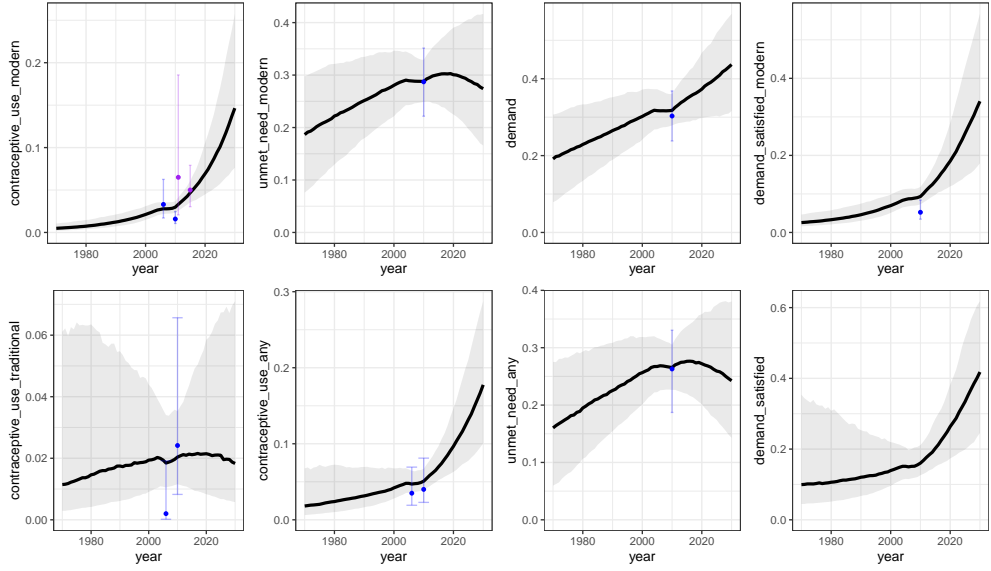

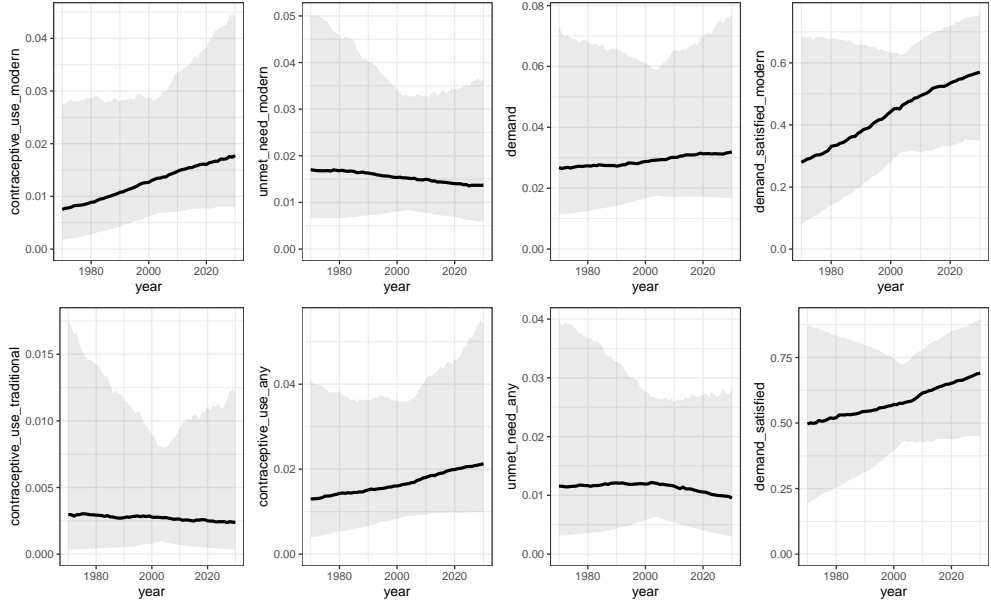

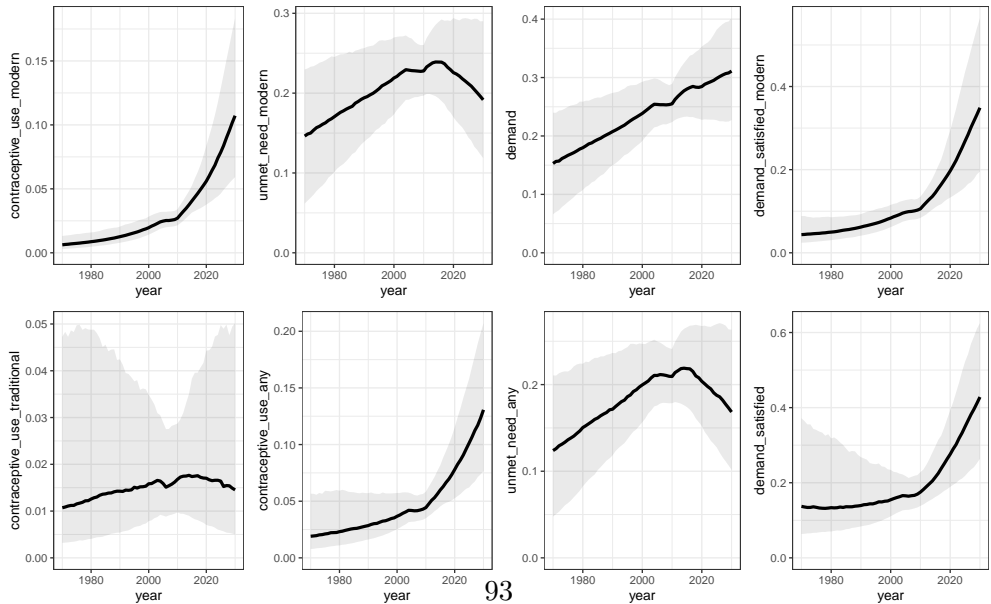

93

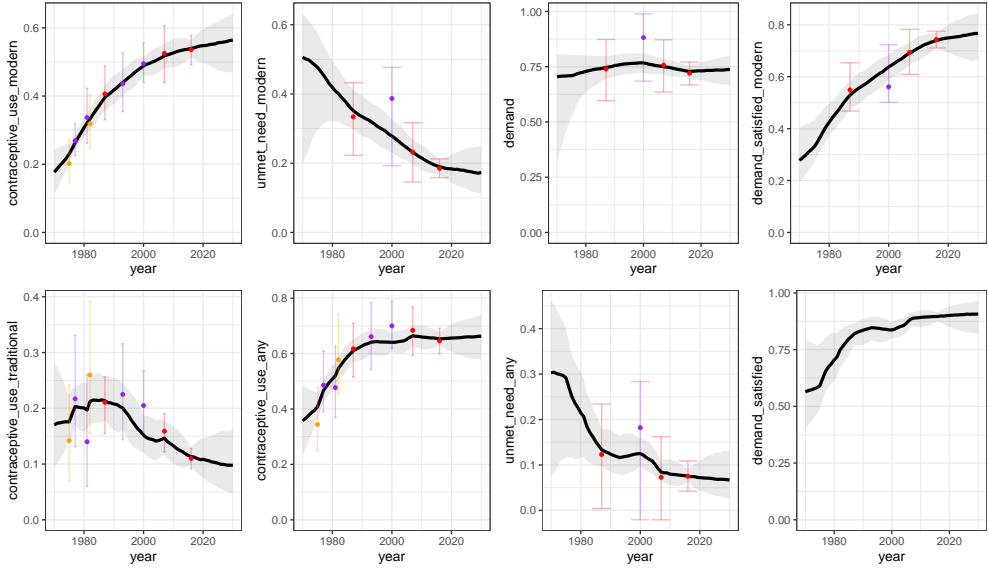

**Sri_Lanka married**

included • Yes   data_series_type ● DHS ● National survey ● Other

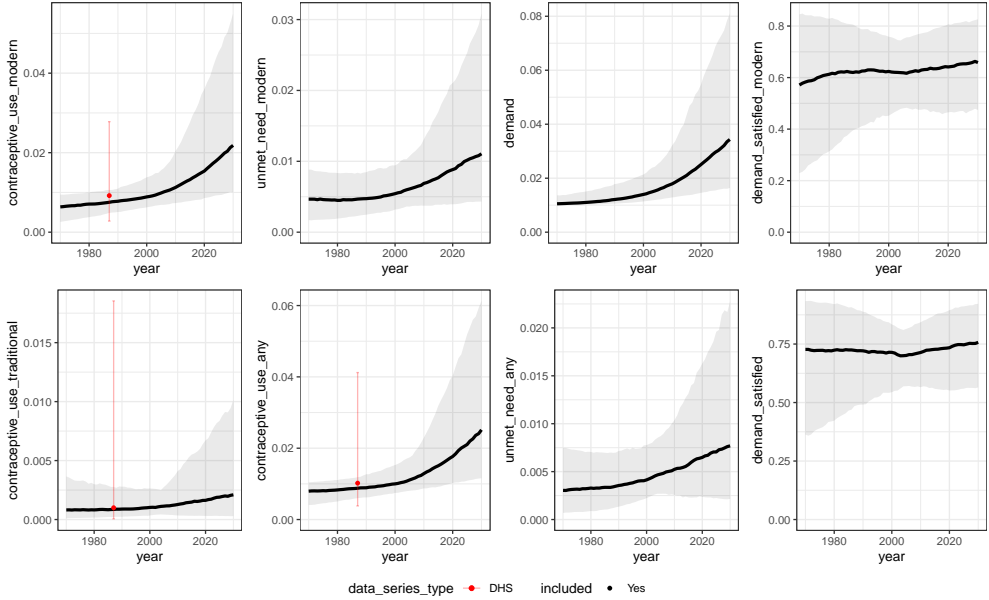

**Sri_Lanka unmarried**

data_series_type ● DHS   included • Yes

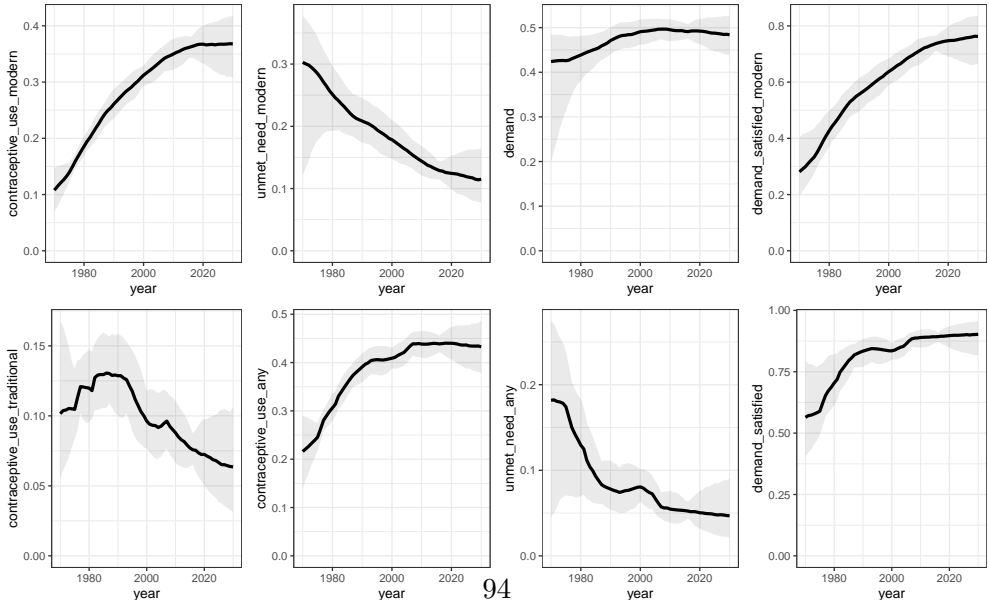

**Sri_Lanka all**

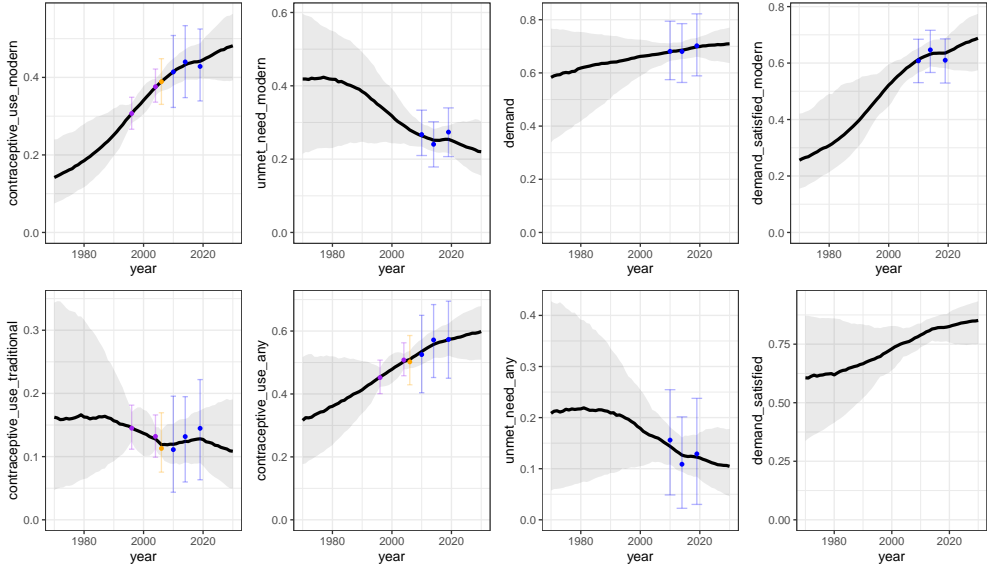

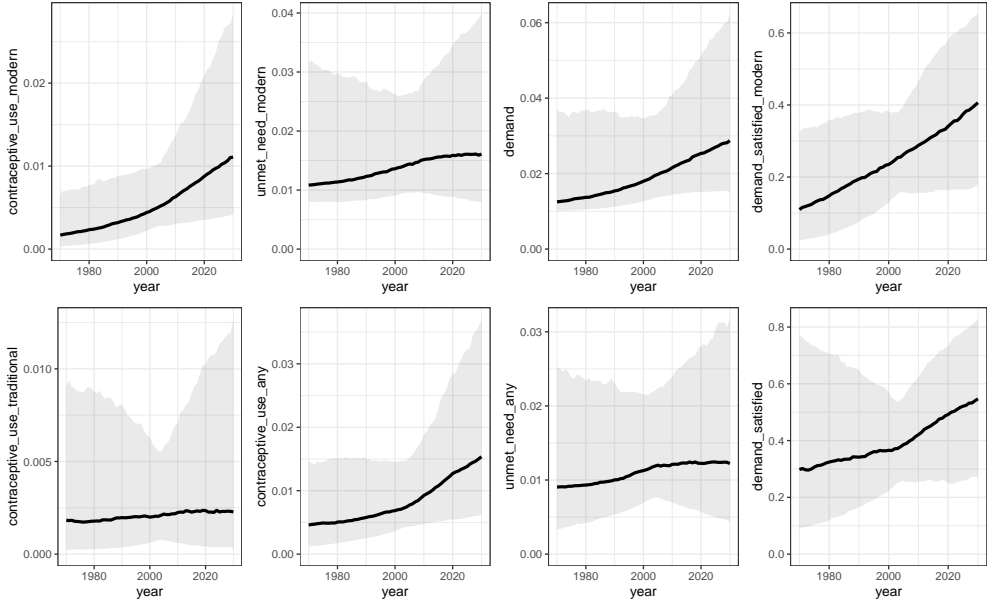

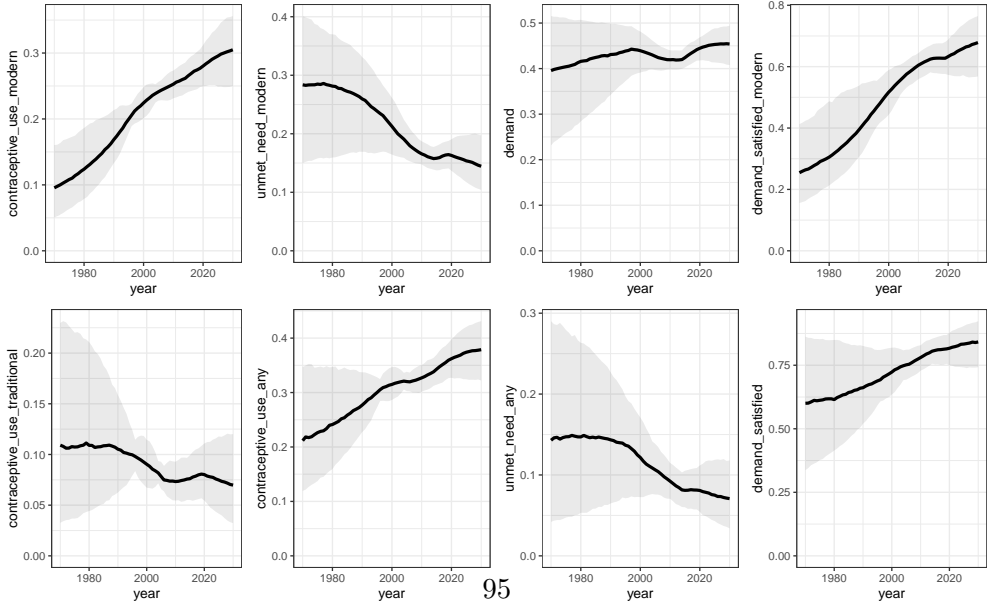



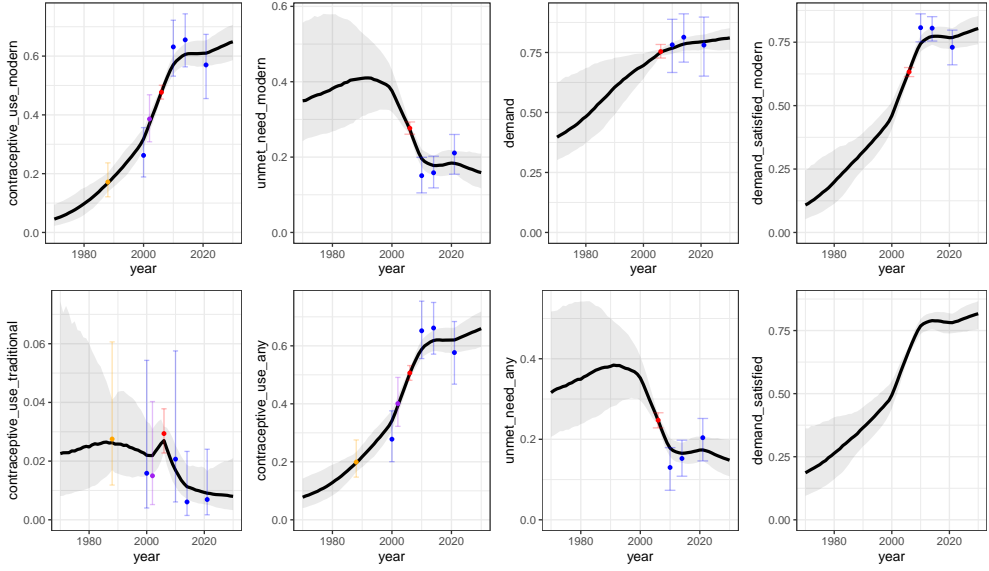

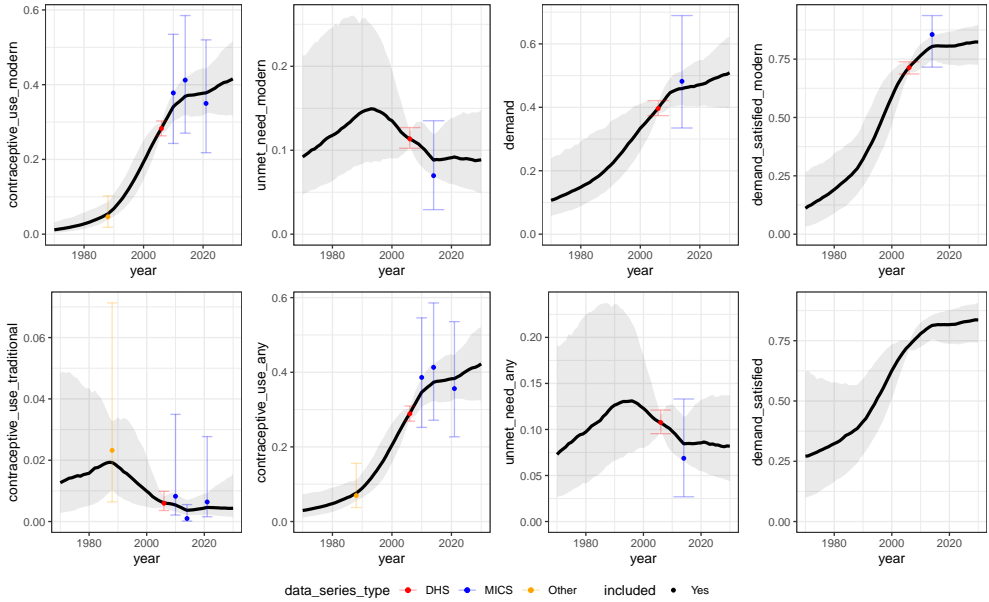

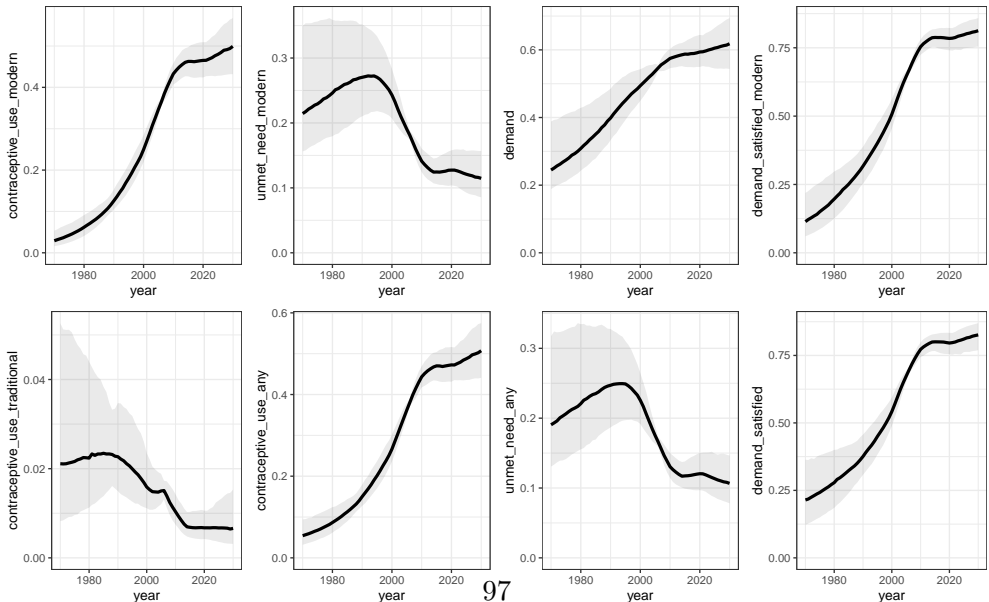

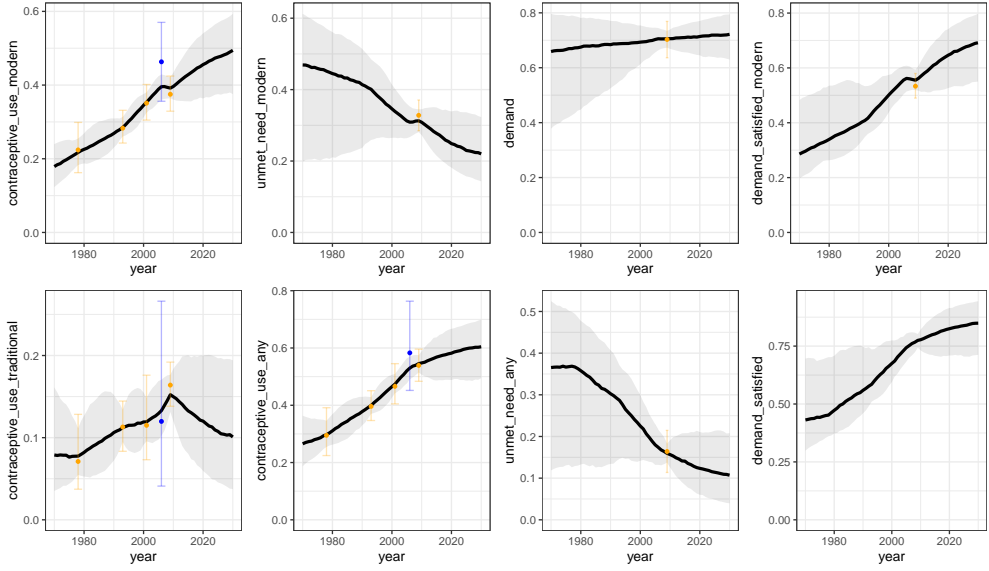

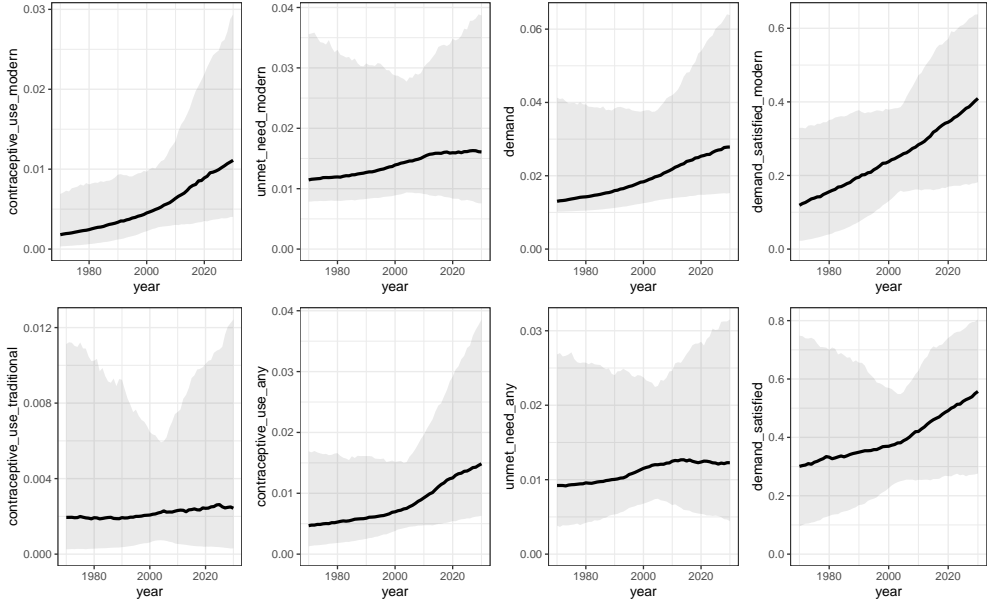

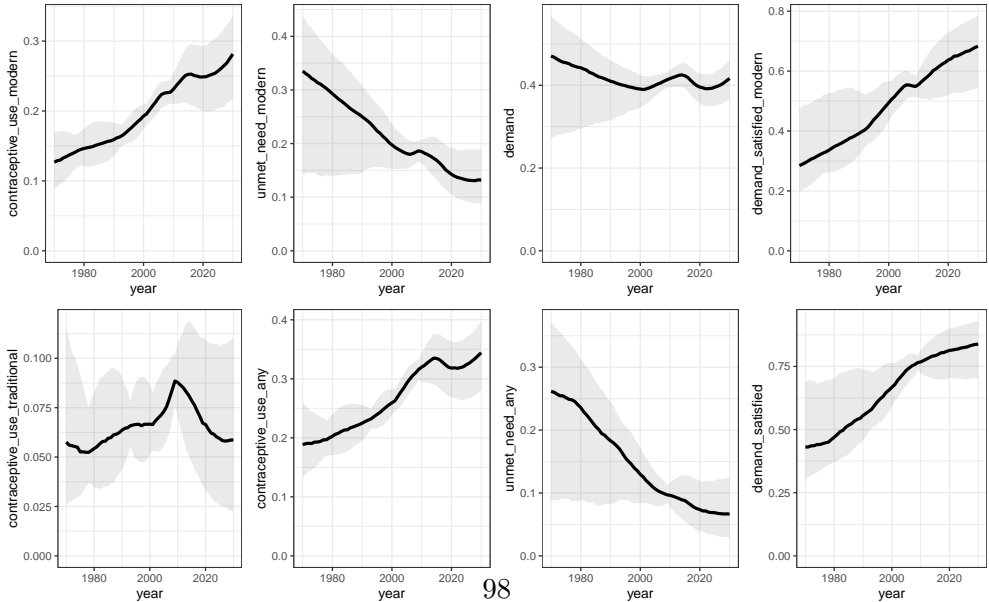

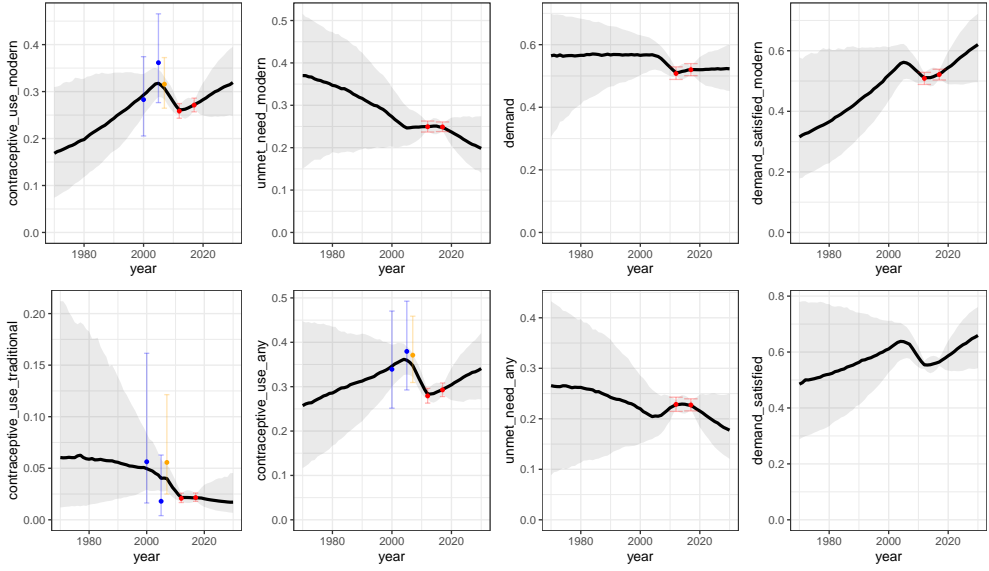

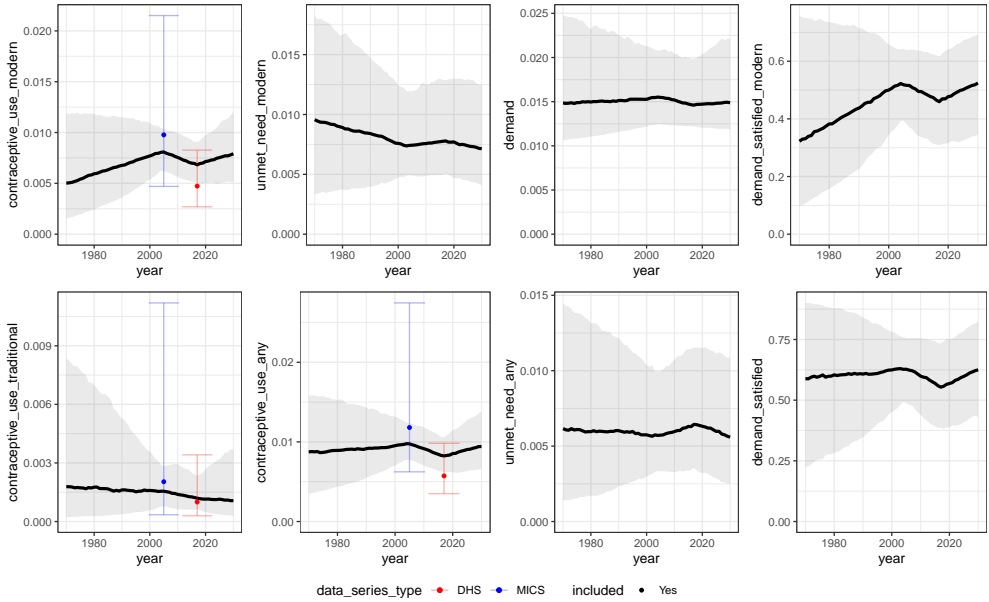

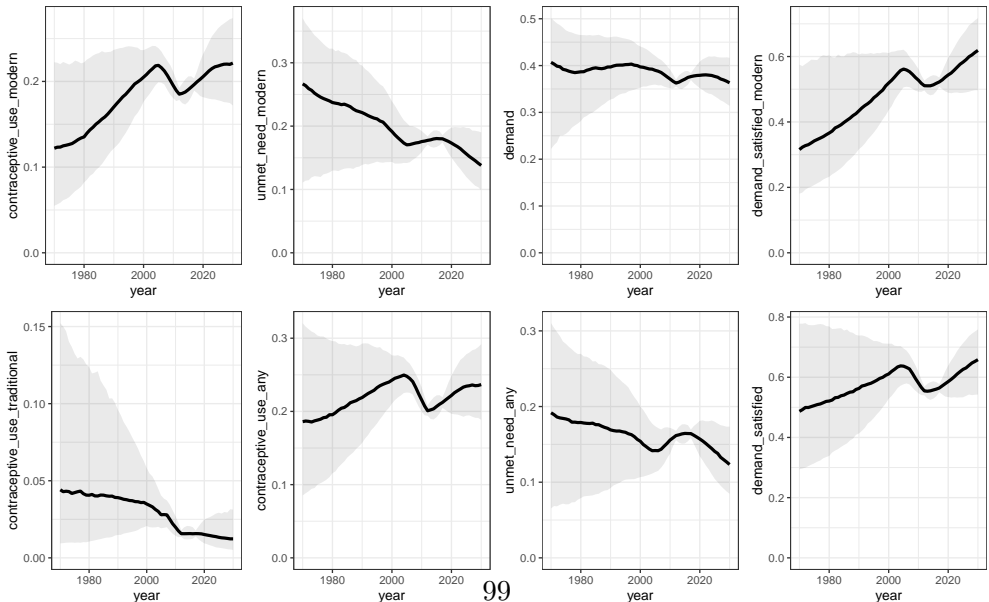

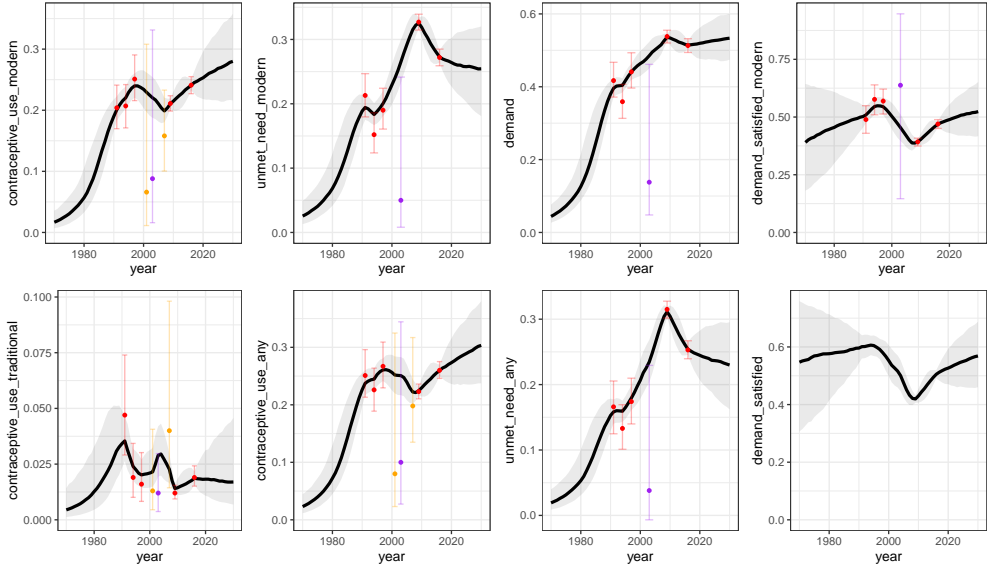

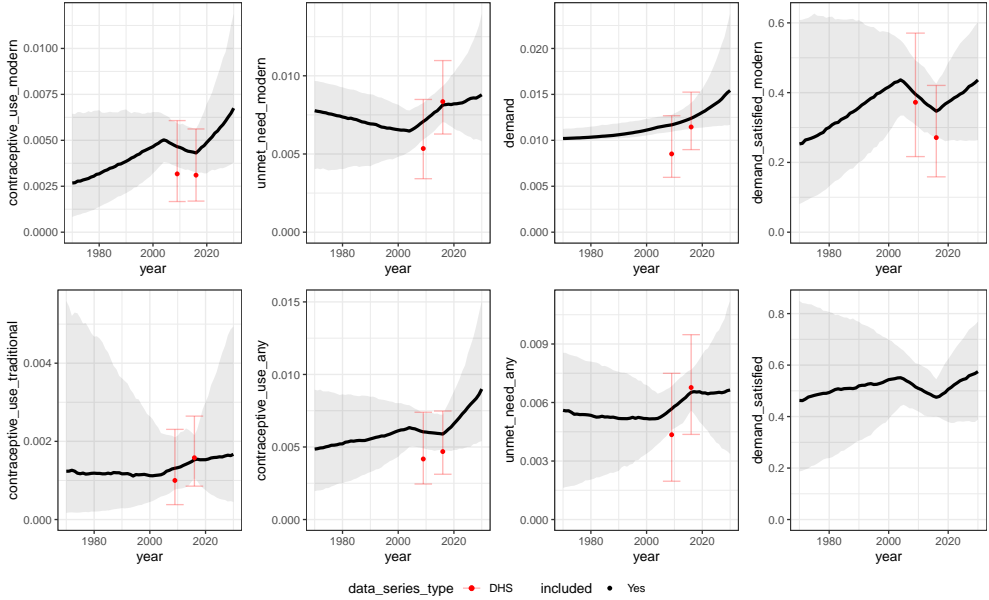

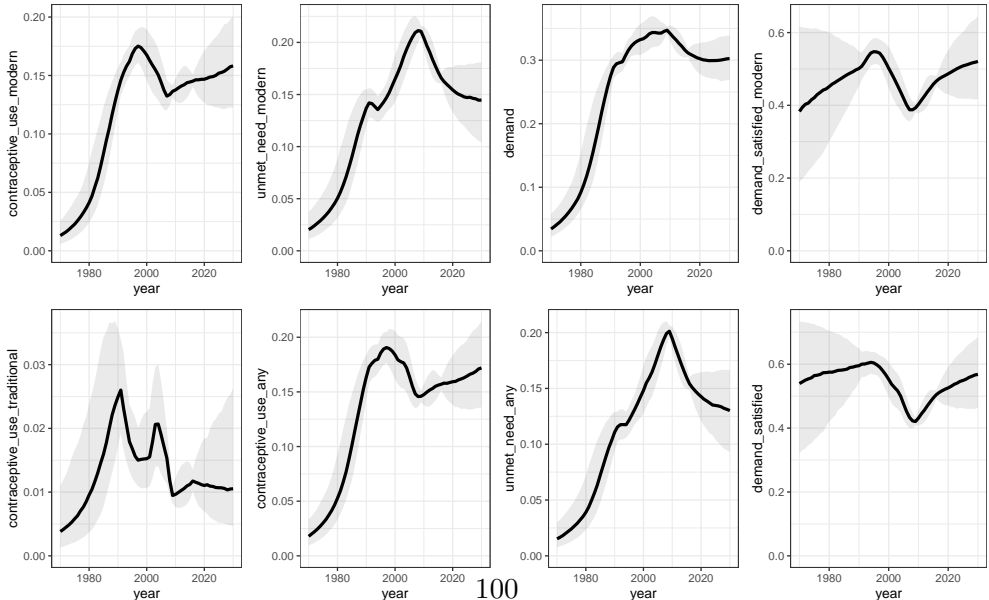

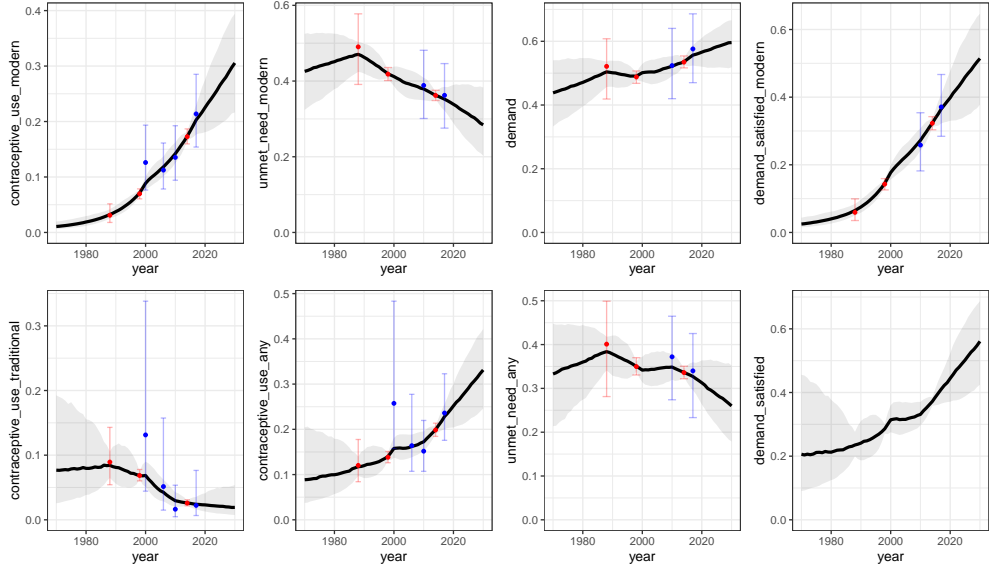

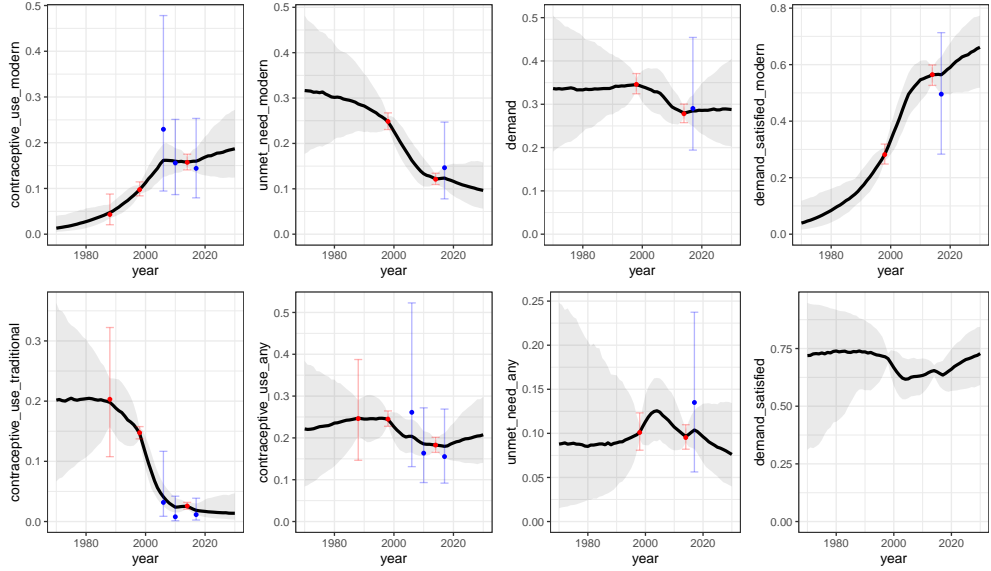

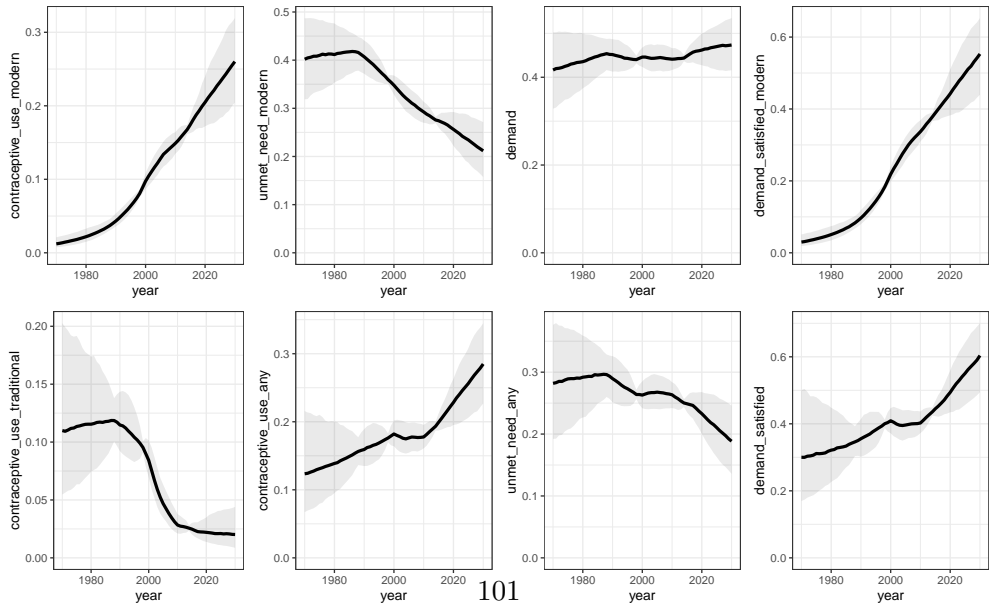

Tunisia married

Tunisia unmarried

Tunisia all

**Uganda married**



**Uganda unmarried**



**Uganda all**

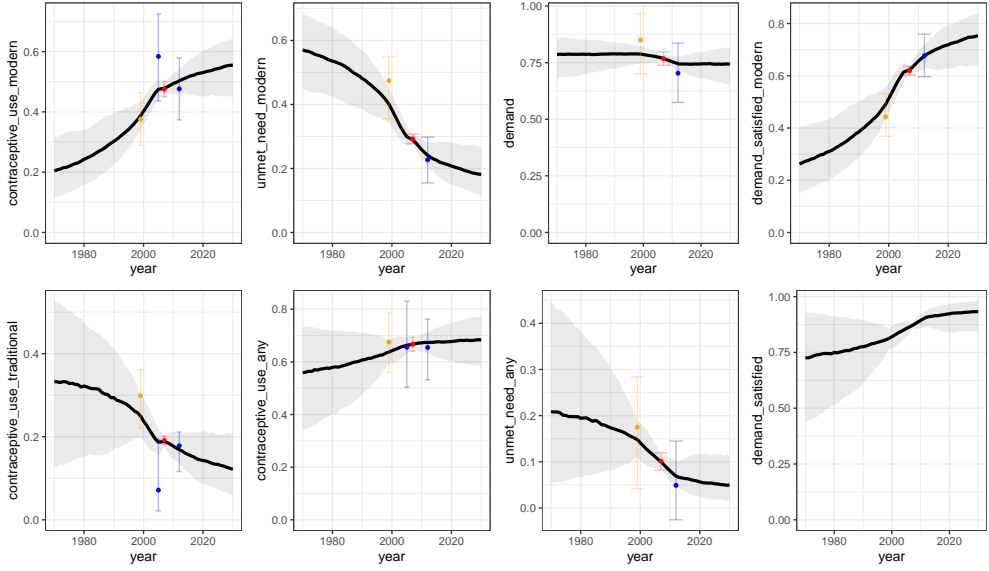

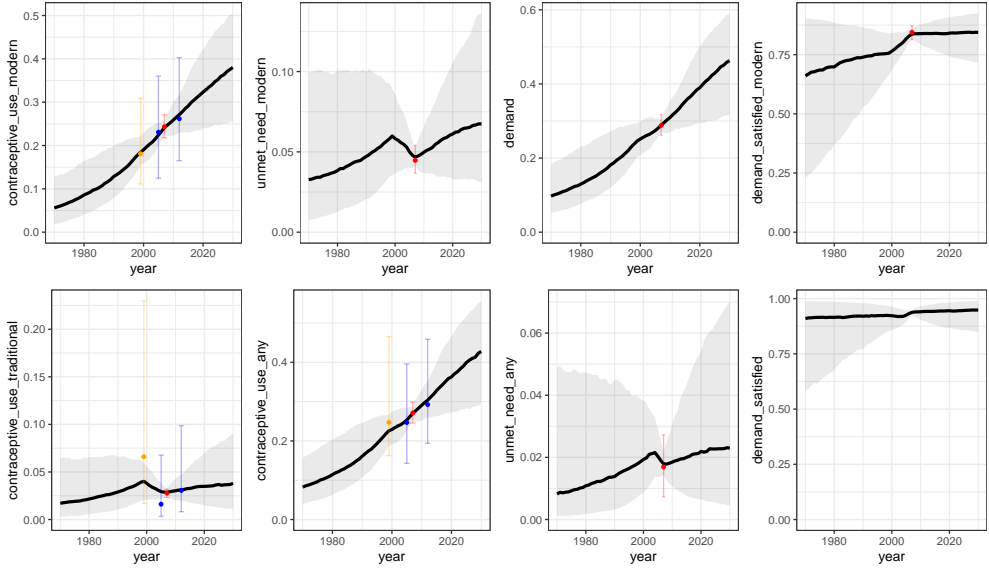

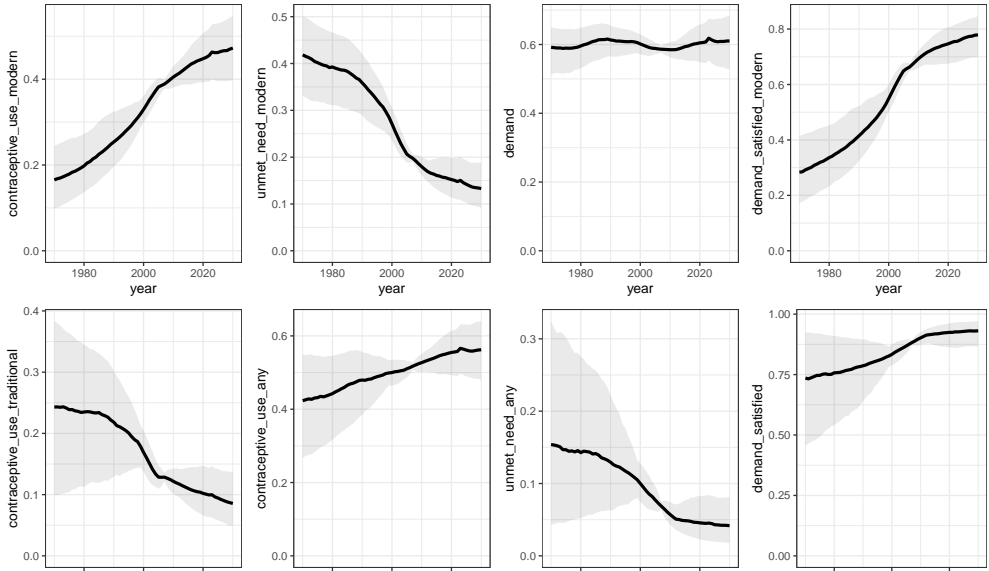

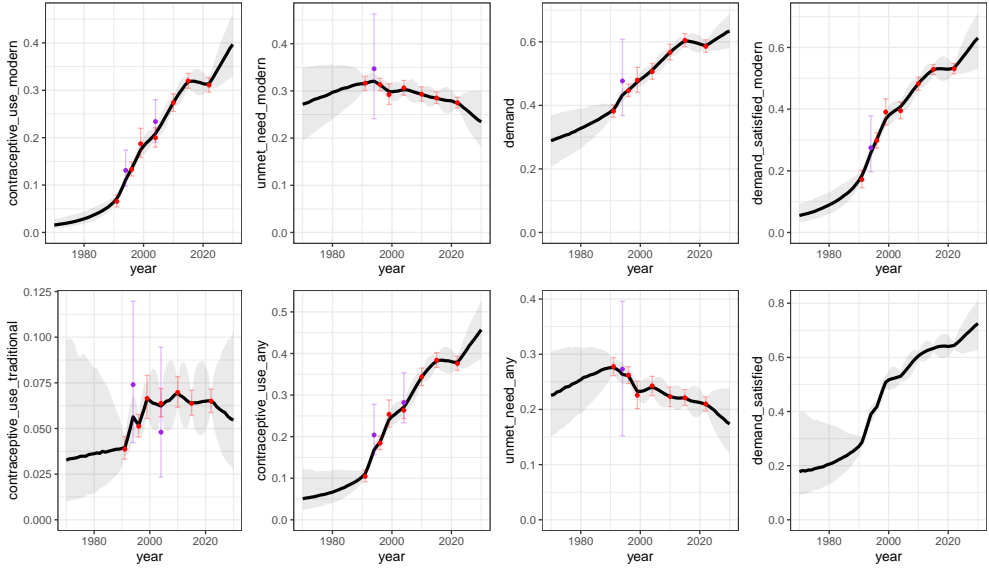

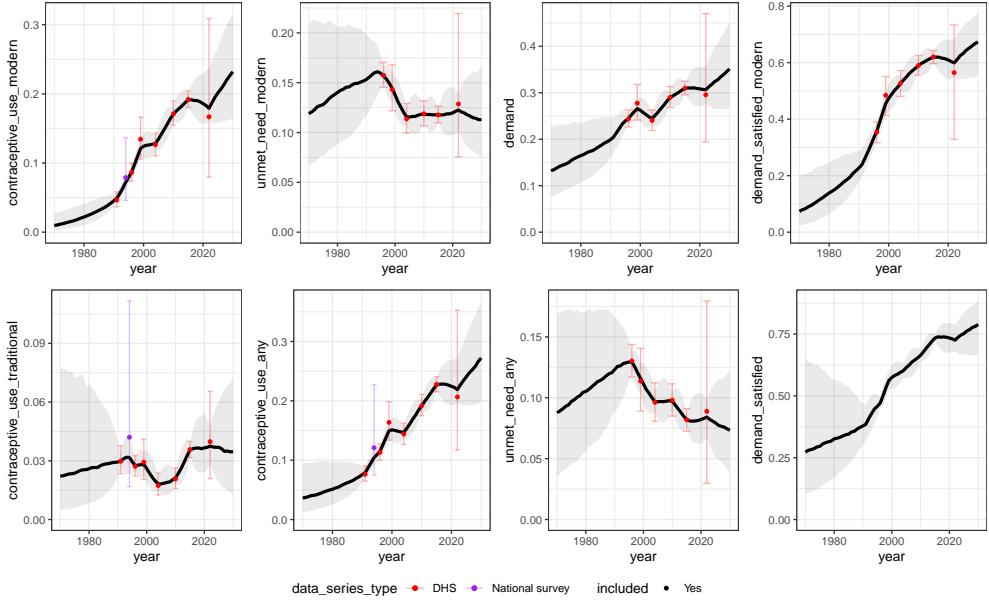

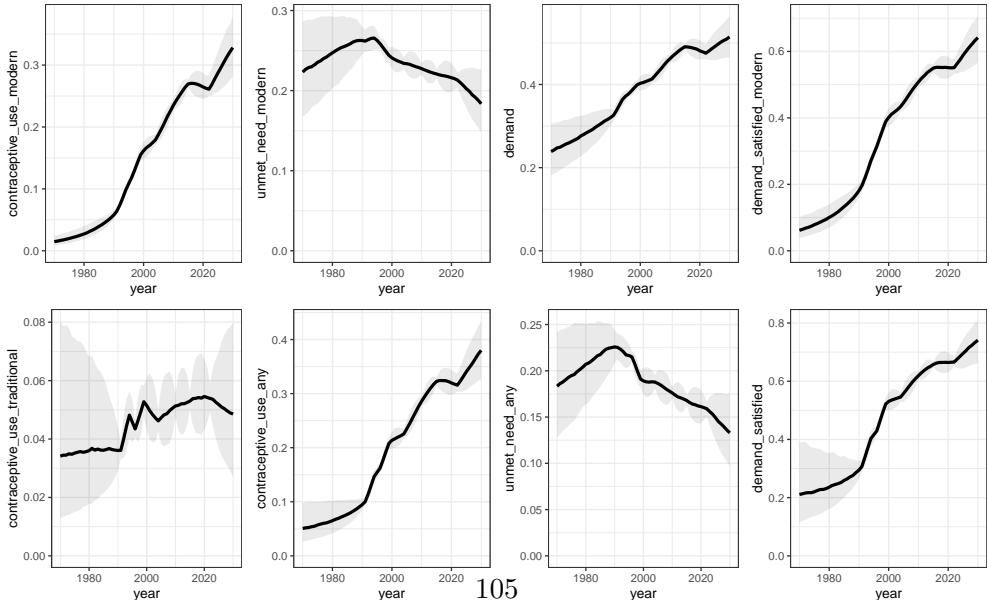

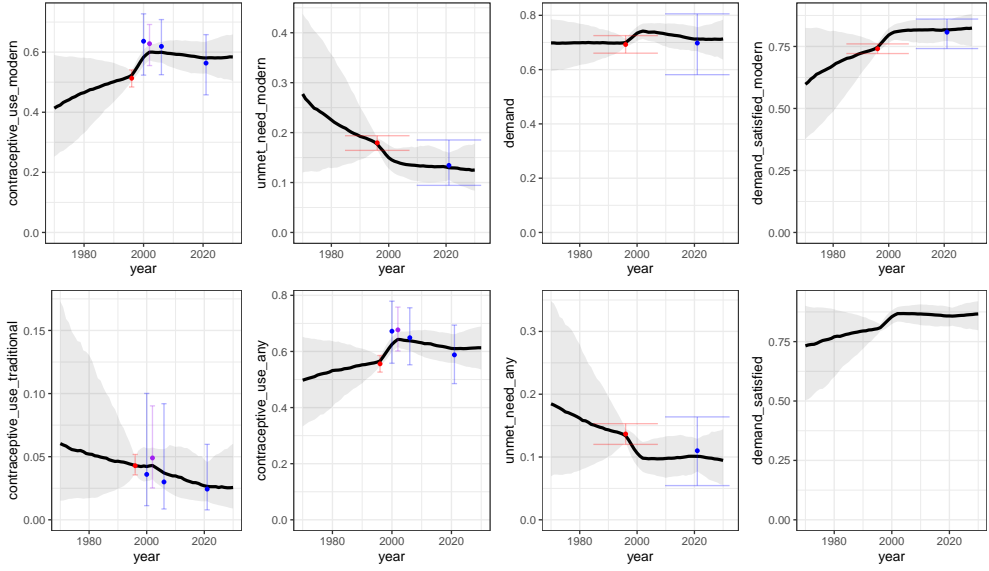

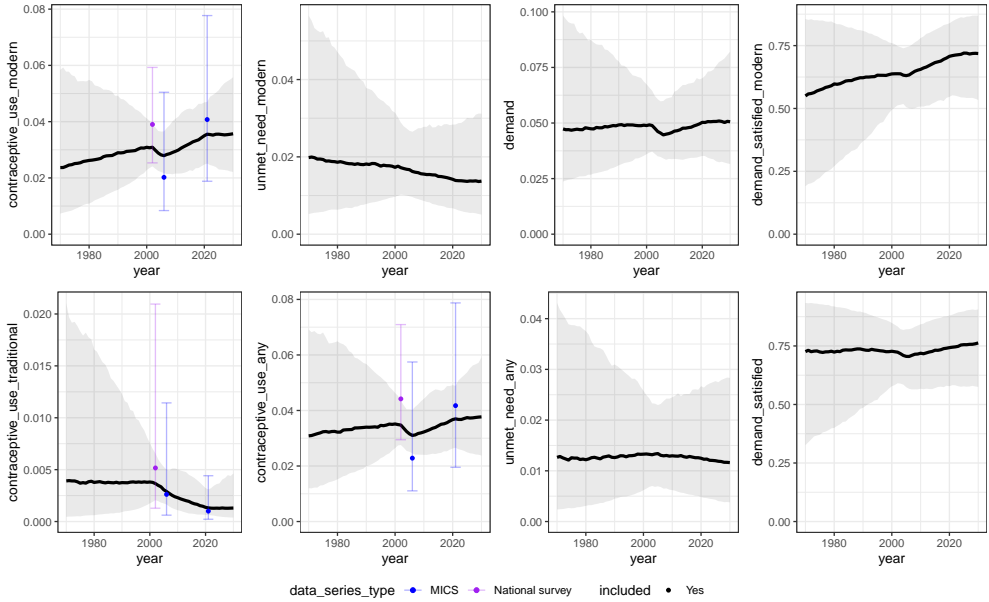

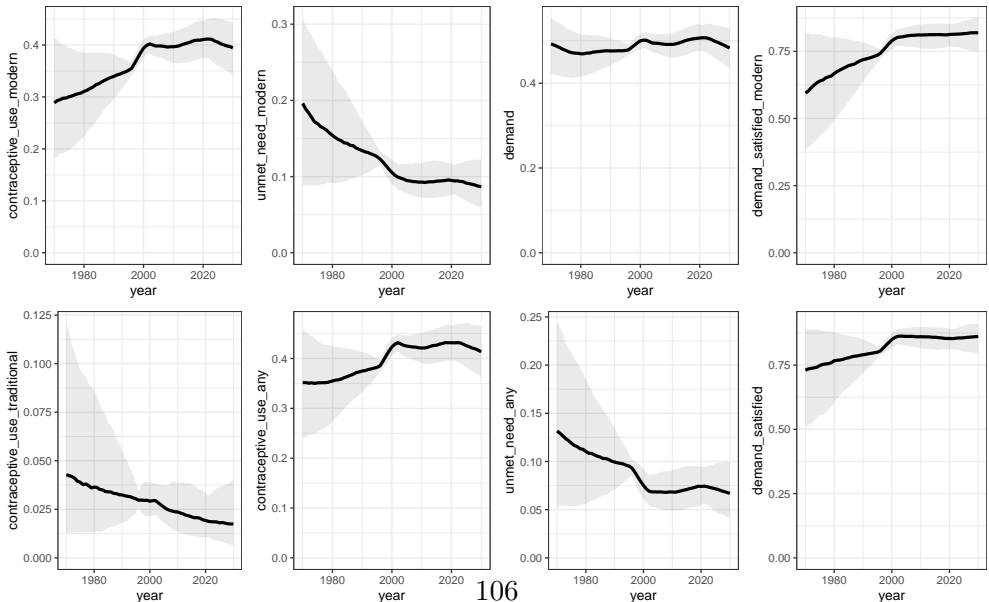

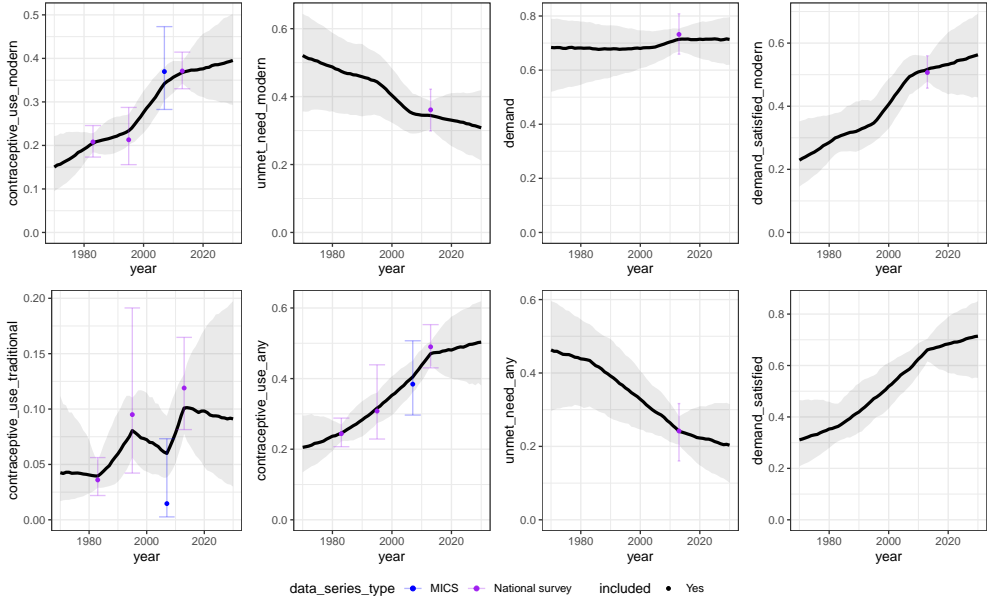

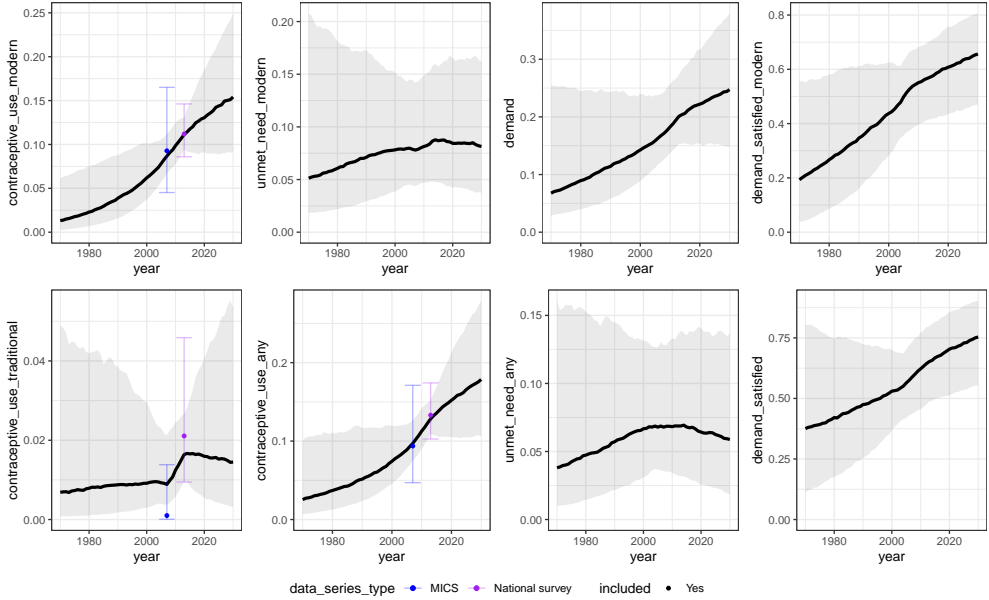

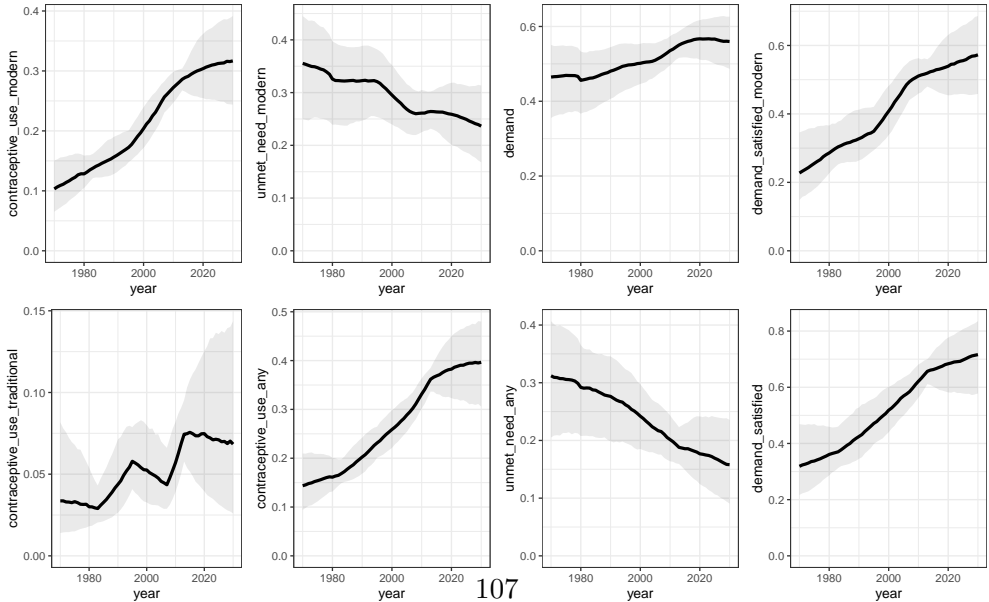

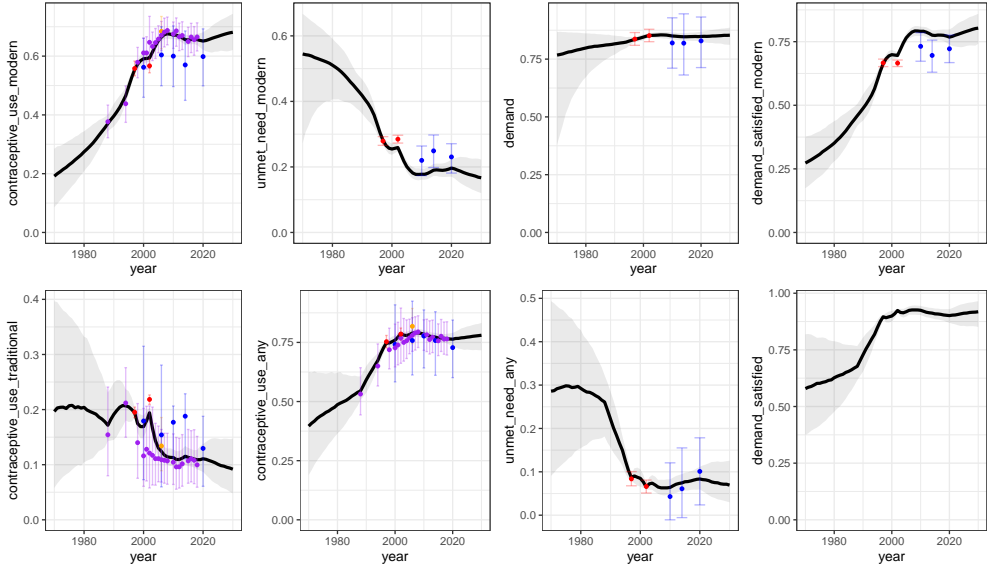

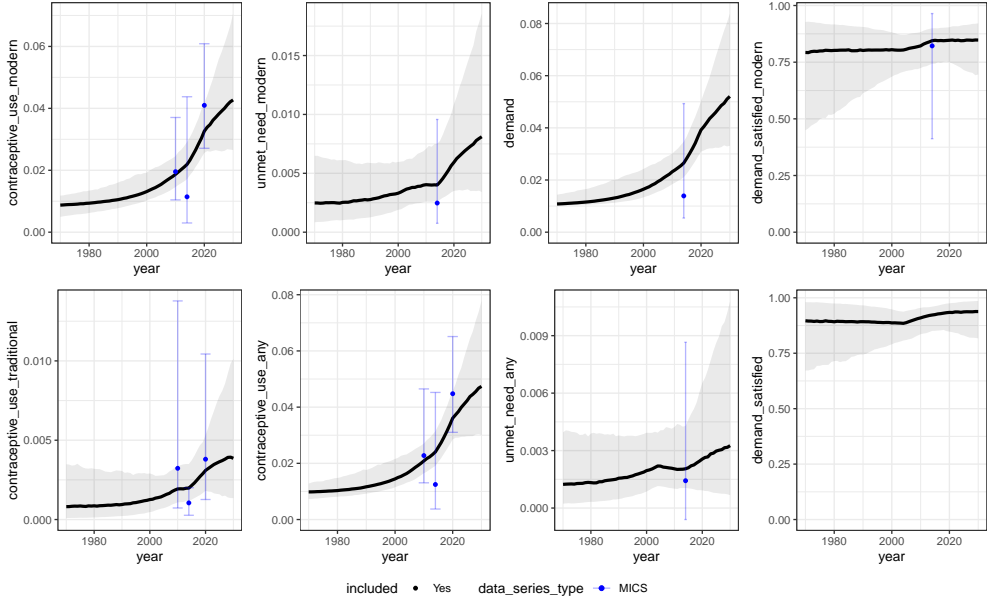

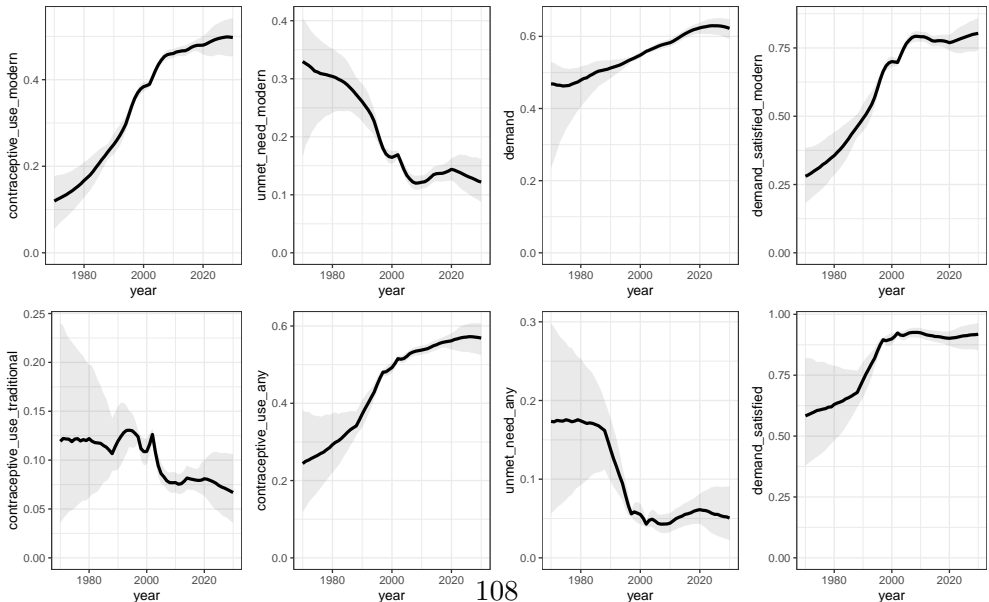

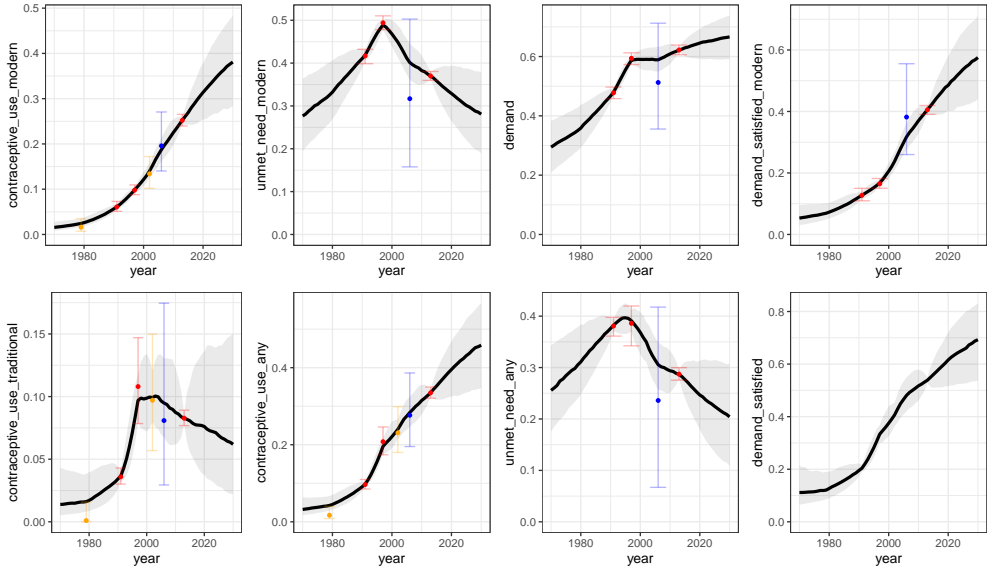

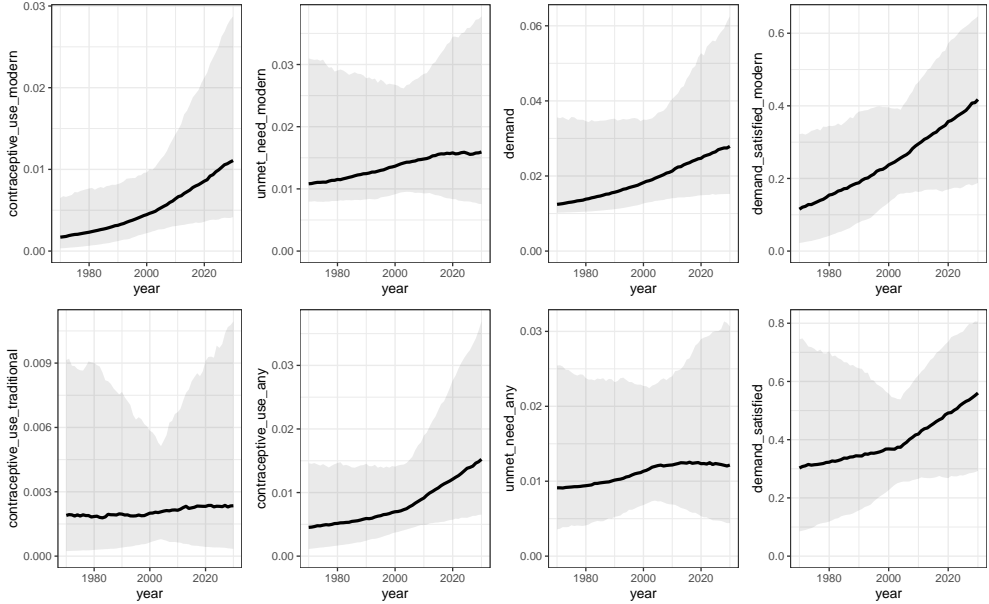

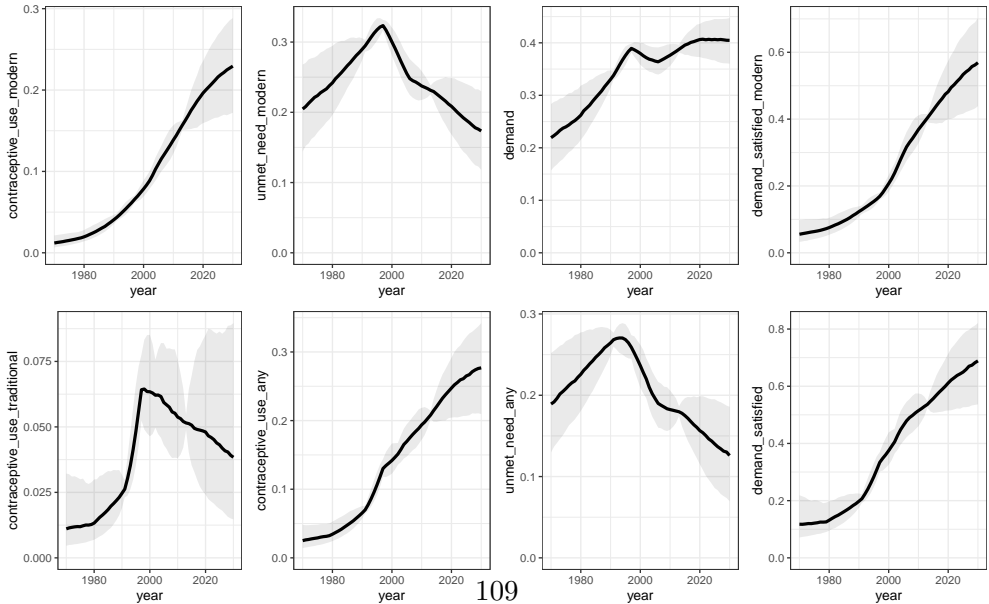

109

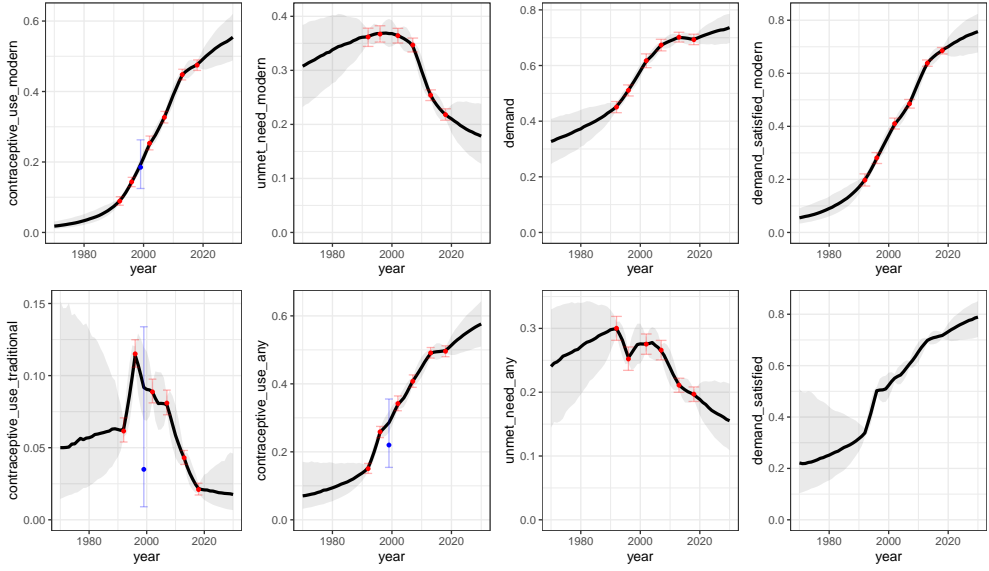

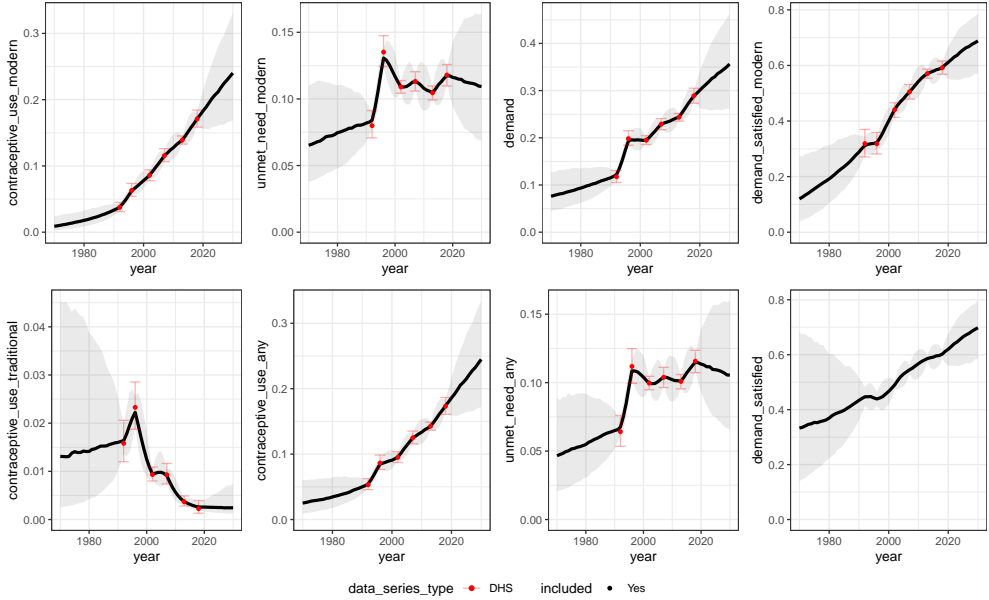

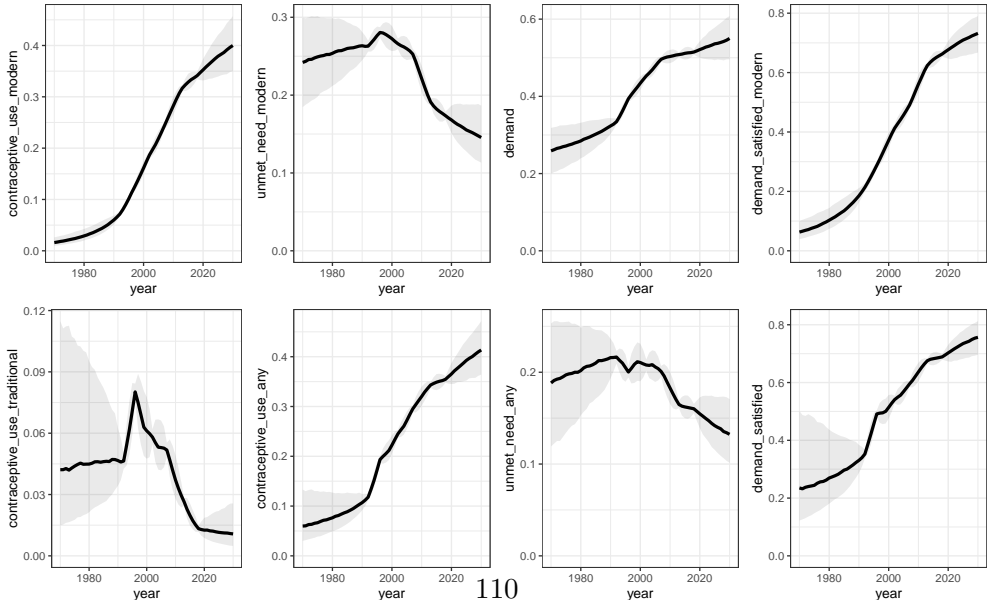

## Zimbabwe married

## Zimbabwe unmarried

## Zimbabwe all

111


\end{document}